\documentclass[10pt]{article}
\usepackage[dvips]{graphicx}
\usepackage{amsmath,amssymb,amsthm,float,arydshln,cite,color}
\usepackage{psfrag,epsfig}
\usepackage[latin1]{inputenc}
\usepackage{dsfont}

\newcommand{\bp}{\begin{proof} \small }
\newcommand{\ep}{\end{proof} \normalsize}
\newcommand{\bpa}{\begin{proofappx} \footnotesize }
\newcommand{\epa}{\end{proofappx} \small }
\newtheorem{lemma}{{\em Lemma}}[section]
\newtheorem{theorem}{{\em Theorem}}[section]
\newtheorem{corollary}{{\em Corollary}}[section]
\newcommand{\addbox}{\addtolength{\fboxsep}{5pt}\boxed}
\newcommand{\Ex}{\mathbb{E}\hspace{0.05cm}}

\newcommand{\bm}[1]{\mbox{\boldmath $#1$}}

\newcommand{\be}{\begin{equation}}
\newcommand{\ee}{\end{equation}}
\newcommand{\bs}{\begin{subequations}}
\newcommand{\es}{\end{subequations}}
\newcommand{\bq}{\begin{eqnarray}}
\newcommand{\eq}{\end{eqnarray}}
\newcommand{\bqn}{\begin{eqnarray*}}
\newcommand{\eqn}{\end{eqnarray*}}
\newcommand{\nn}{\nonumber}
\newcommand{\ba}{\left[ \begin{array}}
\newcommand{\ea}{\\ \end{array} \right]}
\newcommand{\ben}{\begin{enumerate}}
\newcommand{\een}{\end{enumerate}}
\newcommand{\qd}{\hfill{$\square$}}
\newcommand{\define}{\;\stackrel{\Delta}{=}\;}

\def\H{{\boldsymbol{H}}}

\def\R{{\boldsymbol{R}}}

\def\U{{\boldsymbol{U}}}

\def\d{{\boldsymbol{d}}}
\def\e{{\boldsymbol{e}}}

\def\n{{\boldsymbol{n}}}

\def\q{{\boldsymbol{q}}}
\def\r{{\boldsymbol{r}}}
\def\s{{\boldsymbol{s}}}

\def\u{{\boldsymbol{u}}}
\def\v{{\boldsymbol{v}}}
\def\w{{\boldsymbol{w}}}
\def\x{{\boldsymbol{x}}}
\def\y{{\boldsymbol{y}}}
\def\z{{\boldsymbol{z}}}

\def\real{{\mathchoice%
{\hbox{\rm\setbox1=\hbox{I}\copy1\kern-.45\wd1 R}}
{\hbox{\rm\setbox1=\hbox{I}\copy1\kern-.45\wd1 R}}
{\hbox{\scriptsize\rm\setbox1=\hbox{I}\copy1\kern-.45\wd1 R}}
{\hbox{\scriptsize\rm\setbox1=\hbox{I}\copy1\kern-.45\wd1 R}}}}

\def\Zint{{\mathchoice{\setbox1=\hbox{\sf Z}\copy1\kern-.75\wd1\box1}
{\setbox1=\hbox{\sf Z}\copy1\kern-.75\wd1\box1}
{\setbox1=\hbox{\scriptsize\sf Z}\copy1\kern-.75\wd1\box1}
{\setbox1=\hbox{\scriptsize\sf Z}\copy1\kern-.75\wd1\box1}}}
\newcommand{\complex}{ \hbox{\rm C\kern-0.45em\rule[.07em]{.02em}{.58em}%
\kern 0.43em}}

\begin{document}

\title{{\bf DIFFUSION ADAPTATION OVER NETWORKS}\thanks{The cite information for this article is as follows: {\bf A. H. Sayed, ``Diffusion adaptation over networks,'' in {\em E-Reference Signal Processing}, R. Chellapa and S. Theodoridis, {\em editors}, Elsevier, 2013. }
The work was supported in part by NSF grants EECS-060126, EECS-0725441, CCF-0942936, and CCF-1011918. The author is with the Electrical Engineering Department, University of California, Los Angeles, CA 90095, USA. Email: {sayed@ee.ucla.edu}.}}
\author{{\rm Ali H. Sayed}}

\date{}

\maketitle

\begin{center}\vspace{-1cm}

{\rm Electrical Engineering Department}\\
{\rm University of California at Los Angeles}
\end{center}

\bigskip
\bigskip
\bigskip

Adaptive networks are well-suited to perform decentralized
information processing and optimization tasks and to model various types of
self-organized and complex behavior encountered in nature. Adaptive
networks consist of a collection of agents with processing and
learning abilities.  The agents are linked together through a
connection topology, and they cooperate with each other through
local interactions to solve distributed optimization, estimation, and inference problems in
real-time. The continuous diffusion of information across the
network enables agents to adapt their performance in relation to
streaming data and network conditions; it also results in improved
adaptation and learning performance relative to non-cooperative
agents. This article provides an overview of diffusion strategies
for adaptation and learning over networks. The article is divided into several sections:

{ {\small \begin{enumerate}
\item Motivation.
\item Mean-Square-Error Estimation.
\item Distributed Optimization via Diffusion Strategies.
\item Adaptive Diffusion Strategies.
\item Performance of Steepest-Descent Diffusion Strategies.
\item Performance of Adaptive Diffusion Strategies.
\item Comparing the Performance of Cooperative Strategies.
\item Selecting the Combination Weights.
\item Diffusion with Noisy Information Exchanges.
\item Extensions and Further Considerations.
\item Appendix A: Properties of Kronecker Products.
\item Appendix B: Graph Laplacian and Network Connectivity.
\item Appendix C: Stochastic Matrices.
\item Appendix D: Block Maximum Norm.
\item Appendix E: Comparison with Consensus Strategies.
\item References.

\end{enumerate}
}}
\clearpage

\section{Motivation}
Consider a collection of $N$ agents interested in estimating the
same parameter vector, $w^o$, of size $M\times 1$. The vector is
the  minimizer of some global cost
function, denoted by $J^{\rm glob}(w)$, which the agents seek to
optimize, say,
 \be\addbox{\;
w^o = \underset{w}{\operatorname{argmin}} \;\;J^{\rm
glob}(w)\;}\label{adlq79813.a}
 \ee
We are interested in situations where the individual agents have access to partial
information about the global cost function.
In this case, cooperation among the agents becomes beneficial.
For example, by cooperating with their neighbors, and by having these neighbors
cooperate with their neighbors, procedures can be
devised that would  enable all agents in the network to converge towards
the global optimum $w^o$ through local interactions.
The objective of decentralized processing is to allow spatially distributed agents to achieve a global objective by relying solely on local information and on in-network processing. Through a continuous process of cooperation and information sharing with neighbors,
 agents in a network can be made to approach the global performance level despite the localized nature of their interactions.

\subsection{Networks and Neighborhoods}
\begin{figure}[h]
\epsfxsize 8cm \epsfclipon
\begin{center}
\leavevmode \epsffile{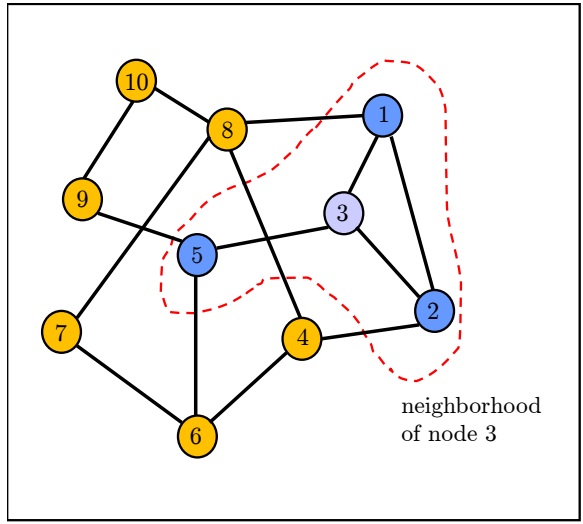} \caption{{\small A network consists of a collection of cooperating nodes. Nodes that
are linked by edges can share information. The neighborhood of any
particular node consists of all nodes that are connected to it by
edges (including the node itself). The figure illustrates the
neighborhood of node $3$, which consists of nodes $\{1,2,3,5\}$.
Accordingly, node $3$ has degree $4$, which is the size of its
neighborhood.}}\label{fig-A.label}
\end{center}
\end{figure}

In this article we focus mainly on {\em connected} networks,
although many of the results hold even if the network graph is
separated into disjoint subgraphs. In a  connected network,
 if we pick any two arbitrary nodes, then
there will  exist at least one path connecting them: the nodes
may be connected directly by an edge if they are neighbors, or
they may be connected by a path that passes through other
intermediate nodes. Figure~\ref{fig-A.label} provides a graphical
representation of a connected network with $N=10$ nodes. Nodes that
are able to share information with each other are connected by
edges. The sharing of information over these edges can be
unidirectional or bi-directional. The neighborhood of any particular
node is defined as the set of nodes that are connected to it by
edges;  we include in this set the node itself.  The figure
illustrates the neighborhood of node $3$, which consists of the
following subset of nodes: ${\cal N}_3=\{1,2,3,5\}$. For each node,
the size of its neighborhood defines its degree. For example, node
$3$ in the figure has degree $|{\cal N}_3|=4$, while node $8$ has
degree $|{\cal N}_8|=5$. Nodes that are well connected have higher degrees. Note
that we are denoting the neighborhood of an arbitrary node $k$ by
${\cal N}_k$ and its size by  $|{\cal N}_k|$.  We shall
also use the notation $n_k$ to refer to $|{\cal N}_k|$.

\begin{figure}[htb]
\epsfxsize 8cm \epsfclipon
\begin{center}
\leavevmode \epsffile{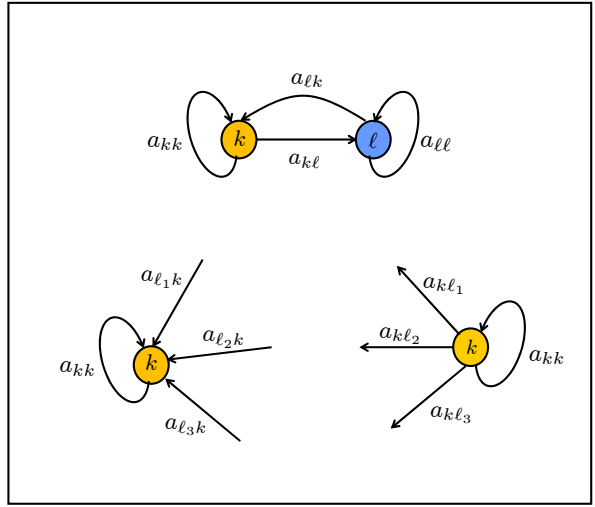} \caption{{\small In the top part, and for emphasis purposes, we are representing the edge between nodes $k$ and $\ell$ by two separate directed links: one moving from $k$ to $\ell$ and the other moving from $\ell$ to $k$. Two nonnegative weights are used to scale the sharing of information over these directed links. The scalar $a_{k\ell}$ denotes  the weight used to scale data sent from node $k$ to $\ell$, while $a_{\ell k}$ denotes the weight used
to scale data sent from node $\ell$ to $k$. The weights $\{a_{k,\ell},a_{\ell k}\}$ can be
different, and one or both of them can be zero, so that the exchange
of information over the edge connecting any two neighboring nodes
need not be symmetric. The bottom part of the figure illustrates directed links
arriving to node $k$ from its neighbors $\{\ell_1,\ell_2,\ell_3,\ldots\}$ (left) and leaving from
node $k$ towards these same neighbors (right).}}\label{fig-Z.label}
\end{center}
\end{figure}

The neighborhood of any node $k$ therefore consists of all nodes with which node $k$ can exchange information.
We assume a symmetric situation in relation to neighbors so
that if node $k$ is a neighbor of node $\ell$, then node $\ell$ is
also a neighbor of node $k$. This does not necessarily mean that the
flow of information between these two nodes is  symmetrical.
For instance,  in future sections, we shall assign pairs of nonnegative
weights to each edge connecting two neighboring nodes --- see
Fig.~\ref{fig-Z.label}. In particular, we will assign the coefficient
$a_{\ell k}$ to denote the weight used by node $k$ to scale the data
it receives from node $\ell$; this scaling can be interpreted as a
measure of trustworthiness or reliability that node $k$ assigns to its interaction with node $\ell$.
Note that we are using
two subscripts, $\ell k$, with the first subscript denoting the
source node (where information originates from) and the second
subscript denoting the sink node (where information moves to) so
that: \be \addbox{\;a_{\ell k}\;\;\equiv \;\;a_{\ell\; \rightarrow
\;k}\;\;\;(\mbox{\rm information flowing from node $\ell$ to node $k$})\;} \ee In this
way, the alternative coefficient $a_{k\ell}$ will denote the weight
used to scale the data sent from node $k$ to $\ell$:
 \be
\addbox{\;a_{k \ell}\;\;\equiv \;\;a_{k \;\rightarrow
\;\ell}\;\;\;(\mbox{\rm information flowing from node $k$ to node $\ell$})\;} \ee
 The weights $\{a_{k\ell},a_{\ell k}\}$ can be
different, and one or both of them can be zero, so that the exchange
of information over the edge connecting the neighboring nodes
$(k,\ell)$ need not be symmetric. When one of the weights is zero,
say, $a_{k\ell}=0$, then this situation means that
even though nodes $(k,\ell)$ are neighbors, node $\ell$ is either not
receiving data from node $k$ or the data emanating
from node $k$ is being annihilated before reaching node $\ell$.
Likewise, when $a_{kk}>0$, then node $k$ scales its own data,
whereas $a_{kk}=0$ corresponds to the situation when node $k$ does
not use its own data. Usually, in graphical representations like
those in Fig.~\ref{fig-A.label}, edges are drawn between
neighboring nodes that can share information.
And, it is understood that the actual sharing of information is controlled by the
values of the scaling weights that are assigned to the edge; these values can
turn off communication in one or both directions and they can also
scale one direction more heavily than the reverse direction, and so
forth.

\subsection{Cooperation Among Agents}
Now, depending on the application under consideration, the solution
vector $w^o$ from (\ref{adlq79813.a}) may admit different
interpretations. For example,
 the entries of $w^o$ may represent the location coordinates of a nutrition source that the agents
 are trying to find, or the location of an accident involving a dangerous chemical
 leak. The nodes may also be interested in locating a predator and tracking its movements over time.
 In these localization applications, the vector $w^o$ is usually two or three-dimensional. In other
 applications, the entries of  $w^o$ may represent the parameters of some model that the network
 wishes to learn, such as identifying the model parameters of a biological process or the occupied frequency bands
 in a shared communications medium. There are also situations where
 different agents in the network may be interested in estimating
 different entries of $w^o$, or even different parameter vectors $w^o$ altogether, say, $\{w_k^o\}$ for node $k$,
 albeit with some relation among the different vectors so that cooperation among the nodes can still  be rewarding. In this article, however, we focus exclusively on the
 special (yet frequent and important) case where all agents are interested in estimating the {\em same} parameter
 vector $w^o$.

Since the agents have a common objective, it is natural to expect
cooperation among them to be beneficial in general. One important question
 is therefore how to develop cooperation strategies that can lead to
better performance than when each agent attempts to solve the
optimization problem individually. Another important question is how
to develop strategies that endow networks with the ability to adapt and learn
in real-time in response to changes in the statistical properties of the data.
This article provides an overview of results in the area of {\em diffusion adaptation} with illustrative examples. Diffusion strategies are powerful methods that enable adaptive learning and cooperation over networks. There have been other useful works in the literature on the use of alternative {\em consensus strategies} to develop distributed optimization solutions over networks. Nevertheless, we explain in App.~\ref{app.C} why diffusion strategies outperform consensus strategies in terms of their mean-square-error stability and performance. For this reason, we focus in the body of the chapter on presenting the theoretical foundations for diffusion strategies and their performance.

\subsection{Notation} In our treatment, we need to distinguish  between random variables and deterministic quantities. For this reason, we use {\bf boldface} letters to represent random
variables and normal font to represent deterministic
(non-random) quantities. For example, the boldface letter $\d$
denotes  a random quantity, while the normal font letter $d$
denotes an observation or realization for it. We also need to distinguish between matrices and vectors. For this purpose, we use CAPITAL
letters to refer to matrices and small letters to refer to both
vectors and scalars; Greek letters always refer to scalars. For
example, we write $R$ to denote a covariance matrix and $w$ to
denote a vector of parameters. We also write $\sigma^2_v$ to refer to the variance of a
random variable. To distinguish between a vector $d$ (small
letter) and a scalar $d$ (also a small letter), we use parentheses
to index scalar quantities and subscripts to index vector
quantities. Thus, we write $d(i)$ to refer to the value of a
scalar quantity $d$ at time $i$, and  $d_i$ to refer to the
value of a vector quantity $d$ at time $i$. Thus, $d(i)$ denotes a
scalar while $d_i$ denotes a vector. All vectors in our presentation
are column vectors, with the exception of the regression vector
(denoted by the letter $u$), which will be taken to be a row vector
for convenience of presentation.  The symbol $T$ denotes
transposition, and the symbol $*$ denotes complex conjugation for
scalars and complex-conjugate transposition for matrices. The
notation $\mbox{\rm col}\{a,b\}$ denotes a column vector with
entries $a$ and $b$ stacked on top of each other, and the notation $\mbox{\rm diag}\{a,b\}$
denotes a diagonal matrix with entries $a$ and $b$. Likewise, the
notation $\mbox{\rm vec}(A)$ vectorizes its matrix argument and
stacks the columns of $A$ on top of each other. The notation $\|x\|$
denotes the Euclidean  norm of its vector argument, while $\|x\|_{b,\infty}$ denotes the block maximum
norm of a block vector (defined in App.~\ref{app.a}). Similarly, the notation $\|x\|_{\Sigma}^2$ denotes the weighted
square value, $x^*\Sigma x$. Moreover, $\|A\|_{b,\infty}$ denotes the block maximum norm
of a matrix (also defined in App.~\ref{app.a}), and $\rho(A)$ denotes the spectral radius of the matrix (i.e., the largest absolute magnitude among its eigenvalues).
 Finally, $I_M$ denotes the
identity matrix of size $M\times M$; sometimes, for simplicity of
notation, we drop the subscript $M$ from $I_M$ when the size of the
identity matrix is obvious from the context. Table~\ref{table9-asas1.label} provides a summary of the symbols used in the article for ease of reference.

{\small
\begin{table}[h]
\begin{center}
\caption{\rm {\small Summary of notation conventions used in the
article}.} {\small
\begin{tabular}{cl}\hline\hline &\\
$\d$ & {\small Boldface notation denotes random variables.}\\
$d$ & {\small Normal font denotes realizations  of random variables.}\\
$A$ & {\small Capital letters denote matrices.}\\
$a$ & {\small Small letters denote vectors or scalars.}\\
$\alpha$& {\small Greek letters denote scalars.}\\
$d(i)$ & {\small Small letters with parenthesis denote scalars.}\\
$d_i$ & {\small Small letters with subscripts denote vectors.}\\
$T$ & {\small Matrix transposition.}\\
$*$ & {\small Complex conjugation for
scalars and complex-conjugate transposition for matrices.}\\
$\mbox{\rm col}\{a,b\}$ & {\small Column vector with
entries $a$ and $b$.}\\
$\mbox{\rm diag}\{a,b\}$ &
{\small Diagonal matrix with entries $a$ and $b$.}\\
$\mbox{\rm vec}(A)$ & {\small Vectorizes matrix $A$ and
stacks its columns on top of each other.} \\
$\|x\|$& {\small Euclidean  norm of its vector argument.}\\
$\;\;\|x\|_{\Sigma}^2$ & {\small Weighted
square value $x^*\Sigma x$.} \\
$\|x\|_{b,\infty}$ & {\small Block maximum norm of a block vector -- see App.~\ref{app.a}}. \\
$\|A\|_{b,\infty}$ & {\small Block maximum norm of a matrix -- see App.~\ref{app.a}.} \\
$\|A\|$& {\small $2-$induced norm of matrix $A$ (its largest singular value).}\\
$\rho(A)$& {\small Spectral radius of its matrix argument.}\\
$I_M$ & {\small Identity matrix of size $M\times M$; sometimes, we drop the subscript $M$.}
\\\hline
\end{tabular} }
\label{table9-asas1.label}
\end{center}
\end{table}
}

\section{Mean-Square-Error Estimation}\label{sec.examples}
Readers interested in the development of the distributed optimization strategies and their adaptive versions can move directly to Sec.~\ref{sec.derivation.1}. The purpose of the current section is to motivate the virtues of distributed in-network processing, and to provide illustrative examples in the context of mean-square-error estimation. Advanced readers can skip this section on a first reading.

We start our development by associating with each agent $k$ an individual cost (or
utility)  function, $J_k(w)$. Although the algorithms presented in
this article apply to more general  situations, we nevertheless assume in our
presentation that the  cost functions $J_k(w)$ are strictly convex so
that each one of them has a unique minimizer.
 We further assume that, for all costs $J_k(w)$, the minimum occurs at the same value $w^o$.
 Obviously, the choice of $J_k(w)$ is limitless and is largely dependent on the application. It is sufficient for our purposes to illustrate the main concepts underlying diffusion adaptation by focusing on the case of mean-square-error (MSE) or quadratic cost functions. In the sequel, we provide several examples to illustrate how such quadratic cost functions arise in applications and how cooperative processing over networks can be beneficial. At the same time, we
note that most of the arguments in this article can be extended beyond MSE optimization to more
general cost functions and to situations where the minimizers of the individual costs $J_k(w)$ need not agree with each other  --- as already shown in \cite{Chen10,Chen10b,ChenSSP2012}; see also Sec.~\ref{sec.diskd8192} for a brief summary.

 In non-cooperative solutions, each agent would operate individually on its own cost function $J_k(w)$ and optimize it to determines $w^o$, without any interaction with the other nodes. However, the analysis and derivations in future sections will reveal that nodes can benefit from cooperation between them in terms of
 better performance (such as converging faster to $w^o$ or tracking a changing $w^o$ more effectively) --- see, e.g.,
 Theorems~\ref{lemaa.cod.2}--\ref{lemaa.cod.2a} and Sec.~\ref{sec.lakdcompa}.

\subsection{Application: Autoregressive Modeling} 
\begin{figure}[h]
\epsfxsize 9cm \epsfclipon
\begin{center}
\leavevmode \epsffile{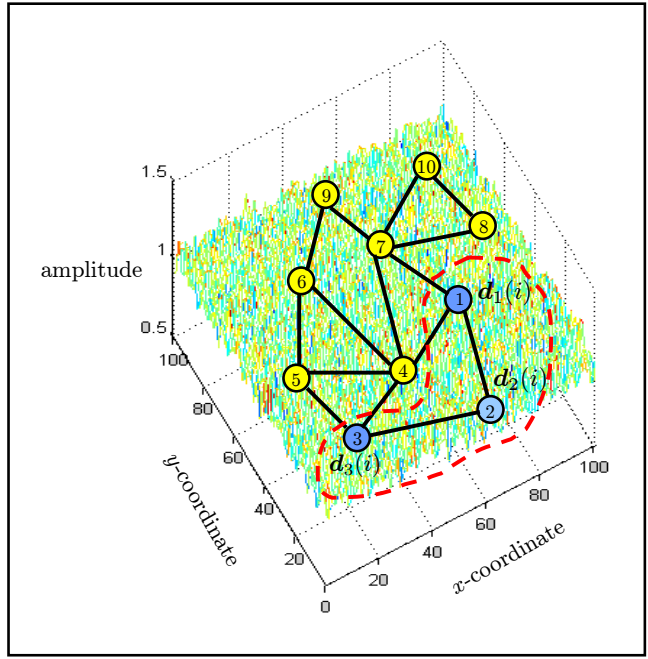} \caption{{\small A collection of
nodes, spread over a geographic region, observes realizations of an
AR random process and cooperates to estimate the underlying model
parameters $\{\beta_m\}$ in the presence of measurement noise. The noise power
profile can vary over space.}}\label{fig-2.label}
\end{center}
\end{figure}

Our first example relates to identifying the parameters of an
auto-regressive (AR) model from noisy data. Thus, consider a
situation where agents are spread over some geographical region and
each agent $k$ is observing realizations $\{d_k(i)\}$ of an AR
zero-mean random process $\{\d_k(i)\}$, which satisfies a model of
the form: \be \d_k(i) = \sum_{m=1}^M \beta_m \d_k(i - m) +
\v_k(i)\label{eq.111a}\ee The scalars $\{\beta_m\}$ represent
the model parameters that the agents wish to identify, and  $\v_k(i)$
represents an additive zero-mean white noise process with power: \be
\sigma_{v,k}^2\define \Ex\left|\v_k(i)\right|^2 \ee It is customary
to assume that the noise process is temporally white {\em and} spatially
independent so that noise terms across different nodes are independent of
each other and, at the same node, successive noise samples are also
independent of each other with a time-independent variance $\sigma_{v,k}^2$: \be
\left\{\begin{array}{l} \Ex \v_{k}(i)\v_{k}^*(j)\;=\;0,\;\;\mbox{\rm
for
all $i\neq j$\;(temporal whiteness)}\\
\Ex \v_{k}(i)\v_{m}^*(j)=\;0,\;\;\mbox{\rm for all $i,j$ whenever
$k\neq m$\;(spatial
whiteness)}\end{array}\right.\label{kaldja.alks}\ee The noise
process $\v_k(i)$ is further assumed to be independent of past
signals $\{\d_{\ell}(i-m), m\geq 1\}$ across all nodes $\ell$. Observe that we are allowing the noise power profile,
$\sigma_{v,k}^2$, to vary with $k$. In this way, the quality of the
measurements is allowed to vary across the network with some nodes
collecting noisier data than other nodes. Figure~\ref{fig-2.label}
illustrates an example of a network consisting of $N=10$ nodes
spread over a region in space. The figure shows the neighborhood of node $2$, which consists of nodes $\{1,2,3\}$.\\

\bigskip
\noindent {\bf {\em Linear Model}}\\
 \noindent To illustrate the difference between cooperative and non-cooperative estimation strategies,
 let us first explain that the data can be represented in terms of a linear model. To do so, we collect
 the model parameters $\{\beta_m\}$ into an $M\times 1$ column vector $w^o$: \be w^o \define
\mbox{\rm col}\left\{\beta_1,\beta_2,\ldots,\beta_M\right\}\ee and
the past data into a $1\times M$ row regression vector $\u_{k,i}$:
\be \u_{k,i} \define \ba{cccc}\d_k(i - 1)& \d_k(i - 2)&\ldots&
\d_k(i - M)\ea\ee Then, we can rewrite the measurement equation
(\ref{eq.111a}) at each node $k$ in the equivalent {\em linear model}
form: \be \addbox{\;\d_k(i) =
\u_{k,i}w^o\;+\;\v_k(i)\;}\label{lkad8912.lakd}\ee  Linear relations of the form (\ref{lkad8912.lakd}) are common in
applications and  arise in many other contexts (as further illustrated by the next three examples in this section).

We assume the stochastic processes $\{\d_k(i),\u_{k,i}\}$ in
(\ref{lkad8912.lakd}) have zero means and are jointly wide-sense stationary. We denote
their second-order moments by: \bq
\sigma_{d,k}^2&\define& \Ex|\d_k(i)|^2\;\;\;\;\;\;\;\;(\mbox{\rm scalar})\\
R_{u,k}&\define& \Ex\u_{k,i}^*\u_{k,i}\;\;\;\;\;\;\;\;(M\times M)\label{rdu.1}\\
r_{du,k}&\define& \Ex\d_k(i)\u_{k,i}^*\;\;\;\;\;\;(M\times 1)
\label{rdu.2}\eq  As was the case with the noise power
profile,   we are allowing the moments
$\{\sigma_{d,k}^2,R_{u,k},r_{du,k}\}$ to depend on the node index
$k$ so that these moments can vary with the spatial dimension as well.
The covariance matrix $R_{u,k}$ is, by definition, always
non-negative definite. However, for convenience of presentation, we
shall assume that $R_{u,k}$ is actually positive-definite (and,
hence, invertible):
\be
R_{u,k}>0
\ee
\bigskip

\noindent {\bf {\em Non-Cooperative Mean-Square-Error Solution}}\\
One immediate result that follows from the linear model
(\ref{lkad8912.lakd}) is that the unknown parameter vector $w^o$ can
be recovered {\em exactly} by each individual node from knowledge of the local moments
$\{r_{du,k},R_{u,k}\}$ alone.  To see this, note
that if we multiply both sides of (\ref{lkad8912.lakd}) by
$\u_{k,i}^*$ and take
 expectations we obtain
\bq
\underbrace{\Ex\u_{k,i}^*\d_k(i)}_{r_{du,k}}&=&\underbrace{\left(\Ex\u_{k,i}^*\u_{k,i}\right)}_{R_{u,k}}w^o\;+\;\underbrace{\Ex\u_{k,i}^*\v_k(i)}_{=0}
\eq so that \be
\addbox{\;r_{du,k}=R_{u,k}\;w^o\;\;\Longleftrightarrow\;\;w^o=R_{u,k}^{-1}\;r_{du,k}\;}\label{lakd891.2lkad}
\ee

\noindent \\

\noindent It is seen from (\ref{lakd891.2lkad}) that  $w^o$ is the
solution to a linear system of equations and that this solution can be computed by every node directly
 from its moments $\{R_{u,k},r_{du,k}\}$. It is useful to  re-interpret construction
(\ref{lakd891.2lkad}) as the solution to a minimum mean-square-error
(MMSE) estimation problem \cite{Sayed03,Sayed08}. Indeed, it can be
verified that the quantity $R_{u,k}^{-1}\;r_{du,k}$ that appears in (\ref{lakd891.2lkad}) is the
unique solution to the following MMSE problem: \be
\addbox{\;\min_{w}\;\Ex\left|\d_k(i)-\u_{k,i}
w\right|^2\;}\label{lkad9812.} \ee  To verify this claim, we denote
the cost function that appears in (\ref{lkad9812.}) by \bq
J_k(w)&\define&\Ex|\d_{k}(i)-\u_{k,i} w|^2\label{lkad89123.12}\eq and
expand it to find that\bq J_k(w)&=&\sigma_{d,k}^2 -w^* r_{du,k} -
r_{du,k}^* w + w^*R_{u,k} w\label{use.lakda} \eq The cost function
$J_k(w)$ is quadratic in $w$ and it has a unique minimizer since
$R_{u,k}>0$. Differentiating $J_k(w)$ with respect to $w$ we find its
gradient vector: \be \nabla_{w} J(w)\;=\;(R_{u,k}w\;-\;r_{du,k})^*
\ee It is seen that this gradient vector is annihilated at the same
value $w=w^o$ given by (\ref{lakd891.2lkad}). Therefore, we can
equivalently state that if each node $k$ solves the MMSE problem
(\ref{lkad9812.}), then the solution vector coincides with the
desired parameter vector $w^o$. This observation explains why it is
often justified to consider mean-square-error cost functions when
dealing with estimation problems that involve  data that satisfy
linear models similar to (\ref{lkad8912.lakd}).

Besides $w^o$, the solution of the MMSE problem (\ref{lkad9812.}) also
conveys information about the noise level in the data.
Note that by substituting $w^o$ from (\ref{lakd891.2lkad}) into
expression (\ref{lkad9812.}) for $J_k(w)$, the resulting minimum mean-square-error
value that is attained by
node $k$ is found to be: \bq \mbox{\rm MSE}_k&\define& J_k(w^o)\nn\\
&=&\Ex|\d_{k}(i)-\u_{k,i} w^o|^2\nn\\
&\stackrel{(\ref{lkad8912.lakd})}{=}&\Ex|\v_k(i)|^2\nn\\
&=&\sigma_{v,k}^2 \label{nois.lkad}\\
&\define& J_{k,\min}\nn\eq We shall use the notation $J_k(w^o)$ and $J_{k,\min}$ interchangeably to denote the
minimum cost value of $J_k(w)$. The above result states that, when each agent $k$ recovers $w^o$ from
knowledge of its moments $\{R_{u,k},r_{du,k}\}$ using
expression (\ref{lakd891.2lkad}), then the agent
attains an MSE performance level that is equal to the noise power
at its location, $\sigma_{v,k}^2$. An alternative useful expression for the
minimum cost can be obtained by
substituting expression (\ref{lakd891.2lkad}) for $w^o$ into
(\ref{use.lakda})  and simplifying the expression to find that \bq
\mbox{\rm MSE}_k
&=&\sigma_{d,k}^2-r^*_{du,k}R_{u,k}^{-1}r_{du,k}\label{kaljkj.a89812}
\eq This second expression is in terms of the data moments
$\{\sigma_{d,k}^2,r_{du,k},R_{u,k}\}$.\\\\

\noindent {\bf {\em Non-Cooperative Adaptive Solution}}\\
The optimal MMSE implementation (\ref{lakd891.2lkad}) for determining
$w^o$ requires knowledge of the statistical information
$\{r_{du,k},R_{u,k}\}$. This information is usually not available
beforehand. Instead, the agents are more likely to have access to successive
time-indexed observations $\{d_k(i),u_{k,i}\}$ of the random processes $\{\d_k(i),\u_{k,i}\}$ for $i\geq 0$.
In this case, it becomes necessary to devise a scheme that would allow each
node to use its measurements to approximate $w^o$. It turns out that an adaptive solution is possible. In this alternative
implementation, each node $k$ feeds its
observations $\{d_k(i),u_{k,i}\}$ into an adaptive filter and
evaluates successive estimates for $w^o$. As time passes by, the
estimates would get closer to $w^o$.

The adaptive solution operates as follows. Let $w_{k,i}$ denote an estimate for $w^o$ that is computed by node
$k$  at time $i$ based on all the observations $\{d_k(j),u_{k,j},\;j\leq i\}$ it has collected up
to that time instant. There are many
 adaptive algorithms that can be used to compute $w_{k,i}$; some filters are more accurate
 than others (usually, at the cost of additional complexity)
 \cite{Sayed03,Sayed08,Haykin01,Widrow85}. It is sufficient for our
 purposes to consider one simple (yet effective)
 filter structure, while noting that most of the discussion in this article can be extended to other structures.
  One of the simplest choices for an adaptive structure is the least-mean-squares
 (LMS) filter, where the data are processed by each node $k$ as follows:
\bq
e_k(i)&=&d_k(i)-u_{k,i}w_{k,i-1}\label{ea.lms}\\
w_{k,i}&=&w_{k,i-1}+\mu_k u_{k,i}^*e_{k}(i),\;\;\;i\geq 0\label{fac.lms} \eq
Starting from some initial condition, say, $w_{k,-1}=0$, the filter
iterates over $i\geq 0$. At each time instant, $i$, the filter uses
the local data $\{d_{k}(i),u_{k,i}\}$ at node $k$ to compute a local
estimation error, $e_k(i)$, via (\ref{ea.lms}). The error is then
used to update the existing estimate from $w_{k,i-1}$ to $w_{k,i}$
via (\ref{fac.lms}). The factor $\mu_k$ that appears in
(\ref{fac.lms}) is a constant positive step-size parameter; usually chosen to
be sufficiently small to ensure mean-square stability and
convergence, as discussed further ahead in the article. The step-size parameter can be
selected to vary with time as well; one popular choice is to replace $\mu_k$ in (\ref{fac.lms}) with the following construction:
\be
\mu_k(i)\define \frac{\mu_k}{\epsilon+\|u_{k,i}\|^2}
\ee
where $\epsilon>0$ is a small positive parameter and $\mu_k>0$. The resulting filter implementation is known as normalized LMS \cite{Sayed08} since the step-size is normalized by the squared norm of the regression vector. Other choices include step-size sequences $\{\mu(i)\geq 0\}$ that satisfy both  conditions:
\be
\sum_{i=0}^{\infty}\mu(i)=\infty,\;\;\;\;\;\;\;\;\;
\sum_{i=0}^{\infty}\mu^2(i)<\infty
\label{kdl8912.alkd}\ee
Such sequences converge slowly towards zero. One example is the choice $\mu_k(i)=\frac{\mu_k}{i+1}$. However, by virtue of the fact that such step-sizes die out as $i\rightarrow\infty$, then these choices  end up turning off adaptation. As such, step-size sequences satisfying (\ref{kdl8912.alkd}) are not generally suitable for applications that require continuous learning, especially under non-stationary environments. For this reason, in this article, we shall focus exclusively on the constant step-size case (\ref{fac.lms}) in order to ensure continuous adaptation and learning.

Equations
(\ref{ea.lms})--(\ref{fac.lms}) are written in terms of the observed
quantities $\{d_{k}(i),u_{k,i}\}$; these are deterministic values
since they correspond to observations of the random processes
$\{\d_{k}(i),\u_{k,i}\}$. Often, when we are interested in
highlighting the random nature of the quantities involved in the
adaptation step, especially when we study the mean-square
performance of adaptive filters, it becomes more useful to rewrite the recursions
using the {\bf boldface}
 notation to highlight the fact that the quantities that appear in (\ref{ea.lms})--(\ref{fac.lms}) are
 actually realizations of random variables. Thus, we also write:
\bq
\e_k(i)&=&\d_k(i)-\u_{k,i}\w_{k,i-1}\label{ea.lms.1}\\
\w_{k,i}&=&\w_{k,i-1}+\mu_k \u_{k,i}^*\e_{k}(i),\;\;\;i\geq 0\label{fac.lms.2}
\eq
where $\{\e_k(i),\w_{k,i}\}$ will be random variables as well.

The performance of adaptive implementations of this kind are
well-understood for both cases of stationary $w^o$ and changing
$w^o$ \cite{Sayed03,Sayed08,Haykin01,Widrow85}. For example, in the
stationary case, if the adaptive implementation
(\ref{ea.lms.1})--(\ref{fac.lms.2}) were to succeed in having its
estimator $\w_{k,i}$ tend to $w^o$ with probability one as
$i\rightarrow \infty$, then we would expect the error signal
$\e_k(i)$ in (\ref{ea.lms.1}) to tend towards the noise signal
$\v_k(i)$  (by virtue of the linear model (\ref{lkad8912.lakd})). This
means that, under this ideal scenario, the variance of the error
signal $\e_k(i)$ would be expected to tend towards the noise variance,
$\sigma_{v,k}^2$, as $i\rightarrow\infty$. Recall from
(\ref{nois.lkad}) that the noise variance is the least cost that the
 MSE solution can attain. Therefore, such limiting behavior
by the adaptive filter would be desirable. However, it is well-known that there is always
some loss in mean-square-error performance when adaptation is
employed due to the effect of gradient noise, which is caused by the algorithm's
 reliance on observations (or realizations) $\{d_k(i),u_{k,i}\}$
  rather than on the actual moments $\{r_{du,k}, R_{u,k}\}$.
In particular, it
is known that for LMS filters, the variance of $\e_k(i)$ in
steady-state will  be larger than $\sigma_{v,k}^2$ by a small
amount, and the size of the offset is proportional to the step-size
parameter $\mu_k$ (so that smaller step-sizes lead to better
mean-square-error (MSE) performance albeit at the expense of slower
convergence). It is easy to see why the variance of $\e_k(i)$
will be larger than $\sigma_{v,k}^2$ from the definition of the error
signal in (\ref{ea.lms.1}). Introduce the weight-error vector \be
\addbox{\widetilde{\w}_{k,i}\define w^o-\w_{k,i}\;} \ee and the
so-called {\em a-priori} error signal \be
\addbox{\;\e_{a,k}(i)\define \u_{k,i}\widetilde{\w}_{k,i-1}\;} \ee
This second error measures the difference between the uncorrupted
term $\u_{k,i}w^o$ and its estimator prior to adaptation,
$\u_{k,i}\w_{k,i-1}$. It then follows from the data model
(\ref{lkad8912.lakd}) and from the defining expression
(\ref{ea.lms.1}) for $\e_k(i)$ that \bq
\e_k(i)&=&\d_k(i)-\u_{k,i}\w_{k,i-1}\nn\\
&=&\u_{k,i}w^o+\v_k(i)-\u_{k,i}\w_{k,i-1}\nn\\
&=&\e_{a,k}(i)\;+\;\v_k(i) \label{rela.erladk}\eq Since the noise
component, $\v_k(i)$, is assumed to be zero-mean and independent of
all other random variables, we conclude that \be
\addbox{\;\Ex|\e_k(i)|^2\;=\;\Ex|\e_{a,k}(i)|^2\;+\;\sigma_{v,k}^2\;}
\label{earlier.alkda}\ee This relation holds for any time instant $i$; it shows that the variance of the output error,
$\e_k(i),$ is larger than $\sigma_{v,k}^2$ by an amount that
is equal to the variance of the {\em a-priori} error, $\e_{a,k}(i)$.
We define the filter mean-square-error (MSE) and
excess-mean-square-error (EMSE) as the
following steady-state measures:
\bq
{\rm MSE}_k&\define& \lim_{i\rightarrow\infty}\Ex|\e_k(i)|^2\\
{\rm EMSE}_k&\define& \lim_{i\rightarrow\infty}\Ex|\e_{a,k}(i)|^2 \eq so that, for the
adaptive implementation (compare with (\ref{nois.lkad})): \be
\addbox{\;{\rm MSE}_k\;=\;{\rm EMSE}_k\;+\;\sigma_{v,k}^2\;} \ee
Therefore, the EMSE term quantifies the size of the offset in the
MSE performance of the adaptive filter. We also define the filter
mean-square-deviation (MSD) as the steady-state measure: \be {\rm
MSD}_k\define \lim_{i\rightarrow\infty}\Ex\|\widetilde{\w}_{k,i}\|^2 \ee which measures
how far $\w_{k,i}$ is from $w^o$ in the mean-square-error sense in
steady-state. It is known that the MSD and EMSE of LMS filters of
the form (\ref{ea.lms.1})--(\ref{fac.lms.2}) can be approximated for
sufficiently small-step sizes by the following expressions
\cite{Sayed03,Sayed08,Haykin01,Widrow85}: \bq
{\rm EMSE}_k &\approx& \mu_k \sigma_{v,k}^2\mbox{\rm Tr}(R_{u,k})/2\label{lkad891.21lk2}\\
{\rm MSD}_k &\approx& \mu_k \sigma_{v,k}^2
M/2\label{lkad891.21lk2.1} \eq It is seen that the smaller the
step-size parameter, the better the performance of the
adaptive solution.\\\\

\noindent {\bf {\em Cooperative Adaptation through Diffusion}}\\
Observe from (\ref{lkad891.21lk2})--(\ref{lkad891.21lk2.1}) that
even if all nodes employ the same step-size, $\mu_k=\mu$, and even
if the regression data are spatially uniform so that $R_{u,k}=R_u$ for all $k$,
the mean-square-error performance across the nodes still varies
in accordance with the variation of the
noise power profile, $\sigma_{v,k}^2$, across the network. Nodes
with larger noise power will perform worse than nodes with smaller
noise power. However, since all nodes are observing data arising
from the {\em same} underlying model $w^o$, it is natural to expect
cooperation among the nodes to be beneficial. By cooperation we
mean that neighboring nodes can share information (such as
measurements or estimates) with each other as permitted by the
network topology. Starting in the next section, we will motivate and describe algorithms that enable nodes to carry
out  adaptation and learning in a cooperative manner to enhance performance.

Specifically, we are going to see that one way to achieve cooperation is by
developing adaptive algorithms that enable the nodes to optimize the
following global cost function in a distributed manner: \be
\addbox{\;\min_{w}\;\sum_{k=1}^N \Ex|\d_{k}(i)-\u_{k,i}
w|^2\;}\label{lkad81.1lk23} \ee where the global cost is the aggregate objective:\be J^{\rm
glob}(w)\define \sum_{k=1}^N \Ex|\d_{k}(i)-\u_{k,i}
w|^2\;=\;\sum_{k=1}^N J_{k}(w) \label{ka8912.ad}\ee  Comparing
(\ref{lkad81.1lk23}) with (\ref{lkad9812.}), we see that we are now
adding the individual costs, $J_{k}(w)$, from across all
nodes. Note that since the desired $w^o$ satisfies
(\ref{lakd891.2lkad}) at every node $k$, then it also satisfies \be
\left(\sum_{k=1}^M R_{u,k}\right)w^o\;=\;\sum_{k=1}^N
r_{du,k}\label{9s0lklad.as} \ee But it can be verified that the optimal
solution to (\ref{lkad81.1lk23}) is given by the same $w^o$ that
satisfies  (\ref{9s0lklad.as}). Therefore, solving the global
optimization problem (\ref{lkad81.1lk23}) also leads to the desired
$w^o$. In future sections, we will show how cooperative and distributed adaptive
schemes for solving (\ref{lkad81.1lk23}), such as
(\ref{Equ:DiffusionAdaptation:ATC.adaptive}) or
(\ref{Equ:DiffusionAdaptation:CTA.adaptive}) further ahead, lead to improved
performance in estimating $w^o$ (in terms of smaller
mean-square-deviation and faster convergence rate) than the
non-cooperative mode (\ref{ea.lms.1})--(\ref{fac.lms.2}), where each
agent runs its own individual adaptive filter --- see, e.g.,
 Theorems~\ref{lemaa.cod.2}--\ref{lemaa.cod.2a} and Sec.~\ref{sec.lakdcompa}.\\

\subsection{Application: Tapped-Delay-Line Models}
\begin{figure}[h]
\epsfxsize 8cm \epsfclipon
\begin{center}
\leavevmode \epsffile{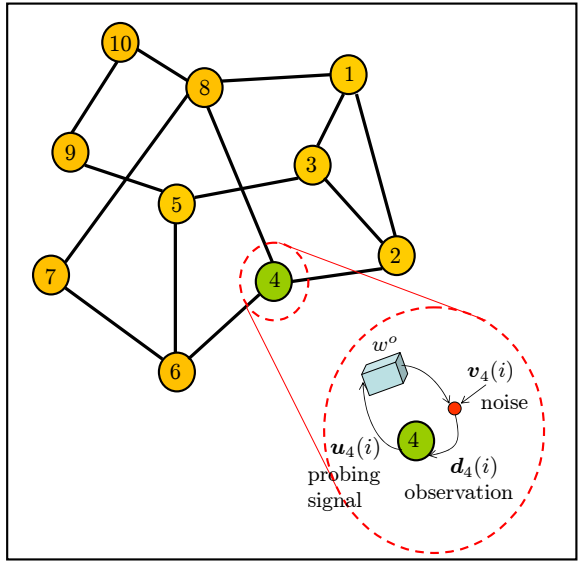} \caption{{\small The network is
interested in estimating the parameter vector $w^o$ that describes
an underlying tapped-delay-line model. The agents are assumed to be
able to independently probe the unknown system, and observe its response to excitations,
under noise, as indicated in the highlighted diagram for node $4$.}}\label{fig-C.label}
\end{center}
\end{figure}

Our second example to motivate MSE cost functions, $J_k(w)$, and linear models relates to identifying the parameters of a
moving-average (MA) model from noisy data. MA models are also known
as finite-impulse-response (FIR) or tapped-delay-line models. Thus,
consider a situation where agents are interested in estimating the
parameters of an FIR model, such as the taps of a communications
channel or the parameters of some (approximate) model of interest in finance or biology.
Assume the agents are able to independently probe the unknown model
and observe its response to excitations in the presence of additive
noise; this situation is illustrated in Fig.~\ref{fig-C.label}, with
the probing operation highlighted for one of the nodes (node $4$).

The schematics inside the enlarged diagram in Fig.~\ref{fig-C.label}
is meant to convey that each node $k$ probes the model with an input
sequence $\{\u_k(i)\}$ and measures the resulting response sequence,
$\{\d_k(i)\}$, in the presence of additive noise. The system
dynamics for each agent $k$ is assumed to be described by a MA model of the form: \be
\d_k(i) = \sum_{m=0}^{M-1} \beta_m \u_k(i - m) +
\v_k(i)\label{eq.111ab}\ee In this model, the term $\v_k(i)$ again
represents an additive zero-mean noise process that is assumed to be
temporally white and spatially independent; it is also assumed to be independent
of the input process, $\{\u_{\ell}(j)\}$, for all $i, j$ and $\ell$. The scalars
$\{\beta_m\}$ represent the model parameters that the agents seek to
identify. If we again collect the model parameters into an $M\times
1$ column vector $w^o$: \be w^o \define \mbox{\rm
col}\left\{\beta_0,\beta_1,\ldots,\beta_{M-1}\right\}\ee and the
input data into a $1\times M$ row regression vector: \be \u_{k,i}
\define  \ba{cccc}\u_k(i)& \u_k(i - 1)&\ldots& \u_k(i - M+1)\ea\label{lkad891.2l1k2}\ee then,
we can again express the measurement equation (\ref{eq.111ab}) at each
node $k$ in the same linear model as
(\ref{lkad8912.lakd}), namely, \be \addbox{\;\d_k(i) =
\u_{k,i}w^o\;+\;\v_k(i)\;}\label{lkad8912.lakdb}\ee As was the case with model
(\ref{lkad8912.lakd}), we
can likewise verify that, in view of (\ref{lkad8912.lakdb}), the
desired parameter vector $w^o$ satisfies the same normal equations (\ref{lakd891.2lkad}), i.e.,
\be
r_{du,k}=R_{u,k}\;w^o\;\;\Longleftrightarrow\;\;w^o=R_{u,k}^{-1}\;r_{du,k}\label{lakd891.2lkadb}
\ee where the moments $\{r_{du,k},R_{u,k}\}$ continue to be defined
by expressions (\ref{rdu.1})--(\ref{rdu.2}) with $\u_{k,i}$ now
defined by (\ref{lkad891.2l1k2}). Therefore, each node $k$ can
determine $w^o$ on its own by solving the same MMSE estimation problem (\ref{lkad9812.}). This
solution method requires knowledge of the moments
$\{r_{du,k},R_{u,k}\}$ and, according to (\ref{nois.lkad}), each
agent $k$ would then attain an MSE level that is equal to
the noise power level at its location.

Alternatively, when the statistical information
$\{r_{du,k},R_{u,k}\}$ is not available, each agent $k$ can estimate
 $w^o$ iteratively by feeding data $\{\d_k(i),\u_{k,i}\}$ into the adaptive
 implementation  (\ref{ea.lms.1})--(\ref{fac.lms.2}). In this way,
 each agent $k$ will achieve the same performance level shown
 earlier in (\ref{lkad891.21lk2})--(\ref{lkad891.21lk2.1}), with the
 limiting performance being again dependent on the local noise power
 level, $\sigma_{v,k}^2$. Therefore, nodes with
larger noise power will perform worse than nodes with
smaller noise power. However, since all nodes are observing data
arising from the same underlying model $w^o$, it is natural to
expect cooperation among the nodes to be beneficial.  As we
are going to see, starting from the next section, one way to achieve
cooperation and improve performance is by developing algorithms that
optimize the same global cost function (\ref{lkad81.1lk23}) in an
adaptive and distributed manner, such as algorithms
(\ref{Equ:DiffusionAdaptation:ATC.adaptive}) and
(\ref{Equ:DiffusionAdaptation:CTA.adaptive}) further ahead.\\

\subsection{Application: Target Localization}
Our third example relates to the problem of locating a destination
of interest (such as the location of a nutrition source or a
chemical leak) or locating and tracking an object of interest (such
as a predator or a projectile). In several such localization
applications, the agents in the network are allowed to move towards the target or away from it, in
which case we would end up with a mobile adaptive network \cite{Tu11}.
Biological networks behave in this manner such as networks
representing fish schools, bird formations, bee swarms, bacteria
motility, and diffusing particles
\cite{Tu11,catmay11,Li11,chenmay11,sasa11}. The
agents may move towards the target (e.g., when it is a nutrition
source) or away from the target (e.g., when it is a predator). In
other applications, the agents may remain static and are simply
interested in locating a target or tracking it (such as tracking a
projectile).

\begin{figure}[h]
\epsfxsize 9cm \epsfclipon
\begin{center}
\leavevmode \epsffile{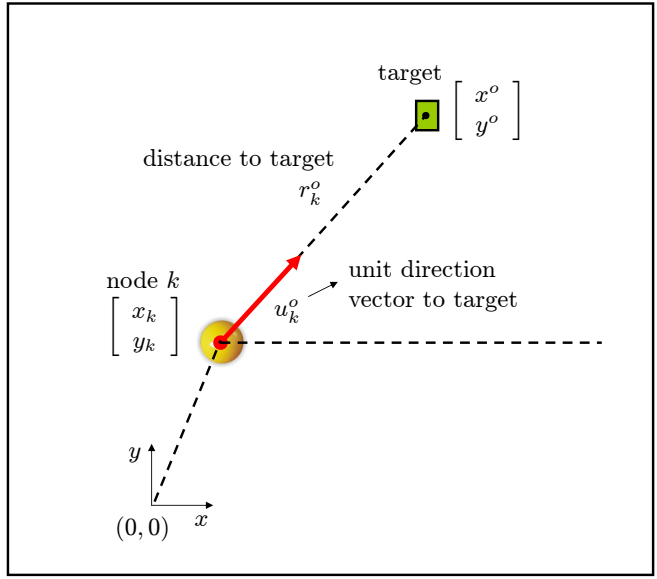} \caption{{\small The
distance from node $k$ to the target is denoted by $r_k^o$ and the
 unit-norm direction vector from the same node to the target
is denoted by $u_{k}^o$. Node $k$ is assumed to have access to noisy
measurements of $\{r_k^o,u_{k}^o\}$}.}\label{fig-D.label}
\end{center}
\end{figure}

To motivate mean-square-error estimation in the context of
localization problems, we start with the situation corresponding to a static target and static
nodes. Thus, assume that the unknown
location of the target in the cartesian plane is represented by the
$2\times 1$  vector $w^o=\mbox{\rm col}\{x^o,y^o\}$. The agents are
spread over the same region of space and are interested in locating
the target. The location of every agent $k$ is denoted by
the $2\times 1$ vector $p_{k}=\mbox{\rm col}\{x_{k},y_{k}\}$ in
terms of its $x$ and $y$ coordinates -- see Fig.~\ref{fig-D.label}. We assume the agents are aware
of their location vectors. The
distance between agent $k$ and the target is denoted by $r_k^o$ and
is equal to: \bq r_k^o&=&\|w^o-p_{k}\|\label{eq.13}\eq The $1\times
2$ unit-norm direction vector pointing from agent $k$ towards the
target is denoted by $u_{k}^o$ and is given by: \bq
u_{k}^o&=&\frac{(w^o-p_{k})^T}{\|w^o-p_{k}\|}\label{eq.16}\eq
 Observe from (\ref{eq.13}) and (\ref{eq.16}) that
$r_k^o$ can be expressed in the following useful
inner-product form:\be r_k^o=u_{k}^o(w^o-p_{k}) \label{eq.use}\ee

\begin{figure}[h]
\epsfxsize 10cm \epsfclipon
\begin{center}
\leavevmode \epsffile{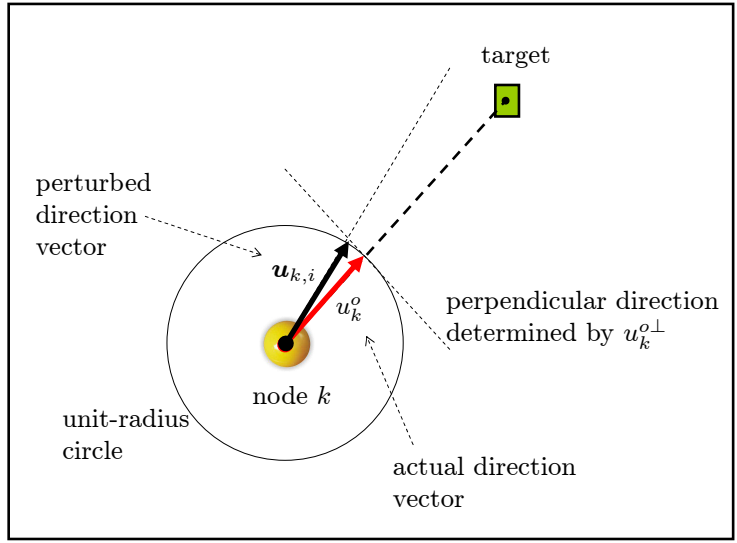} \caption{{\small The tip of the
noisy direction vector is modeled as being approximately perturbed
away from the actual direction by two effects: a larger effect
caused by a deviation along the direction that is perpendicular to
$u_{k}^o$, and a smaller deviation along the direction that is
parallel to $u_{k}^o$.}}\label{fig-E.label}
\end{center}
\end{figure}

In practice, agents have noisy observations of both their
distance and direction vector towards the target. We denote the
noisy distance measurement collected by node $k$ at time $i$ by: \bq
{\r}_k(i) &=& r_k^o\;+\;{\v}_k(i)\label{eq.44}\eq where
 ${\v}_k(i)$ denotes noise and is assumed to be zero-mean, and temporally white and spatially independent with variance
\bq {\sigma}_{v,k}^2&\define&\Ex|{\v}_k(i)|^2\eq We also denote the
noisy direction vector that is measured by node $k$ at time $i$ by
$\u_{k,i}$. This vector is a perturbed version of $u_{k}^o$. We
assume that $\u_{k,i}$ continues to start from the
location of the node at $p_{k}$, but that its tip is perturbed
slightly either to the left or to the right relative to the tip
of $u_{k}^o$ --- see Fig.~\ref{fig-E.label}.  The
perturbation to the tip of $u_{k}^o$ is modeled as
being the result of two effects: a small deviation that occurs along
the direction that is perpendicular to $u_{k}^o$, and a smaller
deviation that occurs along the direction of
$u_{k}^o$.  Since we are assuming that the tip of $\u_{k,i}$ is only
slightly perturbed relative to the tip of
 $u_{k}^o$, then it is reasonable to expect the amount of perturbation along the
parallel direction to be small compared to the amount of perturbation
along the perpendicular direction.

Thus, we write \bq \u_{k,i}&=& u_{k}^o
\;+\;\bm{\alpha}_{k}(i)\;u_{k}^{o \perp}\;+\;\bm{\beta}_k(i)\;
u_{k}^o\;\;\;\;\;\;(1\times 2)\label{eq.uua}\eq
where $u_{k}^{o\perp}$ denotes a unit-norm row vector that lies in
the same plane and whose direction is perpendicular to $u_{k}^o$.
The variables $\bm{\alpha}_k(i)$ and $\bm{\beta}_k(i)$ denote
zero-mean independent random noises that are temporally white and spatially independent with variances: \bq
{\sigma}_{\alpha,k}^2&\define&\Ex|\bm{\alpha}_k(i)|^2,\;\;\;\;\;\;\;\;
{\sigma}_{\beta,k}^2\define \Ex|\bm{\beta}_k(i)|^2
 \eq We assume the contribution of $\bm{\beta}_k(i)$ is small
 compared to the contributions of the other noise sources, $\bm{\alpha}_k(i)$ and $\v_k(i)$, so that
 \be \sigma_{\beta,k}^2\ll \sigma_{\alpha,k}^2,\;\;\;\;\;\;\;
\sigma_{\beta,k}^2\ll \sigma_{v,k}^2\ee
 The random noises $\{\v_k(i),\bm{\alpha}_k(i),\bm{\beta}_k(i)\}$ are further assumed to be
independent of each other.

Using (\ref{eq.use}) we find that the noisy measurements
$\{{\r}_k(i),\u_{k,i}\}$ are related to the unknown $w^o$ via: \be
{\r}_k(i)=\u_{k,i}(w^o-p_{k})\;+\z_k(i) \label{arrive.da}\ee where
the modified noise term $\z_k(i)$ is defined in terms of the noises
in $\r_k(i)$ and $\u_{k,i}$ as follows: \bq \z_k(i)&\define&
{\v}_k(i)\;-\;\bm{\alpha}_k(i)\;
u_{k}^{o\perp}\cdot (w^o-p_{k})\;-\;\bm{\beta}_k(i)\cdot u_{k}^o\cdot(w^o-p_{k})\nn\\
&=& {\v}_k(i)\;-\;\bm{\beta}_k(i)\cdot r_{k}^o\nn\\
&\approx&\v_k(i) \label{eq.vv}\eq since, by construction, \be
u_{k}^{o\perp}\cdot (w^o-p_{k})\;=\;0 \ee and the contribution by
$\bm{\beta}_k(i)$ is assumed to be sufficiently small. If we now
introduce the adjusted signal: \be \d_k(i)\define
{\r}_k(i)\;+\;\u_{k,i}\;p_{k}\label{eq.adj} \ee then we arrive again
from (\ref{arrive.da}) and (\ref{eq.vv}) at the following linear model for
the available measurement variables $\{\d_k(i),\u_{k,i}\}$ in terms
of the target location $w^o$: \be
\addbox{\;\d_k(i)\approx \u_{k,i}w^o+\v_k(i)\;}\label{eq.meas} \ee
There is one important difference in relation to the earlier linear models (\ref{lkad8912.lakd})
and (\ref{lkad8912.lakdb}), namely, the variables $\{\d_k(i),\u_{k,i}\}$ in (\ref{eq.meas}) do not have
zero means any longer. It is nevertheless
straightforward to determine the first and second-order moments of
the variables $\{\d_k(i),\u_{k,i}\}$. First, note from  (\ref{eq.44}), (\ref{eq.uua}), and
(\ref{eq.adj}) that \be \Ex\u_{k,i}=u_{k}^o,\;\;\;\;\;\;\;
\Ex\d_{k}(i)\;=\;r_k^o\;+\;u_{k}^op_{k}\ee
Even in this case of non-zero means, and in view of (\ref{eq.meas}),
the desired parameter vector $w^o$ can still be shown to satisfy the
same normal equations (\ref{lakd891.2lkad}), i.e., \be
r_{du,k}=R_{u,k}w^o\;\;\Longleftrightarrow\;\;w^o=R_{u,k}^{-1}\;r_{du,k}\label{lakd891.2lkadb.xx}
\ee where the moments $\{r_{du,k},R_{u,k}\}$ continue to be defined
as \be R_{u,k}\define \Ex\u_{k,i}^*\u_{k,i},\;\;\;\;r_{du,k}\define
\Ex\u_{k,i}^*\d_k(i)\label{conh.da}\ee To verify that
(\ref{lakd891.2lkadb.xx}) holds, we simply multiply both sides of
(\ref{eq.meas}) by $\u_{k,i}^*$ from the left, compute the expectations of
both sides, and use the fact that $\v_k(i)$ has zero mean and is assumed to be
independent of $\{\u_{\ell,j}\}$ for all times $j$ and nodes $\ell$. However, the difference in relation to the earlier
normal equations (\ref{lakd891.2lkad}) is that the matrix $R_{u,k}$ is not the
actual covariance matrix of $\u_{k,i}$ any longer. When $\u_{k,i}$ is not zero mean, its covariance matrix
is
instead defined as: \bq \mbox{\rm cov}_{u,k}&\define&
\Ex(\u_{k,i}-u_k^o)^*(\u_{k,i}-u_k^o)\nn\\
&=&\Ex\u_{k,i}^*\u_{k,i}\;-\;u_k^{o*}u_k^o\eq so that \be
R_{u,k}\;=\;\mbox{\rm cov}_{u,k}\;+\;u_k^{o*}u_k^o \ee We
conclude from this relation that  $R_{u,k}$ is positive-definite
(and, hence, invertible) so that expression
(\ref{lakd891.2lkadb.xx}) is justified. This is because the
covariance matrix, $\mbox{\rm cov}_{u,k}$, is itself
positive-definite. Indeed, some algebra applied to the difference $\u_{k,i}-u_{k}^o$ from
(\ref{eq.uua})
shows that \bq
 \mbox{\rm
cov}_{u,k}&=&\ba{cc}\left(u_{k}^{o\perp}\right)^*
&\left(u_{k}^{o}\right)^*\ea\ba{cc}\sigma_{\alpha,k}^2&\\&\sigma_{\beta,k}^2\ea\ba{l}u_{k}^{o\perp}\\u_k^o\ea
\eq where the matrix \be \ba{l}u_{k}^{o\perp}\\u_k^o\ea \ee is full
rank since the rows $\{u_k^o,\;u_{k}^{o\perp}\}$ are linearly independent vectors.

Therefore, each node $k$ can determine $w^o$ on its own
by solving the same minimum mean-square-error estimation problem
(\ref{lkad9812.}). This solution method requires knowledge of the
moments $\{r_{du,k},R_{u,k}\}$ and, according to (\ref{nois.lkad}),
each agent $k$ would then attain an MSE level that is
equal to the noise power level, $\sigma_{v,k}^2$, at its location.

Alternatively, when the statistical information
$\{r_{du,k},R_{u,k}\}$ is not available beforehand, each agent $k$
can estimate
 $w^o$ iteratively by feeding data $\{\d_k(i),\u_{k,i}\}$ into the adaptive
 implementation  (\ref{ea.lms.1})--(\ref{fac.lms.2}). In this case,
 each agent $k$ will achieve the performance level shown
 earlier in (\ref{lkad891.21lk2})--(\ref{lkad891.21lk2.1}), with the
 limiting performance being again dependent on the local noise power
 level, $\sigma_{v,k}^2$. Therefore, nodes with
larger noise power will perform worse than nodes with
smaller noise power. However, since all nodes are observing
distances and direction vectors towards the same target location
$w^o$, it is natural to expect cooperation among the nodes to
be beneficial.  As we are going to see, starting from the next
section, one way to achieve cooperation and improve performance is
by developing algorithms that solve the same global cost function
(\ref{lkad81.1lk23}) in an adaptive and distributed manner, by using
algorithms such as (\ref{Equ:DiffusionAdaptation:ATC.adaptive}) and
(\ref{Equ:DiffusionAdaptation:CTA.adaptive}) further ahead.\\

\noindent {\bf {\em Role of Adaptation}}\\
The localization application helps highlight one of the main
advantages of adaptation, namely, the ability of adaptive
implementations to learn and track changing statistical conditions.
For example, in the context of mobile networks, where nodes can move
closer or further away from a target, the location vector for each
agent $k$ becomes time-dependent, say, $p_{k,i}=\mbox{\rm
col}\{x_{k}(i),y_{k}(i)\}$. In this case, the actual distance and
direction vector between agent $k$ and the target also vary with
time and become: \be r_k^o(i)\;=\;\|w^o-p_{k,i}\|,\;\;\;\;\;\;\;\;\;
u_{k,i}^o\;=\;\frac{(w^o-p_{k,i})^T}{\|w^o-p_{k,i}\|}\ee The noisy
distance measurement to the target is then: \bq {\r}_k(i) &=&
r_k^o(i)\;+\;{\v}_k(i)\label{eq.44.abc}\eq where the variance of
$\v_k(i)$ now depends on time as well: \bq
{\sigma}_{v,k}^2(i)&\define&\Ex|{\v}_k(i)|^2\eq In the context of
mobile networks, it is reasonable to assume that the variance of
$\v_k(i)$ varies both with time and with the distance to the target: the
closer the node is to the target, the less noisy the measurement of
the distance is expected to be. Similar remarks hold for the
variances of the noises $\bm{\alpha}_k(i)$ and $\bm{\beta}_k(i)$
that perturb the measurement of the direction vector, say, \be
{\sigma}_{\alpha,k}^2(i)\define\Ex|\bm{\alpha}_k(i)|^2,\;\;\;\;\;\;\;\;
{\sigma}_{\beta,k}^2(i)\define \Ex|\bm{\beta}_k(i)|^2 \ee where now
\bq \u_{k,i}&=& u_{k,i}^o \;+\;\bm{\alpha}_{k}(i)\;u_{k,i}^{o
\perp}\;+\;\bm{\beta}_k(i)\; u_{k,i}^o\label{eq.uua.xx}\eq
 The same
arguments that led to (\ref{eq.meas}) can be repeated to lead to
 the same model, except that now the means of the variables  $\{\d_k(i),\u_{k,i}\}$ become
 time-dependent as well: \be \Ex\u_{k,i}=u_{k,i}^o,\;\;\;\;\;\;\;
\Ex\d_{k}(i)\;=\;r_k^o(i)\;+\;u_{k,i}^o\;p_{k,i} \ee Nevertheless,
 adaptive solutions (whether cooperative or non-cooperative), are
able to track such time-variations because these solutions work
directly with the observations $\{d_k(i),u_{k,i}\}$ and the successive
observations will reflect the changing statistical profile of the data. In
general, adaptive solutions are able to track changes in the
underlying signal statistics rather well \cite{Sayed03,Sayed08}, as
long as the rate of non-stationarity is slow enough for the filter
to be able to follow the changes.\\

\subsection{Application: Collaborative Spectral Sensing}
Our fourth and last example to illustrate the role of
mean-square-error estimation and cooperation relates to spectrum
sensing for cognitive radio applications. Cognitive radio systems
involve two types of users: primary users and secondary users. To
avoid causing harmful interference to incumbent primary users,
unlicensed cognitive radio devices need to detect unused frequency
bands even at low signal-to-noise (SNR) conditions
\cite{Mitola99,Haykin05,Quan11,Zou10}. One way to carry out spectral
sensing is for each secondary user to estimate the aggregated power
spectrum that is transmitted by all active primary users, and to
locate unused frequency bands within the estimated spectrum. This
step can be performed by the secondary users with or without
cooperation.

\begin{figure}[h]
\epsfxsize 10cm \epsfclipon
\begin{center}
\leavevmode \epsffile{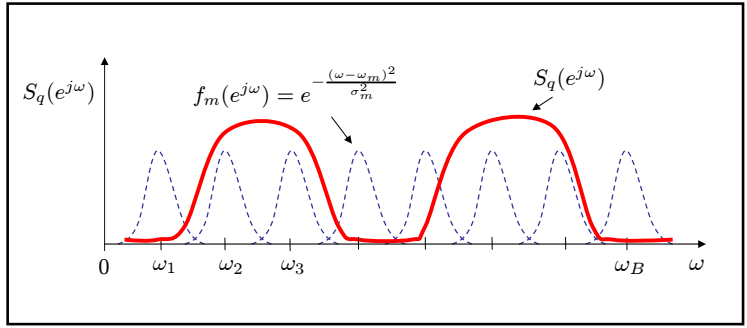} \caption{{\small The interval
$[0,\pi]$ is divided into $B$ sub-intervals of equal width; the
center frequencies of the sub-intervals are denoted by
$\{\omega_m\}$. A power spectrum $S_q(e^{j\omega})$ is approximated
as a linear combination of Gaussian basis functions centered on the
$\{\omega_m\}$.}}\label{fig-F.label}
\end{center}
\end{figure}

Thus, consider a communications environment consisting of $Q$
primary users and $N$ secondary users. Let $S_q(e^{j\omega})$ denote
the power spectrum of the signal transmitted by primary user $q$. To
facilitate estimation of the spectral profile by the secondary users, we assume that each $S_q(e^{j\omega})$
can be represented as a linear combination of basis functions,
$\{f_m(e^{j\omega})\}$, say, $B$ of them \cite{Paolo11}: \bq
S_q(e^{j\omega})&=&\sum_{m=1}^B
\beta_{qm}f_m(e^{j\omega}),\;\;\;q=1,2,\ldots,Q \label{poad.lakd}\eq
In this representation, the scalars $\{\beta_{qm}\}$ denote the
coefficients of the basis expansion for user $q$. The variable
$\omega\in[-\pi,\pi]$ denotes the normalized angular frequency
measured in radians/sample. The power spectrum is often symmetric
about the vertical axis, $\omega=0$, and therefore it is sufficient
to focus on the interval $\omega\in[0,\pi]$. There are many ways by
which the basis functions, $\{f_m(e^{j\omega})\}$, can be selected.
The following is one possible construction for
illustration purposes. We divide the interval $[0,\pi]$ into $B$
identical intervals and denote their center frequencies by
$\{\omega_m\}$. We then place a Gaussian pulse at each location
$\omega_m$ and control its width through the selection of its
standard deviation, $\sigma_m$, i.e., \be f_m(e^{j\omega})\define
e^{-\frac{(\omega-\omega_m)^2}{\sigma_m^2}} \ee
Figure~\ref{fig-F.label} illustrates this construction. The
parameters $\{\omega_m,\sigma_m\}$ are selected by the designer and
are assumed to be known.  For a sufficiently large
  number, $B$, of basis functions, the
  representation (\ref{poad.lakd}) can approximate well a large class of power spectra.

\begin{figure}[h]
\epsfxsize 7cm \epsfclipon
\begin{center}
\leavevmode \epsffile{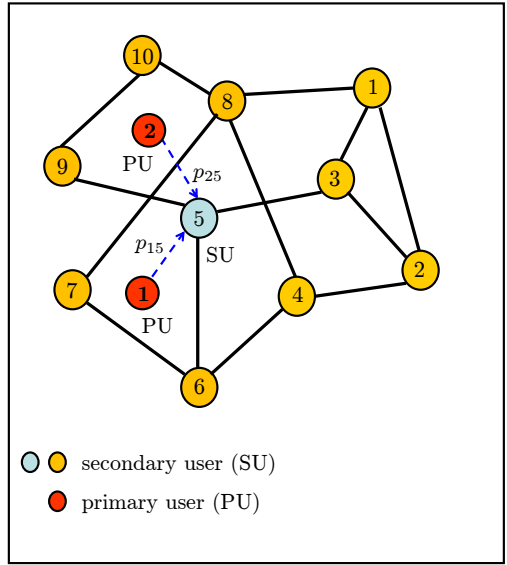} \caption{{\small A network of
secondary users in the presence of two primary users. One of the
secondary users is highlighted and the path loss coefficients from
the primary users to its location are indicated as $p_{15}$ and
$p_{25}$.}}\label{fig-G.label}
\end{center}
\end{figure}

  We collect the combination coefficients $\{\beta_{qm}\}$ for primary user $q$ into a column vector $w_q$:
\be w_q
\define \mbox{\rm col}\left\{\beta_{q1},\beta_{q2},\beta_{q3},\ldots,\beta_{qB}\right\}\;\;\;\;\;\;\;\;\;\;\;\;\;\;\;\;\;\;\;\;(B \times 1)\ee
and collect the basis functions into a row vector: \be f_{\omega}
\define
\ba{cccc}f_1(e^{j\omega})&f_2(e^{j\omega})&\ldots&f_B(e^{j\omega})\ea\;\;\;\;\;\;(1
\times B)\ee Then, the power spectrum (\ref{poad.lakd}) can be
 expressed in the alternative inner-product form:
\be S_q(e^{j\omega})\;=\;f_{\omega} w_q \ee Let $p_{qk}$ denote the
path loss coefficient from primary user $q$ to secondary user $k$.
When the transmitted spectrum $S_q(e^{j\omega})$ travels from
primary user $q$ to secondary user  $k$, the spectrum that is
sensed by node $k$ is $p_{qk}S_q(e^{j\omega})$. We
assume in this example that the path loss factors $\{p_{qk}\}$ are
known and that they have been determined during a prior training stage
involving each of the primary users with each of the secondary
users. The training is usually repeated at regular intervals of time to
accommodate the fact that the path loss coefficients can vary (albeit
slowly) over time. Figure~\ref{fig-G.label} depicts a cognitive
radio system with $2$ primary users and $10$ secondary users. One of
the secondary users (user $5$) is highlighted and the path loss
coefficients from the primary users to its location are indicated;
similar path loss coefficients can be assigned to all other
combinations involving primary and secondary users.

Each user $k$ senses the {\em aggregate} effect of the power spectra that
are transmitted by all active primary users. Therefore, adding the
effect of all primary users, we find that the power spectrum that
arrives at secondary user $k$ is given by: \bq
S_k(e^{j\omega})&=& \sum_{q=1}^Q p_{qk} S_q(e^{j\omega})\;+\;\sigma_{k}^2\nn\\
&=&\sum_{q=1}^Q p_{qk}f_{\omega} w_q\;+\;\sigma_k^2\nn\\
&\define & u_{k,\omega}w^o\;+\;\sigma_k^2 \eq where $\sigma_k^2$
denotes the receiver noise power at node $k$, and where we
introduced the following vector quantities: \bq
w^o&\define&\mbox{col}\{w_1,w_2,\ldots,w_Q\}\;\;\;\;\;\;\;\;\;\;\;\;\;\;\;\;\;\;\;\;\;(BQ\times 1)\\
u_{k,\omega}&\define&\ba{cccc}p_{1k}f_{\omega}&p_{2k}f_{\omega}&\ldots&p_{Qk}f_{\omega}\ea\;\;\;\;(1\times
BQ)\label{row.da} \eq The vector $w^o$ is the collection of all
combination coefficients for all $Q$ primary users. The vector
 $u_{k,\omega}$ contains the path loss coefficients from all primary users to user $k$. Now, at every
 time instant $i$, user $k$ observers its received power spectrum, $S_{k}(e^{j\omega})$, over a discrete
 grid of frequencies, $\{\omega_{r}\}$, in the interval $[0,\pi]$ in the presence of additive measurement noise. We denote these
 measurements by:
\bq \d_{k,{r}}(i)&=&\u_{k,\omega_{r}}
w^o\;+\;\sigma_k^2\;+\;\v_{k,{r}}(i)
\nn\\
&&{r}=1,2,\ldots,R\label{daj897.13lk}\eq The term $\v_{k,{r}}(i)$
denotes sampling noise and is assumed to have zero mean and variance
$\sigma_{v,k}^2$; it is also assumed to be temporally white and spatially
independent; and is also independent of all other random variables. Since the row
vectors $u_{k,\omega}$ in (\ref{row.da}) are defined in terms of the
path loss coefficients $\{p_{qk}\}$, and since these coefficients
are estimated and subject to noisy distortions, we model the
$\u_{k,\omega_{r}}$ as zero-mean random variables in
(\ref{daj897.13lk}) and use the boldface notation for them.

Observe that in this application, each node $k$ collects $R$
measurements at every time instant $i$ and not only a single
measurement, as was the case with the three examples discussed in
the previous sections (AR modeling, MA modeling, and localization).
The implication of this fact is that we now deal with an estimation
problem that involves vector measurements instead of scalar
measurements at each node. The solution structure continues to be the same. We
collect the $R$ measurements at node $k$ at time $i$ into vectors
and introduce the $R\times 1$ quantities: \be \d_{k,i}\define
\ba{c}\d_{k,1}(i)-\sigma_k^2\\\d_{k,2}(i)-\sigma_k^2\\\vdots\\\d_{k,R}(i)-\sigma_k^2\ea,\;\;\;\;\;\;
\v_{k,i}\define
\ba{c}\v_{k,1}(i)\\\v_{k,2}(i)\\\vdots\\\v_{k,R}(i)\ea\ee and the regression matrix:\be
\U_{k,i}\define
\ba{c}\u_{k,\omega_1}\\\u_{k,\omega_2}\\\vdots\\\u_{k,\omega_R}\ea\;\;\;\;\;\;(R\times
QB) \ee The time subscript in $\U_{k,i}$ is used to model the fact
that the path loss coefficients can change over time due to the
possibility of node mobility. With the above notation, expression
(\ref{daj897.13lk}) is equivalent to the linear model:\be
\addbox{\;\d_{k,i}\;=\;\U_{k,i}
w^o\;+\;\v_{k,i}\;}\label{lkad891.21lk} \ee Compared to the earlier examples (\ref{lkad8912.lakd}),
(\ref{lkad8912.lakdb}), and (\ref{eq.meas}), the main difference now
is that each agent $k$ collects a {\em vector} of measurements,
$\d_{k,i}$, as opposed to the scalar $\d_k(i)$, and its regression
data are represented by the matrix quantity, $\U_{k,i}$, as opposed to
the row vector $\u_{k,i}$. Nevertheless, the estimation approach
will continue to be the same. In cognitive network applications, the
secondary users are interested in estimating the aggregate power
spectrum of the primary users in order for the secondary users to
identify vacant frequency bands that can be used by them. In the
context of model (\ref{lkad891.21lk}), this amounts to determining
the parameter vector $w^o$ since knowledge of its entries allows
each secondary user to reconstruct the aggregate power spectrum defined by: \be
S_A(e^{j\omega})\define \sum_{q=1}^Q S_q(e^{j\omega})\;=\;(\mathds{1}_Q^T\otimes
f_{\omega})w^o \ee where the notation $\otimes$ denotes the
Kronecker product operation, and $\mathds{1}_Q$ denotes a $Q\times 1$ vector whose entries are all equal to one.

As before, we can again verify that, in view of
(\ref{lkad891.21lk}), the desired parameter vector $w^o$ satisfies
the same normal equations: \be
R_{dU,k}=R_{U,k}w^o\;\;\Longleftrightarrow\;\;w^o=R_{U,k}^{-1}\;R_{dU,k}\label{lakd891.2lkadbxxxx}
\ee where the moments $\{R_{dU,k},R_{U,k}\}$ are now defined by \bq
R_{dU,k}&\define&\Ex\U_{k,i}^*\d_{k,i}\;\;\;\;\;\;\;\;(QB\times 1)\\
R_{U,k}&\define&\Ex\U_{k,i}^*\U_{k,i}\;\;\;\;\;\;\;(QB\times QB) \eq
Therefore, each secondary user $k$ can determine $w^o$ on its own by
solving the following minimum mean-square-error estimation problem:
 \be
 \min_{w}\;\Ex\left\|\d_{k,i}-\U_{k,i}w\right\|^2
 \ee
This solution method requires knowledge of the moments
$\{R_{dU,k},R_{U,k}\}$ and, in an argument similar to the one that
led to (\ref{nois.lkad}), it can be verified that each agent $k$
would attain an MSE performance level that is equal to the noise
power level, $\sigma_{v,k}^2$,  at its location.

Alternatively, when the statistical information
$\{R_{dU,k},R_{U,k}\}$ is not available, each secondary user $k$ can
estimate
 $w^o$ iteratively by feeding data $\{\d_{k,i},\U_{k,i}\}$ into an adaptive
 implementation  similar to (\ref{ea.lms.1})--(\ref{fac.lms.2}), such as the following vector LMS
 recursion:
\bq
\e_{k,i}&=&\d_{k,i}-\U_{k,i}\w_{k,i-1}\label{ea.lms.1c}\\
\w_{k,i}&=&\w_{k,i-1}+\mu_k \U_{k,i}^*\e_{k,i}\label{fac.lms.2c} \eq
 In this case,
 each secondary user $k$ will achieve the same performance levels shown
 earlier in (\ref{lkad891.21lk2})--(\ref{lkad891.21lk2.1}) with $R_{u,k}$ replaced by $R_{U,k}$. The performance
 will again be dependent on the local noise
 level, $\sigma_{v,k}^2$. As a result, secondary users with
larger noise power will perform worse than secondary users
with smaller noise power. However, since all secondary users are
observing data arising from the same underlying model $w^o$, it is
natural to expect  cooperation among the users to be
beneficial. As we are going to see, starting from the next section,
one way to achieve cooperation and improve performance is by
developing algorithms that solve the following global cost function
in an adaptive and distributed manner:
 \be
 \min_{w}\;\sum_{k=1}^N \Ex\left\|\d_{k,i}-\U_{k,i}w\right\|^2
 \ee

\section{Distributed Optimization via Diffusion
Strategies}\label{sec.derivation.1}
The examples in the previous section were meant to illustrate how MSE cost functions and linear models are useful
design tools and how they arise frequently in applications. We now return to problem
(\ref{adlq79813.a}) and study the distributed optimization of global
cost functions such as (\ref{ka8912.ad}), where  $J^{\rm glob}(w)$ is
assumed to consist of the sum of individual components. Specifically, we are now
interested in solving optimization problems of the type: \be
\min_{w}\;\sum_{k=1}^N J_{k}(w) \label{opt.11}\ee where each
$J_{k}(w)$
 is assumed to be differentiable and convex over $w$. Although the algorithms presented in this article apply to more
general situations,  we shall nevertheless focus on
mean-square-error cost functions of the form: \be
\addbox{\;J_{k}(w)\define \Ex|\d_{k}(i)-\u_{k,i}w|^2\;}
\label{gihjas.alk}\ee where $w$ is an $M\times 1$ column vector, and the random
processes $\{\d_{k}(i),\u_{k,i}\}$ are assumed to be jointly wide-sense stationary with
zero-mean and second-order moments: \bq
\sigma_{d,k}^2&\define&\Ex|\d_{k}(i)|^2\\
R_{u,{k}}&\define& \Ex\u_{{k},i}^*\u_{{k},i}>0\;\;\;\;\;(M\times M)\label{rdu.1.xxx}\\
r_{du,{k}}&\define&
\Ex\d_{k}(i)\u_{{k},i}^*\;\;\;\;\;\;\;\;\;\;(M\times 1)
\label{rdu.2.xxx}\eq It is clear that each $J_{k}(w)$ is quadratic
in $w$ since, after expansion, we get\be
\addbox{\;J_{k}(w)\;=\;\sigma_{d,k}^2 -w^* r_{du,k} - r_{du,k}^*\; w
+ w^*R_{u,k}\; w\;}\label{a90ks.alk}\ee A completion-of-squares
argument shows that $J_k(w)$ can be expressed as the sum of two
squared terms, i.e., \be J_k(w)\;=\;\left(\sigma_{d,k}^2-r_{du,k}^*
R_{u,k}^{-1} r_{du,k}\right)\;+\;(w-w^o)^*R_{u,k}(w-w^o) \ee or,
more compactly, \be \addbox{\;J_k(w)\;=\;
J_{k,\min}\;+\;\|w-w^o\|_{R_{u,k}}^2\;}
 \label{alternakd.alks}\ee
where $w^o$ denotes the minimizer of $J_k(w)$ and is given by \be
w^o\define R_{u,k}^{-1}\;r_{du,k} \label{kad8912.13}\ee and
$J_{k,\min}$ denotes the minimum value of  $J_k(w)$ when evaluated at $w=w^o$: \be
\addbox{J_{k,\min}\define \sigma_{d,k}^2-r_{du,k}^* R_{u,k}^{-1}
r_{du,k}\;=\;J_k(w^o)}\label{factor.1} \ee Observe that this value
is necessarily non-negative since it can be viewed as the Schur
complement of the following covariance matrix: \be
\Ex\left(\ba{c}\d_k^*(i)\\\u^*_{k,i}\ea
\ba{cc}\d_k(i)&\u_{k,i}\ea\right)\;=\;\ba{cc}\sigma_{d,k}^2&r^*_{du,k}\\r_{du,k}&R_{u,k}\ea
 \ee
and covariance matrices are nonnegative-definite.

The choice of the quadratic form (\ref{gihjas.alk}) or (\ref{a90ks.alk}) for $J_{k}(w)$ is
useful for many applications, as was already illustrated in the
previous section for examples involving AR modeling, MA
modeling, localization, and spectral sensing. Other choices for  $J_{k}(w)$ are of course
possible and these choices can
even be different for different nodes. It is sufficient in this
article to illustrate the main concepts underlying diffusion
adaptation by focusing on the useful case of MSE cost functions
 of the form (\ref{a90ks.alk}); still, most of the derivations and arguments in the
coming sections can be extended beyond MSE optimization to more
general cost functions --- as already shown in
\cite{Chen10,Chen10b,ChenSSP2012}; see also Sec.~\ref{sec.diskd8192}.

The positive-definiteness of the covariance matrices $\{R_{u,k}\}$
ensures that each $J_{k}(w)$ in (\ref{a90ks.alk}) is strictly
convex, as well as $J^{\rm glob}(w)$ from (\ref{ka8912.ad}). Moreover, all these cost
functions have a unique minimum at the same $w^o$, which satisfies
the normal equations: \be R_{u,k}\;w^o =
r_{du,k},\;\;\;\;\;\;\mbox{\rm for every}\;k=1,2,\ldots,N
\label{usilka}\ee Therefore, given knowledge of
$\{r_{du,k},R_{u,k}\}$, each node can determine $w^o$
on its own by solving (\ref{usilka}). One then wonders about the need
to seek distributed cooperative and adaptive solutions. There are a couple of reasons:
\begin{enumerate}
\item[(a)] First, even for MSE cost functions, it is often the case that the required
moments $\{r_{du,k}, R_{u,k}\}$ are not known beforehand. In this
case, the optimal $w^o$ cannot be determined from the solution of
the normal equations (\ref{usilka}). The alternative
methods that we shall describe will lead to adaptive techniques that enable
each node $k$ to estimate $w^o$ directly from data realizations.

\item[(b)] Second, since adaptive strategies rely on instantaneous
data, these strategies
possess powerful tracking abilities. Even when the moments vary with
time due to non-stationary behavior (such as $w^o$ changing with
time), these changes will be reflected in the observed data and
will in turn influence the behavior of the adaptive construction.
This is one of the key advantages of adaptive strategies:
they enable learning and tracking in real-time.

\item[(c)] Third, cooperation among nodes is generally beneficial.
When nodes act individually, their performance is limited by
the noise power level at their location. In this way, some nodes can
perform significantly better than other nodes. On the other hand,
when nodes cooperate with their neighbors and share information during the adaptation process, we will see that performance can
be improved across the network.
\end{enumerate}

\subsection{Relating the Global Cost to Neighborhood Costs}
Let us therefore consider the optimization of the
following global cost function: \be \addbox{\;J^{\rm
glob}(w)\;=\;\sum_{k=1}^{N} J_{k}(w)\;} \label{98d.12lk}\ee where
$J_{k}(w)$ is given by (\ref{gihjas.alk}) or (\ref{a90ks.alk}).
Our strategy to optimize $J^{\mathrm{glob}}(w)$ in a distributed
manner is based on two steps, following the developments in
\cite{Cattivelli10,Chen10,Chen10b}. First, using a
completion-of-squares argument (or, equivalently, a second-order Taylor series expansion), we
approximate the global cost
function (\ref{98d.12lk}) by an alternative local cost that is amenable to
distributed optimization. Then, each node will optimize the
alternative cost via a steepest-descent method.

\begin{figure}[h]
\epsfxsize 9cm \epsfclipon
\begin{center}
\leavevmode \epsffile{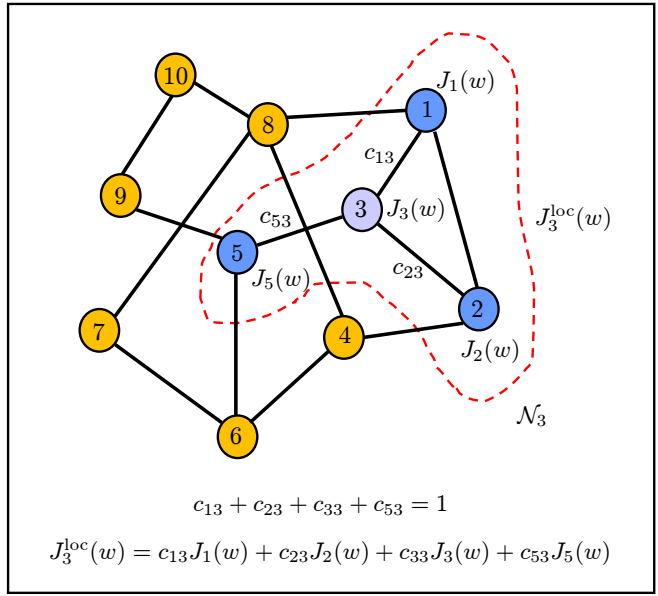} \caption{{\small A network with
$N=10$ nodes. The nodes in the neighborhood of node $3$ are
highlighted with their individual cost functions, and with the
combination weights $\{c_{13},c_{23},c_{53}\}$ along the connecting
edges; there is also a combination weight associated with node $3$
and is denoted by $c_{33}$. The expression for the local cost
function, $J_{3}^{\rm loc}(w)$, is also shown in the
figure.}}\label{fig-H.label}
\end{center}
\end{figure}

To motivate the distributed diffusion-based approach, we start by introducing a set of
nonnegative coefficients $\{c_{k\ell}\}$ that satisfy two
conditions: \bq\mbox{\rm for}\;k=1,2,\ldots,N:&&\nn\\
        c_{k\ell}\geq0,\;\;\;\sum_{\ell=1}^N c_{k\ell} = 1,\;\;\;\mbox{\rm and}\;\;\;  c_{k\ell}=0~\mathrm{if}~\ell \notin
        \mathcal{N}_{k}
\label{Equ:ProblemFormulation:C_Condition} \eq where
$\mathcal{N}_{k}$ denotes the neighborhood of node $k$. Condition
(\ref{Equ:ProblemFormulation:C_Condition}) means that for every node
$k$, the sum of the coefficients $\{c_{k\ell}\}$ that relate it to
its neighbors is one. The coefficients $\{c_{k \ell}\}$ are free
parameters that are chosen by the designer; obviously, as shown
later in Theorem~\ref{lemma.mse}, their selection will have a bearing on the performance of the
resulting algorithms. If we collect the entries $\{c_{k\ell }\}$
into an $N\times N$ matrix $C$, so that the $k-$th row of $C$ is formed of
$\{c_{k\ell},\;\ell=1,2,\ldots,N\}$, then condition
(\ref{Equ:ProblemFormulation:C_Condition}) translates into saying
that each of {\em row} of $C$ adds up to one, i.e., \be
\addbox{\;C\mathds{1}=\mathds{1}\;}\label{columns.1}\ee where the
notation $\mathds{1}$  denotes an $N\times 1$ column vector with all its entries
equal to one: \be \mathds{1}\define \mbox{\rm col}\{1,1,\ldots,1\}
\ee We say that $C$ is a right stochastic matrix. Using the
coefficients $\{c_{k\ell}\}$ so defined, we associate with each node
$\ell$, a local cost function of the following form:
    \be
        \label{Equ:ProblemFormulation:J_k_loc}
        J_{\ell}^{\mathrm{loc}}(w)   \define   \sum_{k \in \mathcal{N}_{\ell}} c_{k\ell} \;J_{k}(w)
\ee This cost consists of a weighted combination of the individual
costs of the neighbors of node $\ell$ (including $\ell$ itself) ---
see Fig.~\ref{fig-H.label}. Since the $\{c_{k\ell}\}$ are all
nonnegative and each $J_{k}(w)$ is strictly convex, then
$J_{\ell}^{\mathrm{loc}}(w)$ is also strictly convex and its
minimizer occurs at the same $w=w^o$. Using the alternative
representation (\ref{alternakd.alks}) for the individual $J_k(w)$, we can
re-express the local cost $J_{\ell}^{\rm loc}(w)$ as \bq
J_{\ell}^{\rm loc}(w)&=&  \sum_{k \in \mathcal{N}_{\ell}} c_{k\ell}
\; J_{k,\min}\;+\;
 \sum_{k \in \mathcal{N}_{\ell}}
c_{k\ell} \;\|w-w^o\|_{R_{u,k}}^2\;\eq or, equivalently, \be
\addbox{J_{\ell}^{\rm loc}(w)\;=\;  J_{\ell,\min}^{\rm loc}\;+\;
\;\|w-w^o\|_{R_{\ell}}^2\;} \label{k8912.12lk}\ee where
$J_{\ell,\min}^{\rm loc}$ corresponds to the minimum value
of $J_{\ell}^{\rm loc}(w)$ at
the minimizer $w=w^o$: \be J_{\ell,\min}^{\rm loc}\define \sum_{k \in
\mathcal{N}_{\ell}} c_{k\ell} \; J_{k,\min} \ee
and $R_{\ell}$ is a positive-definite weighting matrix defined by: \be
R_{\ell}\define \sum_{k\in{\cal N}_{\ell}} c_{k\ell} R_{u,k}
\ee That is,  $R_{\ell}$ is a weighted combination of the covariance
matrices in the neighborhood of node $\ell$.
 Equality (\ref{k8912.12lk}) amounts to a (second-order) Taylor
series expansion of $J_{\ell}^{\rm loc}(w)$ around $w=w^o$. Note
that the right-hand side consists of two terms: the minimum cost and
a weighted quadratic term in the difference $(w-w^o)$.

Now note that we can express $J^{\mathrm{glob}}(w)$ from
\eqref{98d.12lk} as follows: \bq
        J^{\mathrm{glob}}(w)  &\stackrel{(\ref{Equ:ProblemFormulation:C_Condition})}{=}& \sum_{k=1}^{N}\left(\sum_{\ell=1}^N c_{k \ell
        }\right) J_{k}(w)\nn\\
&=&\sum_{\ell=1}^N\left(\sum_{k=1}^{N}
        c_{k \ell}\; J_{k}(w)\right)\nn\\
&\stackrel{(\ref{Equ:ProblemFormulation:J_k_loc})}{=}&\sum_{\ell=1}^N J_{\ell}^{\rm loc}(w)\nn\\
&=&J_{k}^{\mathrm{loc}}(w) \;+\; \sum_{\ell \neq k}^N
J_{\ell}^{\mathrm{loc}}(w) \label{Equ:ProblemFormulation:J_glob}\eq
Substituting (\ref{k8912.12lk}) into the second term on the
right-hand side of the above expression gives: \be J^{\rm
glob}(w)\;=\; J_k^{\rm loc}(w) \;+\;\sum_{\ell \neq
k}\;\|w-w^o\|_{R_{\ell}}^2 \;+\; \sum_{\ell \neq k}
J_{\ell,\min}^{\rm loc} \ee The last term in the above expression
does not depend on $w$. Therefore,
minimizing $J^{\rm glob}(w)$ over $w$ is equivalent to minimizing
the following alternative global cost: \be
\label{Equ:ProblemFormulation:J_glob_prime}
   \addbox{     J^{\mathrm{glob}'}(w)\;=\;
                            J_k^{\rm loc}(w)
                            \;+\;
                            \sum_{\ell \neq k}
                            \|w-w^{o}\|_{R_{\ell}}^2}
\ee
 Expression
\eqref{Equ:ProblemFormulation:J_glob_prime} relates the optimization
of the original global cost function, $J^{\rm glob}(w)$ or its
equivalent $J^{\rm glob'}(w)$, to the newly-introduced local
cost function $J_{k}^{\mathrm{loc}}(w)$. The relation is through the
second term on the right-hand side of
\eqref{Equ:ProblemFormulation:J_glob_prime}, which corresponds to a
sum of quadratic factors involving the minimizer $w^o$; this term tells us how the
local cost $J_{k}^{\rm loc}(w)$ can be corrected to the global cost $J^{\rm glob'}(w)$. Obviously, the
minimizer $w^o$ that appears in the correction term is not known since the nodes wish to determine its value. Likewise, not all the weighting matrices
$R_{\ell}$ are available to node $k$; only those matrices that
originate from its neighbors can be assumed to be available. Still,
expression \eqref{Equ:ProblemFormulation:J_glob_prime} suggests a
useful way to replace $J_{k}^{\rm loc}$ by another local cost that is closer to
$J^{\rm glob'}(w)$. This alternative cost  will be shown to lead to
a powerful distributed solution to optimize $J^{\rm
glob}(w)$ through localized interactions.

 Our first step is to limit the summation on the right-hand side of
 \eqref{Equ:ProblemFormulation:J_glob_prime} to the neighbors of node $k$ (since every
 node $k$ can only have access to information from its neighbors).
 We thus introduce the modified cost function at node $k$:
\be        \label{Equ:DiffusionAdaptation:J_glob_prime_local}
        J_k^{\mathrm{glob}'}(w) \define  \displaystyle J_k^{\mathrm{loc}}(w)
                            +
                            \sum_{\ell\in{\mathcal{N}}_k\backslash\{k\}} \|w-w^o\|^2_{R_{\ell}}
   \ee
The cost functions $J_{k}^{\rm loc}(w)$ and
$J_k^{\mathrm{glob}'}(w)$ are both associated with node $k$; the
difference between them is that the expression for the latter is
closer  to the
 global cost function \eqref{Equ:ProblemFormulation:J_glob_prime}
 that we want to optimize.

The weighting matrices $\{R_{\ell}\}$ that appear in \eqref{Equ:DiffusionAdaptation:J_glob_prime_local} may or may not be available because the second-order moments $\{R_{u,\ell}\}$ may or may not be known beforehand. If these moments are known, then we can proceed with the analysis by assuming knowledge of the $\{R_{\ell}\}$. However, the more interesting case is when these moments are not known. This is generally the case in practice, especially in the context of adaptive solutions and problems involving non-stationary data. Often, nodes can only observe realizations $\{u_{\ell,i}\}$ of the regression data $\{\u_{\ell,i}\}$ arising from distributions whose covariance matrices are the unknown $\{R_{u,\ell}\}$. One way to address the difficulty is to replace each of the weighted norms $\|w-w^o\|^2_{R_{\ell}}$ in \eqref{Equ:DiffusionAdaptation:J_glob_prime_local} by a scaled multiple of the un-weighted norm, say,
\be
\|w-w^o\|^2_{R_{\ell}}\;\approx\;  b_{\ell k}\cdot\|w-w^o\|^2   \label{replaeXX.alkda}
\ee
where $b_{\ell k}$ is some nonnegative coefficient; we are even allowing its value to change with the node index $k$. The above substitution amounts to having each node $k$ approximate the $\{R_{\ell}\}$ from its neighbors by multiples of the identity matrix
\be R_{\ell} \approx b_{\ell k}\; I_M\label{kd8901.alkd}\ee
Approximation (\ref{replaeXX.alkda}) is reasonable in view of the fact that all vector norms are equivalent \cite{golub,horn,kre}; this norm property ensures that we can bound the weighted norm $\|w-w^o\|^2_{R_{\ell}}$ by some constants multiplying the un-weighted norm $\|w-w^o\|^2,$ say, as:
\be
r_1\|w-w^o\|^2 \;\leq \;\|w-w^o\|^2_{R_{\ell}} \;\leq \;r_2 \|w-w^o\|^2
\ee
for some positive constants $(r_1,r_2)$. Using the fact that the $\{R_{\ell}\}$ are Hermitian positive-definite matrices, and calling upon the Rayleigh-Ritz characterization of eigenvalues \cite{golub,horn}, we can be more specific and replace the above inequalities by
\be
\lambda_{\min}(R_{\ell})\cdot\|w-w^o\|^2 \;\leq \;\|w-w^o\|^2_{R_{\ell}} \;\leq \; \lambda_{\max}(R_{\ell}) \cdot \|w-w^o\|^2
\ee
We note that approximations similar to (\ref{kd8901.alkd}) are common in stochastic approximation theory and they mark the difference between using a Newton's iterative method or a stochastic gradient method
\cite{Sayed08,Poljak87}; the former uses Hessian matrices as approximations
for $R_{\ell}$ and the latter uses multiples of the identity matrix. Furthermore, as the derivation will reveal, we do not need
to worry at this stage about how to select the scalars $\{b_{\ell k}\}$; they
will end up being embedded into another set of coefficients $\{a_{\ell k}\}$ that will be set
by the designer or adjusted by the algorithm --- see (\ref{Equ:AdaptiveDiffusion:Condition_a}) further ahead.

Thus, we replace \eqref{Equ:DiffusionAdaptation:J_glob_prime_local}
by
    \begin{align}
        \label{Equ:DiffusionAdaptation:J_glob_prime_prime}
        J_k^{\mathrm{glob}''}(w)    =    \displaystyle J_k^{\mathrm{loc}}(w)
                                \;+\;
                                \sum_{\ell \in{\mathcal N}_k\backslash\{k\}} b_{\ell k}\; \|w-w^o\|^2
    \end{align}
The argument so far has suggested how to modify $J_k^{\mathrm{loc}}(w)$ from (\ref{Equ:ProblemFormulation:J_k_loc})
and replace it by the cost
\eqref{Equ:DiffusionAdaptation:J_glob_prime_prime} that is closer in
form to the global cost function
\eqref{Equ:ProblemFormulation:J_glob_prime}. If we replace
$J_k^\mathrm{loc}(w)$ by its definition
\eqref{Equ:ProblemFormulation:J_k_loc}, we can rewrite
\eqref{Equ:DiffusionAdaptation:J_glob_prime_prime} as
 \be
    \label{Equ:DiffusionAdaptation:J_glob_prime_prime_final}
   \addbox{\;     J_k^{\mathrm{glob}''}(w) \;=\;    \displaystyle
                                \sum_{\ell \in{\cal N}_k} \! c_{\ell
                                k}\;J_{\ell}(w)\;+\;
                                \sum_{\ell \in{\cal N}_k\backslash\{k\}} b_{\ell k}\;\|w-w^o\|^2
                                \!\;}
\ee With the exception of the variable $w^o$, this approximate cost
at node $k$ relies solely on information that is available to node $k$ from its neighborhood. We
will soon explain how to handle the fact that $w^o$ is not known
beforehand to node $k$.

\subsection{Steepest-Descent Iterations}\label{sec.steepest.sec}
Node $k$ can apply a steepest-descent iteration to minimize
$J_k^{\mathrm{glob}''}(w)$. Let $w_{k,i}$ denote the estimate for
the minimizer $w^o$ that is evaluated by node $k$ at time $i$. Starting from an initial condition $w_{k,-1}$, node
$k$ can compute successive estimates iteratively as follows: \be
w_{k,i} \;=\;   w_{k,i-1} - \mu_k\;\left[\nabla_{w}
J_k^{\mathrm{glob}''}(w_{k,i-1})\right]^*,\;\;\;\;\;i\geq 0\label{eq119.11}\ee where $\mu_k$ is a
small positive step-size parameter, and the notation $\nabla_w
J(a)$ denotes the gradient vector of the function $J(w)$ relative to
 $w$ and evaluated at $w=a$.  The step-size parameter $\mu_k$ can be
selected to vary with time as well. One choice that is common in the optimization literature
\cite{Poljak87,Ber97,Sayed08} is to replace $\mu_k$ in (\ref{eq119.11}) by step-size sequences $\{\mu(i)\geq 0\}$ that satisfy the two conditions (\ref{kdl8912.alkd}). However, such step-size sequences are not suitable for applications that require continuous learning because they turn off adaptation as $i\rightarrow\infty$; the steepest-descent iteration (\ref{eq119.11}) would stop updating since $\mu_k(i)$ would be tending towards zero.  For this reason, we shall focus mainly on the constant step-size case described by (\ref{eq119.11}) since we are interested in developing distributed algorithms that will endow networks with continuous adaptation abilities.

 Returning to (\ref{eq119.11}) and computing the gradient vector of
(\ref{Equ:DiffusionAdaptation:J_glob_prime_prime_final}) we get: \be
w_{k,i} =   w_{k,i-1} - \mu_k\; \sum_{\ell \in \mathcal{N}_k}
c_{\ell k}\;\left[ \nabla_w J_{\ell}(w_{k,i-1})\right]^*
                   \;-\; \mu_k \sum_{\ell \in \mathcal{N}_k\backslash \{k\}} b_{\ell k}\;(w_{k,i-1}
                                -w^{o})\ee
Using the expression for $J_{\ell}(w)$ from (\ref{a90ks.alk}) we
arrive at \be w_{k,i} \;=\;   w_{k,i-1} + \mu_k\;
\sum_{\ell \in \mathcal{N}_k} c_{\ell k}\;(r_{du,\ell}-R_{u,\ell }\;
w_{k,i-1})
                   \;+\; \mu_k \sum_{\ell \in \mathcal{N}_k\backslash \{k\}} b_{\ell
                   k}\;(w^o-w_{k,i-1})
                                        \label{Equ:DiffusionAdaptation:GradientDescent}\ee
This iteration indicates that the update from $w_{k,i-1}$ to
$w_{k,i}$ involves adding two correction terms to $w_{k,i-1}$. Among
many other forms, we can implement the update in two successive
steps by adding one correction term at a time, say, as follows: \bq
\label{Equ:DiffusionAdaptation:Adaptation_intermedieate}
        \psi_{k,i}  &=&   \displaystyle w_{k,i-1} \;+\; \mu_k\;
\sum_{\ell \in \mathcal{N}_k} c_{\ell k}\;(r_{du,\ell}-R_{u,\ell}
\;w_{k,i-1})
\\\nn\\
        \label{Equ:DiffusionAdaptation:Combination_intermediate}
        w_{k,i} &=&   \displaystyle \psi_{k,i}  \;+\; \mu_k \sum_{\ell \in \mathcal{N}_k\backslash \{k\}} b_{\ell
                   k}\;(w^o-w_{k,i-1})
\eq Step \eqref{Equ:DiffusionAdaptation:Adaptation_intermedieate}
updates $w_{k,i-1}$ to an intermediate value $\psi_{k,i}$ by using
local gradient vectors from the neighborhood of node $k$. Step
\eqref{Equ:DiffusionAdaptation:Combination_intermediate} further
updates $\psi_{k,i}$ to $w_{k,i}$. However, this second step is not
realizable since $w^o$ is not known and the nodes are actually
trying to estimate it. Two issues stand
out from examining
\eqref{Equ:DiffusionAdaptation:Combination_intermediate}:
    \begin{enumerate}
        \item[(a)]
        First, iteration \eqref{Equ:DiffusionAdaptation:Combination_intermediate}
        requires knowledge of the minimizer $w^o$.
        Neither node $k$ nor its neighbors know the value of the minimizer; each of these nodes is actually
        performing steps similar to \eqref{Equ:DiffusionAdaptation:Adaptation_intermedieate}
        and \eqref{Equ:DiffusionAdaptation:Combination_intermediate} to estimate the minimizer.
        However,  each node $\ell$ has a readily available
        approximation for $w^o$, which is its local intermediate estimate $\psi_{\ell,i}$.
        Therefore,
        we replace $w^{o}$ in \eqref{Equ:DiffusionAdaptation:Combination_intermediate} by
        $\psi_{\ell,i}$.  This step helps diffuse information throughout the
        network. This is because each neighbor of node $k$
        determines its estimate $\psi_{\ell,i}$ by processing
        information from its own neighbors, which process information from their neighbors, and so forth.

        \item[(b)]
        Second, the intermediate value $\psi_{k,i}$ at node $k$ is generally a better estimate for $w^o$
        than $w_{k,i-1}$ since it is obtained by incorporating information from the neighbors
        through the first step \eqref{Equ:DiffusionAdaptation:Adaptation_intermedieate}.
        Therefore, we further replace $w_{k,i-1}$ in \eqref{Equ:DiffusionAdaptation:Combination_intermediate}
        by $\psi_{k,i}$. This step is reminiscent of
        incremental-type approaches to optimization, which have been widely
        studied in the literature
        \cite{Bertsekas97,Nedic01,Rabbat05,Lopes07}.\\
    \end{enumerate}

\noindent With the substitutions described in items (a) and (b) above, we
replace the second step
\eqref{Equ:DiffusionAdaptation:Combination_intermediate} by
 \bq
        w_{k,i} &=&  \psi_{k,i}  \;+\; \mu_k \sum_{\ell \in \mathcal{N}_k\backslash \{k\}} b_{\ell
                  k}\;(\psi_{\ell,i}-\psi_{k,i})\nn\\
 &=&   \left(1-\mu_k \sum_{\ell \in \mathcal{N}_k\backslash \{k\}} b_{\ell
                  k}\right)\psi_{k,i}  \;+\; \mu_k \sum_{\ell \in \mathcal{N}_k\backslash \{k\}} b_{\ell
                  k}\;\psi_{\ell,i}
        \label{Equ:DiffusionAdaptation:Combination_intermediate.1}\eq
Introduce the weighting coefficients: \bq a_{kk}&\define&1-\mu_k
\sum_{\ell \in \mathcal{N}_k\backslash \{k\}} b_{\ell
                  k}\\
                  a_{\ell k}&\define& \mu_{k} b_{\ell
                  k},\;\;\;\ell \in \mathcal{N}_k\backslash
                  \{k\}\\
                  a_{\ell k}&\define&0,\;\;\;\;\;\;\;\;\;\ell\notin {\cal N}_k
\eq and observe that, for sufficiently small step-sizes $\mu_k$,
these coefficients are nonnegative and, moreover, they satisfy the
conditions:  \bq\mbox{\rm for}\;k=1,2,\ldots,N:&&\nn\\
        a_{\ell k}\geq0,\;\;\;\;\sum_{\ell=1}^N a_{\ell k} = 1,\;\;\;\mbox{\rm and}\;\;\;  a_{\ell k}=0~\mathrm{if}~\ell \notin
        \mathcal{N}_{k}
 \label{Equ:AdaptiveDiffusion:Condition_a} \eq  Condition
(\ref{Equ:AdaptiveDiffusion:Condition_a}) means that for every node
$k$, the sum of the coefficients $\{a_{\ell k}\}$ that relate it to
its neighbors is one. Just like the $\{c_{\ell k}\}$, from now on, we will
treat the coefficients $\{a_{\ell k}\}$ as free weighting parameters
that are chosen by the designer according to (\ref{Equ:AdaptiveDiffusion:Condition_a});  their selection will also have a
bearing on the performance of the resulting algorithms --- see  Theorem~\ref{lemma.mse}. If we
collect the entries $\{a_{\ell k }\}$ into an $N\times N$ matrix
$A$, such that the $k-$th column of $A$ consists of $\{a_{\ell k},\;\ell=1,2,\ldots,N\}$,
then condition (\ref{Equ:AdaptiveDiffusion:Condition_a})
translates into saying that each {\em column} of $A$ adds up to one: \be
\addbox{\;A^T\mathds{1}=\mathds{1}\;}\label{rows.1}\ee We say that
$A$ is a left stochastic matrix.

\subsection{Adapt-then-Combine (ATC) Diffusion Strategy}
\begin{figure}[h]
\epsfxsize 9cm \epsfclipon
\begin{center}
\leavevmode \epsffile{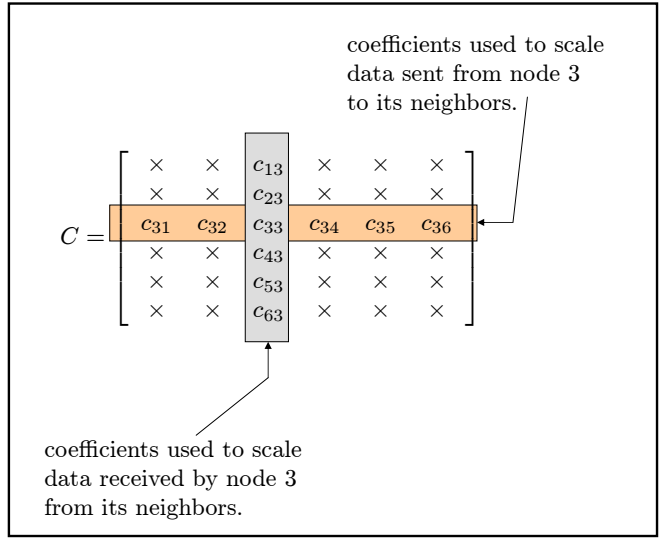} \caption{{\small Interpretation of the columns and rows of
combination matrices. The pair of entries $\{c_{k\ell},c_{\ell k}\}$ correspond to
weighting coefficients used over the edge
connecting nodes $k$ and $\ell$. When nodes $(k,\ell)$ are not neighbors, then
these weights are zero.}}\label{fig-L.label}
\end{center}
\end{figure}

 Using the
coefficients $\{a_{\ell k}\}$ so defined, we replace
\eqref{Equ:DiffusionAdaptation:Adaptation_intermedieate} and \eqref{Equ:DiffusionAdaptation:Combination_intermediate.1}
by the following recursions for $i\geq 0$:    \be
        \mbox{\rm (ATC strategy)}\;\;\;
        \addbox{
                \label{Equ:DiffusionAdaptation:ATC}
                \begin{array}{l}
                    \psi_{k,i}  =   \displaystyle w_{k,i-1} + \mu_k\; \sum_{\ell \in \mathcal{N}_k}
c_{\ell k}\;(r_{du,\ell}-R_{u,\ell}\; w_{k,i-1})\\
                    w_{k,i} =   \displaystyle \sum_{\ell \in \mathcal{N}_k} a_{\ell k}\; \psi_{\ell,i}
                \end{array}
            }
       \ee
for some nonnegative coefficients $\{c_{\ell k},a_{\ell k}\}$ that satisfy
conditions (\ref{columns.1}) and (\ref{rows.1}), namely,
\be
\addbox{\;C\mathds{1}=\mathds{1},\;\;\;\;\;A^T\mathds{1}=\mathds{1}\;}
\label{satisyud}\ee
or, equivalently,
\be
\begin{array}{r}
\mbox{\rm for}\;k=1,2,\ldots,N:\\
c_{\ell k}\geq 0,\;\;\;\;\displaystyle\sum_{k=1}^N c_{\ell k} = 1,\;\;\;\;  c_{\ell k}=0~\mathrm{if}~\ell \notin \mathcal{N}_{k}\\
a_{\ell k}\geq 0,\;\;\;\;\displaystyle\sum_{\ell=1}^N a_{\ell k} = 1,\;\;\;\;  a_{\ell k}=0~\mathrm{if}~\ell \notin \mathcal{N}_{k}
\end{array}\label{run.run}
\ee
 To run algorithm
\eqref{Equ:DiffusionAdaptation:ATC}, we only need to select the
coefficients $\{a_{\ell k},c_{\ell k}\}$ that satisfy (\ref{satisyud}) or (\ref{run.run});
there is no need to worry
about the intermediate coefficients $\{b_{\ell k}\}$ any longer since they
have been blended into the $\{a_{\ell k}\}$. The scalars $\{c_{\ell k}, a_{\ell k}\}$ that
appear in (\ref{Equ:DiffusionAdaptation:ATC}) correspond to weighting coefficients over the edge
linking node $k$ to its neighbors $\ell\in{\cal N}_k$. Note that two sets of coefficients are
used to scale the data that are being received by node $k$: one set of coefficients, $\{c_{\ell k}\},$ is used in the
first step of (\ref{Equ:DiffusionAdaptation:ATC}) to scale the moment data $\{r_{du,\ell},R_{u,\ell}\}$, and a
second set of coefficients, $\{a_{\ell k}\},$ is used in the second step of (\ref{Equ:DiffusionAdaptation:ATC}) to scale the
estimates $\{\psi_{\ell,i}\}$. Figure~\ref{fig-L.label} explains what the entries on the columns and rows
of the combination matrices $\{A,C\}$ stand for using an example with $N=6$ and the matrix $C$ for illustration. When the combination
matrix is right-stochastic (as is the case with $C$), each of its rows would add up to one. On the other hand, when
the matrix is left-stochastic (as is the case with $A$), each of its columns would add up to one.

At every time instant $i$, the ATC strategy
\eqref{Equ:DiffusionAdaptation:ATC} performs two steps. The first
step is an {\em information exchange} step where node $k$ receives from
its neighbors their moments $\{R_{u,\ell},r_{du,\ell}\}$. Node
$k$ combines this information and uses it to update its existing
estimate $w_{k,i-1}$ to an intermediate value $\psi_{k,i}$. All
other nodes in the network are performing a similar step and
updating their existing estimates $\{w_{\ell,i-1}\}$ into intermediate estimates $\{\psi_{\ell,i}\}$ by
using information from their neighbors. The second step in
\eqref{Equ:DiffusionAdaptation:ATC} is an {\em aggregation} or consultation step where
node $k$ combines the intermediate estimates of its neighbors to
obtain its update estimate $w_{k,i}$. Again, all other nodes in the
network are simultaneously performing a similar step. The reason for the name Adapt-then-Combine (ATC) strategy
is that the first step in \eqref{Equ:DiffusionAdaptation:ATC} will be shown to lead to an adaptive step, while the
second step in \eqref{Equ:DiffusionAdaptation:ATC} corresponds to a combination step. Hence, strategy \eqref{Equ:DiffusionAdaptation:ATC}
involves adaptation followed by combination or ATC for short. The reason for the qualification ``diffusion'' is that the combination
step in \eqref{Equ:DiffusionAdaptation:ATC} allows information to diffuse through the network in real time. This is because
each of the estimates $\psi_{\ell,i}$ is influenced by data beyond the immediate neighborhood of node $k$.

In the special case when $C=I$,
so that no information exchange is performed but only the aggregation step, the ATC
strategy \eqref{Equ:DiffusionAdaptation:ATC} reduces to:
 \be
 \begin{array}{l}\mbox{\rm (ATC strategy without}\\\mbox{\rm information exchange)}\end{array}\;\;\;
\addbox{\;                \begin{array}{l}
                    \psi_{k,i}  =   \displaystyle w_{k,i-1} + \mu_k\; (r_{du,k}-R_{u,k}\;
                    w_{k,i-1})\\\\
                    w_{k,i} =   \displaystyle \sum_{\ell \in \mathcal{N}_k} a_{\ell k}\; \psi_{\ell,i}
                \end{array}\;}
\label{without.atc}       \ee
where the first step relies solely on the information
$\{R_{u,k},r_{du,k}\}$ that is available locally at node $k$.

\begin{figure}[h]
\epsfxsize 9cm \epsfclipon
\begin{center}
\leavevmode \epsffile{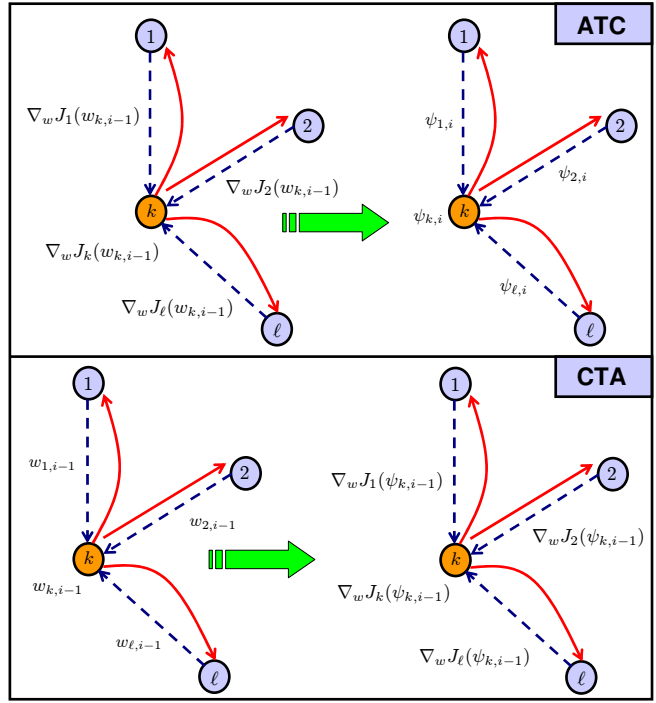} \caption{{\small Illustration
of the ATC and CTA strategies for a node $k$ with three other
neighbors $\{1,2,\ell\}$. The updates involve two steps: information
exchange followed by aggregation in ATC and aggregation followed by
information exchange in CTA. The dotted blue arrows represent the
data received from the neighbors of node $k$, and the solid red arrows
represent the data sent from node $k$ to its
neighbors.}}\label{fig-33.label}
\end{center}
\end{figure}

Observe in passing that the term that appears in the information exchange step of
\eqref{Equ:DiffusionAdaptation:ATC} is related to the gradient
vectors of the local costs $\{J_{\ell}(w)\}$ evaluated at $w_{k,i-1}$, i.e., it holds that
\be
r_{du,\ell}-R_{u,\ell}\; w_{k,i-1}\;=\;-\left[\nabla_{w}J_{\ell} (w_{k,i-1})\right]^*
\ee
so that the ATC strategy \eqref{Equ:DiffusionAdaptation:ATC} can also be written in the
following equivalent form:

\be \mbox{\rm (ATC strategy)}\;\;\;
\addbox{\;
                \label{Equ:DiffusionAdaptation:ATC.2}
                \begin{array}{l}
                    \psi_{k,i}  =   \displaystyle w_{k,i-1} - \mu_k\; \sum_{\ell \in \mathcal{N}_k}
c_{\ell k}\;\left[\nabla_{w}J_{\ell} (w_{k,i-1})\right]^*\\
                    w_{k,i} =   \displaystyle \sum_{\ell \in \mathcal{N}_k} a_{\ell k}\; \psi_{\ell,i}
                \end{array}\;}
       \ee
The significance of this general form is that it is applicable to
optimization problems involving more general local costs
$J_{\ell}(w)$ that are not necessarily quadratic in $w$, as detailed in
\cite{Chen10,Chen10b,ChenSSP2012} --- see also Sec.~\ref{sec.diskd8192}. The top part of Fig.~\ref{fig-33.label}
illustrates the two steps involved in the ATC procedure for a
situation where node $k$ has three other neighbors labeled
$\{1,2,\ell\}$. In the first step, node $k$ evaluates the gradient
vectors of its neighbors at $w_{k,i-1}$, and subsequently aggregates
the estimates $\{\psi_{1,i},\psi_{2,i},\psi_{\ell,i}\}$ from its
neighbors. The dotted arrows represent flow of information towards node $k$ from its neighbors. The solid arrows
represent flow of information from node $k$ to its neighbors. The CTA diffusion strategy is discussed next.

\subsection{Combine-then-Adapt (CTA) Diffusion Strategy}
 Similarly, if we return to (\ref{Equ:DiffusionAdaptation:GradientDescent}) and add the second correction term first, then
 \eqref{Equ:DiffusionAdaptation:Adaptation_intermedieate}--\eqref{Equ:DiffusionAdaptation:Combination_intermediate} are replaced by:
 \bq
\label{Equ:DiffusionAdaptation:Adaptation_intermedieate.r}
\psi_{k,i-1} &=&   \displaystyle w_{k,i-1}  \;+\; \mu_k \sum_{\ell
\in \mathcal{N}_k\backslash \{k\}} b_{\ell
k}\;(w^o-w_{k,i-1})\\\nn\\
w_{k,i}  &=&   \displaystyle \psi_{k,i-1} + \mu_k\; \sum_{\ell \in
\mathcal{N}_k} c_{\ell k}\;(r_{du,\ell}-R_{u,\ell}\; w_{k,i-1})
        \label{Equ:DiffusionAdaptation:Combination_intermediate.r}
\eq Following similar reasoning to what we did before in the ATC
case, we replace $w^o$ in step
\eqref{Equ:DiffusionAdaptation:Adaptation_intermedieate.r} by
$w_{\ell,i-1}$ and replace $w_{k,i-1}$ in
\eqref{Equ:DiffusionAdaptation:Combination_intermediate.r} by
$\psi_{k,i-1}$. We then introduce the same coefficients $\{a_{\ell
k}\}$ and arrive at the following combine-then-adapt (CTA) strategy:
\be
   \mbox{\rm (CTA strategy)}\;\;\;\;\;
        \addbox{
                \label{Equ:DiffusionAdaptation:CTA}
                \begin{array}{l}
                    \psi_{k,i-1}    =   \displaystyle \sum_{\ell \in \mathcal{N}_k} a_{\ell k} \;w_{\ell,i-1}
                    \\
                    w_{k,i} =    \displaystyle\psi_{k,i-1}
                                + \mu_k \sum_{\ell \in \mathcal{N}_k} c_{\ell
                                k}\left(r_{du,\ell}-R_{u,\ell}\;\psi_{k,i-1}\right)
                \end{array}
            }
   \ee
where the nonnegative coefficients $\{c_{\ell k},a_{\ell k}\}$  satisfy the same
conditions (\ref{columns.1}) and (\ref{rows.1}), namely,
\be
\addbox{\;C\mathds{1}=\mathds{1},\;\;\;\;\;A^T\mathds{1}=\mathds{1}\;}
\label{satisyud.3}\ee
or, equivalently,
\be
\begin{array}{r}
\mbox{\rm for}\;k=1,2,\ldots,N:\\
c_{\ell k}\geq 0,\;\;\;\;\displaystyle\sum_{k=1}^N c_{\ell k} = 1,\;\;\;\;  c_{\ell k}=0~\mathrm{if}~\ell \notin \mathcal{N}_{k}\\
a_{\ell k}\geq 0,\;\;\;\;\displaystyle\sum_{\ell=1}^N a_{\ell k} = 1,\;\;\;\;  a_{\ell k}=0~\mathrm{if}~\ell \notin \mathcal{N}_{k}
\end{array}\label{run.run.2}
\ee
At every time instant $i$, the CTA strategy
\eqref{Equ:DiffusionAdaptation:CTA} also consists of two steps. The
first step is an aggregation step where node $k$ combines the
existing estimates of its neighbors to obtain the intermediate
estimate $\psi_{k,i-1}$. All other nodes in the network are
simultaneously performing a similar step and aggregating the
estimates of their neighbors. The second step in
\eqref{Equ:DiffusionAdaptation:CTA} is an information exchange step
where node $k$ receives from its neighbors their moments
$\{R_{du,\ell},r_{du,\ell}\}$ and uses this information to update
its intermediate estimate to $w_{k,i}$.  Again, all other nodes in
the network are simultaneously performing a similar information
exchange step. The reason for the name Combine-then-Adapt (CTA) strategy
is that the first step in \eqref{Equ:DiffusionAdaptation:CTA} involves a combination step, while the
second step will be shown to lead to an adaptive step. Hence, strategy \eqref{Equ:DiffusionAdaptation:CTA}
involves combination followed by adaptation or CTA for short. The reason for the qualification ``diffusion''
is that the combination step of \eqref{Equ:DiffusionAdaptation:CTA} allows information to diffuse through the network
in real time.

In the special case when $C=I$,
so that no information exchange is performed but only the aggregation step, the CTA
strategy \eqref{Equ:DiffusionAdaptation:CTA} reduces to:
\be
 \begin{array}{l}\mbox{\rm (CTA strategy without}\\\mbox{\rm information exchange)}\end{array}\;\;\;
\addbox{\;
                \begin{array}{l}
                    \psi_{k,i-1}    =   \displaystyle \sum_{\ell \in \mathcal{N}_k} a_{\ell k} \;w_{\ell,i-1}
                    \\
                    w_{k,i} =    \displaystyle\psi_{k,i-1}
                                + \mu_k \left(r_{du,k}-R_{u,k}\;\psi_{k,i-1}\right)
                \end{array}\;}
\label{without.cta}      \ee
where the second step relies solely on the information
$\{R_{u,k},r_{du,k}\}$ that is available locally at node $k$.
Again, the CTA strategy \eqref{Equ:DiffusionAdaptation:CTA}  can be rewritten in terms of the gradient vectors of the local costs
$\{J_{\ell}(w)\}$ as follows:
\be \mbox{\rm (CTA strategy)}\;\;\;
\addbox{\;
                \label{Equ:DiffusionAdaptation:CTA.2}
                \begin{array}{l}
                    \psi_{k,i-1}    =   \displaystyle \sum_{\ell \in \mathcal{N}_k} a_{\ell k} \;w_{\ell,i-1}
                    \\
                    w_{k,i} =    \displaystyle\psi_{k,i-1}
                                - \mu_k \sum_{\ell \in \mathcal{N}_k} c_{\ell
                                k}\;\left[\nabla_{w}J_{\ell} (\psi_{k,i-1})\right]^*
                \end{array}\;}
   \ee
The bottom part of Fig.~\ref{fig-33.label} illustrates the two steps
involved in the CTA procedure for a situation where node $k$ has
three other neighbors labeled $\{1,2,\ell\}$. In the first step,
node $k$ aggregates the estimates
$\{w_{1,i-1},w_{2,i-1},w_{\ell,i-1}\}$ from its neighbors, and
subsequently performs information exchange by evaluating the
gradient vectors of its neighbors at $\psi_{k,i-1}$.

\subsection{Useful Properties of Diffusion Strategies}
Note that the structure of the ATC and CTA diffusion strategies
\eqref{Equ:DiffusionAdaptation:ATC} and \eqref{Equ:DiffusionAdaptation:CTA} are fundamentally the same: the difference between the implementations lies in which variable we choose to correspond to the updated weight estimate $w_{k,i}$. In the ATC case, we choose the result of the {\em combination} step to be $w_{k,i}$, whereas in the CTA case we choose the result of the {\em adaptation} step to be $w_{k,i}$.

For ease of reference, Table~\ref{table9-1.label.xxxas} lists the steepest-descent diffusion algorithms derived in the previous sections. The derivation of the ATC and CTA strategies
\eqref{Equ:DiffusionAdaptation:ATC} and \eqref{Equ:DiffusionAdaptation:CTA}
followed the approach proposed in \cite{Cattivelli08,Cattivelli10}.  CTA estimation schemes
 were first proposed in the works
\cite{Lopes06,Sayed07,Lopes07a,Lopes07b,Lopes08}, and later extended in
\cite{Cattivelli08,Cattivelli10,Sayed09,Cattivelli10b}. The earlier
versions of CTA in \cite{Lopes06,Sayed07,Lopes07a} used the choice $C=I$. This
form of the algorithm with $C=I$, and with the additional constraint that the step-sizes $\mu_k$ should be time-dependent
and decay towards zero as time progresses, was later applied by
\cite{ramdistributed,bianchi}  to solve distributed optimization problems that
require all nodes to reach consensus or agreement. Likewise, special cases of the ATC estimation scheme
\eqref{Equ:DiffusionAdaptation:ATC}, involving an information exchange step followed by an
aggregation step, first appeared in the work \cite{Cattivelli07} on diffusion least-squares
schemes and subsequently in the works \cite{Cattivelli10,Cattivelli08a,Cattivelli08b,Cattivelli08c,Sayed09,Cattivelli10b} on
distributed mean-square-error and state-space estimation methods. A special case of the ATC strategy \eqref{Equ:DiffusionAdaptation:ATC} corresponding to the choice $C=I$ with decaying step-sizes was adopted in \cite{Stankovic11} to ensure convergence towards a consensus state. Diffusion strategies of the form
\eqref{Equ:DiffusionAdaptation:ATC} and
\eqref{Equ:DiffusionAdaptation:CTA} (or, equivalently,
\eqref{Equ:DiffusionAdaptation:ATC.2} and \eqref{Equ:DiffusionAdaptation:CTA.2}) are general in several
respects:

\begin{table}[h]
\begin{center}
\caption{\rm {\small Summary of steepest-descent diffusion strategies for the distributed optimization
of general problems of the form (\ref{opt.11}), and their specialization to the
case of mean-square-error (MSE) individual cost functions given by (\ref{gihjas.alk}).}}\noindent \\ {\small
\begin{tabular}{|l|l|c|}\hline\hline
{\sc Algorithm}& {\sc Recursions} & {\sc Reference}\\\hline\hline &&\\
$\begin{array}{l}\mbox{\rm {\bf ATC strategy}}\\(\mbox{\rm general case})\end{array}$ & $
                \begin{array}{l}
                    \psi_{k,i}  =   \displaystyle w_{k,i-1} - \mu_k\; \sum_{\ell \in \mathcal{N}_k}
c_{\ell k}\;\left[\nabla_{w}J_{\ell} (w_{k,i-1})\right]^*\\
                    w_{k,i} =   \displaystyle \sum_{\ell \in \mathcal{N}_k} a_{\ell k}\; \psi_{\ell,i}
                \end{array}$ & (\ref{Equ:DiffusionAdaptation:ATC.2})\\&&\\\hline &&\\
$\begin{array}{l}\mbox{\rm {\bf ATC strategy}}\\(\mbox{\rm MSE costs})\end{array}$
&$                \begin{array}{l}
                    \psi_{k,i}  =   \displaystyle w_{k,i-1} + \mu_k\; \sum_{\ell \in \mathcal{N}_k}
c_{\ell k}\;(r_{du,\ell}-R_{u,\ell}\; w_{k,i-1})\\
                    w_{k,i} =   \displaystyle \sum_{\ell \in \mathcal{N}_k} a_{\ell k}\; \psi_{\ell,i}
                \end{array}$ & (\ref{Equ:DiffusionAdaptation:ATC})\\&&\\\hline\hline&&\\
$\begin{array}{l}\mbox{\rm {\bf CTA strategy}}\\(\mbox{\rm general case})\end{array}$ &$
                \begin{array}{l}
                    \psi_{k,i-1}    =   \displaystyle \sum_{\ell \in \mathcal{N}_k} a_{\ell k} \;w_{\ell,i-1}
                    \\
                    w_{k,i} =    \displaystyle\psi_{k,i-1}
                                - \mu_k \sum_{\ell \in \mathcal{N}_k} c_{\ell
                                k}\;\left[\nabla_{w}J_{\ell} (\psi_{k,i-1})\right]^*
                \end{array}$
                & (\ref{Equ:DiffusionAdaptation:CTA.2})\\&&\\\hline&&\\
$\begin{array}{l}\mbox{\rm {\bf CTA strategy}}\\(\mbox{\rm MSE costs})\end{array}$&
 $  \begin{array}{l}
                    \psi_{k,i-1}    =   \displaystyle \sum_{\ell \in \mathcal{N}_k} a_{\ell k} \;w_{\ell,i-1}
                    \\
                    w_{k,i} =    \displaystyle\psi_{k,i-1}
                                + \mu_k \sum_{\ell \in \mathcal{N}_k} c_{\ell
                                k}\left(r_{du,\ell}-R_{u,\ell}\;\psi_{k,i-1}\right)
                \end{array}$ &(\ref{Equ:DiffusionAdaptation:CTA})\\&&\\\hline
\end{tabular} }
\label{table9-1.label.xxxas}
\end{center}
\end{table}

\begin{enumerate}
\item[(1)] These strategies do not only diffuse the local weight
estimates, but they can also diffuse the local gradient vectors. In
other words, two sets of combination coefficients
$\{a_{\ell k},c_{\ell k}\}$ are used.

\item[(2)] In the derivation that led to the diffusion strategies, the combination
matrices $C$ and $A$ are only required to be
right-stochastic (for $C$) and left-stochastic (for $A$). In comparison,
it is common in consensus-type strategies to
require the corresponding combination matrix $A$ to be doubly stochastic (i.e., its rows and columns should add
up to one) --- see, e.g., App.~\ref{app.C} and \cite{ramdistributed,groot74,berger,Tsi84}.

\item[(3)] As the analysis in Sec.~\ref{sec.mse} will reveal, ATC and CTA
strategies do {\em not} force  nodes to converge to an agreement about the desired
parameter vector $w^o$, as is common in consensus-type strategies (see App.~\ref{app.C} and
 \cite{ramdistributed,moore03,Saber04,Saber05,Saber07,Xiao04,Xiao05,Khan08}). Forcing nodes to reach agreement on
$w^o$ ends up limiting the adaptation and learning abilities of these nodes, as well as
their ability to react to information in real-time. Nodes in diffusion networks enjoy more flexibility in the learning
process, which allows their individual estimates, $\{w_{k,i}\}$, to tend to values that
lie within a reasonable mean-square-deviation (MSD) level from the optimal solution, $w^o$.
Multi-agent systems in nature
behave in this manner; they do not require exact agreement among their agents (see, e.g.,
\cite{Tu11,catmay11,Li11}).

\item[(4)] The step-size parameters $\{\mu_k\}$  are not required
to depend on the time index $i$ and are not required to vanish as
$i\rightarrow\infty$ (as is common in many works on distributed optimization, e.g.,
\cite{Poljak87,ramdistributed,Ber97,Nedic2010}). Instead, the step-sizes can assume constant values,
which is a critical property to endow networks with continuous adaptation
and learning abilities. An important contribution in the study of diffusion strategies is to show that
distributed optimization is still possible even for constant step-sizes, in addition to the ability to
perform adaptation, learning, and tracking. Sections~\ref{sec.sd} and~\ref{sec.mse}
highlight the convergence properties of the diffusion strategies --- see also \cite{Chen10,Chen10b,ChenSSP2012} for results pertaining to
more general cost functions.

\item[(5)] Even the combination weights $\{a_{\ell k},c_{\ell k}\}$ can be adapted, as we shall discuss later
in Sec.~\ref{sec7.2}. In this way, diffusion strategies allow multiple layers of adaptation: the nodes
perform adaptive processing, the combination weights can be adapted, and even the topology can be
adapted especially for mobile networks \cite{Tu11}.
\end{enumerate}

\section{Adaptive Diffusion Strategies}\label{sec.adaptive.dd}
The distributed ATC and CTA steepest-descent strategies
\eqref{Equ:DiffusionAdaptation:ATC} and \eqref{Equ:DiffusionAdaptation:CTA}
 for determining the $w^o$ that solves (\ref{opt.11})--(\ref{gihjas.alk})
 require knowledge of the statistical information
$\{R_{u,k},r_{du,k}\}$.  These moments are needed in order to be
able to evaluate the gradient vectors that appear in
(\ref{Equ:DiffusionAdaptation:ATC}) and
\eqref{Equ:DiffusionAdaptation:CTA}, namely, the terms: \bq
-\left[\nabla_{w} J_{\ell}(w_{k,i-1})\right]^* &=&(r_{du,\ell}-R_{u,\ell}\;w_{k,i-1})\\
-\left[\nabla_{w} J_{\ell}(\psi_{k,i-1})\right]^*&=&(r_{du,\ell}-R_{u,\ell}\;\psi_{k,i-1}) \eq for all
$\ell\in{\cal N}_k$. However, the moments
$\{R_{u,\ell},r_{du,\ell}\}$ are often not available beforehand, which
means that the true gradient vectors are generally not available.
Instead, the agents have access to observations $\{d_k(i),u_{k,i}\}$
of the random processes $\{\d_k(i),\u_{k,i}\}$.  There are many ways by which the
true gradient vectors can
be approximated by using these observations.  Recall that, by definition, \be R_{u,\ell}\define
\Ex \u_{\ell,i}^*\u_{\ell,i},\;\;\;\;\;\; r_{du,\ell}\define \Ex
\d_{\ell}(i)\u_{\ell,i}^* \ee One common stochastic approximation
method is to drop the expectation operator
from the definitions of $\{R_{u,\ell},r_{du,\ell}\}$ and to use the
following instantaneous approximations instead
\cite{Sayed03,Sayed08,Haykin01,Widrow85}: \be R_{u,\ell}\approx
u_{\ell,i}^*u_{\ell,i},\;\;\;\;\; r_{du,\ell}\approx
d_{\ell}(i)u_{\ell,i}^* \label{instaldlakd.laprad}\ee In this case, the approximate gradient
vectors become: \bq (r_{du,\ell}-R_{u,\ell}\;w_{k,i-1})&\approx&
u_{\ell,i}^*\left[d_{\ell}(i)-u_{\ell,i}\;w_{k,i-1}\right]\label{approxalksa}\\
(r_{du,\ell}-R_{u,\ell}\;\psi_{k,i-1})&\approx&
u_{\ell,i}^*\left[d_{\ell}(i)-u_{\ell,i}\;\psi_{k,i-1}\right]\label{approxalksa.1}\eq
Substituting into the ATC and CTA steepest-descent strategies
(\ref{Equ:DiffusionAdaptation:ATC}) and
\eqref{Equ:DiffusionAdaptation:CTA}, we
arrive at the following adaptive implementations of the diffusion
strategies for $i\geq 0$: \be
        \mbox{\rm (adaptive ATC strategy)}\;\;\;
        \addbox{
                \label{Equ:DiffusionAdaptation:ATC.adaptive}
                \begin{array}{l}
                    \psi_{k,i}  =   \displaystyle w_{k,i-1} + \mu_k\; \sum_{\ell \in \mathcal{N}_k}
c_{\ell k}\;u_{\ell,i}^*\left[d_{\ell}(i)-u_{\ell,i} w_{k,i-1}\right]\\
                    w_{k,i} =   \displaystyle \sum_{\ell \in \mathcal{N}_k} a_{\ell k}\; \psi_{\ell,i}
                \end{array}
            }
       \ee
and \be
   \mbox{\rm (adaptive CTA strategy)}\;\;\;\;\;
        \addbox{
                \label{Equ:DiffusionAdaptation:CTA.adaptive}
                \begin{array}{l}
                    \psi_{k,i-1}    =   \displaystyle \sum_{\ell \in \mathcal{N}_k} a_{\ell k} \;w_{\ell,i-1}
                    \\
                    w_{k,i} =    \displaystyle\psi_{k,i-1}
                                + \mu_k \sum_{\ell \in \mathcal{N}_k} c_{\ell
                                k}\;u_{\ell,i}^*\left[d_{\ell}(i)-u_{\ell,i}\;\psi_{k,i-1}\right]
                \end{array}
            }
   \ee
where the coefficients $\{a_{\ell k},c_{\ell k}\}$ are chosen to satisfy:
\be\begin{array}{r}
\mbox{\rm for}\;k=1,2,\ldots,N:\\
c_{\ell k}\geq 0,\;\;\;\;\displaystyle\sum_{k=1}^N c_{\ell k} = 1,\;\;\;\;  c_{\ell k}=0~\mathrm{if}~\ell \notin \mathcal{N}_{k}\\
a_{\ell k}\geq 0,\;\;\;\;\displaystyle\sum_{\ell=1}^N a_{\ell k} = 1,\;\;\;\;  a_{\ell k}=0~\mathrm{if}~\ell \notin \mathcal{N}_{k}
\end{array}\label{run.run.3}
\ee
The adaptive implementations usually start from the initial conditions
$w_{\ell,-1}=0$ for all $\ell$, or from some other convenient
initial values. Clearly, in view of the approximations
(\ref{approxalksa})--(\ref{approxalksa.1}), the successive iterates
$\{w_{k,i},\psi_{k,i},\psi_{k,i-1}\}$  that are generated by the
above adaptive implementations are different from the iterates that
result from the steepest-descent implementations
(\ref{Equ:DiffusionAdaptation:ATC}) and
(\ref{Equ:DiffusionAdaptation:CTA}).
Nevertheless, we shall continue to use the same notation for these
variables for ease of reference. One key advantage of the adaptive implementations (\ref{Equ:DiffusionAdaptation:ATC.adaptive})--(\ref{Equ:DiffusionAdaptation:CTA.adaptive})
is that they
enable the agents to react to changes in the underlying statistical
information $\{r_{du,\ell},R_{u,\ell}\}$ and to changes in $w^o$.
This is because these changes end up being reflected in the data realizations $\{d_k(i),u_{k,i}\}$. Therefore, adaptive implementations have an innate tracking and
learning ability that is of paramount significance in practice.

We say that the stochastic
gradient approximations (\ref{approxalksa})--(\ref{approxalksa.1})
introduce gradient noise into each step of the recursive updates
(\ref{Equ:DiffusionAdaptation:ATC.adaptive})--(\ref{Equ:DiffusionAdaptation:CTA.adaptive}).
This is because the updates
(\ref{Equ:DiffusionAdaptation:ATC.adaptive})--(\ref{Equ:DiffusionAdaptation:CTA.adaptive})
can be interpreted as corresponding to the following forms: \be \mbox{\rm (adaptive
ATC strategy)}\;\;\;
               \label{Equ:DiffusionAdaptation:ATC.2x}
               \addbox{\; \begin{array}{l}
                    \psi_{k,i}  =   \displaystyle w_{k,i-1} - \mu_k\; \sum_{\ell \in \mathcal{N}_k}
c_{\ell k}\;\widehat{\left[{\nabla_{w}J_{\ell}} (w_{k,i-1})\right]}^*\\
                    w_{k,i} =   \displaystyle \sum_{\ell \in \mathcal{N}_k} a_{\ell k}\; \psi_{\ell,i}
                \end{array}\;}
       \ee
and \be \hspace{-0.2cm} \mbox{\rm (adaptive CTA strategy)}\;\;\;
                \label{Equ:DiffusionAdaptation:CTA.2x}
               \addbox{\; \begin{array}{l}
                    \psi_{k,i-1}    =   \displaystyle \sum_{\ell \in \mathcal{N}_k} a_{\ell k} \;w_{\ell,i-1}
                    \\
                    w_{k,i} =    \displaystyle\psi_{k,i-1}
                                - \mu_k \sum_{\ell \in \mathcal{N}_k} c_{\ell
                                k}\;\widehat{\left[{\nabla_{w}J_{\ell}} (\psi_{k,i-1})\right]}^*
                \end{array}\;}
   \ee
\noindent \\

\noindent where the true gradient vectors, $\{\nabla_{w}
J_{\ell}(\cdot)\}$, have been replaced by approximations,
$\{\widehat{\nabla_{w} J_{\ell}(\cdot)}\}$ --- compare with \eqref{Equ:DiffusionAdaptation:ATC.2} and \eqref{Equ:DiffusionAdaptation:CTA.2}.
The significance of
the alternative forms (\ref{Equ:DiffusionAdaptation:ATC.2x})--(\ref{Equ:DiffusionAdaptation:CTA.2x}) is that they are
 applicable to optimization problems involving more general local costs $J_{\ell}(w)$ that
are not necessarily quadratic, as detailed in \cite{Chen10b,ChenSSP2012}; see also Sec.~\ref{sec.diskd8192}. In the
next section, we  examine how gradient noise affects the
performance of the diffusion strategies and how close the successive
estimates $\{w_{k,i}\}$ get to the desired optimal solution $w^o$. Table~\ref{table9-1.label.alkalkdaxxxas} lists
several of the adaptive diffusion algorithms derived in this section.

\begin{table}[htp]
\begin{center}
\caption{\rm {\small Summary of adaptive diffusion strategies for the distributed optimization
of general problems of the form (\ref{opt.11}), and their specialization to the
case of mean-square-error (MSE) individual cost functions given by (\ref{gihjas.alk}). These adaptive solutions
rely on stochastic approximations.}}\noindent \\ {\small
\begin{tabular}{|l|l|c|}\hline\hline
{\sc Algorithm}& {\sc Recursions} & {\sc Reference}\\\hline\hline &&\\
$\begin{array}{l}\mbox{\rm {\bf Adaptive ATC strategy}}\\(\mbox{\rm general case})\end{array}$&
$\begin{array}{l}
                    \psi_{k,i}  =   \displaystyle w_{k,i-1} - \mu_k\; \sum_{\ell \in \mathcal{N}_k}
c_{\ell k}\;\widehat{\left[{\nabla_{w}J_{\ell}} (w_{k,i-1})\right]}^*\\
                    w_{k,i} =   \displaystyle \sum_{\ell \in \mathcal{N}_k} a_{\ell k}\; \psi_{\ell,i}
                \end{array}$ & (\ref{Equ:DiffusionAdaptation:ATC.2x}) \\&&\\\hline &&\\
$\begin{array}{l}\mbox{\rm {\bf Adaptive ATC strategy}}\\(\mbox{\rm MSE costs})\end{array}$ &
 $                \begin{array}{l}
                    \psi_{k,i}  =   \displaystyle w_{k,i-1} + \mu_k\; \sum_{\ell \in \mathcal{N}_k}
c_{\ell k}\;u_{\ell,i}^*\left[d_{\ell}(i)-u_{\ell,i} w_{k,i-1}\right]\\
                    w_{k,i} =   \displaystyle \sum_{\ell \in \mathcal{N}_k} a_{\ell k}\; \psi_{\ell,i}
                \end{array}$& (\ref{Equ:DiffusionAdaptation:ATC.adaptive})\\&&\\\hline &&\\

$\begin{array}{l}\mbox{\rm {\bf Adaptive ATC strategy}}\\(\mbox{\rm MSE costs})\\(\mbox{\rm no information exchange})\end{array}$
&$
                \begin{array}{l}
                    \psi_{k,i}  =   \displaystyle w_{k,i-1} + \mu_k\; u_{k,i}^*\left[d_{k}(i)-u_{k,i}
                    w_{k,i-1}\right]\\\\
                    w_{k,i} =   \displaystyle \sum_{\ell \in \mathcal{N}_k} a_{\ell k}\; \psi_{\ell,i}
                \end{array}$& (\ref{Equ:DiffusionAdaptation:ATC.adaptive.2})\\&&\\\hline\hline&&\\
$\begin{array}{l}\mbox{\rm {\bf Adaptive CTA strategy}}\\(\mbox{\rm general case})\end{array}$&
$\begin{array}{l}
                    \psi_{k,i-1}    =   \displaystyle \sum_{\ell \in \mathcal{N}_k} a_{\ell k} \;w_{\ell,i-1}
                    \\
                    w_{k,i} =    \displaystyle\psi_{k,i-1}
                                - \mu_k \sum_{\ell \in \mathcal{N}_k} c_{\ell
                                k}\;\widehat{\left[{\nabla_{w}J_{\ell}} (\psi_{k,i-1})\right]}^*
                \end{array}$ & (\ref{Equ:DiffusionAdaptation:CTA.2x})\\&&\\\hline&&\\
$\begin{array}{l}\mbox{\rm {\bf Adaptive CTA strategy}}\\(\mbox{\rm MSE costs})\end{array}$ &$
                \begin{array}{l}
                    \psi_{k,i-1}    =   \displaystyle \sum_{\ell \in \mathcal{N}_k} a_{\ell k} \;w_{\ell,i-1}
                    \\
                    w_{k,i} =    \displaystyle\psi_{k,i-1}
                                + \mu_k \sum_{\ell \in \mathcal{N}_k} c_{\ell
                                k}\;u_{\ell,i}^*\left[d_{\ell}(i)-u_{\ell,i}\;\psi_{k,i-1}\right]
                \end{array}$
                & (\ref{Equ:DiffusionAdaptation:CTA.adaptive})\\&&\\\hline&&\\
$\begin{array}{l}\mbox{\rm {\bf Adaptive CTA strategy}}\\(\mbox{\rm MSE costs})\\(\mbox{\rm no information exchange})\end{array}$&
 $
                \begin{array}{l}
                    \psi_{k,i-1}    =   \displaystyle \sum_{\ell \in \mathcal{N}_k} a_{\ell k} \;w_{\ell,i-1}
                    \\
                    w_{k,i} =    \displaystyle\psi_{k,i-1}
                                + \mu_k \;u_{k,i}^*\left[d_{k}(i)-u_{k,i}\;\psi_{k,i-1}\right]
                \end{array}$ &(\ref{Equ:DiffusionAdaptation:CTA.adaptive.2})\\&&\\\hline
\end{tabular} }
\label{table9-1.label.alkalkdaxxxas}
\end{center}
\end{table}

\begin{figure}[h]
\epsfxsize 10cm \epsfclipon
\begin{center}
\leavevmode \epsffile{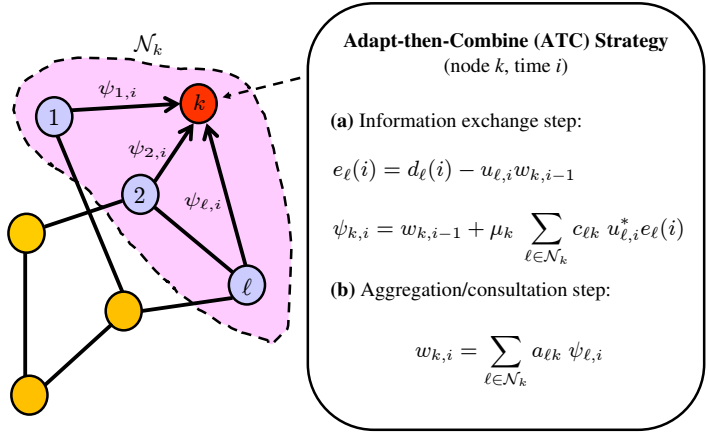} \caption{{\small Illustration
of the adaptive ATC strategy, which involves two steps: information
exchange followed by aggregation.}}\label{fig-11.label}
\end{center}
\end{figure}

The operation of the adaptive diffusion strategies is similar to the
operation of the steepest-descent diffusion strategies of the
previous section. Thus, note that at every time instant $i$, the ATC
strategy \eqref{Equ:DiffusionAdaptation:ATC.adaptive} performs two
steps; as illustrated in Fig.~\ref{fig-11.label}. The first step is
an {\em information exchange} step where node $k$ receives from its
neighbors their information $\{d_{\ell}(i),u_{\ell,i}\}$. Node $k$
combines this information and uses it to update its existing
estimate $w_{k,i-1}$ to an intermediate value $\psi_{k,i}$. All
other nodes in the network are performing a similar step and
updating their existing estimates $\{w_{\ell,i-1}\}$ into intermediate estimates $\{\psi_{\ell,i}\}$ by
using information from their neighbors. The second step in
\eqref{Equ:DiffusionAdaptation:ATC.adaptive} is an {\em aggregation}
or consultation step where node $k$ combines the intermediate
estimates $\{\psi_{\ell,i}\}$ of its neighbors to obtain its update estimate $w_{k,i}$.
Again, all other nodes in the network are simultaneously performing
a similar step. In the special case when $C=I$, so that no
information exchange is performed but only the aggregation step, the
ATC strategy \eqref{Equ:DiffusionAdaptation:ATC.adaptive} reduces
to:

\be
      \begin{array}{l}  \mbox{\rm (adaptive ATC strategy}\\\mbox{\rm without information exchange})\end{array}\;\;\;\;\;
        \addbox{
                \label{Equ:DiffusionAdaptation:ATC.adaptive.2}
                \begin{array}{l}
                    \psi_{k,i}  =   \displaystyle w_{k,i-1} + \mu_k\; u_{k,i}^*\left[d_{k}(i)-u_{k,i}
                    w_{k,i-1}\right]\\\\
                    w_{k,i} =   \displaystyle \sum_{\ell \in \mathcal{N}_k} a_{\ell k}\; \psi_{\ell,i}
                \end{array}
            }
       \ee

Likewise, at every time instant $i$, the CTA strategy
\eqref{Equ:DiffusionAdaptation:CTA.adaptive} also consists of two
steps -- see Fig.~\ref{fig-22.label}. The first step is an
aggregation step where node $k$ combines the existing estimates of
its neighbors to obtain the intermediate estimate $\psi_{k,i-1}$.
All other nodes in the network are simultaneously performing a
similar step and aggregating the estimates of their neighbors. The
second step in \eqref{Equ:DiffusionAdaptation:CTA.adaptive} is an
information exchange step where node $k$ receives from its neighbors
their information $\{d_{\ell}(i),u_{\ell,i}\}$ and uses this
information to update its intermediate estimate to $w_{k,i}$. Again,
all other nodes in the network are simultaneously performing a
similar information exchange step. In the special case when $C=I$,
so that no information exchange is performed but only the
aggregation step, the CTA strategy
\eqref{Equ:DiffusionAdaptation:CTA.adaptive} reduces to:

\begin{figure}[h]
\epsfxsize 10cm \epsfclipon
\begin{center}
\leavevmode \epsffile{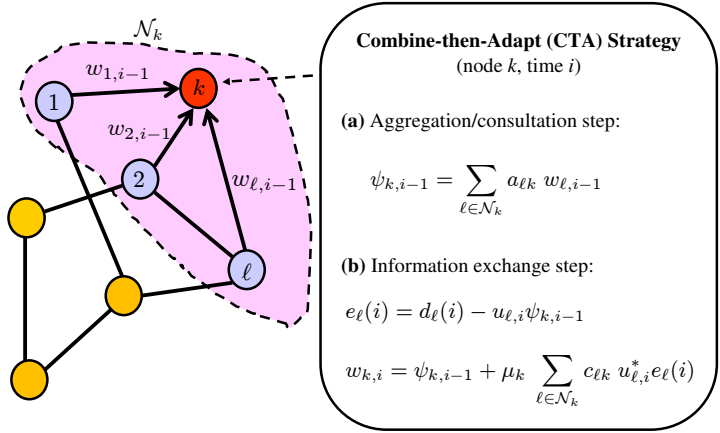} \caption{{\small Illustration
of the adaptive CTA strategy, which involves two steps: aggregation
followed by information exchange.}}\label{fig-22.label}
\end{center}
\end{figure}

\be
 \begin{array}{l}  \mbox{\rm (adaptive CTA strategy}\\\mbox{\rm without information exchange})\end{array}\;\;\;\;\;
        \addbox{
                \label{Equ:DiffusionAdaptation:CTA.adaptive.2}
                \begin{array}{l}
                    \psi_{k,i-1}    =   \displaystyle \sum_{\ell \in \mathcal{N}_k} a_{\ell k} \;w_{\ell,i-1}
                    \\
                    w_{k,i} =    \displaystyle\psi_{k,i-1}
                                + \mu_k \;u_{k,i}^*\left[d_{k}(i)-u_{k,i}\;\psi_{k,i-1}\right]
                \end{array}
            }
   \ee

\noindent We further note that the adaptive ATC and CTA strategies
(\ref{Equ:DiffusionAdaptation:ATC.adaptive})--(\ref{Equ:DiffusionAdaptation:CTA.adaptive})
reduce to the non-cooperative adaptive solution
(\ref{ea.lms})--(\ref{fac.lms}), where each node $k$ runs its own individual LMS filter,
when the coefficients $\{a_{\ell
k},c_{\ell k}\}$ are selected as \be a_{\ell k}=\delta_{\ell
k}=c_{\ell k}\;\;\;\;\;(\mbox{\rm non-cooperative case}) \ee where
$\delta_{\ell k}$ denotes the Kronecker delta function: \be
\delta_{\ell k}\define
\left\{\begin{array}{ccl}1,&&\ell=k\\0,&&\mbox{\rm
otherwise}\end{array}\right. \ee
 In
terms of the combination matrices $A$ and $C$, this situation
corresponds to setting \be
\addbox{\;A=I_{N}=C\;}\;\;\;\;\;(\mbox{\rm non-cooperative case})
\ee

\section{Performance of Steepest-Descent Diffusion Strategies}\label{sec.sd}
Before studying in some detail the mean-square performance of the adaptive
diffusion implementations
(\ref{Equ:DiffusionAdaptation:ATC.adaptive})--(\ref{Equ:DiffusionAdaptation:CTA.adaptive}),
and the influence of gradient noise, we examine first the
convergence behavior of the steepest-descent diffusion strategies
(\ref{Equ:DiffusionAdaptation:ATC}) and
(\ref{Equ:DiffusionAdaptation:CTA}), which employ the true gradient vectors. Doing so, will help
introduce the necessary notation and highlight some features of the analysis in preparation for
the more challenging treatment of the adaptive strategies in Sec.~\ref{sec.mse}.

\subsection{General Diffusion Model}
Rather than study
the performance of the ATC and CTA steepest-descent strategies (\ref{Equ:DiffusionAdaptation:ATC}) and
(\ref{Equ:DiffusionAdaptation:CTA}) separately, it is useful to
introduce a more general description that includes the ATC and CTA
recursions as special cases. Thus, consider a distributed
steepest-descent diffusion implementation of the following general
form for $i\geq 0$:  \bq
                {\phi}_{k,i-1}   &=&  \displaystyle\sum_{\ell\in{\cal N}_k} a_{1,\ell k}\; w_{\ell,i-1}   \label{general.1}\\
                {\psi}_{k,i}      &=&   {\phi}_{k,i-1} +
                \mu_k \displaystyle\sum_{\ell\in{\cal N}_k} c_{\ell
                k}\;\left[r_{du,\ell}-R_{u,\ell}\;{\phi}_{k,i-1}\right]\\
                w_{k,i}     &=&    \displaystyle\sum_{\ell\in{\cal N}_k} a_{2,\ell k}\;
                {\psi}_{\ell,i}\label{general.3}
  \eq
where the scalars $\{a_{1,\ell k},c_{\ell k},a_{2,\ell k}\}$ denote three sets of
non-negative real coefficients corresponding to the $(\ell,k)$
entries of $N\times N$ combination matrices $\{A_1,C,A_2\}$,
respectively. These matrices are assumed to satisfy the conditions:
\be \addbox{
        A_1^T \mathds{1}=\mathds{1}, \quad
        C\mathds{1}=\mathds{1}, \quad
        A_2^T \mathds{1}=\mathds{1}
        }
\label{dlajs8.12}\ee so that $\{A_1,A_2\}$ are left stochastic and $C$ is right-stochastic, i.e.,
\be
\begin{array}{llr}
&&\mbox{\rm for}\;k=1,2,\ldots,N:\\
c_{\ell k}\geq 0,&\;\;\;\displaystyle\sum_{k=1}^N c_{\ell k} = 1,&  c_{\ell k}=0~\mathrm{if}~\ell \notin \mathcal{N}_{k}\\
a_{1,\ell k}\geq 0,&\;\;\;\displaystyle\sum_{\ell=1}^N a_{1,\ell k} = 1,&  a_{1,\ell k}=0~\mathrm{if}~\ell \notin \mathcal{N}_{k}\\
a_{2,\ell k}\geq 0,&\;\;\;\displaystyle\sum_{\ell=1}^N a_{2,\ell k} = 1,&  a_{2,\ell k}=0~\mathrm{if}~\ell \notin \mathcal{N}_{k}
\end{array}\label{run.run.4}
\ee
Different choices for $\{A_1,C,A_2\}$
correspond to different cooperation modes. For example, the choice
$A_1 = I_N$ and $A_2=A$ corresponds to the ATC implementation
(\ref{Equ:DiffusionAdaptation:ATC}), while the choice $A_1 = A$ and
$A_2 = I_N$ corresponds to the CTA implementation
(\ref{Equ:DiffusionAdaptation:CTA}).
Likewise, the choice $C=I_N$ corresponds to the case in which the
nodes only share weight estimates and the distributed diffusion
recursions (\ref{general.1})--(\ref{general.3}) become
 \bq
                {\phi}_{k,i-1}   &=&  \displaystyle\sum_{\ell\in{\cal N}_k} a_{1,\ell k}\; w_{\ell,i-1}   \label{general.1x}\\
                {\psi}_{k,i}      &=&   {\phi}_{k,i-1} +
                \mu_k \;\left(r_{du,k}-R_{u,k}\;{\phi}_{k,i-1}\right)\\
                w_{k,i}     &=&    \displaystyle\sum_{\ell\in{\cal N}_k} a_{2,\ell k}\;
                {\psi}_{\ell,i}\label{general.3x}
  \eq
Furthermore, the choice $A_1=A_2=C=I_N$ corresponds to the
non-cooperative mode of operation, in which case the recursions
reduce to the classical (stand-alone) steepest-descent recursion \cite{Sayed03,Sayed08,Haykin01,Widrow85},
where each node minimizes individually its own quadratic cost
$J_k(w)$, defined earlier in (\ref{a90ks.alk}):\bq
                w_{k,i}      &=&    w_{k,i-1} +
                \mu_k \;\left[r_{du,k}-R_{u,k}\;w_{k,i-1}\right],\;\;\;i\geq 0\label{no.general.3}
  \eq

\begin{table}[h]
\begin{center}
\caption{\rm {\small Different choices for the combination matrices
$\{A_1,A_2,C\}$ in (\ref{general.1})--(\ref{general.3}) correspond
to different cooperation strategies.}} {\small
\begin{tabular}{|c|c|c|l|}\hline\hline
$A_1$ & $A_2$ & $C$ & {\sc Cooperation Mode}\\\hline\hline$I_N$ & $A$
&
$C$ & {\rm ATC strategy} (\ref{Equ:DiffusionAdaptation:ATC}).\\
$I_N$ & $A$
&
$I_N$ & {\rm ATC strategy} (\ref{without.atc}) without information exchange.\\
$A$ & $I_N$ &
$C$ & {\rm CTA strategy} (\ref{Equ:DiffusionAdaptation:CTA}).\\
$A$ & $I_N$ &
$I_N$ & {\rm CTA strategy} (\ref{without.cta}) without information exchange.\\
$I_N$ & $I_N$ & $I_N$ & {\rm non-cooperative steepest-descent} (\ref{no.general.3}).
\\\hline\hline
\end{tabular} }
\label{table9-1.label}
\end{center}
\end{table}

\subsection{Error Recursions}
\noindent Our objective is to examine whether, and how fast, the weight
estimates $\{w_{k,i}\}$ from the distributed implementation
(\ref{general.1})--(\ref{general.3}) converge towards the solution $w^o$ of (\ref{opt.11})--(\ref{gihjas.alk}). To do
so, we introduce  the $M\times 1$ error vectors: \bq
\widetilde{\phi}_{k,i}&
\define &w^o-\phi_{k,i}\\
        \widetilde{\psi}_{k,i}  &\define&   w^o-\psi_{k,i}\\
        \widetilde{w}_{k,i} &\define &  w^o - w_{k,i}\eq
Each of these error vectors measures the residual relative to the
desired minimizer $w^o$. Now recall from (\ref{kad8912.13}) that \be
r_{du,k}=R_{u,k}\;w^o \label{jakd8127}\ee
 Then, subtracting $w^o$ from both sides of the relations
in \eqref{general.1}--\eqref{general.3} we get \bq
                \widetilde{\phi}_{k,i-1}   &=&  \displaystyle\sum_{\ell\in{\cal N}_k} a_{1,\ell k}\; \widetilde{w}_{\ell,i-1}   \label{general.1a}\\
                \widetilde{\psi}_{k,i}      &=&   \left(I_M-\mu_k \sum_{\ell\in{\cal N}_k} c_{\ell
                k}\;R_{u,\ell}\right)\;\widetilde{\phi}_{k,i-1}\\
                \widetilde{w}_{k,i}     &=&    \displaystyle\sum_{\ell\in{\cal N}_k} a_{2,\ell k}\;
                \widetilde{\psi}_{\ell,i}\label{general.3a}
  \eq
We can describe these relations more compactly by collecting the
information from across the network into block vectors and matrices.
We collect the error vectors from across all
nodes into the following $N\times 1$ {\em block} vectors, whose
individual entries are of size $M\times 1$ each:\be
\widetilde{\psi}_i\define
\ba{c}\widetilde{\psi}_{1,i}\\
\widetilde{\psi}_{2,i}\\\vdots\\
\widetilde{\psi}_{N,i}\ea,\quad\quad \widetilde{\phi}_i\define
\ba{c}\widetilde{\phi}_{1,i}\\
\widetilde{\phi}_{2,i}\\\vdots\\
\widetilde{\phi}_{N,i}\ea,\quad\quad \widetilde{w}_i\define
\ba{c}\widetilde{w}_{1,i}\\
\widetilde{w}_{2,i}\\\vdots\\ \widetilde{w}_{N,i}\ea\ee The block
quantities
$\{\widetilde{\psi}_i,\widetilde{\phi}_i,\widetilde{w}_i\}$
represent the state of the errors across the network at time $i$.
Likewise, we introduce the following $N\times N$ {\em block}
diagonal matrices, whose individual entries are of size $M\times M$
each: \bq {\cal M}&\define& \mbox{\rm diag}\{\;\mu_1
I_M,\;\mu_2I_M,\;\ldots,\;\mu_N I_M\;\}\label{r.dklajda.11}\\\nn\\
 {\cal R}&\define&\mbox{\rm diag}\left\{\;\sum_{\ell\in{\cal N}_1} c_{\ell
                1}\;R_{u,\ell},\;\displaystyle\sum_{\ell\in{\cal N}_2} c_{\ell
                2}\;R_{u,\ell},\;\ldots,\;\sum_{\ell\in{\cal N}_N} c_{\ell
                N}\;R_{u,\ell}\;\right\}\label{r.dklajda}
\eq Each block diagonal entry of ${\cal R}$, say, the $k$-th entry,
contains the combination of the covariance matrices in the
neighborhood of node $k$. We can simplify the notation by denoting
these neighborhood combinations as follows: \be \addbox{\;R_k\define
\sum_{\ell\in{\cal N}_k} c_{\ell
                k}\;R_{u,\ell}\;}
\label{defkajs812}\ee so that ${\cal R}$ becomes \be \addbox{\;{\cal
R}\define\mbox{\rm
diag}\left\{\;R_1,\;R_2,\;\ldots,R_N\;\right\}\;}\;\;\;\;\;(\mbox{\rm
when $C\neq I$})\label{lkad912.as}\ee In the special case when $C=I_N$, the matrix
${\cal R}$ reduces to \be \addbox{\;{\cal R}_u=\mbox{\rm
diag}\{R_{u,1},R_{u,2},\ldots,R_{u,N}\}\;}\;\;\;\;\;(\mbox{\rm when
$C=I$}) \label{replaced}\ee with the individual covariance matrices
appearing on its diagonal; we denote ${\cal R}$ by ${\cal R}_u$ in
this special case. We further introduce the Kronecker products \be {\cal A}_1\define
A_1\otimes I_M,\quad\quad{\cal A}_2\define A_2\otimes I_M
\label{defioas.a}\ee The matrix ${\cal A}_1$ is an $N\times N$ {\em block}
matrix whose $(\ell,k)$ block is equal to $a_{1,\ell k} I_M$.
Likewise, for ${\cal A}_2$. In other words, the Kronecker
transformation defined above simply replaces the matrices
$\{A_1,A_2\}$ by block matrices $\{{\cal A}_1,{\cal A}_2\}$ where
each entry $\{a_{1,\ell k},a_{2,\ell k}\}$ in the original matrices
is replaced by the diagonal matrices $\{a_{1,\ell k} I_M,\;a_{2,\ell
k} I_M\}$. For ease of reference, Table~\ref{table9-xxxx.label} lists the various symbols that have been
defined so far, and others that will be defined in the sequel.

\begin{table}[h]
\begin{center}
\caption{{\small \rm Definitions of network variables used throughout the analysis.}}
\noindent \\
 {\small
\begin{tabular}{l|c}\hline\hline
{\sc Variable} & {\sc Equation}\\\hline&\\
${\cal A}_1=A_1\otimes I_M$& (\ref{defioas.a})\\
${\cal A}_2=A_2\otimes I_M$& (\ref{defioas.a})\\
${\cal C}=C\otimes I_M$& (\ref{cclakdla})\\&\\
$R_k=
\displaystyle \sum_{\ell\in{\cal N}_k} c_{\ell
                k}\;R_{u,\ell}$& (\ref{defkajs812})\\&\\
${\cal R}=
\mbox{\rm diag}\;\left\{R_1,\;R_2,\;\ldots,R_N\;\right\}$&(\ref{lkad912.as})\\
${\cal R}_u=\mbox{\rm diag}\{
R_{u,1},\;R_{u,2},\;\ldots,\;R_{u,N}
\}$& (\ref{defl.sa})\\
$R_v=\mbox{\rm diag}\{\sigma_{v,1}^2,\;\sigma_{v,2}^2,\ldots,\sigma_{v,N}^2\}$& (\ref{kdla.salkd})\\
${\cal M}=\mbox{\rm diag}\{\;\mu_1
I_M,\;\mu_2I_M,\;\ldots,\;\mu_N I_M\;\}$ & (\ref{r.dklajda.11})\\
${\cal S}=\mbox{\rm diag}\{\sigma_{v,1}^2
R_{u,1},\;\sigma_{v,2}^2 R_{u,2},\;\ldots,\sigma_{v,N}^2 R_{u,N}
\}$& (\ref{defl.sa})\\&\\
${\cal G}={\cal A}_2^T{\cal M}{\cal C}^T$ & (\ref{g.dlakd})\\
${\cal B} ={\cal
A}_2^T\left(I_{NM}-{\cal M} {\cal R}\right){\cal
A}_1^T$ & (\ref{kdjalk12})\\
${\cal Y}={\cal G}{\cal S}{\cal G}^T$& (\ref{dlkas9812.1})\\
${\cal F}\approx {\cal B}^T\otimes {\cal B}^*$ & (\ref{ka8912.lkad})\\&\\
${\cal J}_k= \mbox{\rm
diag}\{\;0_M,\ldots,0_M,I_M,0_M,\ldots,0_M\;\}$ & (\ref{901asl})\\
${\cal
T}_k= \mbox{\rm
diag}\{\;0_M,\ldots,0_M,R_{u,k},0_M,\ldots,0_M\;\}$ & (\ref{dj8192})\\&\\\hline
\end{tabular} }
\label{table9-xxxx.label}
\end{center}
\end{table}

Returning to (\ref{general.1a})--(\ref{general.3a}), we
conclude that the following relations hold for the block
quantities:\bq
                \widetilde{\phi}_{i-1}   &=&  {\cal A}_1^T \widetilde{w}_{i-1}   \label{general.1b}\\
                \widetilde{\psi}_{i}      &=&   \left(I_{NM}-{\cal M}{\cal R}\right)\;\widetilde{\phi}_{i-1}\\
                \widetilde{w}_{i}     &=&    {\cal A}_2^T\;
                \widetilde{\psi}_{i}\label{general.3b}
  \eq
so that the network weight error vector, $\widetilde{w}_i$, ends up
evolving according to the following dynamics:\be
\addbox{\;\widetilde{w}_i={\cal A}_2^T\left(I_{NM}-{\cal M}{\cal
R}\right){\cal A}_1^T\widetilde{w}_{i-1},\;\;i\geq
0\;}\;\;\;\;\;(\mbox{\rm diffusion strategy}) \label{like.lakd}\ee
For comparison purposes, if each node in the network minimizes its own
cost function, $J_k(w)$, separately from the other nodes and uses the
non-cooperative steepest-descent
strategy (\ref{no.general.3}), then the weight error vector across
all $N$ nodes would evolve according to the following alternative
dynamics: \be \addbox{\;\widetilde{w}_i=\;\left(I_{NM}-{\cal M}{\cal
R}_u\right)\widetilde{w}_{i-1},\;\;i\geq
0\;}\;\;\;\;\;(\mbox{\rm non-cooperative strategy})
\label{like.lakd.no}\ee where the matrices ${\cal A}_1$ and ${\cal
A}_2$ do not appear, and ${\cal R}$ is replaced by ${\cal R}_u$ from
(\ref{replaced}). This recursion is a special case of (\ref{like.lakd}) when
$A_1=A_2=C=I_N$.

\subsection{Convergence Behavior}
Note from (\ref{like.lakd}) that the evolution of the weight error
vector involves block vectors and block matrices; this will be
characteristic of the distributed implementations that we
consider in this article. To examine the stability and convergence
properties of recursions that involve such block quantities, it
becomes useful to rely on a certain block vector norm. In App.~\ref{app.a},
we describe a so-called {\em block maximum block} and establish some
of its useful properties. The results of the appendix will be used
extensively in our exposition. It is therefore
advisable for the reader to review the properties stated in the
appendix at this stage.

Using the result of Lemma~\ref{Lemma:BlockMaximumNorm.block.left}, we
can establish the following useful statement about the convergence of the steepest-descent
diffusion strategy (\ref{general.1})--(\ref{general.3}). The result establishes that all nodes
end up converging to the optimal solution $w^o$ if the nodes employ positive step-sizes
$\mu_k$ that are small enough; the lemma provides a sufficient bound on the $\{\mu_k\}$.

\begin{theorem} ({\rm {\bf Convergence to Optimal Solution}}) \label{thm.1xx} Consider
the problem of optimizing the global cost (\ref{opt.11}) with the
individual cost functions given by (\ref{gihjas.alk}). Pick a right
stochastic matrix $C$ and left stochastic matrices $A_1$ and $A_2$
satisfying (\ref{dlajs8.12}) or (\ref{run.run.4}); these matrices define the network
topology and how information is shared over neighborhoods. Assume
each node in the network runs the (distributed)
steepest-descent diffusion algorithm (\ref{general.1})--(\ref{general.3}).
Then, all estimates $\{w_{k,i}\}$ across the network converge to the
optimal solution $w^o$ if the positive step-size parameters $\{\mu_k\}$
satisfy \be \addbox{\;\mu_{k}< \frac{2}{\lambda_{\max}(R_k)}\;}
\label{KD8912.ALKD}\ee where the neighborhood covariance matrix
$R_k$ is defined by (\ref{defkajs812}).
\end{theorem}

\bp The weight error vector $\widetilde{w}_i$ converges to zero if,
and only if, the coefficient matrix ${\cal A}_2^T\left(I_{NM}-{\cal
M}{\cal R}\right){\cal A}_1^T$ in (\ref{like.lakd}) is a stable matrix (meaning that all
its eigenvalues lie strictly inside the unit disc). From property
(\ref{kajd8912.xxas}) established in App.~\ref{app.a}, we know that ${\cal
A}_2^T\left(I_{NM}-{\cal M}{\cal R}\right){\cal A}_1^T$ is stable if
the block diagonal matrix $\left(I_{NM}-{\cal M}{\cal R}\right)$ is
stable. It is now straightforward to verify that condition
(\ref{KD8912.ALKD}) ensures the stability of $\left(I_{NM}-{\cal
M}{\cal R}\right)$. It follows that \be
\widetilde{w}_i\longrightarrow 0 \;\;\;\mbox{\rm
as}\;\;\;i\longrightarrow \infty \ee

\ep

\noindent Observe that the stability condition (\ref{KD8912.ALKD})
does not depend on the specific combination matrices $A_1$ and $A_2$. Thus, as long as these matrices are chosen to be left-stochastic, the weight-error vectors will converge to zero under condition (\ref{KD8912.ALKD}) no matter what $\{A_1,A_2\}$ are. Only the combination matrix $C$ influences the condition on the step-size through the neighborhood covariance
matrices $\{R_k\}$. Observe further that the statement of the lemma
does not require the network to be connected. Moreover, when $C=I$,
in which case the nodes only share weight estimates and do not share
the neighborhood moments $\{r_{du,\ell},R_{u,\ell}\}$, as in
(\ref{general.1x})--(\ref{general.3x}), condition
(\ref{KD8912.ALKD}) becomes \be\addbox{\;
\mu_k<\frac{2}{\lambda_{\max}(R_{u,k})}\;}\;\;\;\;\;\;\;\;\;\;\;(\mbox{\rm
cooperation with $C=I$}) \label{90alkad.231}\ee  in terms of the actual covariance matrices
$\{R_{u,k}\}$. Results
(\ref{KD8912.ALKD}) and (\ref{90alkad.231}) are reminiscent of a
classical result for stand-alone steepest-descent algorithms, as in
the non-cooperative case (\ref{no.general.3}), where it is known
that the estimate by each individual node in this case will
converge to $w^o$ if, and only if, its positive step-size satisfies \be
\addbox{\;\mu_k<\frac{2}{\lambda_{\max}(R_{u,k})}\;}\;\;\;\;\;\;\;\;(\mbox{\rm
non-cooperative case (\ref{no.general.3}) with $A_1=A_2=C=I_N$}) \label{90alkad.1}\ee This is the same
condition as (\ref{90alkad.231}) for the case $C=I$.

The following statement provides a {\em bi-directional} statement that
ensures convergence of the (distributed)
steepest-descent diffusion strategy (\ref{general.1})--(\ref{general.3}) for {\em any} choice of left-stochastic
combination matrices $A_1$ and $A_2$.

\begin{theorem} ({\rm {\bf Convergence for Arbitrary Combination Matrices}})\label{thm.2xx} Consider
the problem of optimizing the global cost (\ref{opt.11}) with the
individual cost functions given by (\ref{gihjas.alk}). Pick a right
stochastic matrix $C$ satisfying (\ref{dlajs8.12}). Then, the
estimates $\{w_{k,i}\}$ generated by (\ref{general.1})--(\ref{general.3})
converge to $w^o$, for all choices of left-stochastic
matrices $A_1$ and $A_2$ satisfying (\ref{dlajs8.12}) if,
and only if, \be \addbox{\;\mu_{k}< \frac{2}{\lambda_{\max}(R_k)}\;}
\label{KD8912.ALKD.2}\ee
\end{theorem}
\bp The result follows from property (b) of
Corollary~\ref{coruais.as}, which is established in App.~\ref{app.a}.

\ep

More importantly, we can verify that under fairly general
conditions, employing the steepest-descent diffusion strategy
(\ref{general.1})--(\ref{general.3}) enhances the convergence rate
of the error vector towards zero relative to the non-cooperative strategy
(\ref{no.general.3}). The next three results establish this fact when
$C$ is a doubly stochastic matrix, i.e., it has non-negative entries and satisfies \be
C\mathds{1}=\mathds{1},\quad\quad\quad C^T\mathds{1}=\mathds{1} \label{c.doubdlkajs}\ee
with both its rows and columns adding up to one. Compared to the
earlier right-stochastic condition on $C$ in
(\ref{Equ:ProblemFormulation:C_Condition}), we are now requiring \be
\sum_{\ell\in {\cal N}_k} c_{k\ell} = 1,\quad\quad\quad
\sum_{\ell\in {\cal N}_k} c_{\ell k} = 1 \ee For example, these
conditions are satisfied when $C$ is right stochastic and symmetric.
They are also satisfied for $C=I$, when only weight estimates are
shared as in (\ref{general.1x})--(\ref{general.3x}); this latter case covers
the ATC and CTA diffusion strategies (\ref{without.atc}) and (\ref{without.cta}), which do not
involve information exchange.

\begin{theorem} ({\rm {\bf Convergence Rate is Enhanced: Uniform Step-Sizes}})
\label{lemaa.cod} Consider
the problem of optimizing the global cost (\ref{opt.11}) with the
individual cost functions given by (\ref{gihjas.alk}). Pick a doubly
stochastic matrix $C$ satisfying (\ref{c.doubdlkajs}) and left stochastic matrices $A_1$ and $A_2$
satisfying (\ref{dlajs8.12}). Consider two modes of operation. In
one mode, each node in the network runs the (distributed)
steepest-descent diffusion algorithm
(\ref{general.1})--(\ref{general.3}). In the second mode, each node
operates individually and runs the non-cooperative steepest-descent
algorithm (\ref{no.general.3}). In both cases, the positive step-sizes used
by all nodes are assumed to be the same, say, $\mu_k=\mu$ for all
$k$, and the value of $\mu$ is chosen to satisfy the required
stability conditions (\ref{KD8912.ALKD}) and (\ref{90alkad.1}), which are met by selecting
\be \mu
\;<\;\min_{1\leq k\leq N}\;\left\{\frac{2}{\lambda_{\max}(R_{u,k})}\right\}
\label{condiuad}\ee It then holds that the magnitude of the error
vector, $\|\widetilde{w}_i\|,$ in the diffusion case decays to zero
more rapidly than in the non-cooperative case. In other words,
diffusion cooperation enhances convergence rate.
\end{theorem}

\bp Let us first establish that any positive step-size $\mu$ satisfying (\ref{condiuad}) will satisfy both
stability conditions (\ref{KD8912.ALKD}) and (\ref{90alkad.1}). It is obvious that
(\ref{90alkad.1}) is satisfied. We verify that (\ref{KD8912.ALKD}) is also satisfied when $C$ is doubly stochastic.
In this case, each neighborhood covariance matrix, $R_k$, becomes a convex
combination of individual covariance matrices $\{R_{u,\ell}\}$, i.e.,
\[
R_k=\sum_{\ell\in{\cal N}_k}c_{\ell k}R_{u,\ell}
\]
where now
\[
\sum_{\ell\in{\cal N}_k} c_{\ell
k}\;=\;1\quad\quad\;\;\;\;(\mbox{\rm when $C$ is doubly stochastic})
\]
To proceed, we
recall that the spectral norm (maximum singular value) of any matrix $X$
is a convex function of $X$ \cite{boyd}. Moreover, for Hermitian
matrices $X$, their spectral norms coincide with their spectral radii
(largest eigenvalue magnitude). Then, Jensen's inequality
 \cite{boyd} states that for any convex function $f(\cdot)$ it holds
that
\[
f\left(\sum_m \theta_m X_m\right)\;\leq \;\sum_m \theta_m f(X_m)
\]
for Hermitian matrices $X_m$ and nonnegative scalars $\theta_m$ that satisfy
\[
\sum_{m}\theta_m =1
\]
Choosing $f(\cdot)$ as the spectral radius function, and applying it to the
definition of $R_k$ above, we get
\bqn
\rho(R_k)&=&\rho\left(\sum_{\ell\in{\cal N}_k} c_{\ell k}R_{u,\ell}\right)\\
&\leq &\sum_{\ell\in{\cal N}_k} c_{\ell k}\cdot\rho(R_{u,\ell})\\
&\leq &\sum_{\ell\in{\cal N}_k} c_{\ell k}\cdot\left[\max_{1\leq \ell\leq N}\rho(R_{u,\ell})\right]\\
&= &\max_{1\leq \ell\leq N}\rho(R_{u,\ell})
\eqn
In other words,
\[
\lambda_{\max}(R_k)\;\leq\;\max_{1\leq k\leq N}\;\left\{\lambda_{\max}(R_{u,k})\right\}
\]
It then follows from (\ref{condiuad})  that
\[
\mu\;<\;\frac{2}{\lambda_{\max}(R_k)},\;\;\;\;\mbox{\rm for all $k=1,2,\ldots,N$}
\]
so that (\ref{KD8912.ALKD}) is satisfied as well.

Let us now examine the convergence rate. To begin with, we note that the matrix $\left(I_{NM}-{\cal M}{\cal
R}\right)$ that appears in the weight-error recursion (\ref{like.lakd})
is block diagonal:
\[
\left(I_{NM}-{\cal M}{\cal R}\right)=\mbox{\rm diag}\{(I_{M}-\mu
R_{1}),\;(I_{M}-\mu R_{2}),\ldots,\;(I_{M}-\mu R_{N})\}
\]
and each individual block entry, $(I_M-\mu
R_{k})$, is a stable matrix since  $\mu$ satisfies
(\ref{KD8912.ALKD}). Moreover,  each of these entries can be written as
\[
I_M-\mu R_k\;=\;\sum_{\ell\in{\cal N}_k} c_{\ell k}(I_M-\mu
R_{u,\ell})
\]
which expresses $(I_M-\mu R_k)$  as a convex combination of stable terms
$(I_M-\mu R_{u,\ell})$. Applying  Jensen's inequality again we get
\[
\rho\left(\sum_{\ell\in{\cal N}_k} c_{\ell k}(I_M-\mu
R_{u,\ell})\right)\;\leq\; \sum_{\ell\in{\cal N}_k} c_{\ell
k}\;\rho(I_M-\mu R_{u,\ell})
\]
Now, we know from (\ref{like.lakd}) that the rate of decay of
$\widetilde{w}_i$ to zero in the diffusion case is determined by the
spectral radius of the coefficient matrix ${\cal
A}_2^T\left(I_{NM}-{\cal M}{\cal R}\right){\cal A}_1^T$. Likewise,
we know from (\ref{like.lakd.no}) that the rate of decay of
$\widetilde{w}_i$ to zero in the non-cooperative case is determined
by the spectral radius of the coefficient matrix
$\left(I_{NM}-{\cal M}{\cal R}_u\right)$. Then, note that

\bqn \rho\left({\cal A}_2^T\left(I_{NM}-{\cal M}{\cal
R}\right){\cal
A}_1^T\;\right)&\stackrel{(\ref{kajd8912.xxas})}{\leq}&
\rho(I_{NM}-{\cal M}{\cal R})\nn\\
&=&
\max_{1\leq k\leq N}\; \rho\left(I_{M}-\mu R_k\right)\nn\\
&=& \max_{1\leq k\leq N} \rho\left(\sum_{\ell\in{\cal N}_k} c_{\ell
k}(I_M-\mu R_{u,\ell})\right)\nn\\
&\leq & \max_{1\leq k\leq N} \sum_{\ell\in{\cal N}_k} c_{\ell
k}\;\rho(I_M-\mu R_{u,\ell})\nn\\
&\leq & \max_{1\leq k\leq N} \sum_{\ell\in{\cal N}_k} c_{\ell
k}\;\left(\max_{1\leq \ell\leq N}\rho(I_M-\mu R_{u,\ell})\right)\nn\\
 &= & \max_{1\leq k\leq N}\left\{
\left(\max_{1\leq \ell\leq N} \rho(I_M-\mu R_{u,\ell})\right)\cdot \sum_{\ell\in{\cal N}_k} c_{\ell k}\right\} \nn\\
 &= & \max_{1\leq k\leq N}
\left(\max_{1\leq \ell\leq N} \rho(I_M-\mu R_{u,\ell})\right)\nn\\
 &= &\max_{1\leq \ell\leq N} \rho(I_M-\mu R_{u,\ell})\\
  &= &\rho(I_{NM}-{\cal M}{\cal R}_u)
 \eqn
Therefore, the spectral radius of ${\cal A}_2^T\left(I_{NM}-{\cal
M}{\cal R}\right){\cal A}_1^T$ is at most as large as the largest
individual spectral radius in the non-cooperative case.

\ep

\noindent The argument can be modified to handle different
step-sizes across the nodes if we assume uniform covariance data across the network, as stated below.

\begin{theorem} ({\rm {\bf Convergence Rate is Enhanced: Uniform Covariance Data}}) \label{lemaa.cod.2x} Consider
the same setting of Theorem~\ref{lemaa.cod}.  Assume the covariance data are uniform across all nodes, say,
$R_{u,k}=R_u$ is independent of $k$. Assume further that the
nodes in both modes of operation employ steps-sizes $\mu_k$ that are chosen to satisfy
the required stability conditions (\ref{KD8912.ALKD}) and (\ref{90alkad.1}), which in this case
are met by:
\be
\mu_k \;<\;\frac{2}{\lambda_{\max}(R_{u})},\;\;\;\;\;k=1,2,\ldots,N
\label{condiuad.2}\ee  It then holds that the magnitude of the error
vector, $\|\widetilde{w}_i\|,$ in the diffusion case decays to zero
more rapidly than in the non-cooperative case. In other words,
diffusion enhances convergence rate.
\end{theorem}

\bp Since $R_{u,\ell}=R_u$ for all $\ell$ and $C$ is doubly stochastic, we get $R_k=R_u$ and $
I_{NM}-{\cal M}{\cal R}=I_{NM}-{\cal M}{\cal R}_u$. Then,
\bqn \rho\left({\cal A}_2^T\left(I_{NM}-{\cal M}{\cal
R}\right){\cal
A}_1^T\;\right)&\stackrel{(\ref{kajd8912.xxas})}{\leq}&
\rho(I_{NM}-{\cal M}{\cal R})\nn\\
&=&
\rho(I_{NM}-{\cal M}{\cal R}_u)
 \eqn

\ep

\noindent The next statement considers the case of ATC and CTA strategies
 (\ref{without.atc}) and (\ref{without.cta}) without information exchange, which correspond to
 the case $C=I_N$. The result establishes that these strategies always enhance the convergence rate over the non-cooperative case,
 without the need to assume uniform step-sizes or uniform covariance data.

\begin{theorem} ({\rm {\bf Convergence Rate is Enhanced when $\bm{C}=\bm{I}$}}) \label{lemaa.cod.3}
Consider the problem of optimizing the global cost (\ref{opt.11}) with the
individual cost functions given by (\ref{gihjas.alk}). Pick left stochastic matrices $A_1$ and $A_2$
satisfying (\ref{dlajs8.12}) and set $C=I_N$. This situation covers the ATC and CTA strategies
 (\ref{without.atc}) and (\ref{without.cta}), which do not involve information exchange.
Consider two modes of operation. In
one mode, each node in the network runs the (distributed)
steepest-descent diffusion algorithm
(\ref{general.1x})--(\ref{general.3x}). In the second mode, each node
operates individually and runs the non-cooperative steepest-descent
algorithm (\ref{no.general.3}). In both cases, the positive step-sizes are chosen
 to satisfy the required
stability conditions (\ref{90alkad.231}) and (\ref{90alkad.1}), which in this case are met by
 \be \mu_k
\;<\;\frac{2}{\lambda_{\max}(R_{u,k})},\;\;\;\;\;k=1,2,\ldots,N
\label{condiuad.3}\ee It then holds that the magnitude of the error
vector, $\|\widetilde{w}_i\|,$ in the diffusion case decays to zero
more rapidly than in the non-cooperative case. In other words,
diffusion cooperation enhances convergence rate.
\end{theorem}

\bp When $C=I_N$, we get $R_k=R_{u,k}$ and, therefore, ${\cal R}={\cal R}_u$ and  $I_{NM}-{\cal M}{\cal R}=
I_{NM}-{\cal M}{\cal R}_u$. Then,
\bqn \rho\left({\cal A}_2^T\left(I_{NM}-{\cal M}{\cal
R}\right){\cal
A}_1^T\;\right)&\stackrel{(\ref{kajd8912.xxas})}{\leq}&
\rho(I_{NM}-{\cal M}{\cal R})\nn\\
&=&
\rho(I_{NM}-{\cal M}{\cal R}_u)
 \eqn
\ep

\noindent The results of the previous theorems highlight the
following important facts about the role of the combination matrices
$\{A_1,A_2,C\}$ in the convergence behavior of the diffusion strategy
(\ref{general.1})--(\ref{general.3}):
\begin{itemize}
\item[(a)] The matrix $C$ influences the stability of the network
through its influence on the bound
in (\ref{KD8912.ALKD}). This is because the matrices $\{R_k\}$ depend on the entries of $C$.
The matrices $\{A_1,A_2\}$ do not influence network stability.

\item[(b)] The matrices $\{A_1,A_2,C\}$ influence the
rate of convergence of the network since they influence the spectral
radius of the matrix ${\cal A}_2^T\left(I_{NM}-{\cal M}{\cal
R}\right){\cal A}_1^T$, which controls the dynamics of the weight
error vector in (\ref{like.lakd}).

\end{itemize}

\section{Performance of Adaptive Diffusion
Strategies}\label{sec.mse}
We now move on to examine the behavior of the {\em adaptive} diffusion
implementations
(\ref{Equ:DiffusionAdaptation:ATC.adaptive})--(\ref{Equ:DiffusionAdaptation:CTA.adaptive}),
and the influence of both gradient noise and measurement noise on
convergence and steady-state performance. Due to the random nature
of the perturbations, it becomes necessary to evaluate the behavior
of the algorithms on average, using mean-square convergence
analysis. For this reason, we shall study the convergence of the
weight estimates both in the mean and mean-square
sense. To do so, we will again consider a general diffusion
structure that includes the ATC and CTA strategies
(\ref{Equ:DiffusionAdaptation:ATC.adaptive})--(\ref{Equ:DiffusionAdaptation:CTA.adaptive})
as special cases. We shall further resort to the boldface
notation to refer to the measurements and weight estimates in order to
highlight the fact that they are now being treated as random
variables. In this way, the update equations becomes stochastic
updates. Thus, consider the following general adaptive diffusion
strategy for $i\geq 0$:
 \bq
                \bm{\phi}_{k,i-1}   &=&  \displaystyle\sum_{\ell\in{\cal N}_k} a_{1,\ell k}\; \w_{\ell,i-1}   \label{general.1ada}\\
                \bm{\psi}_{k,i}      &=&    \bm{\phi}_{k,i-1}\; +\;
                \mu_k \displaystyle\sum_{\ell\in{\cal N}_k} c_{\ell
                k}\;\u_{\ell,i}^*\left[\d_{\ell}(i)-\u_{\ell,i}\;\bm{\phi}_{k,i-1}\right]\label{general.2ada}\\
                \w_{k,i}     &=&    \displaystyle\sum_{\ell\in{\cal N}_k} a_{2,\ell k}\;
                \bm{\psi}_{\ell,i}\label{general.3ada}
  \eq
As before,  the scalars $\{a_{1,\ell k},c_{\ell k},a_{2,\ell k}\}$
are non-negative real coefficients corresponding to the $(\ell,k)$
entries of $N\times N$ combination matrices $\{A_1,C,A_2\}$,
respectively. These matrices are assumed to satisfy the same
conditions (\ref{dlajs8.12}) or (\ref{run.run.4}). Again, different choices for
$\{A_1,C,A_2\}$ correspond to different cooperation modes. For
example, the choice $A_1 = I_N$ and $A_2=A$ corresponds to the
adaptive ATC implementation
(\ref{Equ:DiffusionAdaptation:ATC.adaptive}), while the choice $A_1
= A$ and $A_2 = I_N$ corresponds to the adaptive CTA implementation
(\ref{Equ:DiffusionAdaptation:CTA.adaptive}). Likewise, the choice
$C=I_N$ corresponds to the case in which the nodes only share weight
estimates and the distributed diffusion recursions
(\ref{general.1ada})--(\ref{general.3ada}) become
 \bq
                \bm{\phi}_{k,i-1}   &=&  \displaystyle\sum_{\ell\in{\cal N}_k} a_{1,\ell k}\; \w_{\ell,i-1}   \label{general.1adax}\\
                \bm{\psi}_{k,i}      &=&    \bm{\phi}_{k,i-1} +
                \mu_k \u_{k,i}^*\left[\d_{k}(i)-\u_{k,i}\;\bm{\phi}_{k,i-1}\right]\\
                \w_{k,i}     &=&    \displaystyle\sum_{\ell\in{\cal N}_k} a_{2,\ell k}\;
                \bm{\psi}_{\ell,i}\label{general.3adax}
  \eq
Furthermore, the choice $A_1=A_2=C=I_N$ corresponds to the
non-cooperative mode of operation, where each node runs the classical (stand-alone)
least-mean-squares (LMS)
filter independently of the other nodes: \cite{Sayed03,Sayed08,Haykin01,Widrow85}:\bq
                \w_{k,i}      &=&    \w_{k,i-1} +
                \mu_k \u_{k,i}\left[\d_{k}(i)-\u_{k,i}\;\w_{k,i-1}\right],\;\;\;i\geq 0\label{no.general.3xx}
  \eq

\subsection{Data Model}\label{sec.daada}
When we studied the performance of the steepest-descent diffusion
strategy (\ref{general.1})--(\ref{general.3}) we exploited result
(\ref{jakd8127}), which indicated how the moments
$\{r_{du,k},R_{u,k}\}$ that appeared in the recursions
related to the optimal solution $w^o$.
Likewise, in order to be able to analyze the performance of the
{\em adaptive} diffusion strategy
(\ref{general.1ada})--(\ref{general.3ada}),  we need to
know how the data $\{\d_k(i),\u_{k,i}\}$ across the network relate to
$w^o$. Motivated
by the several examples presented earlier in Sec.~\ref{sec.examples}, we shall assume that the
data satisfy a linear model of the form:\be \addbox{\;\d_k(i)
= \u_{k,i}w^o\;+\;\v_k(i)\;}\label{lkad8912.lakd.ada}\ee where
$\v_k(i)$ is measurement noise with variance $\sigma_{v,k}^2$: \be
\sigma_{v,k}^2\define \Ex|\v_k(i)|^2 \ee
and where the stochastic processes $\{\d_k(i),\u_{k,i}\}$
are assumed to be jointly wide-sense stationary with moments:
\bq
\sigma_{d,k}^2&\define& \Ex|\d_k(i)|^2\;\;\;\;\;\;\;\;\;\;\;\;\;\;\;(\mbox{\rm scalar})\\
R_{u,k}&\define& \Ex\u_{k,i}^*\u_{k,i}>0\;\;\;\;\;\;\;\;(M\times M)\label{rdu.1.ada}\\
r_{du,k}&\define&
\Ex\d_k(i)\u_{k,i}^*\;\;\;\;\;\;\;\;\;\;\;\;\;(M\times 1)
\label{rdu.2.ada}\eq All variables are assumed to be zero-mean.
Furthermore, the noise process $\{\v_k(i)\}$ is assumed to be
temporally white and spatially independent, as described earlier by (\ref{kaldja.alks}),
namely, \be \left\{\begin{array}{l} \Ex
\v_{k}(i)\v_{k}^*(j)\;=\;0,\;\;\mbox{\rm for
all $i\neq j$\;(temporal whiteness)}\\
\Ex \v_{k}(i)\v_{m}^*(j)=\;0,\;\;\mbox{\rm for all $i,j$ whenever
$k\neq m$\;(spatial
whiteness)}\end{array}\right.\label{kaldja.ada.alks}\ee The noise
process $\v_k(i)$ is further assumed to be independent of the
regression data $\u_{m,j}$ for all $k,m$ and $i,j$ so that: \be
\Ex\v_{k}(i)\u_{m,j}^*=0,\quad\;\;\;\;\mbox{\rm for all $k,m,i,j$}
\ee We shall also assume that the regression data are temporally white
and spatially independent so that: \be
\Ex\u_{k,i}^*\u_{\ell,j}\;=\;R_{u,k}\delta_{k\ell}\delta_{ij} \label{lkad8912.lakd.ada.1}\ee
Although we are going to derive performance measures for the network
under this independence assumption on the regression data, it turns
out that the resulting expressions continue to match well with
simulation results for sufficiently small step-sizes, even when the
independence assumption does not hold (in a manner similar to the behavior of stand-alone adaptive filters)
\cite{Sayed03,Sayed08}.

\subsection{Performance Measures}
Our objective is to analyze whether, and how fast, the weight estimates
$\{\w_{k,i}\}$ from the adaptive diffusion implementation
(\ref{general.1ada})--(\ref{general.3ada}) converge towards $w^o$.
To do so, we again introduce  the $M\times 1$ weight error vectors: \bq
\widetilde{\bm{\phi}}_{k,i}&
\define &w^o-\bm{\phi}_{k,i}\\
        \widetilde{\bm{\psi}}_{k,i}  &\define&   w^o-\bm{\psi}_{k,i}\\
        \widetilde{\w}_{k,i} &\define &  w^o - \w_{k,i}\eq
Each of these error vectors measures the residual relative to the
desired  $w^o$ in (\ref{lkad8912.lakd.ada}). We further introduce two scalar error
measures: \bq \e_k(i)&\define& \d_k(i)-\u_{k,i}\w_{k,i-1}\;\;\;\;\;(\mbox{\rm output error})\\
\e_{a,k}(i)&\define
&\u_{k,i}\widetilde{\w}_{k,i-1}\;\;\;\;\;\;\;\;\;\;\;\;\;\;\;\;\;(\mbox{\rm
{\em a-priori} error})\eq The first error measures how well the term
$\u_{k,i}\w_{k,i-1}$ approximates the measured data, $\d_k(i)$; in view of (\ref{lkad8912.lakd.ada}),
this
error can be interpreted as an estimator for the noise term
$\v_k(i)$. If node $k$ is able to estimate $w^o$ well, then $\e_k(i)$ would
get close to $\v_k(i)$. Therefore, under ideal conditions, we would
expect the variance of $\e_k(i)$ to tend towards the variance of
$\v_k(i)$. However, as remarked earlier in (\ref{earlier.alkda}),
there is generally an offset term for adaptive implementations that is
captured by the variance of the {\em
a-priori} error, $\e_{a,k}(i)$. This second error measures how well
 $\u_{k,i}\w_{k,i-1}$ approximates the uncorrupted term
$\u_{k,i}w^o$. Using  the data model (\ref{lkad8912.lakd.ada}), we
can relate $\{\e_k(i),\e_{a,k}(i)\}$ as \bq \e_k(i)&=& \e_{a,k}+\v_k(i)\eq
Since the noise component, $\v_k(i)$, is assumed to be zero-mean and
independent of all other random variables, we recover (\ref{earlier.alkda}): \be
\addbox{\;\Ex|\e_k(i)|^2\;=\;\Ex|\e_{a,k}(i)|^2\;+\;\sigma_{v,k}^2\;}
\ee This relation confirms that the variance of the output error,
$\e_k(i),$ is always at least as large as $\sigma_{v,k}^2$ and away
from it by an amount that is equal to the variance of the {\em
a-priori} error, $\e_{a,k}(i)$. Accordingly,
in order to quantify the performance of any particular node in the
network, we define the mean-square-error (MSE) and
excess-mean-square-error (EMSE) for node $k$ as
the following steady-state measures: \bq
{\rm MSE}_k&\define& \lim_{i\rightarrow\infty}\Ex|\e_k(i)|^2\\
{\rm EMSE}_k&\define& \lim_{i\rightarrow\infty}\Ex|\e_{a,k}(i)|^2 \eq Then, it holds
that \be \addbox{\;{\rm MSE}_k\;=\;{\rm
EMSE}_k\;+\;\sigma_{v,k}^2\;} \ee Therefore, the EMSE term
quantifies the size of the offset in the MSE performance of each node.
We also define the mean-square-deviation (MSD) of each node as the
steady-state measure: \be {\rm MSD}_k\define
\lim_{i\rightarrow\infty}\Ex\|\widetilde{\w}_{k,i}\|^2 \ee which measures how far
$\w_{k,i}$ is from $w^o$ in the mean-square-error sense.

We indicated earlier in
(\ref{lkad891.21lk2})--(\ref{lkad891.21lk2.1}) how the MSD and EMSE
of stand-alone LMS filters in the non-cooperative case depend on
$\{\mu_k,\sigma_{v}^2,R_{u,k}\}$. In this section, we examine how
cooperation among the nodes influences their performance. Since cooperation
couples the operation of the nodes, with data originating from one node
influencing the behavior of its neighbors and their neighbors, the study of the network
performance requires more effort than in the non-cooperative case.
Nevertheless, when all is said and done, we will arrive at
expressions that approximate well the network performance and reveal some interesting
conclusions.

\subsection{Error Recursions}\label{msnkajd}
Using the data model (\ref{lkad8912.lakd.ada})  and subtracting
$w^o$ from both sides of the relations in
\eqref{general.1ada}--\eqref{general.3ada} we get \bq
                \widetilde{\bm{\phi}}_{k,i-1}   &=&  \displaystyle\sum_{\ell\in{\cal N}_k} a_{1,\ell k}\; \widetilde{\w}_{\ell,i-1}   \label{general.1a.ada}\\
                \widetilde{\bm{\psi}}_{k,i}      &=&   \left(I_M-\mu_k \sum_{\ell\in{\cal N}_k} c_{\ell
                k}\;\u_{\ell,i}^*\u_{\ell,i}\right)\;\widetilde{\bm{\phi}}_{k,i-1}\;-\;\mu_k\sum_{\ell\in{\cal N}_k} c_{\ell k}\;\u_{\ell,i}^*\v_{\ell}(i)\\
                \widetilde{\w}_{k,i}     &=&    \displaystyle\sum_{\ell\in{\cal N}_k} a_{2,\ell k}\;
                \widetilde{\bm{\psi}}_{\ell,i}\label{general.3a.ada}
  \eq
Comparing the second recursion with the corresponding recursion in
the steepest-descent case (\ref{general.1a})--(\ref{general.3a}), we
see that two new effects arise: the effect of gradient noise, which
replaces the covariance matrices $R_{u,\ell}$ by the instantaneous
approximation $\u_{\ell,i}^*\u_{\ell,i}$, and the effect of
measurement noise, $\v_{\ell}(i)$.

We again describe the above relations more compactly by collecting
the information from across the network in block vectors and
matrices. We collect the error vectors from across all nodes into
the following $N\times 1$ {\em block} vectors, whose individual
entries are of size $M\times 1$ each:\be
\widetilde{\bm{\psi}}_i\define
\ba{c}\widetilde{\bm{\psi}}_{1,i}\\
\widetilde{\bm{\psi}}_{2,i}\\\vdots\\
\widetilde{\bm{\psi}}_{N,i}\ea,\quad\quad
\widetilde{\bm{\phi}}_i\define
\ba{c}\widetilde{\bm{\phi}}_{1,i}\\
\widetilde{\bm{\phi}}_{2,i}\\\vdots\\
\widetilde{\bm{\phi}}_{N,i}\ea,\quad\quad \widetilde{\w}_i\define
\ba{c}\widetilde{\w}_{1,i}\\
\widetilde{\w}_{2,i}\\\vdots\\ \widetilde{\w}_{N,i}\ea\label{ahdak819aldka}\ee The block
quantities
$\{\widetilde{\bm{\psi}}_i,\widetilde{\bm{\phi}}_i,\widetilde{\w}_i\}$
represent the state of the errors across the network at time $i$.
Likewise, we introduce the following $N\times N$ {\em block}
diagonal matrices, whose individual entries are of size $M\times M$
each: \bq {\cal M}&\define& \mbox{\rm diag}\{\;\mu_1
I_M,\;\mu_2I_M,\;\ldots,\;\mu_N I_M\;\}\label{m.ladlk}\\
 \bm{\cal R}_i&\define&\mbox{\rm diag}\left\{\;\sum_{\ell\in{\cal N}_1} c_{\ell
                1}\;\u_{\ell,i}^*\u_{\ell,i},\;\displaystyle\sum_{\ell\in{\cal N}_2} c_{\ell
                2}\;\u_{\ell,i}^*\u_{\ell,i},\;\ldots,\;\sum_{\ell\in{\cal N}_N} c_{\ell
                N}\;\u_{\ell,i}^*\u_{\ell,i}\;\right\}
\eq Each block diagonal entry of $\bm{\cal R}_i$, say, the $k$-th
entry, contains a combination of rank-one regression terms collected from
 the neighborhood of node $k$. In this way, the matrix
$\bm{\cal R}_i$ is now stochastic {\em and} dependent on time, in
contrast to the matrix ${\cal R}$ in the steepest-descent case in
(\ref{r.dklajda}), which was a constant matrix. Nevertheless, it holds that
 \be \addbox{\;\Ex\bm{\cal R}_i\;=\;{\cal R}\;} \ee
so that, on average, $\bm{\cal R}_i$ agrees with ${\cal R}$. We can simplify the
notation by denoting the neighborhood combinations as follows: \be
\R_{k,i}\define \sum_{\ell\in{\cal N}_k} c_{\ell
                k}\;\u_{\ell,i}^*\u_{\ell,i}
\label{defkajs812.ada}\ee so that $\bm{\cal R}_i$ becomes \be
\bm{\cal R}_i\define\mbox{\rm
diag}\left\{\;\R_{1,i},\;\R_{2,i},\;\ldots,\R_{N,i}\;\right\}\;\;\;\;\;(\mbox{\rm
when $C\neq I$})\label{compadlkas.12}\ee Again, compared with the
matrix $R_k$ defined in (\ref{defkajs812}), we find that $\R_{k,i}$
is now both stochastic and time-dependent. Nevertheless, it again holds
that \be \addbox{\;\Ex\R_{k,i}\;=\;R_k\;} \ee  In the special case when $C=I$, the matrix
$\bm{\cal R}_i$ reduces to \be \bm{\cal R}_{u,i}\define \mbox{\rm
diag}\{\u_{1,i}^*\u_{1,i},\;\u_{2,i}^*\u_{2,i},\;\ldots,\;\u_{N,i}^*\u_{N,i}\}\;\;\;\;\;(\mbox{\rm
when $C=I$}) \label{replaced.ada}\ee with  \be \addbox{\;\Ex\bm{\cal
R}_{u,i}\;=\;{\cal R}_u \;}\ee where ${\cal R}_u$ was defined earlier in
(\ref{replaced}).

We further introduce the following $N\times 1$ block column vector,
whose entries are of size $M\times 1$ each: \bq {\s}_i&\define&\mbox{\rm
col}\{\;\u_{1,i}^*\v_1(i),\;\u_{2,i}^*\v_2(i),\ldots,\;\u_{N,i}^*\v_{N}(i)\;\}
\label{dkja891.lakdlk}\eq Obviously, given that the regression data
and measurement noise are zero-mean and independent of each other,
we have \be \addbox{\;\Ex{\s}_i\;=\;0\;} \ee and the covariance
matrix of $\s_i$ is $N\times N$ block diagonal with blocks of size
$M\times M$:\be \addbox{\;{\cal S}\define
\Ex\s_i\s_i^*\;=\;\mbox{\rm diag}\{\sigma_{v,1}^2
R_{u,1},\;\sigma_{v,2}^2 R_{u,2},\;\ldots,\sigma_{v,N}^2 R_{u,N}
\}\;} \label{defl.sa} \ee Returning to
(\ref{general.1a.ada})--(\ref{general.3a.ada}), we conclude that the
following relations hold for the block quantities:\bq
                \widetilde{\bm{\phi}}_{i-1}   &=&  {\cal A}_1^T \widetilde{\w}_{i-1}   \label{general.1bx}\\
                \widetilde{\bm{\psi}}_{i}      &=&   \left(I_{NM}-{\cal M}\bm{\cal R}_i\right)\;\widetilde{\bm{\phi}}_{i-1}\;-\;
{\cal M} {\cal C}^T\s_i
                \\
                \widetilde{\w}_{i}     &=&    {\cal A}_2^T
                \widetilde{\bm{\psi}}_{i}\label{general.3bx}
  \eq
  where
  \be
\addbox{\;{\cal C}\define C\otimes I_M\;} \label{cclakdla}
  \ee
so that the network weight error vector, $\widetilde{\w}_i$, ends up
evolving according to the following {\em stochastic} recursion:\be
\addbox{\;\widetilde{\w}_i={\cal A}_2^T\left(I_{NM}-{\cal M}\bm
{\cal R}_i\right){\cal A}_1^T\widetilde{\w}_{i-1}\;-\;{\cal
A}_2^T{\cal M}{\cal C}^T\s_i,\;\;i\geq 0\;}\;\;\;\;\;(\mbox{\rm
diffusion strategy}) \label{like.lakd.ada}\ee For comparison
purposes, if each node operates individually and uses the
non-cooperative LMS recursion (\ref{no.general.3xx}), then the
weight error vector across all $N$ nodes would evolve according to
the following stochastic recursion: \be
\addbox{\;\widetilde{\w}_i=\;\left(I_{NM}-{\cal M}\bm{\cal
R}_{u,i}\right)\widetilde{\w}_{i-1}\;-\;{\cal M}\s_i,\;\;i\geq
0\;}\;\;\;\;\;(\mbox{\rm non-cooperative strategy})
\label{like.lakd.no.ada}\ee where the matrices ${\cal A}_1$ and
${\cal A}_2$ do not appear, and $\bm{\cal R}_i$ is replaced by
$\bm{\cal R}_{u,i}$ from (\ref{replaced.ada}).

\subsection{Convergence in the Mean}
 Taking expectations of both sides of (\ref{like.lakd.ada}) we find that:
 \be
\addbox{\;\Ex\widetilde{\w}_i={\cal A}_2^T\left(I_{NM}-{\cal M}
{\cal R}\right){\cal
A}_1^T\cdot\Ex\widetilde{\w}_{i-1},\;\;i\geq
0\;}\;\;\;\;\;(\mbox{\rm diffusion strategy})
\label{like.lakd.ada.2}\ee where we used the fact that
$\widetilde{\w}_{i-1}$ and $\bm{\cal R}_{i}$ are independent of each
other in view of our earlier assumptions on the regression data and noise in Sec.~\ref{sec.daada}.
Comparing with the error recursion (\ref{like.lakd}) in the
steepest-descent case, we find that both recursions are identical with
$\widetilde{w}_i$ replaced by $\Ex\widetilde{\w}_i$.
Therefore, the convergence statements from the steepest-descent case
can be extended to the adaptive case to provide conditions on the
step-size to ensure stability in the mean, i.e., to ensure \be
\Ex\widetilde{\w}_i\longrightarrow 0\;\;\;\mbox{\rm
as}\;\;\;\;i\longrightarrow \infty \label{fguadl12}\ee When (\ref{fguadl12}) is guaranteed,
we would say that the adaptive
diffusion solution is asymptotically unbiased. The following statements restate
 the results of  Theorems~\ref{thm.1xx}--\ref{lemaa.cod.3}
 in the context of mean error analysis.

\begin{theorem} ({\rm {\bf Convergence in the Mean}}) \label{kald8912.lkas} Consider
the problem of optimizing the global cost (\ref{opt.11}) with the
individual cost functions given by (\ref{gihjas.alk}). Pick a right
stochastic matrix $C$ and left stochastic matrices $A_1$ and $A_2$
satisfying (\ref{dlajs8.12}) or (\ref{run.run.4}). Assume each node in the network measures data that satisfy
the conditions described in Sec.~\ref{sec.daada}, and runs
the adaptive diffusion algorithm
(\ref{general.1ada})--(\ref{general.3ada}). Then, all estimators
$\{\w_{k,i}\}$ across the network converge in the mean to the
optimal solution $w^o$ if the positive step-size parameters $\{\mu_k\}$
satisfy \be \addbox{\;\mu_{k}< \frac{2}{\lambda_{\max}(R_k)}\;}
\label{KD8912.ALKD.ada}\ee where the neighborhood covariance matrix
$R_k$ is defined by (\ref{defkajs812}). In other words, $\Ex\w_{k,i}\rightarrow w^o$ for all
nodes $k$ as $i\rightarrow\infty$.
\end{theorem}

\qd

 \noindent Observe again that the mean stability condition
(\ref{KD8912.ALKD.ada}) does not depend on the specific combination
matrices $A_1$ and $A_2$ that are being used. Only the combination
matrix $C$ influences the condition on the step-size through the
neighborhood covariance matrices $\{R_k\}$. Observe further that the
statement of the lemma does not require the network to be connected.
Moreover, when $C=I_N$, in which case the nodes only share weight
estimators and do not share neighborhood data
$\{\d_{\ell}(i),\u_{\ell,i}\}$ as in
(\ref{general.1adax})--(\ref{general.3adax}), condition
(\ref{KD8912.ALKD.ada}) becomes \be\addbox{\;
\mu_k<\frac{2}{\lambda_{\max}(R_{u,k})}\;}\;\;\;\;\;\;\;\;(\mbox{\rm
adaptive cooperation with $C=I_N$}) \label{90alkad.231.ada}\ee  Results (\ref{KD8912.ALKD.ada})
and (\ref{90alkad.231.ada}) are reminiscent
of a classical result for the stand-alone LMS algorithm, as in the
non-cooperative case (\ref{no.general.3xx}), where it is known that
the estimator by each individual node in this case would converge in
the mean to $w^o$ if, and only if, its step-size satisfies \be
\addbox{\;\mu_k<\frac{2}{\lambda_{\max}(R_{u,k})}\;}\;\;\;\;\;\;\;\;(\mbox{\rm
non-cooperative adaptation}) \label{90alkad.1.ada}\ee
The following statement provides a bi-directional result that
ensures the mean convergence of the adaptive diffusion strategy for {\em
any} choice of left-stochastic combination matrices $A_1$ and $A_2$.

\begin{theorem} ({\rm {\bf Mean Convergence for Arbitrary Combination Matrices}}) Consider
the problem of optimizing the global cost (\ref{opt.11}) with the
individual cost functions given by (\ref{gihjas.alk}). Pick a right
stochastic matrix $C$ satisfying (\ref{dlajs8.12}). Assume each node in the network measures data that satisfy
the conditions described in Sec.~\ref{sec.daada}. Then, the
estimators $\{\w_{k,i}\}$ generated by the adaptive diffusion strategy
(\ref{general.1ada})--(\ref{general.3ada}),
converge in the mean to $w^o$, for all
choices of left stochastic matrices $A_1$ and $A_2$ satisfying
(\ref{dlajs8.12}) if, and only if, \be \addbox{\;\mu_{k}<
\frac{2}{\lambda_{\max}(R_k)}\;} \label{KD8912.ALKD.2.ada}\ee
\end{theorem}
\qd

\noindent As was the case with steepest-descent diffusion strategies,
the adaptive diffusion strategy
(\ref{general.1ada})--(\ref{general.3ada}) also enhances the
convergence rate of the mean of the error vector towards zero relative to
the non-cooperative strategy (\ref{no.general.3xx}). The next results
restate Theorems~\ref{lemaa.cod}--\ref{lemaa.cod.3}; they assume $C$ is a
doubly stochastic matrix.

\begin{theorem} ({\rm {\bf Mean Convergence Rate is Enhanced: Uniform Step-Sizes}})
\label{lemaa.cod.2} Consider
the problem of optimizing the global cost (\ref{opt.11}) with the
individual cost functions given by (\ref{gihjas.alk}). Pick a doubly
stochastic matrix $C$ satisfying (\ref{c.doubdlkajs}) and left stochastic matrices $A_1$ and $A_2$
satisfying (\ref{dlajs8.12}). Assume each node in the network measures data that satisfy
the conditions described in Sec.~\ref{sec.daada}. Consider two modes of operation. In
one mode, each node in the network runs the adaptive diffusion
 algorithm (\ref{general.1ada})--(\ref{general.3ada}). In
the second mode, each node operates individually and runs the
non-cooperative LMS algorithm (\ref{no.general.3xx}). In both cases,
the positive step-sizes used by all nodes are assumed to be the same, say,
$\mu_k=\mu$ for all $k$, and the value of $\mu$ is chosen to satisfy
the required mean stability conditions (\ref{KD8912.ALKD.ada}) and
(\ref{90alkad.1.ada}), which are met by
selecting
\be  \mu
\;<\;\min_{1\leq k\leq N}\;\left\{\frac{2}{\lambda_{\max}(R_{u,k})}\right\}
\label{condiuad.ada}\ee It then holds that the magnitude of the mean
error vector, $\|\Ex\widetilde{\w}_i\|$ in the diffusion case decays
to zero more rapidly than in the non-cooperative case. In other
words, diffusion enhances convergence rate.
\end{theorem}
\qd

\smallskip
\begin{theorem} ({\rm {\bf Mean Convergence Rate is Enhanced: Uniform Covariance Data}})
 \label{lemaa.cod.2xa} Consider
the same setting of Theorem~\ref{lemaa.cod.2}.  Assume the covariance data are uniform across all nodes, say,
$R_{u,k}=R_u$ is independent of $k$. Assume further that the
nodes in both modes of operation employ steps-sizes $\mu_k$ that are chosen to satisfy
the required stability conditions (\ref{KD8912.ALKD.ada}) and (\ref{90alkad.1.ada}),
which in this case
are met by:
\be
\mu_k \;<\;\frac{2}{\lambda_{\max}(R_{u})},\;\;\;\;\;k=1,2,\ldots,N
\label{condiuad.2aaa}\ee  It then holds that the magnitude of the mean error
vector, $\|\Ex\widetilde{\w}_i\|,$ in the diffusion case also decays to zero
more rapidly than in the non-cooperative case. In other words,
diffusion enhances convergence rate.
\end{theorem}

\qd

\noindent The next statement considers the case of ATC and CTA strategies
 (\ref{general.1adax})--(\ref{general.3adax}) without information exchange, which correspond to
 the choice $C=I_N$. The result establishes that these strategies always enhance the convergence rate over the non-cooperative case,
 without the need to assume uniform step-sizes or uniform covariance data.

\begin{theorem} ({\rm {\bf Mean Convergence Rate is Enhanced when $\bm{C}=\bm{I}$}}) \label{lemaa.cod.2a}
Consider the problem of optimizing the global cost (\ref{opt.11}) with the
individual cost functions given by (\ref{gihjas.alk}). Pick left stochastic matrices $A_1$ and $A_2$
satisfying (\ref{dlajs8.12}) and set $C=I_N$. This situation covers the ATC and CTA strategies
 (\ref{general.1adax})--(\ref{general.3adax}) that do not involve information exchange.
Assume each node in the network measures data that satisfy
the conditions described in Sec.~\ref{sec.daada}. Consider two modes of operation.  In
one mode, each node in the network runs the adaptive diffusion
 algorithm (\ref{general.1adax})--(\ref{general.3adax}). In
the second mode, each node operates individually and runs the
non-cooperative LMS algorithm (\ref{no.general.3xx}).  In both cases, the positive step-sizes are chosen
 to satisfy the required
stability conditions (\ref{90alkad.231.ada}) and (\ref{90alkad.1.ada}), which in this case are met by
 \be \mu_k
\;<\;\frac{2}{\lambda_{\max}(R_{u,k})},\;\;\;\;\;k=1,2,\ldots,N
\label{condiuad.3.ada}\ee It then holds that the magnitude of the mean error
vector, $\|\Ex\widetilde{\w}_i\|,$ in the diffusion case decays to zero
more rapidly than in the non-cooperative case. In other words,
diffusion cooperation enhances convergence rate.
\end{theorem}
\qd

\smallskip

\noindent The results of the previous theorems again highlight the
following important facts about the role of the combination matrices
$\{A_1,A_2,C\}$ in the convergence behavior of the adaptive diffusion strategy
(\ref{general.1ada})--(\ref{general.3ada}):
\begin{itemize}
\item[(a)] The matrix $C$ influences the mean stability of the network
through its influence on the bound
in (\ref{KD8912.ALKD.ada}). This is because the matrices $\{R_k\}$ depend on the entries of $C$.
The matrices $\{A_1,A_2\}$ do not influence network mean stability.

\item[(b)] The matrices $\{A_1,A_2,C\}$ influence the
rate of convergence of the mean weight-error vector over the network since
they influence the spectral
radius of the matrix ${\cal A}_2^T\left(I_{NM}-{\cal M}{\cal
R}\right){\cal A}_1^T$, which controls the dynamics of the weight
error vector in (\ref{like.lakd.ada.2}).

\end{itemize}

\subsection{Mean-Square Stability}
It is not sufficient to ensure the stability of the weight-error
vector in the mean sense. The error vectors, $\tilde{\w}_{k,i}$, may be converging on average to
zero but they may have large fluctuations around the zero value. We therefore
need to examine how small the error vectors get. To do so, we  perform a mean-square-error
analysis. The purpose of the analysis is to evaluate how the variances
$\Ex\|\widetilde{\w}_{k,i}\|^2$ evolve with time and what their
steady-state values are, for each node $k$.

In this section, we are particularly interested
in evaluating the evolution of two mean-square-errors, namely, \be
\Ex\|\widetilde{\w}_{k,i}\|^2\;\;\;\;\;\;\mbox{\rm
and}\;\;\;\;\;\;\;\;\;\Ex|\e_{a,k}(i)|^2 \label{lkad891.alk}\ee The steady-state values of these quantities determine the
MSD and EMSE performance levels at node $k$ and, therefore, convey critical information about the
performance of the network. Under the independence assumption
on the regression data from Sec.~\ref{sec.daada}, it can be verified that the EMSE variance
can be written as: \bq \Ex|\e_{a,k}(i)|^2&\define&
\Ex|\u_{k,i}\widetilde{\w}_{k,i-1}|^2\nn\\
&=&\Ex \widetilde{\w}_{k,i-1}^*\u_{k,i}^*\u_{k,i}\widetilde{\w}_{k,i-1}\nn\\
&=&\Ex\left[\Ex (\widetilde{\w}_{k,i-1}^*\u_{k,i}^*\u_{k,i}\widetilde{\w}_{k,i-1}|\widetilde{\w}_{k,i})\right]\nn\\
&=&\Ex \widetilde{\w}_{k,i-1}^*\left[\Ex \u_{k,i}^*\u_{k,i}\right]\widetilde{\w}_{k,i-1}\nn\\
&=&\Ex \widetilde{\w}_{k,i-1}^*R_{u,k}\widetilde{\w}_{k,i-1}\nn\\
&=&\Ex\|\widetilde{\w}_{k,i-1}\|^2_{R_{u,k}} \eq in terms of a
weighted square measure with weighting matrix $R_{u,k}$. Here we are using the notation
$\|x\|^2_{\Sigma}$ to denote the weighted square quantity $x^*\Sigma x$, for any column vector $x$ and matrix $\Sigma$.
Thus, we can evaluate mean-square-errors of the form (\ref{lkad891.alk}) by
evaluating the means of
weighted square quantities of  the following form:\be \Ex\|\widetilde{\w}_{k,i}\|_{\Sigma_k}^2\ee for an
arbitrary Hermitian nonnegative-definite weighting matrix $\Sigma_k$ that we are free to
choose. By setting $\Sigma_{k}$ to different values  (say, $\Sigma_k=I$ or $\Sigma_k=R_{u,k}$), we can extract various types of information about the nodes and the network, as the discussion will reveal.
The approach we follow is based on the energy conservation framework of \cite{Sayed03,Sayed08,naffuori}.

So, let $\Sigma$ denote an arbitrary $N\times N$ block Hermitian
nonnegative-definite matrix that we are free to choose, with $M\times
M$ block entries $\{\Sigma_{\ell k}\}$. Let $\sigma$ denote the
$(NM)^2\times 1$ vector that is obtained by stacking the columns of
$\Sigma$ on top of each other, written as \be \sigma \define \mbox{\rm
vec}(\Sigma)\ee
In the sequel, it will
become more convenient to work with the vector representation
$\sigma$ than with the matrix $\Sigma$ itself.

We start from the weight-error vector recursion
(\ref{like.lakd.ada}) and re-write it more compactly as: \be
\addbox{\;\widetilde{\w}_i\;=\;\bm{\cal
B}_i\widetilde{\w}_{i-1}\;-\;{\cal G}\s_i,\;\;i\geq 0\;}
\label{like.lakd.ada.1}\ee where the coefficient matrices $\bm{\cal
B}_i$ and ${\cal G}$ are short-hand representations for \be
\addbox{\;\bm{\cal B}_i\define {\cal A}_2^T\left(I_{NM}-{\cal
M}\bm {\cal R}_i\right){\cal A}_1^T\;} \label{89.1lksk}\ee and \be
\addbox{\;{\cal G}\define {\cal A}_2^T{\cal M}{\cal C}^T\;} \label{g.dlakd}\ee Note
that $\bm{\cal B}_i$ is stochastic and time-variant, while ${\cal
G}$ is constant. We denote the mean of $\bm{\cal B}_i$ by \be
\addbox{\;{\cal B}\define \Ex\bm{\cal B}_i\;=\; {\cal
A}_2^T\left(I_{NM}-{\cal M} {\cal R}\right){\cal
A}_1^T\;}\label{kdjalk12}\ee where ${\cal R}$ is defined by
(\ref{r.dklajda}). Now equating weighted square measures on both
sides of (\ref{like.lakd.ada.1}) we get \be
\|\widetilde{\w}_i\|_{\Sigma}^2\;=\;\left\|\bm{\cal
B}_i\widetilde{\w}_{i-1}\;-\;{\cal G}\s_i\right\|^2_{\Sigma}
\label{like.lakd.ada.1asdc}\ee
Expanding the right-hand side we find that \bq
\|\widetilde{\w}_i\|_{\Sigma}^2&=& \widetilde{\w}_{i-1}^*\bm{\cal
B}_i^*\Sigma \bm{\cal B}_i \widetilde{\w}_{i-1}\;+\;\s_i^* {\cal G}^T\Sigma {\cal G}\s_i-\nn\\
&& \widetilde{\w}_{i-1}^*\bm{\cal B}_i^*\Sigma {\cal G}\s_i\;-\;
\s_i^*{\cal G}^T\Sigma \bm{\cal B}_i \widetilde{\w}_{i-1} \eq Under
expectation, the last two terms on the right-hand side evaluate to
zero so that \bq \Ex\|\widetilde{\w}_i\|_{\Sigma}^2&=&
\Ex\left(\widetilde{\w}_{i-1}^*\bm{\cal B}_i^*\Sigma \bm{\cal B}_i
\widetilde{\w}_{i-1}\right)\;+\;\Ex\left(\s_i^* {\cal G}^T\Sigma
{\cal G}\s_i\right)\label{ridoaka}\eq Let us evaluate each of the
expectations on the right-hand side. The last expectation is given
by \bq \Ex\left(\s_i^* {\cal G}^T \Sigma {\cal G}\s_i\right)&=&
\mbox{\rm
Tr}\left({\cal G}^T \Sigma {\cal G}\;\Ex\s_i\s_i^* \right)\nn\\
&\stackrel{(\ref{defl.sa})}{=}& \mbox{\rm Tr}\left({\cal G}^T \Sigma {\cal G}{\cal S} \right)\nn\\
&=& \mbox{\rm Tr}\left(\Sigma {\cal G}{\cal S}{\cal G}^T  \right)
 \eq
 where ${\cal S}$ is defined by (\ref{defl.sa}) and where we used the fact that $\mbox{\rm Tr}(AB)=\mbox{\rm
 Tr}(BA)$ for any two matrices $A$ and $B$ of compatible dimensions.
 With regards to the first expectation on the right-hand side of
 (\ref{ridoaka}), we have
 \bq
\Ex\left(\widetilde{\w}_{i-1}^*\bm{\cal B}_i^*\Sigma \bm{\cal B}_i
\widetilde{\w}_{i-1}\right)&=&\Ex\left[\Ex\left(\widetilde{\w}_{i-1}^*\bm{\cal
B}_i^*\Sigma \bm{\cal B}_i
\widetilde{\w}_{i-1}|\widetilde{\w}_{i-1}\right)\right]\nn\\
&=&\Ex\widetilde{\w}_{i-1}^*\left[\Ex\left(\bm{\cal B}_i^*\Sigma
\bm{\cal B}_i\right)\right]\widetilde{\w}_{i-1}\nn\\
&\define&\Ex\widetilde{\w}_{i-1}^*\Sigma'\widetilde{\w}_{i-1}\nn\\
&=&\Ex\|\widetilde{\w}_{i-1}\|_{\Sigma'}^2
 \eq
 where we introduced the nonnegative-definite weighting matrix
 \bq
\Sigma'&\define& \Ex\bm{\cal B}_i^*\Sigma \bm{\cal B}_i\nn\\
&\stackrel{(\ref{89.1lksk})}{=}&\Ex {\cal A}_1\left(I_{NM}-\bm
{\cal R}_i{\cal M}\right){\cal A}_2\Sigma {\cal
A}_2^T\left(I_{NM}-{\cal M}\bm {\cal R}_i\right){\cal A}_1^T\nn\\
&=& {\cal A}_1{\cal A}_2\Sigma{\cal A}_2^T{\cal A}_1^T - {\cal
A}_1{\cal A}_2\Sigma{\cal A}_2^T {\cal M}{\cal R}{\cal A}_1^T- {\cal
A}_1 {\cal R}{\cal M}{\cal A}_2\Sigma{\cal A}_2^T{\cal
A}_1^T\;+\;O({\cal M}^2)
 \eq
where ${\cal R}$ is defined by (\ref{r.dklajda}) and the term
$O({\cal M}^2)$ denotes the following factor, which depends on the
square of the step-sizes, $\{\mu_k^2\}$: \be O({\cal M}^2)\;=\; \Ex
\left({\cal A}_1\bm{\cal R}_i{\cal M}{\cal A}_2\Sigma{\cal
A}_2^T{\cal M}\bm{\cal R}_i{\cal A}_1^T\right)\label{kd8912.adlk}\ee
The evaluation of the above expectation depends on higher-order
moments of the regression data. While we can continue with the analysis
by taking this factor into account, as was done in
\cite{Cattivelli10,Sayed03,Sayed08,naffuori}, it is sufficient for the exposition in this
article to focus on the case of sufficiently small step-sizes where terms involving
higher powers of the step-sizes can be ignored.   Therefore, we continue our discussion by letting \be
\addbox{\;\Sigma'\define {\cal A}_1{\cal A}_2\Sigma{\cal A}_2^T{\cal
A}_1^T - {\cal A}_1{\cal A}_2\Sigma{\cal A}_2^T {\cal M}{\cal
R}{\cal A}_1^T- {\cal A}_1 {\cal R}{\cal M}{\cal A}_2\Sigma{\cal
A}_2^T{\cal A}_1^T\;} \label{kdj8912.alkd}\ee The weighting matrix
$\Sigma'$ is fully defined in terms of the step-size matrix, ${\cal
M}$, the network topology through the matrices $\{{\cal A}_1,{\cal
A}_2,{\cal C}\}$, and the regression statistical profile
through ${\cal R}$. Expression (\ref{kdj8912.alkd}) tells us how to
construct $\Sigma'$ from $\Sigma$. The expression can be transformed
into a more compact and revealing form if we instead relate the vector forms
$\sigma'=\mbox{\rm vec}(\Sigma')$ and $\sigma=\mbox{\rm vec}(\Sigma)$. Using the following equalities for arbitrary
matrices $\{U,W,\Sigma\}$ of compatible dimensions
\cite{Sayed08}:\bq \mbox{\rm
vec}(U\Sigma W)&=&(W^T\otimes U)\sigma\\
\mbox{\rm Tr}(\Sigma W)&=&\left[\mbox{\rm
vec}(W^T)\right]^T\sigma\label{jd8912.}
 \eq
and applying the vec operation to both sides of (\ref{kdj8912.alkd})
we get \bq \sigma'&=& \left({\cal A}_1{\cal A}_2\otimes {\cal
A}_1{\cal A}_2\right)\sigma\;-\; \left({\cal A}_1{\cal R}^T{\cal
M}{\cal A}_2\otimes {\cal A}_1{\cal A}_2\right)\sigma\;-\;
\left({\cal A}_1{\cal A}_2\otimes {\cal A}_1{\cal R}{\cal M}{\cal
A}_2\right)\sigma\nn\eq That is, \be\addbox{\; \sigma' \define {\cal
F}\sigma\;} \label{defajsa}\ee where we are introducing the coefficient
matrix of size $(NM)^2\times (NM)^2$:\be \addbox{\;{\cal F}\define
\left({\cal A}_1{\cal A}_2\otimes {\cal A}_1{\cal A}_2\right)\;-\;
\left({\cal A}_1{\cal R}^T{\cal M}{\cal A}_2\otimes {\cal A}_1{\cal
A}_2\right)\;-\; \left({\cal A}_1{\cal A}_2\otimes {\cal A}_1{\cal
R}{\cal M}{\cal A}_2\right)\;} \label{jdk8912.l.kas}\ee A reasonable
approximate expression for ${\cal F}$ for sufficiently small
step-sizes is \be \addbox{\;{\cal F}\approx {\cal B}^T\otimes {\cal
B}^* \;}\label{ka8912.lkad}\ee Indeed, if we replace ${\cal B}$ from
(\ref{kdjalk12}) into (\ref{ka8912.lkad}) and expand
terms, we obtain the same factors that appear in
(\ref{jdk8912.l.kas}) plus an additional term that depends on the
square of the step-sizes, $\{\mu_k^2\}$, whose effect can be ignored
for sufficiently small step-sizes.

 In this way, using in addition property (\ref{jd8912.}), we find that relation
(\ref{ridoaka}) becomes: \be
\addbox{\;\Ex\|\widetilde{\w}_i\|_{\Sigma}^2\;=\;
\Ex\|\widetilde{\w}_{i-1}\|_{\Sigma'}^2\;+\;\left[\mbox{\rm
vec}\left({\cal G}{\cal S}^T{\cal G}^T\right)\right]^T
\sigma\;}\label{kdl8912.1}\ee The last term is dependent on the
network topology through the matrix ${\cal G}$, which is defined in
terms of $\{{\cal A}_2,{\cal C},{\cal M}\}$, and the noise and
regression data statistical profile through ${\cal S}$. It is
convenient to introduce the alternative notation $\|x\|^2_{\sigma}$
to refer to the weighted square quantity $\|x\|^2_{\Sigma}$, where
$\sigma=\mbox{\rm vec}(\Sigma)$. We shall use these
two notations interchangeably. The convenience of the vector notation
is that it allows us to exploit the simpler linear relation
(\ref{defajsa}) between $\sigma'$ and $\sigma$ to rewrite
(\ref{kdl8912.1}) as shown in (\ref{kdl8912.12}) below, with the {\em same} weight vector $\sigma$
appearing on both sides.

\begin{theorem} ({\rm {\bf Variance Relation}})\label{90alasdlk.thm} Consider the data model of Sec.~\ref{sec.daada}
and the independence statistical conditions imposed on the noise and regression data, including
(\ref{lkad8912.lakd.ada})--(\ref{lkad8912.lakd.ada.1}). Assume further sufficiently small step-sizes are used so that
terms that depend on higher-powers of the step-sizes can be ignored.
Pick left stochastic matrices $A_1$ and $A_2$ and a right stochastic matrix $C$
satisfying (\ref{dlajs8.12}). Under these conditions, the
weight-error vector $\widetilde{\w}_i=\mbox{\rm col}\{\widetilde{\w}_{k,i}\}_{k=1}^N$ associated with a network running the adaptive diffusion strategy
(\ref{general.1ada})--(\ref{general.3ada}) satisfies the following variance relation
\be
\addbox{\;\Ex\|\widetilde{\w}_i\|_{\sigma}^2\;=\;
\Ex\|\widetilde{\w}_{i-1}\|_{{\cal F}\sigma}^2\;+\;\left[\mbox{\rm
vec}\left({\cal Y}^T\right)\right]^T
\sigma\;}\label{kdl8912.12}\ee
for any Hermitian nonnegative-definite matrix $\Sigma$ with $\sigma=\mbox{\rm vec}(\Sigma)$, and where
$\{ {\cal S}, {\cal G}, {\cal F}\}$ are defined by (\ref{defl.sa}), (\ref{g.dlakd}), and
(\ref{ka8912.lkad}), and
\be \addbox{\;{\cal Y}\define {\cal G}{\cal
S}{\cal G}^T\;} \label{dlkas9812.1}\ee
\end{theorem}

\qd

\noindent Note that relation (\ref{kdl8912.12})
is not an actual recursion; this is because the weighting matrices $\{\sigma, {\cal F}\sigma\}$
on both sides of the equality are different. The relation can be
transformed into a true recursion by expanding it into a convenient
state-space model; this argument was pursued in
\cite{Cattivelli10,Sayed03,Sayed08,naffuori} and is not necessary for the
exposition here, except to say that stability of the matrix ${\cal
F}$ ensures the mean-square stability of the filter --- this fact is
also established further ahead through
relation (\ref{reflalkda.a}). By
mean-square stability we mean that each term
$\Ex\|\widetilde{\w}_{k,i}\|^2$ remains bounded over time and
converges to a steady-state ${\rm MSD}_k$ value. Moreover, the spectral radius
of ${\cal F}$ controls the rate of convergence of
$\Ex\|\widetilde{\w}_{i}\|^2$ towards its steady-state value.\\

\begin{theorem} ({\rm {\bf Mean-Square Stability}}) \label{escxlasdk} Consider the same setting
of Theorem~\ref{90alasdlk.thm}. The adaptive
diffusion strategy (\ref{general.1ada})--(\ref{general.3ada}) is
mean-square stable if, and only if, the matrix ${\cal F}$ defined by
(\ref{jdk8912.l.kas}), or its approximation (\ref{ka8912.lkad}),
 is stable (i.e., all its
eigenvalues lie strictly inside the unit disc). This condition is
satisfied by sufficiently small positive step-sizes $\{\mu_k\}$ that also satisfy:
\be \mu_{k}< \frac{2}{\lambda_{\max}(R_k)}
\label{KD8912.ALKD.ada.3}\ee where the neighborhood covariance
matrix $R_k$ is defined by (\ref{defkajs812}). Moreover, the
convergence rate of the algorithm is determined by the value
$[\rho({\cal B})]^2$ (the square of the spectral radius of ${\cal B}$).
\end{theorem}

\bp Recall that, for two arbitrary matrices $A$ and $B$ of
compatible dimensions, the eigenvalues of the Kronecker product
$A\otimes B$ is formed of all product combinations
$\lambda_i(A)\lambda_j(B)$ of the eigenvalues of $A$ and $B$
\cite{golub}. Therefore, using expression (\ref{ka8912.lkad}), we
have that $\rho({\cal F})=\left[\rho({\cal B})\right]^2$.
It follows that  ${\cal F}$ is stable if, and only if, ${\cal B}$ is
stable. We already noted earlier in Theorem~\ref{kald8912.lkas} that
condition (\ref{KD8912.ALKD.ada.3}) ensures the
stability of ${\cal B}$. Therefore, step-sizes that ensure stability
in the mean and are sufficiently small will also ensure mean-square stability.

\ep

\noindent {\bf {\em Remark}}. More generally, had we not ignored the
second-order term (\ref{kd8912.adlk}), the expression for ${\cal F}$
would have been the following. Starting from the definition $\Sigma'=\Ex\bm{\cal
B}_i^*\Sigma \bm{\cal B}_i$, we would get
\[
\sigma'\;=\;\left(\Ex\bm{\cal B}_i^T\otimes \bm{\cal
B}_i^*\right)\sigma
\]
so that \bq {\cal F}&\define& \Ex\left(\bm{\cal B}_i^T\otimes
\bm{\cal B}_i^*\right)\;\;\;\;\;(\mbox{\rm for general
step-sizes})\\
&=&({\cal A}_1\otimes {\cal A}_1)\cdot\left\{I\;-\;\left({\cal
R}^T{\cal M}\otimes I\right)\;-\;\left(I\otimes {\cal R}{\cal
M}\right)\;+\;\Ex\left(\bm{\cal R}_i^T{\cal M}\otimes \bm{\cal
R}_i{\cal M}\right)\right\}\cdot ({\cal A}_2\otimes {\cal A}_2)
\nn\eq Mean-square stability of the filter would then require the
step-sizes $\{\mu_k\}$ to be chosen such that they ensure the
stability of this matrix ${\cal F}$ (in addition to condition (\ref{KD8912.ALKD.ada.3}) to ensure
mean stability).

\qd

\subsection{Network Mean-Square Performance}
We can now use the variance relation (\ref{kdl8912.12}) to evaluate
the network performance, as well as the performance of the
individual nodes, in steady-state. Since the dynamics is mean-square
stable for sufficiently small step-sizes, we take the limit of (\ref{kdl8912.12})
as $i\rightarrow\infty$ and write:
\be
\lim_{i\rightarrow\infty}\Ex\|\widetilde{\w}_{i}\|_{\sigma}^2\;=\;
\lim_{i\rightarrow\infty}\Ex\|\widetilde{\w}_{i-1}\|_{{\cal
F}\sigma}^2\;+\;\left[\mbox{\rm vec}\left({\cal Y}^T\right)\right]^T \sigma\label{kdl8912.12.as}
\ee
Grouping terms leads to the following result.

\begin{corollary} ({\rm {\bf Steady-State Variance Relation}})
Consider the same setting of Theorem~\ref{90alasdlk.thm}. The weight-error vector,
$\widetilde{\w}_i=\mbox{\rm col}\{\widetilde{\w}_{k,i}\}_{k=1}^N,$ of the adaptive
diffusion strategy (\ref{general.1ada})--(\ref{general.3ada}) satisfies
the following relation in steady-state:
\be\addbox{\; \lim_{i\rightarrow\infty}\;\Ex\|\widetilde{\w}_{i}\|_{(I-{\cal
F})\sigma}^2\;=\; \left[\mbox{\rm vec}\left({\cal
Y}^T\right)\right]^T \sigma\;}\label{kdl8912.12.asas}\ee
for any Hermitian nonnegative-definite matrix $\Sigma$ with $\sigma=\mbox{\rm vec}(\Sigma)$,
and where $\{{\cal F}, {\cal Y}\}$ are defined by
(\ref{ka8912.lkad}) and (\ref{dlkas9812.1}).
\end{corollary}

\qd

\smallskip

\noindent Expression (\ref{kdl8912.12.asas}) is a very useful
relation; it allows us to evaluate the network MSD and EMSE through
proper selection of the weighting vector $\sigma$ (or, equivalently,
the weighting matrix $\Sigma$). For example, the network MSD is
defined as the average value: \be {\rm MSD}^{\rm network}\define
\lim_{i\rightarrow\infty}\;\frac{1}{N}\sum_{k=1}^N\;\Ex\|\widetilde{\w}_{k,i}\|^2 \ee
which amounts to averaging the MSDs of the individual
nodes. Therefore, \be {\rm MSD}^{\rm network}\;=\;
\lim_{i\rightarrow\infty}\;\frac{1}{N}\Ex\|\widetilde{\w}_{i}\|^2\;=\;
\lim_{i\rightarrow\infty}\;\Ex\|\widetilde{\w}_{i}\|_{1/N}^2 \ee This means that in order
to recover the network MSD from relation (\ref{kdl8912.12.asas}), we
should select the weighting vector $\sigma$ such that
\[
(I-{\cal F})\sigma\;=\;\frac{1}{N}\mbox{\rm
vec}\left(I_{NM}\right)
\]
Solving for $\sigma$ and substituting back into (\ref{kdl8912.12.asas})
we arrive at the following expression for the network MSD:
\be \addbox{\;{\rm MSD}^{\rm network}\;=\;
\frac{1}{N}\cdot\left[\mbox{\rm vec}\left({\cal
Y}^T\right)\right]^T\cdot(I-{\cal F})^{-1}\cdot\mbox{\rm
vec}\left(I_{NM}\right)\;} \label{lkad8912.13}\ee
Likewise,  the network EMSE is defined as the average value \bq {\rm
EMSE}^{\rm network}&\define&
\lim_{i\rightarrow\infty}\;\frac{1}{N}\sum_{k=1}^N\;\Ex|\e_{a,k}(i)|^2\nn\\
&=&\lim_{i\rightarrow\infty}\;
\frac{1}{N}\sum_{k=1}^N\;\Ex\|\widetilde{\w}_{k,i}\|^2_{R_{u,k}}
\eq which amounts to averaging the EMSEs of the
individual nodes. Therefore, \be {\rm EMSE}^{\rm network}\;=\;
\lim_{i\rightarrow\infty}\;\frac{1}{N}\Ex\|\widetilde{\w}_{i}\|^2_{\mbox{{\small \rm
diag}}\{R_{u,1},R_{u,2},\ldots,R_{u,N}\}}\;=\;
\lim_{i\rightarrow\infty}\;\frac{1}{N}\Ex\|\widetilde{\w}_{i}\|^2_{{\cal R}_u}
 \ee where ${\cal R}_u$ is the matrix defined earlier by (\ref{replaced}), and which we repeat below for
 ease of reference:
 \be {\cal R}_u=\mbox{\rm diag}\{R_{u,1},R_{u,2},\ldots,R_{u,N}\}\label{90als.asdl}\ee
 This means that
in order to recover the network EMSE from relation
(\ref{kdl8912.12.asas}), we should select the weighting vector
$\sigma$ such that
\be
(I-{\cal F})\sigma\;=\;\frac{1}{N}\mbox{\rm vec}\left({\cal
R}_u\right)
\ee
Solving for $\sigma$ and substituting into (\ref{kdl8912.12.asas})
we arrive at the following expression for the network EMSE:
\be \addbox{\;{\rm EMSE}^{\rm network}\;=\;
\frac{1}{N}\cdot\left[\mbox{\rm vec}\left({\cal
Y}^T\right)\right]^T\cdot(I-{\cal F})^{-1}\cdot\mbox{\rm
vec}\left({\cal R}_u\right)\;} \ee

\subsection{Mean-Square Performance of Individual Nodes}
We can also assess the mean-square performance of the individual
nodes in the network from (\ref{kdl8912.12.asas}). For instance, the
MSD of any particular node $k$ is defined by \be {\rm MSD}_k\define
\lim_{i\rightarrow\infty}\Ex\|\widetilde{\w}_{k,i}\|^2 \ee Introduce the $N\times N$
block diagonal matrix with blocks of size $M\times M$, where all
blocks on the diagonal are zero except for an identity matrix on the
diagonal block of index $k$, i.e., \be {\cal J}_k\define \mbox{\rm
diag}\{\;0_M,\ldots,0_M,I_M,0_M,\ldots,0_M\;\} \label{901asl}\ee
Then, we can express the node MSD as follows: \be {\rm
MSD}_k\define \lim_{i\rightarrow\infty}\;\Ex\|\widetilde{\w}_{i}\|^2_{{\cal J}_k} \ee
The same argument that was used to obtain the network MSD then leads
to \be \addbox{\;{\rm MSD}_k\;=\; \left[\mbox{\rm vec}\left({\cal
Y}^T\right)\right]^T\cdot(I-{\cal F})^{-1}\cdot\mbox{\rm
vec}\left({\cal J}_k\right)\;} \ee
Likewise,  the EMSE of node $k$ is defined by\bq {\rm
EMSE}_k&\define&
\lim_{i\rightarrow\infty}\Ex|\e_{a,k}(i)|^2\nn\\
&=& \lim_{i\rightarrow\infty}\Ex\|\widetilde{\w}_{k,i}\|^2_{R_{u,k}} \eq Introduce the
$N\times N$ block diagonal matrix with blocks of size $M\times M$,
where all blocks on the diagonal are zero except for the diagonal
block of index $k$ whose value is $R_{u,k}$, i.e., \be {\cal
T}_k\define \mbox{\rm
diag}\{\;0_M,\ldots,0_M,R_{u,k},0_M,\ldots,0_M\;\} \label{dj8192}\ee
Then, we can express the node EMSE as follows: \be {\rm
EMSE}_k\define \lim_{i\rightarrow\infty}\;\Ex\|\widetilde{\w}_{i}\|^2_{{\cal T}_k} \ee
The same argument that was used to obtain the network EMSE then
leads to \be \addbox{\;{\rm EMSE}_k\;=\; \left[\mbox{\rm
vec}\left({\cal Y}^T\right)\right]^T\cdot(I-{\cal
F})^{-1}\cdot\mbox{\rm vec}\left({\cal T}_k\right)\;} \ee
We summarize the results in the following statement.

\begin{theorem} ({\rm {\bf Network Mean-Square Performance}}) \label{lemma.mse} Consider the same setting of Theorem~\ref{90alasdlk.thm}.
 Introduce the
$1\times (NM)^2$ row vector $h^T$ defined by \be h^T\define
\left[\mbox{\rm vec}\left({\cal Y}^T\right)\right]^T\cdot(I-{\cal
F})^{-1}
 \ee
 where $\{{\cal F}, {\cal Y}\}$ are defined by
(\ref{ka8912.lkad}) and (\ref{dlkas9812.1}). Then the
network MSD and EMSE and the individual node
 performance measures are given by
 \bq
{\rm MSD}^{\rm network}&=& h^T\cdot\mbox{\rm vec}\left(I_{NM}\right)/N\label{kdla8912.12}\\
{\rm EMSE}^{\rm network}&=&h^T\cdot\mbox{\rm vec}\left({\cal R}_{u}\right)/N\\
{\rm MSD}_k&=& h^T\cdot\mbox{\rm vec}\left({\cal J}_k\right)\\
{\rm EMSE}_k&=&h^T\cdot\mbox{\rm vec}\left({\cal T}_k\right)
 \eq
where $\{{\cal J}_k,{\cal T}_k\}$ are defined by (\ref{901asl}) and
(\ref{dj8192}).
\end{theorem}
\qd

\noindent We can obviously recover from the above expressions the performance
of the nodes in the non-cooperative implementation
(\ref{no.general.3xx}), where each node performs its adaptation
individually, by setting ${A}_1={A}_2={C}=I_N$.

We can express the network MSD, and its EMSE if desired, in an alternative useful form
involving a series representation.

\begin{corollary} ({\rm {\bf Series Representation for Network MSD}}) \label{lemma.mse2} Consider the same setting of Theorem~\ref{90alasdlk.thm}.
The network MSD can be expressed in the following alternative series expansion form:
\be \addbox{\;{\rm MSD}^{\rm network}\;=\;
\frac{1}{N}\sum_{j=0}^{\infty}\;\mbox{\rm Tr}\left({\cal B}^j {\cal
Y}{\cal B}^{*j}\right)\;}\label{imajdka}\ee
where
\bq
{\cal Y}&=&{\cal G}{\cal S}{\cal G}^T\label{yuadlas}\\
{\cal G}&=&{\cal A}_2^T{\cal M}{\cal C}^T\\
{\cal B}&=&{\cal A}_2^T(I-{\cal M}{\cal R}){\cal A}_1^T\label{lA912.a}
 \eq
\end{corollary}
\bp Since ${\cal F}$ is stable when the filter is mean-square
stable, we can expand $(I-{\cal F})^{-1}$ as \bq (I-{\cal
F})^{-1}&=&I\;+\;{\cal F}\;+\;{\cal
F}^2+\ldots\nn\\
&\stackrel{(\ref{ka8912.lkad})}{=}& I \;+\;\left({\cal B}^T\otimes
{\cal B}^*\right)\;+\; \left({\cal B}^T\otimes {\cal
B}^*\right)^2\;+\;\ldots\nn
 \eq Substituting into (\ref{lkad8912.13}) and using property (\ref{jd8912.}),  we obtain the desired result.
 \ep

\subsection{Uniform Data Profile}
We can simplify expressions (\ref{yuadlas})--(\ref{lA912.a})  for $\{{\cal Y}, {\cal G}, {\cal B}\}$ in the
case when the regression covariance matrices are uniform across the network and all nodes
employ the same step-size, i.e., when
\bq
R_{u,k}&=&R_u,\;\;\;\;\mbox{\rm for all $k$}\;\;\;\;\;\;(\mbox{\rm uniform covariance profile})\label{cond.1}\\
\mu_k&=&\mu,\;\;\;\;\;\;\mbox{\rm for all $k$}\;\;\;\;\;\;(\mbox{\rm uniform step-sizes})
\label{cond.2}\eq
and when the combination matrix $C$ is doubly stochastic, so that
 \be
\addbox{\; C\mathds{1}=\mathds{1},\;\;\;\;\;C^T\mathds{1}=\mathds{1}\;}\label{cond.3}
 \ee
We refer to conditions (\ref{cond.1})--(\ref{cond.3}) as corresponding to a {\em uniform data profile} environment.
The noise variances, $\{\sigma_{v,k}^2\}$, do not need to be uniform so that the signal-to-noise ratio (SNR) across the network can still vary from node to node. The
simplified expressions derived in the sequel will be useful in Sec.~\ref{sec.comapre} when we compare the performance of various
cooperation strategies.

Thus, under conditions (\ref{cond.1})--(\ref{cond.3}), expressions
(\ref{r.dklajda.11}), (\ref{r.dklajda}), and (\ref{g.dlakd}) for $\{{\cal M}, {\cal R},  {\cal G}\}$ simplify to
\bq
{\cal M}&=&\mu I_{NM}\\
{\cal R}&=& I_N\otimes R_u\\
{\cal G}&=&\mu {\cal A}_2^T {\cal C}^T
\eq
Substituting these values into expression (\ref{lA912.a}) for ${\cal B}$ we get
\bq
{\cal B}&=&{\cal A}_2^T(I-{\cal M}{\cal R}){\cal A}_1^T\nn\\
&=&(A_2^T\otimes I)\cdot(I-\mu(I\otimes R_u))\cdot(A_1^T\otimes I)\nn\\
&=&(A_2^T\otimes I)(A_1^T\otimes I)\;-\;\mu (A_2^T\otimes I)(I\otimes R_u)(A_1^T\otimes I)\nn\\
&=&(A_2^TA_1^T\otimes I)\;-\;\mu(A_2^TA_1^T\otimes R_u)\nn\\
&=&A_2^TA_1^T\otimes (I-\mu R_u)
\eq
where we used the useful Kronecker product identities:
\bq
(X+Y)\otimes Z&=&(X\otimes Z)\;+\;(Y\otimes Z)\label{use.property}\\
(X\otimes Y)(W\otimes Z)&=&(XW\otimes YZ)
\eq
for any matrices $\{X,Y,Z,W\}$ of compatible dimensions. Likewise, introduce the $N\times N$ diagonal matrix with noise variances:
\be
\addbox{\;{R_v}\define \mbox{\rm diag}\{\sigma_{v,1}^2,\;\sigma_{v,2}^2,\ldots,\sigma_{v,N}^2\}\;}
\label{kdla.salkd}\ee
Then, expression (\ref{defl.sa}) for ${\cal S}$ becomes
\bq
{\cal S}&=&\mbox{\rm diag}\{\sigma_{v,1}^2 R_u,\;\sigma_{v,2}^2 R_{u},\ldots,\;\sigma_{v,N}^2 R_u\}\nn\\
&=&{R_v}\otimes R_u
\eq
It then follows that we can simplify expression (\ref{yuadlas}) for ${\cal Y}$ as:
\bq
{\cal Y}&=&\mu^2 {\cal A}_2^T{\cal C}^T{\cal S}{\cal C}{\cal A}_2\nn\\
&=&\mu^2\cdot(A_2^T\otimes I)\cdot(C^T\otimes I)\otimes (R_v\otimes R_u)\cdot(C\otimes I)\cdot (A_2\otimes I)\nn\\
&=&\mu^2 (A_2^T C^T R_v CA_2\otimes R_u)
\eq

\begin{corollary} ({\rm {\bf Network MSD for Uniform Data Profile}}) \label{lemma.mseasx}
Consider the same setting of Theorem~\ref{90alasdlk.thm} with the additional requirement that
conditions (\ref{cond.1})--(\ref{cond.3}) for a uniform data profile hold.
The network MSD is still given by the same series representation
(\ref{imajdka}) where now
\bq
{\cal Y}&=&\mu^2 (A_2^T C^T R_v CA_2\otimes R_u)\label{exp.aa}\\
{\cal B}&=&A_2^TA_1^T\otimes (I-\mu R_u)\label{exp.aa.2}
 \eq
Using these expressions, we can decouple the network MSD expression (\ref{imajdka})
into two separate factors: one is
dependent on the step-size and data covariance $\{\mu,R_u\}$,
and the other is dependent on the combination matrices and noise profile
$\{A_1,A_2,C,R_v\}$:

{\small
\be
\addbox{\;{\rm MSD}^{\rm network}=
\dfrac{\mu^2}{N}\displaystyle \sum_{j=0}^{\infty}\;\mbox{\rm Tr}\left(
\left[\left(A_2^TA_1^T\right)^j
\left(A_2^T C^T R_v CA_2\right)(A_1 A_2)^j\right]\otimes\left[
(I-\mu R_u)^j R_u (I-\mu R_u)^j \right]\right)\;}\label{resu.t313}
\ee
}

\end{corollary}

\bp Using (\ref{imajdka}) and the given expressions (\ref{exp.aa})--(\ref{exp.aa.2})
for $\{{\cal Y},{\cal B}\}$, we get
\[{\rm MSD}^{\rm network}=\dfrac{\mu^2}{N}\sum_{j=0}^{\infty}\;\mbox{\rm Tr}\left(
\left[\left(A_2^TA_1^T\right)^j \otimes (I-\mu R_u)^j\right] (A_2^T C^T R_v CA_2\otimes R_u)
\left[(A_1 A_2)^j\otimes(I-\mu R_u)^j \right]\right)\nn
\]
Result (\ref{resu.t313}) follows from property (\ref{use.property}).

\ep

\subsection{Transient Mean-Square Performance}
Before comparing the mean-square performance of various cooperation strategies, we
pause to comment that the variance relation (\ref{kdl8912.12}) can also
be used to characterize the transient behavior of the network,
and not just its steady-state performance. To see this, iterating (\ref{kdl8912.12}) starting from $i=0$, we find that
\be
\Ex\|\widetilde{\w}_{i}\|_{\sigma}^2\;=\;\Ex\|\widetilde{\w}_{-1}\|^2_{{\cal F}^{i+1}\sigma}\;+
\;\left[\mbox{\rm vec}\left({\cal Y}^T\right)\right]^T\cdot\left(\sum_{j=0}^{i}{\cal F}^j \sigma\right)
\label{coad.ad}\ee
where
\be
\widetilde{\w}_{-1}\define w^o-\w_{-1}
\ee
in terms of the initial condition, $\w_{-1}$. If this initial condition happens to
 be $\w_{-1}=0$, then $\widetilde{\w}_{-1}=w^o$.
Comparing expression (\ref{coad.ad}) at time instants $i$ and $i-1$ we can relate
$\Ex\|\widetilde{\w}_i\|_{\sigma}^2$ and $\Ex\|\widetilde{\w}_{i-1}\|_{\sigma}^2$ as follows:
\be
\addbox{\;
\Ex\|\widetilde{\w}_{i}\|_{\sigma}^2\;=\;\Ex\|\widetilde{\w}_{i-1}\|_{\sigma}^2\;+
\;\left[\mbox{\rm vec}\left({\cal Y}^T\right)\right]^T
\cdot {\cal F}^i \sigma\;-\;\Ex\|\widetilde{\w}_{-1}\|_{(I-{\cal F}){\cal F}^i\sigma}^2
\;}
\label{reflalkda.a}\ee
This recursion relates the same weighted square measures of the error
vectors $\{\widetilde{\w}_{i},\widetilde{\w}_{i-1}\}$.
It therefore describes how these weighted square measures evolve over time. It is clear from this relation that, for mean-square stability,
the matrix ${\cal F}$ needs to be stable
so that the terms involving ${\cal F}^i$ do not grow unbounded.

The learning curve of the network is the curve that describes the evolution of the network EMSE over time.
At any time $i$, the network EMSE is denoted by $\zeta(i)$ and measured as: \bq
\zeta(i)&\define&
\frac{1}{N}\sum_{k=1}^N\;\Ex|\e_{a,k}(i)|^2\label{define.zeta}\\
&=&
\frac{1}{N}\sum_{k=1}^N\;\Ex\|\widetilde{\w}_{k,i}\|^2_{R_{u,k}}\nn
\eq The above expression indicates that $\zeta(i)$ is obtained
by averaging the  EMSE of the
individual nodes at time $i$. Therefore, \be \zeta(i)\;=\;
\frac{1}{N}\Ex\|\widetilde{\w}_{i}\|^2_{\mbox{{\small \rm
diag}}\{R_{u,1},R_{u,2},\ldots,R_{u,N}\}}\;=\;
\frac{1}{N}\Ex\|\widetilde{\w}_{i}\|^2_{{\cal R}_u}
 \ee where ${\cal R}_u$ is the matrix defined by (\ref{90als.asdl}).
 This means that
in order to evaluate the evolution of the network EMSE from relation
(\ref{reflalkda.a}), we simply select the weighting vector
$\sigma$ such that
\be
\sigma\;=\;\frac{1}{N}\mbox{\rm vec}\left({\cal
R}_u\right)
\ee
Substituting into (\ref{reflalkda.a})
we arrive at the learning curve for the network.

\begin{corollary} ({\rm {\bf Network Learning Curve}}) \label{lemma.Xamseasx}
Consider the same setting of Theorem~\ref{90alasdlk.thm}. Let $\zeta(i)$ denote the network EMSE at
time $i$, as defined by (\ref{define.zeta}). Then, the learning curve of the network corresponds to the evolution of $\zeta(i)$ with time and is described by the
following recursion over $i\geq 0$:
\be \addbox{\;\zeta(i)\;=\; \zeta(i-1)\;+\;
\frac{1}{N}\left[\mbox{\rm vec}\left({\cal Y}^T\right)\right]^T\cdot {\cal F}^i \cdot \mbox{\rm vec}\left({\cal
R}_u\right)\;-\;\frac{1}{N}\Ex\|\tilde{\w}_{-1}\|_{(I-{\cal F}){\cal F}^i{\footnotesize \mbox{\rm vec}}\left({\cal
R}_u\right)}^2
\;} \ee
where $\{{\cal F}, {\cal Y}, {\cal R}_u\}$ are defined by
(\ref{ka8912.lkad}), (\ref{dlkas9812.1}), and (\ref{90als.asdl}).
\end{corollary}

\qd

\section{Comparing the Performance of Cooperative Strategies}\label{sec.comapre}
Using the expressions just derived for the MSD of the network,
we can compare the performance of various cooperative and non-cooperative strategies. Table~\ref{table.compare} further ahead
summarizes the results derived in this section and the conditions under which they hold.\\

\subsection{Comparing ATC and CTA Strategies}
We first compare the
performance of the adaptive ATC and CTA diffusion strategies
(\ref{Equ:DiffusionAdaptation:ATC.adaptive}) and (\ref{Equ:DiffusionAdaptation:CTA.adaptive})
when they employ a {\em doubly} stochastic
combination matrix $A$. That is, let us consider the two
scenarios:\bq C,\;\; A_1=A, \;\;A_2=I_N\;\;\;&&(\mbox{\rm adaptive CTA
strategy})\\
C,\;\; A_1=I_N, \;\;A_2=A\;\;\;&&(\mbox{\rm adaptive ATC strategy})\eq
where $A$ is now assumed to be doubly stochastic, i.e., \be A\mathds{1}=\mathds{1},\quad\quad
A^T\mathds{1}=\mathds{1}
 \ee with its rows and columns adding
up to one.  For example, these conditions are satisfied when $A$ is
left stochastic and symmetric. Then, expressions
(\ref{yuadlas}) and (\ref{lA912.a}) give:
\bq {\cal B}_{\rm
cta}&=&(I-{\cal M}{\cal R}){\cal A}^T,\;\;\;\;\;\;\;{\cal Y}_{\rm
cta}\;=\;{\cal M}{\cal C}^T{\cal S}{\cal C}{\cal M}\\
{\cal B}_{\rm atc}&=&{\cal A}^T(I-{\cal M}{\cal R}),\;\;\;\;\;\;\;{\cal
Y}_{\rm atc}\;=\;{\cal A}^T{\cal M}{\cal C}^T{\cal S}{\cal C}{\cal
M}{\cal A}\eq where \be {\cal A}=A\otimes I_M \ee Following
\cite{Cattivelli10}, introduce the auxiliary nonnegative-definite
matrix \be {\cal H}_j\define \left[(I-{\cal M}{\cal R}){\cal
A}^T\right]^j\cdot {\cal M}{\cal C}^T{\cal S}{\cal C}{\cal
M}\cdot\left[(I-{\cal M}{\cal R}){\cal A}^T\right]^{*j}  \ee Then,
it is immediate to verify from (\ref{imajdka}) that \bq {\rm
MSD}^{\rm network}_{\rm
cta}&=&\frac{1}{N}\sum_{j=0}^{\infty}\mbox{\rm Tr}({\cal H}_j)\\
{\rm MSD}^{\rm network}_{\rm
atc}&=&\frac{1}{N}\sum_{j=0}^{\infty}\mbox{\rm Tr}({\cal A}^T\;{\cal
H}_j\;{\cal A})
 \eq
so that \bq {\rm MSD}^{\rm network}_{\rm cta}- {\rm MSD}^{\rm
network}_{\rm atc}&=&\frac{1}{N}\sum_{j=0}^{\infty}\mbox{\rm
Tr}\left({\cal H}_j-{\cal A}^T{\cal H}_j{\cal A}\right)\label{jdka.aldk12}
 \eq
Now, since $A$ is doubly stochastic, it also holds that the enlarged
matrix ${\cal A}$ is doubly stochastic. Moreover, for any doubly
stochastic matrix ${\cal A}$ and any nonnegative-definite matrix
${\cal H}$ of compatible dimensions, it holds that (see part (f) of
Theorem~\ref{thm.adasxxa}): \be \mbox{\rm Tr}({\cal A}^T{\cal
H}{\cal A})\;\leq\;\mbox{\rm Tr}({\cal H}) \label{jdk8912.1}\ee
 Applying result
(\ref{jdk8912.1}) to (\ref{jdka.aldk12}) we conclude that
\be \addbox{\;{\rm MSD}^{\rm network}_{\rm
atc}\;\leq\; {\rm MSD}^{\rm network}_{\rm cta}\;}\;\;\;\;(\mbox{\rm
doubly stochastic $A$}) \label{jakd.ka9831}\ee so that the adaptive ATC strategy
(\ref{Equ:DiffusionAdaptation:ATC.adaptive}) outperforms the adaptive CTA strategy
 (\ref{Equ:DiffusionAdaptation:CTA.adaptive})
for doubly stochastic combination matrices $A$.

\subsection{Comparing Strategies with and without Information Exchange}
\noindent We now examine the effect of information exchange $(C\neq I)$ on the performance of
the adaptive ATC and CTA diffusion strategies
(\ref{Equ:DiffusionAdaptation:ATC.adaptive})--(\ref{Equ:DiffusionAdaptation:CTA.adaptive})
under conditions (\ref{cond.1})--(\ref{cond.3}) for
 {\em uniform data profile}.\\

\noindent {\bf {\em CTA Strategies}}\\
\noindent We start with the adaptive CTA strategy (\ref{Equ:DiffusionAdaptation:CTA.adaptive}),
and consider two scenarios with and without information exchange. These scenarios correspond to the
following selections in the general description (\ref{general.1ada})--(\ref{general.3ada}):
\bq C\neq I,\;\; A_1=A, \;\;A_2=I_N\;\;\;&&(\mbox{\rm adaptive CTA with information exchange
})\\
C=I,\;\; A_1=A, \;\;A_2=I_N\;\;\;&&(\mbox{\rm adaptive CTA without information exchange})\eq
Then, expressions
(\ref{exp.aa}) and (\ref{exp.aa.2}) give: \bq {\cal B}_{\rm
cta, C\neq I}&=&A^T\otimes (I-\mu R_u),\;\;\;{\cal Y}_{\rm
cta,C\neq I}\;=\;\mu^2 (C^T R_v C\otimes R_u)\\
{\cal B}_{\rm
cta, C=I}&=&A^T\otimes (I-\mu R_u),\;\;\;{\cal Y}_{\rm
cta,C=I}\;=\;\mu^2 (R_v\otimes R_u)\eq  where the matrix $R_v$ is defined by (\ref{kdla.salkd}).
Note that ${\cal B}_{\rm
cta, C\neq I}={\cal B}_{\rm
cta, C= I}$, so we denote them simply by ${\cal B}$ in the derivation that follows. Then, from expression (\ref{imajdka})
for the network MSD we get:
\bq
{\rm MSD}_{{\rm cta}, C= I}^{\rm network} - {\rm MSD}_{{\rm cta}, C\neq I}^{\rm network}
&=&\frac{\mu^2}{N}\sum_{j=0}^{\infty} \mbox{\rm Tr}\left({\cal B}^j\left[(R_v-C^TR_v C)\otimes
R_u\right]{\cal B}^{*j}\right)
\eq
It follows that the difference in performance between both CTA implementations depends on how the matrices
$R_v$ and $C^TR_v C$ compare to each other:
\begin{enumerate}
\item[(1)] When $R_v-C^TR_v C\geq 0$, we obtain
\be\addbox{\;
{\rm MSD}_{{\rm cta}, C= I}^{\rm network} \;\geq\; {\rm MSD}_{{\rm cta}, C\neq I}^{\rm network}\;}
\;\;\;\;\;\;\;\;(\mbox{\rm when $C^TR_v C\leq R_v$})
\ee
so that a CTA implementation with information exchange performs better than a CTA implementation without information
exchange. Note that the condition on $\{R_v,C\}$ corresponds to requiring
\be
C^T R_v C \leq R_v
\ee
which can be interpreted to mean that the cooperation matrix $C$ should be such that it
does not amplify the effect of measurement noise. For example, this situation occurs
when the noise profile is uniform across the network, in which case $R_v=\sigma_v^2 I_M$. This is because it would
then hold that
\bq
R_v-C^TR_v C&=&\sigma_v^2(I-C^T C) \geq 0
\eq
in view of the fact that $(I-C^TC)\geq 0$ since $C$ is doubly stochastic (cf.~property (e) in Lemma~\ref{thm.adasxxa}).

\item[(2)] When $R_v-C^TR_v C\leq 0$, we obtain
\be
\addbox{\;{\rm MSD}_{{\rm cta}, C= I}^{\rm network} \;\leq\; {\rm MSD}_{{\rm cta}, C\neq I}^{\rm
 network}\;}\;\;\;\;\;\;\;\;(\mbox{\rm when $C^TR_v C\geq R_v$})\ee
so that a CTA implementation without information exchange performs better than a CTA implementation with information exchange.
In this case, the condition on $\{R_v,C\}$ indicates that the combination matrix $C$ ends up
amplifying the
effect of noise.

\end{enumerate}

\bigskip

\noindent {\bf {\em ATC Strategies}}\\
We can repeat the argument for the adaptive ATC strategy
(\ref{Equ:DiffusionAdaptation:ATC.adaptive}),
and consider two scenarios with and without information exchange. These
scenarios correspond to the
following selections in the general description (\ref{general.1ada})--(\ref{general.3ada}):
\bq C\neq I,\;\; A_1=I_N, \;\;A_2=A\;\;\;&&(\mbox{\rm adaptive ATC with information exchange
})\\
C=I,\;\; A_1=I_N, \;\;A_2=A\;\;\;&&(\mbox{\rm adaptive ATC without information exchange})\eq
Then, expressions
(\ref{exp.aa}) and (\ref{exp.aa.2}) give: \bq {\cal B}_{\rm
atc, C\neq I}&=&A^T\otimes (I-\mu R_u),\;\;\;{\cal Y}_{\rm
atc,C\neq I}\;=\;\mu^2 (A^TC^TR_vCA\otimes R_u)\\
{\cal B}_{\rm
atc, C=I}&=&A^T\otimes(I-\mu R_u),\;\;\;{\cal Y}_{\rm
atc,C=I}\;=\;\mu^2 (A^TR_v A\otimes R_u)\eq
Note again that ${\cal B}_{\rm
atc, C\neq I}={\cal B}_{\rm
atc, C= I}$, so we denote them simply by ${\cal B}$. Then,
\bq
{\rm MSD}_{{\rm atc}, C= I}^{\rm network} - {\rm MSD}_{{\rm atc}, C\neq I}^{\rm network}&=&
\frac{\mu^2}{N}\sum_{j=0}^{\infty} \mbox{\rm Tr}\left({\cal B}^j\left[A^T(R_v-C^TR_v C)A\otimes
 R_u\right]{\cal B}^{*j}\right)
\eq
 It again follows that the difference in performance between both ATC
  implementations depends on how the matrices
$R_v$ and $C^TR_v C$ compare to each other and we obtain:
\be\addbox{\;
{\rm MSD}_{{\rm atc}, C= I}^{\rm network} \;\geq\; {\rm MSD}_{{\rm atc}, C\neq I}^{\rm network}\;}
\;\;\;\;\;\;\;\;(\mbox{\rm when $C^TR_v C\leq R_v$})
\ee
and\be
\addbox{\;{\rm MSD}_{{\rm atc}, C= I}^{\rm network} \;\leq\; {\rm MSD}_{{\rm atc}, C\neq I}^{\rm
 network}\;}\;\;\;\;\;\;\;\;(\mbox{\rm when $C^TR_v C\geq R_v$})\ee

\smallskip
{\small
\begin{table}[h]
\begin{center}
\caption{\rm {\small Comparison of the MSD performance of various cooperative strategies. }} {\small
\begin{tabular}{|l|l|}\hline\hline &\\
{\sc Comparison} & {\sc Conditions}\\\hline\hline &\\
${\rm MSD}^{\rm network}_{\rm
atc}\;\leq\; {\rm MSD}^{\rm network}_{\rm cta}$ & {\small $A$ doubly stochastic, $C$ right stochastic.}\\\hline&\\
${\rm MSD}_{{\rm cta}, C\neq I}^{\rm network} \;\leq\; {\rm MSD}_{{\rm cta}, C= I}^{\rm network}$&
$C^TR_v C\leq R_v$, $C$ doubly stochastic, $R_{u,k}=R_u$, $\mu_k=\mu$.\\\hline&\\
${\rm MSD}_{{\rm cta}, C= I}^{\rm network} \;\leq\; {\rm MSD}_{{\rm cta}, C\neq I}^{\rm network}$&
$C^TR_v C\geq R_v$, $C$ doubly stochastic, $R_{u,k}=R_u$, $\mu_k=\mu$.\\\hline&\\
${\rm MSD}_{{\rm atc}, C\neq I}^{\rm network} \;\leq\; {\rm MSD}_{{\rm atc}, C= I}^{\rm network}$&
$C^TR_v C\leq R_v$, $C$ doubly stochastic, $R_{u,k}=R_u$, $\mu_k=\mu$.\\\hline&\\
${\rm MSD}_{{\rm atc}, C= I}^{\rm network} \;\leq\; {\rm MSD}_{{\rm atc}, C\neq I}^{\rm network}$&
$C^TR_v C\geq R_v$, $C$ doubly stochastic, $R_{u,k}=R_u$, $\mu_k=\mu$.\\\hline&\\
${\rm MSD}_{{\rm atc}}^{\rm network}\;\leq\;
{\rm MSD}_{{\rm cta}}^{\rm network}\;\leq\;
{\rm MSD}_{{\rm lms}}^{\rm network}$ & $\{A,C\}$ doubly stochastic, $R_{u,k}=R_u$, $\mu_k=\mu$.
\\\hline\hline
\end{tabular} }
\label{table.compare}
\end{center}
\end{table}
}

\subsection{Comparing Diffusion Strategies with the Non-Cooperative Strategy}\label{sec.lakdcompa}
\noindent We now compare the performance of the adaptive CTA
strategy (\ref{Equ:DiffusionAdaptation:CTA.adaptive}) to the
non-cooperative LMS strategy (\ref{no.general.3xx}) assuming conditions
(\ref{cond.1})--(\ref{cond.3}) for uniform data profile.
 These
scenarios correspond to the
following selections in the general description (\ref{general.1ada})--(\ref{general.3ada}):
 \bq C,\;\; A_1=A,
\;\;A_2=I&&(\mbox{\rm adaptive CTA})\\
C=I,\;\; A_1=I,\;\; A_2=I&&(\mbox{\rm non-cooperative LMS})\eq
where $A$ is further assumed to be doubly stochastic (along with $C$) so that
\be A\mathds{1}=\mathds{1},\quad\quad
A^T\mathds{1}=\mathds{1}
 \ee  Then, expressions
(\ref{exp.aa}) and (\ref{exp.aa.2}) give: \bq
{\cal B}_{\rm cta}&=&A^T\otimes (I-\mu R_u),\;\;\;\;\;{\cal Y}_{\rm
cta}\;=\;\mu^2(C^TR_v C\otimes R_u)\\
{\cal B}_{\rm lms}&=&I\otimes (I-\mu R_u),\;\;\;\;\;\;\;\;{\cal Y}_{\rm
lms}\;=\;\mu^2 (R_v\otimes R_u)\eq
Now recall that
\be
{\cal C}=C\otimes I_M
\ee so that, using the Kronecker product property (\ref{use.property}),
\bq
{\cal Y}_{\rm cta}&=& \mu^2(C^TR_v C\otimes R_u)\nn\\
&=&
\mu^2(C^T\otimes I_M)(R_v \otimes R_u)(C\otimes I_M)\nn\\
&=&
\mu^2{\cal C}^T(R_v \otimes R_u){\cal C}\nn\\
&=&{\cal C}^T {\cal Y}_{\rm lms} \;{\cal C}\eq
Then,
\bq
{\rm MSD}_{{\rm lms}}^{\rm network} - {\rm MSD}_{{\rm cta}}^{\rm network}
&=&\frac{1}{N}\sum_{j=0}^{\infty} \mbox{\rm Tr}\left({\cal B}_{\rm lms}^j{\cal Y}_{\rm lms}{\cal B}_{\rm lms}^{*j}\right)\;-\;\frac{1}{N}\sum_{j=0}^{\infty} \mbox{\rm Tr}\left({\cal B}_{\rm cta}^j{\cal C}^T{\cal Y}_{\rm lms}{\cal C}
{\cal B}_{\rm cta}^{*j}\right)\nn\\
&=&\frac{1}{N}\sum_{j=0}^{\infty} \mbox{\rm Tr}\left({\cal B}_{\rm lms}^{*j}{\cal B}_{\rm lms}^j
{\cal Y}_{\rm lms}\right)\;-\;\frac{1}{N}\sum_{j=0}^{\infty} \mbox{\rm Tr}\left(
{\cal C}{\cal B}_{\rm cta}^{*j}
{\cal B}_{\rm cta}^j{\cal C}^T{\cal Y}_{\rm lms}\right)\nn\\
&=&\frac{1}{N}\sum_{j=0}^{\infty} \mbox{\rm Tr}\left[\left({\cal B}_{\rm lms}^{*j}{\cal B}_{\rm lms}^j
\;-\;{\cal C}{\cal B}_{\rm cta}^{*j}{\cal B}_{\rm cta}^j{\cal C}^T\right){\cal Y}_{\rm lms}\right]\label{condkja.13}
\eq
Let us examine the difference:
\bq
{\cal B}_{\rm lms}^{*j}{\cal B}_{\rm lms}^j\;-\;{\cal C}{\cal B}_{\rm cta}^{*j}{\cal B}_{\rm cta}^j{\cal C}^T
&=&
\left(I\otimes (I-\mu R_u)^{2j}\right)\;-\;\left(CA^{j}\otimes (I-\mu R_u)^{j}\right)
\left(A^{j T}C^T\otimes (I-\mu R_u)^{j}\right)\nn\\
&\stackrel{(\ref{use.property})}{=}&
\left(I\otimes (I-\mu R_u)^{2j}\right)\;-\;\left(CA^{j}A^{j T}C^T\otimes (I-\mu R_u)^{2j}\right)\nn\\
&=&(I-\;CA^{j}A^{j T}C^T)\otimes (I-\mu R_u)^{2j}\eq
Now, due to the even power, it always holds that  $(I-\mu R_u)^{2j}\geq 0$.
Moreover, since $A^j$ and $C$ are doubly stochastic, it follows that $CA^j A^{j T}C^T$ is
also doubly stochastic.
Therefore, the matrix $(I-CA^{j}A^{j T}C^T)$ is nonnegative-definite as well (cf.~property (e) of Lemma~\ref{thm.adasxxa}).
It follows that
\be
{\cal B}_{\rm lms}^{*j}{\cal B}_{\rm lms}^j\;-\;{\cal C}{\cal B}_{\rm cta}^{*j}{\cal B}_{\rm cta}^j{\cal C}^T\;\geq\;0\ee
But since ${\cal Y}_{\rm lms}\geq 0$, we conclude from (\ref{condkja.13}) that
\bq
\addbox{\;{\rm MSD}_{{\rm lms}}^{\rm network} \geq {\rm MSD}_{{\rm cta}}^{\rm network}\;}\eq
This is because for any two Hermitian nonnegative-definite matrices $A$ and $B$ of compatible dimensions, it holds that
$\mbox{\rm Tr}(AB)\geq 0$; indeed if we factor $B=XX^*$ with $X$ full rank, then $\mbox{\rm Tr}(AB)=\mbox{\rm Tr}(X^*AX)\geq 0$. We conclude from this analysis that
adaptive CTA diffusion performs better than non-cooperative LMS under uniform data profile conditions {\em and} doubly stochastic $A$.
If we refer to the earlier result (\ref{jakd.ka9831}), we conclude that the following relation holds:

\bq
\addbox{\;{\rm MSD}_{{\rm atc}}^{\rm network}\;\leq\;
{\rm MSD}_{{\rm cta}}^{\rm network}\;\leq\;
{\rm MSD}_{{\rm lms}}^{\rm network} \;}\eq
\noindent Table~\ref{table.compare} lists the comparison results derived in this section and lists the
conditions under which the conclusions hold.

\section{Selecting the Combination Weights}\label{sec.comb.rules}
The adaptive diffusion strategy
(\ref{general.1ada})--(\ref{general.3ada}) employs combination
weights $\{a_{1,\ell k},a_{2,\ell k},c_{\ell k}\}$ or, equivalently,
combination matrices $\{A_1,A_2,C\}$, where $A_1$ and $A_2$ are
left-stochastic matrices and $C$ is a right-stochastic matrix.
There are several ways by which these matrices can be selected. In
this section, we describe constructions that result in
left-stochastic or doubly-stochastic combination matrices, $A$. When
a right-stochastic combination matrix is needed, such as $C$, then it can be obtained by transposition of the left-stochastic constructions shown below.

\subsection{Constant Combination Weights}
Table~\ref{tablecons-1.label} lists a couple of common choices for
selecting constant combination weights for a network with $N$ nodes.
Several of these constructions appeared originally in
the literature on graph theory. In the
table, the symbol $n_k$ denotes the degree of node $k$, which refers
to the size of its neighborhood. Likewise, the symbol
$n_{\max}$ refers to the maximum degree across the network, i.e.,
\be n_{\max}\;=\;\max_{1\leq k\leq N}\;\{n_k\} \ee

The Laplacian
rule, which appears in the second line of the table, relies on the
use of the Laplacian matrix ${\cal L}$ of the network and a positive
scalar $\gamma$. The Laplacian matrix is defined by (\ref{define.laplacian})
in App.~\ref{appendix.B}, namely,  it is a symmetric matrix whose
entries are constructed as
follows \cite{cve98,bel98,kocay}: \be \left[{\cal
L}\right]_{k\ell}=\left\{\begin{array}{rl}n_k-1,&\mbox{\rm if}\;
k=\ell\\
-1,&\mbox{\rm if}\;k\neq \ell\;\mbox{\rm and nodes}\;k\;{\rm
and}\;\ell\;\mbox{\rm are neighbors}\\ 0,&\mbox{\rm
otherwise}\end{array}\right. \label{xx.define.laplacian}\ee
The Laplacian rule
can be reduced to other forms through the selection of the positive
parameter $\gamma$. One choice is $\gamma=1/n_{\max},$ while another
choice is $\gamma=1/N$ and leads to the maximum-degree rule.
Obviously, it always holds that $n_{\max}\leq N$ so that
$1/n_{\max}\geq 1/N$. Therefore, the choice $\gamma=1/n_{\max}$ ends
up assigning larger weights to neighbors than the choice
$\gamma=1/N$. The averaging rule in the first row of the table is one of the simplest combination rules whereby
nodes simply average data from their neighbors.

{\small
\begin{table}[ht]
\begin{center}
\caption{{\small \rm Selections for combination matrices}
$A=[a_{\ell k}]$.} {\small
\begin{tabular}{|l|l|}\hline\hline &\\
{\sc Entries of Combination Matrix $A$} & {\sc Type of} $A$  \\\hline &\\
\;\;{\bf 1. Averaging rule} \cite{blondel}:& \\&\\
$\;\;a_{\ell k} =\left\{
\begin{array}{ll}1/n_{k},&\mbox{\rm if $k\neq \ell$ are
neighbors or $k=\ell$}\\
0,&\mbox{\rm otherwise}
\end{array}
\right.$ & $\begin{array}{l}
\mbox{\rm left-stochastic}\end{array}$ \\\hline&\\
\;\;{\bf 2. Laplacian rule} \cite{scherber04_ipsn,Xiao04}:&\\
$\;\;A=I_N-\gamma {\cal L},\;\gamma>0$ &$\begin{array}{l}\mbox{\rm symmetric and} \\
\mbox{\rm doubly-stochastic}\end{array}$
\\\hline&\\
$\begin{array}{l}\mbox{\rm {\bf 3. Laplacian rule} using}\;
 \gamma=1/n_{\max}:\end{array}$ &\\&\\ $\;\;a_{\ell k}=\left\{
\begin{array}{ll}1/n_{\max},&\mbox{\rm if $k\neq \ell$ are
neighbors}\\
1-(n_k-1)/n_{\max},&\mbox{\rm $k=\ell$ }\\
0,&\mbox{\rm otherwise}
\end{array}
\right.$&$\begin{array}{l}\mbox{\rm symmetric and} \\
\mbox{\rm doubly-stochastic}\end{array}$\\ \hline&\\
$\begin{array}{l}\mbox{\rm {\bf 4. Laplacian rule} using}\;
 \gamma=1/N (\mbox{\rm maximum-degree rule \cite{Xiao05}}):\end{array}$
 &\\&\\ $\;\;a_{\ell
k}=\left\{
\begin{array}{ll}1/N,&\mbox{\rm if $k\neq \ell$ are
neighbors}\\
1-(n_k-1)/N,&\mbox{\rm $k=\ell$ }\\
0,&\mbox{\rm otherwise}
\end{array}
\right.$& $\begin{array}{l}\mbox{\rm symmetric and} \\
\mbox{\rm doubly-stochastic}\end{array}$\\\hline&\\
\;\;{\bf 5. Metropolis rule} \cite{Xiao04,metropolis,hastings}:&\\&\\
$\;\;a_{\ell k}\;=\;\left\{\begin{array}{ll}
1/\max\{n_k,n_\ell\},&\mbox{\rm if  $k\neq \ell$ are neighbors}\\
1-\displaystyle \sum_{\ell\in{\cal{N}}_k\backslash\{k\}}a_{\ell k},
&\mbox{\rm
  $k=\ell$}\\
   0, &\mbox{\rm otherwise}\\
\end{array}\right.
$ & $\begin{array}{l}\mbox{\rm symmetric and} \\
\mbox{\rm doubly-stochastic}\end{array}$\\\hline&\\
\;\;{\bf 6. Relative-degree rule} \cite{Cattivelli08a}: &\\&\\ $\;\;a_{\ell k}=\left\{
\begin{array}{ll}n_{\ell} / \displaystyle \left(\sum_{m\in\mathcal{N}_k} n_m\right),&\mbox{\rm if $k$
and $\ell$ are neighbors or $k=\ell$}\\
0,&\mbox{\rm otherwise}
\end{array}
\right. $ & $\begin{array}{l} \mbox{\rm
left-stochastic}\end{array}$\\\hline\hline
\end{tabular} }
\label{tablecons-1.label}
\end{center}
\end{table}
}

In the constructions in Table~\ref{tablecons-1.label}, the
values of the weights $\{a_{\ell k}\}$ are largely dependent on
the degree of the nodes. In this way, the number of connections
that each node has influences the combination weights with its
neighbors. While such selections may be appropriate in some
applications, they can nevertheless degrade the performance of
adaptation  over networks \cite{Takahashi10}. This is because such weighting schemes
ignore the noise profile across the network. And since
some nodes can be noisier than others, it is not sufficient to rely solely on the amount of
connectivity that nodes have to determine the combination weights
to their neighbors. It is important to take into
account the amount of noise that is present at the nodes as well.
Therefore, designing combination rules that are aware of the
variation in noise profile across the network is an important task.
It is also important to devise strategies that are able to {\em adapt} these combination
weights in response to variations in network topology and data statistical profile.
For this reason, following \cite{TS11,zhaoxiao}, we describe in the next subsection
one adaptive procedure for adjusting the combination weights.
This procedure allows  the network to assign more or less relevance
to nodes according to the quality of their data.

\subsection{Optimizing the Combination Weights}\label{sec7.1}
Ideally, we would like to select $N\times N$ combination matrices
$\{A_1,A_2,C\}$ in order to minimize the network MSD given by
(\ref{kdla8912.12}) or (\ref{imajdka}). In \cite{Cattivelli10}, the selection of the
combination weights was formulated as the following
optimization problem:
\begin{equation}\label{eq19}\boxed{\;\;\;\;\\
\begin{aligned}\\
    &\min_{\{A_1,A_2,C\}} \quad \text{MSD}^{\rm network}\;\;\mbox{\rm given by}\;(\ref{kdla8912.12})\;\mbox{\rm or}\;
    (\ref{imajdka})\\
    &\;\;\;\text{over left and right-stochastic matrices with nonnegative entries:}\\
    &\qquad A_1^T\mathds{1}=\mathds{1},\quad
    a_{1,\ell k}=0 \text{ if $\ell\notin \mathcal{N}_k$}\\
    &\qquad A_2^T\mathds{1}=\mathds{1},\quad
    a_{2,\ell k}=0 \text{ if $\ell\notin \mathcal{N}_k$}\\
    &\qquad C\mathds{1}=\mathds{1},\quad\quad
    c_{\ell k}=0 \;\text{ if $\ell\notin \mathcal{N}_k$\;\;\;\;}\\
\end{aligned}}
\end{equation}

 \noindent We can pursue a numerical solution to (\ref{eq19}) in order to
search for optimal combination matrices, as was done in
\cite{Cattivelli10}. Here, however, we are interested in an adaptive
solution that becomes part of the learning process so that the
network can adapt the weights on the fly in response to network
conditions.  We illustrate an
approximate approach from \cite{TS11,zhaoxiao} that leads to one adaptive solution that
performs reasonably well in practice.

We illustrate the construction by considering the ATC strategy (\ref{Equ:DiffusionAdaptation:ATC.adaptive.2}) without information
exchange where $A_1=I_N,$  $A_2=A$, and $C=I$. In this case, recursions
(\ref{general.1adax})--(\ref{general.3adax}) take the form:
 \bq
                \bm{\psi}_{k,i}      &=&    \w_{k,i-1} +
                \mu_k \u_{k,i}^*\left[\d_{k}(i)-\u_{k,i}\w_{k,i-1}\right]\label{cokajd.2}\\
                \w_{k,i}     &=&    \displaystyle\sum_{\ell\in{\cal N}_k} a_{\ell k}
                \bm{\psi}_{\ell,i}\label{cokajd}
  \eq
and, from (\ref{imajdka}),  the corresponding network MSD
performance is: \be {\rm MSD}^{\rm network}_{\rm atc}\;=\;
\frac{1}{N}\sum_{j=0}^{\infty}\;\mbox{\rm Tr}\left({\cal B}_{\rm
atc}^j {\cal Y}_{\rm atc}{\cal B}_{\rm
atc}^{*j}\right)\label{imajdka.act}\ee where \bq {\cal B}_{\rm
atc}&=&{\cal A}^T(I-{\cal M}{\cal R}_u)\\
{\cal Y}_{\rm
atc}&=&{\cal A}^T{\cal M}{\cal S}{\cal M}{\cal A}\label{usialdalk}\\
{\cal R}_u&=&\mbox{\rm diag}\{R_{u,1},R_{u,2},\ldots,R_{u,N}\}\\
{\cal S}&=&\mbox{\rm diag}\{\sigma_{v,1}^2 R_{u,1},\;\sigma_{v,2}^2
R_{u,2},\ldots,\sigma_{v,N}^2 R_{u,N}\}\\
{\cal M}&=&\mbox{\rm diag}\{\mu_1 I_M,\;\mu_2 I_M,\ldots,\mu_N I_M\}\\
{\cal A}&=&A\otimes I_M \eq Minimizing the MSD expression
(\ref{imajdka.act}) over left-stochastic matrices $A$ is generally
non-trivial. We pursue an approximate solution.

To begin with, for compactness of notation, let $r$ denote the spectral radius of the $N\times N$ block matrix
$I-{\cal M}{\cal R}_u$:
\be
r\define\;\rho(I-{\cal M}{\cal R}_u)
\ee
We already know, in view of the mean and mean-square stability of the network, that $|r|<1$. Now,
consider the series that appears in (\ref{imajdka.act}) and whose trace we wish to minimize over $A$.
Note that its block maximum norm can be bounded as follows:
\bq
\left\|\sum_{j=0}^{\infty} {\cal B}_{\rm
atc}^j {\cal Y}_{\rm atc}{\cal B}_{\rm
atc}^{*j}\right\|_{b,\infty}&\leq&
\sum_{j=0}^{\infty}\; \left\|{\cal B}_{\rm
atc}^j\right\|_{b,\infty}\cdot \left\|{\cal Y}_{\rm atc}\right\|_{b,\infty}\cdot\left\|{\cal B}_{\rm
atc}^{*j}\right\|_{b,\infty}\nn\\
&\stackrel{(a)}{\leq}&
N\cdot \left(\sum_{j=0}^{\infty} \;\left\|{\cal B}_{\rm
atc}^{j}\right\|_{b,\infty}^2\cdot \left\|{\cal Y}_{\rm atc}\right\|_{b,\infty}\right)\nn\\
&\leq&
N\cdot \left(\sum_{j=0}^{\infty}\; \left\|{\cal B}_{\rm
atc}\right\|_{b,\infty}^{2j}\cdot \left\|{\cal Y}_{\rm atc}\right\|_{b,\infty}\right)\nn\\
&\stackrel{(b)}{\leq}&
N\cdot \left(\sum_{j=0}^{\infty}\; r^{2j}\cdot
\left\|{\cal Y}_{\rm atc}\right\|_{b,\infty}\right)\nn\\
&=&
\frac{N}{1-r^2}\cdot
\left\|{\cal Y}_{\rm atc}\right\|_{b,\infty}\label{ser.bd}
\eq
where for step (b) we use result (\ref{hdk912l}) to conclude that
\bq \|{\cal B}_{\rm atc}\|_{b,\infty}&=&
\|{\cal A}^T(I-{\cal M}{\cal R}_u)\|_{b,\infty}\nn\\
&\leq &
\|{\cal A}^T\|_{b,\infty}\cdot \|I-{\cal M}{\cal R}_u\|_{b,\infty}\nn\\
&= &
\|I-{\cal M}{\cal R}_u\|_{b,\infty}\nn\\
&\stackrel{(\ref{hdk912l})}{=} &
r\eq
To justify step (a), we use result (\ref{89akd.uasuyef})
 to relate the norms of ${\cal B}_{\rm atc}^j$ and its
complex conjugate, $\left[{\cal B}_{\rm atc}^j\right]^*$, as
\be
\|{\cal B}_{\rm atc}^{*j}\|_{b,\infty}\;\leq\;N\cdot \|{\cal B}_{\rm atc}^{j}\|_{b,\infty}
\ee
Expression (\ref{ser.bd}) then shows that the norm of the series appearing in (\ref{imajdka.act}) is bounded by
a scaled multiple of the norm of ${\cal Y}_{\rm atc}$, and the scaling constant is independent of $A$. Using
property (\ref{use.kajd}) we conclude that there exists a positive constant $c$, also independent of $A$, such that
\be
\mbox{\rm Tr}\left(\sum_{j=0}^{\infty} {\cal B}_{\rm
atc}^j {\cal Y}_{\rm atc}{\cal B}_{\rm
atc}^{*j}\right)\;\leq\; c\cdot \mbox{\rm Tr}({\cal Y}_{\rm atc})
\ee
Therefore,
instead of attempting to
minimize the trace of the series, the above result motivates us to minimize an upper bound to the trace.
Thus, we consider the alternative problem of minimizing
the first term of the series (\ref{imajdka.act}), namely,
\be\label{reduasa}
\begin{aligned}
    &\min_{A} \quad \text{Tr}({\cal Y}_{\rm atc})\\
    &\text{subject to}\;\; A^T\mathds{1}=\mathds{1},\;\;a_{\ell k}\geq 0,\;\;
    a_{\ell k}=0 \text{ if $\ell \notin \mathcal{N}_k$}
\end{aligned}
\ee Using (\ref{usialdalk}), the trace of ${\cal Y}_{\rm atc}$ can be
expressed in terms of the combination coefficients as follows: \be
\mbox{\rm Tr}({\cal Y}_{\rm atc})\;=\;\sum_{k=1}^N
\sum_{\ell=1}^N\;\mu_\ell^2\; a_{\ell k}^2\;\sigma_{v,\ell}^2\;\mbox{\rm
Tr}(R_{u,\ell}) \ee so that problem (\ref{reduasa}) can be decoupled
into $N$ separate optimization problems of the form:
\begin{equation}\label{eq20.xx}\boxed{\;\;\;
\begin{aligned}\\
    &\min_{\{a_{\ell k}\}_{\ell =1}^N} \quad \sum_{\ell =1}^N \mu_{\ell}^2\; a^2_{\ell
    k}\;\sigma^2_{v,\ell}\;\mbox{\rm Tr}(R_{u,\ell})
    \text{,\;\;\;  $k=1,\ldots,N$}\\
    &\text{subject to}\\
    &\qquad a_{\ell k}\geq 0,\qquad \sum_{\ell =1}^Na_{\ell k}=1,\quad
    a_{\ell k}=0 \text{ if $\ell \notin \mathcal{N}_k$\;\;\;}\\
\end{aligned}}
\end{equation}
With each node $\ell$, we associate the following nonnegative
noise-data-dependent measure: \be \addbox{\;\gamma_{\ell}^2\define
\mu_{\ell}^2\cdot \sigma^{2}_{v,\ell}\cdot \mbox{\rm Tr}(R_{u,\ell})\;} \label{defi.ahdas}\ee This
measure amounts to scaling the noise variance at node $\ell$ by $\mu_{\ell}^2$ and by the
power of the regression data (measured through the trace of its
covariance matrix). We shall refer to $\gamma_{\ell}^2$ as the
noise-data variance product (or {\em variance product}, for simplicity) at node $\ell$. Then, the solution of
(\ref{eq20.xx}) is given by:
\begin{equation} \label{eq11.xx}
\addbox{\;    a_{\ell k}=
    \begin{cases}
    \frac{\gamma_{\ell}^{-2}}{\sum_{m\in\mathcal{N}_k}\gamma_{m}^{-2}},
    &\text{if $\ell \in\mathcal{N}_k$}\\
    0, &\text{otherwise}
    \end{cases}\;}\;\;\;\;\;(\mbox{\rm relative-variance
    rule})
\end{equation}
We refer to this combination rule as the {\em relative-variance
combination rule} \cite{TS11}; it leads to a left-stochastic matrix
$A$. In this construction, node $k$ combines the intermediate estimates
$\{\bm{\psi}_{\ell,i}\}$ from its neighbors in (\ref{cokajd}) in
proportion to the inverses of their variance products, $\{\gamma_{m}^{-2}\}$. The result
is physically meaningful. Nodes with smaller  variance products
will generally be given larger weights. In comparison, the following
{\em relative-degree-variance rule} was proposed in
\cite{Cattivelli10} (a typo appears in Table III in
\cite{Cattivelli10}, where the noise variances appear written in the
table instead of their inverses):
\begin{equation} \label{eq12}
\addbox{\;a_{\ell k}=
    \begin{cases}
    \frac{n_\ell  \sigma_{v,\ell }^{-2}}{\sum_{m\in{\cal N}_k} n_m
    \sigma_{v,m}^{-2}},
    &\text{if $\ell\in\mathcal{N}_k$}\\
    0, &\text{otherwise}
    \end{cases}\;}\;\;\;\;\;\;(\mbox{\rm relative degree-variance rule})
\end{equation}
This second form also leads to a left-stochastic combination matrix
$A$. However, rule (\ref{eq12}) does not take into account the covariance
matrices of the regression data across the network. Observe that in the special case
when step-sizes, the regression covariance matrices, and the noise variances are uniform across the
network, i.e., $\mu_k=\mu$, $R_{u,k}=R_u$, and $\sigma_{v,k}^2=\sigma_v^2$ for all
$k$, expression (\ref{eq11.xx}) reduces to the simple averaging rule
(first line of Table~\ref{tablecons-1.label}). In contrast,
expression (\ref{eq12}) reduces the relative degree rule (last line
of Table~\ref{tablecons-1.label}).

\subsection{Adaptive Combination Weights}\label{sec7.2}
To evaluate the combination weights (\ref{eq11.xx}), the nodes need to
know the variance products, $\{\gamma_{m}^2\}$, of their
neighbors.  According to (\ref{defi.ahdas}), the factors
 $\{\gamma_{m}^2\}$ are defined in terms of the noise
variances, $\{\sigma_{v,m}^2\}$, and the regression covariance
matrices, $\{\mbox{\rm Tr}(R_{u,m})\}$, and these quantities are not
known beforehand. The nodes only have access to realizations of
$\{\d_{m}(i),\u_{m,i}\}$. We now describe one procedure that allows
every node $k$ to learn the variance products of its neighbors in
an adaptive manner. Note that if a particular node $\ell$ happens to
belong to two neighborhoods, say, the neighborhood of node $k_1$ and
the neighborhood of node $k_2$, then each of $k_1$ and $k_2$ need to
evaluate the variance product, $\gamma_{\ell}^2$, of node $\ell$.
The procedure described below allows each node in the network to
estimate the variance products of its neighbors in a recursive
manner.

To motivate the algorithm, we refer to the ATC recursion
(\ref{cokajd.2})--(\ref{cokajd}) and use the data model
(\ref{lkad8912.lakd.ada}) to write for node $\ell$: \be
\boldsymbol{\psi}_{\ell,i} =\w_{\ell,i-1}\;+\;\mu_{\ell}
\u_{\ell,i}^*\left[\u_{\ell,i}\widetilde{\w}_{\ell,i-1}+\v_{\ell}(i)\right]\ee
so that, in view of our earlier assumptions on the regression data
and noise in Sec.~\ref{sec.daada}, we obtain in the limit as
$i\rightarrow\infty$: \be \lim_{i\rightarrow\infty}\;
\Ex\left\|\boldsymbol{\psi}_{\ell,i} -\w_{\ell,i-1}\right\|^2\;=\;
\mu_{\ell}^2\cdot
\left(\lim_{i\rightarrow\infty}\Ex\|\widetilde{\w}_{i-1}\|^2_{\Ex\left(\u_{\ell,i}^*
\|\u_{\ell,i}\|^2\u_{\ell,i}\right)}\right)\;+\;\mu_{\ell}^2\cdot
\sigma_{v,\ell}^2\cdot \mbox{\rm Tr}(R_{u,\ell}) \label{89ajd712}\ee
We can evaluate the limit on the right-hand side by using the
steady-state result (\ref{kdl8912.12.asas}). Indeed, we select the
vector $\sigma$ in (\ref{kdl8912.12.asas}) to satisfy \be (I-{\cal
F})\sigma\;=\;\mbox{\rm vec}\left[\Ex\left(\u_{\ell,i}^*
\|\u_{\ell,i}\|^2\u_{\ell,i}\right)\right] \ee Then, from
(\ref{kdl8912.12.asas}),
\be\lim_{i\rightarrow\infty}\;\Ex\|\widetilde{\w}_{i-1}\|_{\Ex\left(\u_{\ell,i}^*
\|\u_{\ell,i}\|^2\u_{\ell,i}\right)}^2\;=\; \left[\mbox{\rm
vec}\left({\cal Y}^T\right)\right]^T \cdot (I-{\cal
F})^{-1}\cdot\mbox{\rm vec}\left[\Ex\left(\u_{\ell,i}^*
\|\u_{\ell,i}\|^2\u_{\ell,i}\right)\right]\ee Now recall from
expression (\ref{usialdalk}) for ${\cal Y}$ that for the ATC
algorithm  under consideration we have \be {\cal Y}={\cal A}^T{\cal
M}{\cal S}{\cal M}{\cal A} \ee so that the entries of ${\cal Y}$
depend on combinations of the squared step-sizes,
$\{\mu_m^2,\;m=1,2,\ldots,N\}$.
 This fact implies that the first term on the right-hand side of (\ref{89ajd712}) depends on products of
 the form $\{\mu_{\ell}^2\mu_{m}^2\}$; these fourth-order factors can be ignored in comparison to the second-order factor $\mu_{\ell}^2$
 for small step-sizes so that
 \bq
\lim_{i\rightarrow\infty}\;
\Ex\left\|\boldsymbol{\psi}_{\ell,i} -\w_{\ell,i-1}\right\|^2&\approx&
\mu_{\ell}^2\cdot \sigma_{v,\ell}^2\cdot \mbox{\rm Tr}(R_{u,\ell})\nn\\
&=&\gamma_{\ell}^2\eq
in terms of the desired variance product, $\gamma_{\ell}^2$. Using
the following instantaneous approximation at node $k$ (where $w_{\ell,i-1}$ is replaced by $w_{k,i-1}$):
\begin{align}
   \Ex\|\boldsymbol{\psi}_{\ell,i}-\w_{\ell,i-1}\|^2\;\approx\;\|\psi_{\ell,i}-w_{k,i-1}\|^2
\end{align}
we can motivate an algorithm that enables node $k$ to estimate the
variance product, $\gamma_{\ell}^2,$ of its neighbor $\ell$. Thus,
let ${\gamma}_{\ell k}^2(i)$ denote an estimate for
$\gamma_{\ell}^2$ that is computed by node $k$ at time $i$. Then,
one way to evaluate ${\gamma}_{\ell k}^2(i)$ is through the
recursion:
\begin{equation} \label{eq20}
\begin{aligned}
 \addbox{\;   {\gamma}^2_{\ell k}(i)=(1-\nu_k)\cdot {\gamma}^2_{\ell
 k}(i-1)\;+\;\nu_k\cdot\|{\psi}_{\ell,i}-w_{k,i-1}\|^2\;}
\end{aligned}
\end{equation}
where $0<\nu_k\ll 1$ is a positive coefficient smaller than one. Note that under expectation, expression (\ref{eq20}) becomes \be
\Ex\boldsymbol{\gamma}^2_{\ell k}(i)=
    (1-\nu_k)\cdot \Ex\boldsymbol{\gamma}^2_{\ell k}(i-1)\;+\;
    \nu_k\cdot
    \Ex\|\boldsymbol{\psi}_{\ell,i}-\boldsymbol{w}_{k,i-1}\|^2
    \ee
so that in steady-state, as $i\rightarrow\infty$, \be
\lim_{i\rightarrow\infty}\;\Ex\boldsymbol{\gamma}^2_{\ell k}(i)   \; \approx\;
(1-\nu_k)\cdot \lim_{i\rightarrow\infty}\;\Ex\boldsymbol{\gamma}^2_{\ell k}(i-1)\;+\;
    \nu_k\cdot \gamma^2_{\ell}
\end{equation}
Hence, we obtain
\begin{equation}
    \lim_{i\rightarrow\infty}\Ex\boldsymbol{\gamma}^2_{\ell k}(i)
    \approx \gamma^2_{\ell}
\end{equation}
That is, the estimator $\bm{\gamma}^2_{\ell k}(i)$ converges on
average close to the desired variance product $\gamma^2_{\ell}$. In this
way, we can replace the optimal weights (\ref{eq11.xx}) by the adaptive
construction:
\begin{equation}\label{eq15.xx}
\addbox{\;    a_{\ell k}(i)=
    \begin{cases}
    \frac{\gamma^{-2}_{\ell k}(i)}{\sum_{m\in\mathcal{N}_k}\gamma^{-2}_{m k}(i)},
    &\text{if $\ell\in\mathcal{N}_k$}\\
    0, &\text{otherwise}
    \end{cases}\;}
\end{equation}
Equations (\ref{eq20}) and (\ref{eq15.xx}) provide one adaptive
construction for the combination weights $\{a_{\ell k}\}$.

\section{Diffusion with Noisy Information Exchanges}
The adaptive diffusion strategy
(\ref{general.1ada})--(\ref{general.3ada}) relies on the fusion of
local information collected from neighborhoods through the use of
combination matrices $\{A_1,A_2,C\}$. In the previous section, we
described several constructions for selecting such combination
matrices. We also motivated and developed an adaptive scheme for the ATC mode of operation (\ref{cokajd.2})--(\ref{cokajd})
that
computes combination weights in a manner that is aware of the
variation of the variance-product  profile across the network.
Nevertheless, in addition to the measurement noises $\{\v_{k}(i)\}$
at the individual nodes, we also need to consider the effect of
perturbations that are introduced during the exchange of information
among neighboring nodes. Noise over the communication links can be
due to various factors including thermal noise and imperfect channel
information. Studying the degradation in mean-square performance
that results from these noisy exchanges can be pursued by
straightforward extension of the mean-square analysis of
Sec.~\ref{sec.mse}, as we proceed to illustrate.
Subsequently, we shall use the
results to show how the combination weights can also be adapted in the presence of
noisy exchange links.

\subsection{Noise Sources over Exchange Links}\label{sec9.1}
To model noisy links, we introduce an additive noise component into
each of the steps of the diffusion strategy
(\ref{general.1ada})--(\ref{general.3ada}) during the operations of
 information exchange among the nodes. The notation becomes a bit cumbersome
because we need to account for both the source and destination of
the information that is being exchanged. For example, the same
signal $\d_{\ell}(i)$ that is generated by node $\ell$ will be
broadcast to all the neighbors of node $\ell$. When this is done, a
different noise will interfere with the exchange of $\d_{\ell}(i)$ over each of the
edges that link node $\ell$ to its neighbors. Thus, we will need to
use a notation of the form $\d_{\ell k}(i)$, with two subscripts
$\ell$ and $k$, to indicate that this is the noisy version of
$\d_{\ell}(i)$ that is received by node $k$ from node $\ell$. The
subscript $\ell k$ indicates that $\ell$ is the source and $k$ is
the sink, i.e., information is moving from $\ell$ to $k$. For the
reverse situation where information flows from node $k$ to $\ell$,
we would use instead the subscript $k \ell$.

\begin{figure}[h]
\epsfxsize 9cm \epsfclipon
\begin{center}
\leavevmode \epsffile{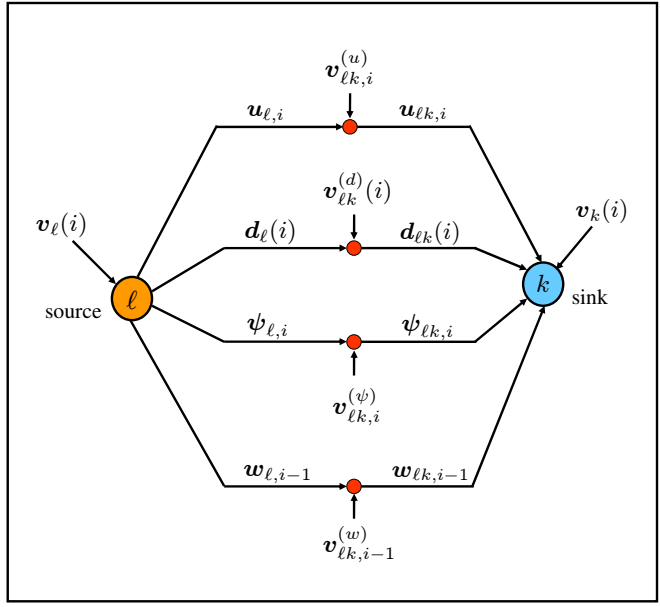} \caption{{\small Additive noise
sources perturb the exchange of information from node $\ell$ to node
$k$. The subscript $\ell k$ in this illustration indicates that $\ell$ is
the source node and
$k$ is  the sink node so that information is flowing from $\ell$ to $k$.
}}\label{fig-J.label}
\end{center}
\end{figure}

With this notation in mind, we model the noisy data received by node
$k$ from its neighbor $\ell$ as follows: \bq \label{eqn:noisyw}
{\bm{w}}_{\ell k,i-1}&=&{\bm{w}}_{\ell,i-1}+{\bm{v}}_{\ell k,i-1}^{(w)}\\
\label{eqn:noisypsi}
{\bm{\psi}}_{\ell k,i}&=&{\bm{\psi}}_{\ell,i}+{\bm{v}}_{\ell k,i}^{(\psi)}\\
\label{eqn:noisyu}
{\bm{u}}_{\ell k,i}&=&{\bm{u}}_{\ell,i}+{\bm{v}}_{\ell k,i}^{(u)}\\
\label{eqn:noisyd} {\bm{d}}_{\ell
k}(i)&=&{\bm{d}}_{\ell}(i)+{\bm{v}}_{\ell k}^{(d)}(i)\eq
 where
${\bm{v}}_{\ell k,i-1}^{(w)} \;(M\times 1)$, ${\bm{v}}_{\ell k,i}^{(\psi)}\; (M\times 1)$, and
${\bm{v}}_{\ell k,i}^{(u)}\; (1\times M)$ are vector noise signals, and
${\bm{v}}_{\ell k}^{(d)}(i)$ is a scalar noise signal. These are the
noise signals that perturb exchanges over the edge linking source
$\ell$ to sink $k$ (i.e., for data sent from node $\ell$ to node
$k$). The superscripts $\{(w),\;(\psi),\;(u),\;(d)\}$ in each case
refer to the variable that these noises perturb.
Figure~\ref{fig-J.label} illustrates the various noise sources that
perturb the exchange of information from node $\ell$ to node $k$.
The figure also shows the measurement noises
$\{\v_{\ell}(i),\v_{k}(i)\}$ that exist locally at the nodes in view
of the data model (\ref{lkad8912.lakd.ada}).

We assume that the following noise signals, which
influence the data received by node $k$, \be
\left\{{\boldsymbol{v}}_{k}(i),\;\;{\bm{v}}_{\ell
k}^{(d)}(i),\;\;{\bm{v}}_{\ell k,i-1}^{(w)}, \;\;{\bm{v}}_{\ell
k,i}^{(\psi)},\;\; {\bm{v}}_{\ell
k,i}^{(u)}\right\}\label{noise.eqaa}\ee are temporally white and spatially
independent random processes with zero mean and variances or covariances given by \be
\addbox{\;\left\{\sigma_{v,k}^{2},\;\;\sigma_{v,\ell k}^{2},\;\;
R_{v,\ell k}^{(w)},\;\; R_{v,\ell k}^{(\psi)},\;\;R_{v,\ell
k}^{(u)}\right\}\;}\ee Obviously, the quantities \be
\left\{\sigma_{v,\ell k}^{2},\;\; R_{v,\ell k}^{(w)},\;\; R_{v,\ell
k}^{(\psi)},\;\;R_{v,\ell k}^{(u)}\right\} \ee are all zero if
$\ell\notin{\mathcal{N}}_k$ or when $\ell=k$. We further assume that
the noise processes (\ref{noise.eqaa}) are independent of each other
and of the regression data $\u_{m,j}$ for all $k, \ell, m $ and
$i,j$.

\subsection{Error Recursion}
 Using the perturbed data (\ref{eqn:noisyw})--(\ref{eqn:noisyd}), the adaptive
diffusion strategy (\ref{general.1ada})--(\ref{general.3ada})
becomes \bq
\label{eqn:noisyDiffusionPriorDiff1}
{\bm{\phi}}_{k,i-1}&=&\sum_{\ell\in{\cal N}_k} a_{1,\ell k}\;{\bm{w}}_{\ell k,i-1}\\
\label{eqn:noisyDiffusionIncremental1}
{\bm{\psi}}_{k,i}&=&{\bm{\phi}}_{k,i-1}+\mu_k\sum_{\ell\in{\cal
N}_k}c_{\ell k}\;{\bm{u}}_{\ell k,i}^*\;
[{\bm{d}}_{\ell k}(i)-{\bm{u}}_{\ell k,i}{\bm{\phi}}_{k,i-1}]\\
\label{eqn:noisyDiffusionPostDiff1}
{\bm{w}}_{k,i}&=&\sum_{\ell\in{\cal N}_k}a_{2,\ell
k}\;{\bm{\psi}}_{\ell k,i} \eq Observe that the perturbed quantities
$\{\w_{\ell k,i-1}, \u_{\ell k,i}, \d_{\ell k}(i),\bm{\psi}_{\ell
k,i}\}$, with subscripts $\ell k$, appear in
(\ref{eqn:noisyDiffusionPriorDiff1})--(\ref{eqn:noisyDiffusionPostDiff1})
in place of the original quantities $\{\w_{\ell,i-1}, \u_{\ell,i},
\d_{\ell}(i),\bm{\psi}_{\ell,i}\}$ that appear in
(\ref{general.1ada})--(\ref{general.3ada}). This is because these
quantities are now subject to exchange noises. As before, we are
still  interested in examining the evolution of the weight-error
vectors: \be \widetilde{\w}_{k,i}\define w^o-\w_{k,i},\;\;\;\;k=1,2,\ldots,N \ee For this
purpose, we again introduce the following  $N\times 1$ block vector,
whose entries are of size $M\times 1$ each: \be
\widetilde{\w}_i\define
\ba{c}\widetilde{\w}_{1,i}\\
\widetilde{\w}_{2,i}\\\vdots\\\widetilde{\w}_{N,i}\ea \ee and
proceed to determine a recursion for its evolution over time. The
arguments are largely similar to what we already did before in
Sec.~\ref{msnkajd} and, therefore, we shall emphasize the differences that arise. The main
deviation is that we now need to account for
the presence of the new noise signals; they will contribute
additional terms to the recursion for  $\widetilde{\w}_i$ --- see
(\ref{eqn:noisyerrorrecursion1}) further ahead. We may note that some
studies on the effect of imperfect data exchanges on the performance
of adaptive diffusion algorithms are considered in \cite{abdolee,khalili,TS11b}. However, these earlier investigations were limited to particular cases in which only noise in the exchange of $\w_{\ell,i-1}$  was considered (as in (\ref{eqn:noisyw})), in addition to setting $C=I$ (in which case there is {\em no} exchange of $\{\d_{\ell}(i),\u_{\ell,i}\}$),
and by focusing on the CTA case for which $A_2=I$. Here, we consider instead
the general case that accounts
for the additional sources of imperfections shown in (\ref{eqn:noisypsi})--(\ref{eqn:noisyd}), in addition to the general
diffusion strategy (\ref{general.1ada})--(\ref{general.3ada}) with
combination matrices $\{A_1,A_2,C\}$.

To begin with, we introduce the aggregate $M\times 1$ zero-mean
noise signals: \be \label{eqn:vkwpsidef}
{\bm{v}}_{k,i-1}^{(w)}\;\define \;\sum_{\ell\in{\cal N}_k}a_{1,\ell
k}\;{\bm{v}}_{\ell k,i-1}^{(w)},\;\;\;\;\;\;\;\;\;\;
{\bm{v}}_{k,i}^{(\psi)}\;\define\;\sum_{\ell\in{\cal N}_k}a_{2,\ell
k}\;{\bm{v}}_{\ell k,i}^{(\psi)} \ee These noises represent the
aggregate effect on node $k$ of all exchange noises from the
neighbors of node $k$ while exchanging the estimates
$\{\w_{\ell,i-1},\bm{\psi}_{\ell,i}\}$ during the two combination
steps (\ref{general.1ada}) and (\ref{general.3ada}). The $M\times M$
covariance matrices of these noises are given by\be
\label{eqn:Rvkwpsidef} \addbox{\;R_{v,k}^{(w)}\define
\sum_{\ell\in{\cal N}_k}a_{1,\ell k}^2 \;R_{v,\ell
k}^{(w)},\;\;\;\;\;\;\;\;\;\; R_{v,k}^{(\psi)}\define \sum_{\ell\in{\cal
N}_k}a_{2,\ell k}^2\;R_{v,\ell k}^{(\psi)}\;} \ee These expressions
aggregate the exchange noise covariances in the neighborhood of node
$k$; the covariances are scaled by the squared coefficients
$\{a_{1,\ell k}^2,\;a_{2,\ell k}^2\}$. We collect these noise
signals, and their covariances, from across the network into
$N\times 1$ block vectors and $N\times N$ block diagonal matrices as
follows: \bq \label{eqn:vwdef}
{\bm{v}}_{i-1}^{(w)}&\define&{\mathrm{col}}\left\{{\bm{v}}_{1,i-1}^{(w)},\;{\bm{v}}_{2,i-1}^{(w)},\dots,\;{\bm{v}}_{N,i-1}^{(w)}\right\}\\
\label{eqn:vpsidef}
{\bm{v}}_i^{(\psi)}&\define&{\mathrm{col}}\left\{{\bm{v}}_{1,i}^{(\psi)},\;{\bm{v}}_{2,i}^{(\psi)},\dots,\;{\bm{v}}_{N,i}^{(\psi)}\right\}\\
\label{eqn:Rvwdef}
R_v^{(w)}&\define&{\mathrm{diag}}\left\{R_{v,1}^{(w)},\;R_{v,2}^{(w)},\;\dots,\;R_{v,N}^{(w)}\right\}\\
\label{eqn:Rvpsidef}
R_v^{(\psi)}&\define&{\mathrm{diag}}\left\{R_{v,1}^{(\psi)},\;R_{v,2}^{(\psi)},\;\dots,\;R_{v,N}^{(\psi)}\right\}
\eq We further introduce the following scalar zero-mean noise
signal: \be \v_{\ell k}(i)\define \v_{\ell}(i)\;+\;\v_{\ell
k}^{(d)}(i)\;-\;\v_{\ell k,i}^{(u)}\; w^o \label{lkad012.lkad}\ee
whose variance is \be \addbox{\;\sigma_{\ell k}^{2}=
\sigma_{v,\ell}^2\;+\;\sigma_{v,\ell k}^2\;+\;w^{o*} R_{v,\ell
k}^{(u)}w^o\;}
 \ee
In the absence of exchange noises for the data
$\{\d_{\ell}(i),\u_{\ell,i}\}$, the signal $\v_{\ell k}(i)$ would
coincide with the measurement noise $\v_{\ell}(i)$. Expression
(\ref{lkad012.lkad}) is simply a reflection of the aggregate effect of the
noises in exchanging $\{\d_{\ell}(i),\u_{\ell,i}\}$ on node $k$. Indeed,
starting from the data model (\ref{lkad8912.lakd.ada}) and using
(\ref{eqn:noisyu})--(\ref{eqn:noisyd}), we can easily verify that
the noisy data $\{\d_{\ell k}(i),\u_{\ell k,i}\}$ are related via:
\be \d_{\ell k}(i)\;=\;\u_{\ell k,i} w^o\;+\;\v_{\ell k}(i)
 \ee
We also define (compare with
(\ref{defkajs812.ada})--(\ref{compadlkas.12}) and note that we are now using the perturbed regression vectors
$\{\u_{\ell k,i}\}$):
 \bq
\label{eqn:Ukiprimedef} {\bm{R}}_{k,i}'&\define&\sum_{\ell\in{\cal
N}_k}c_{\ell k}{\bm{u}}_{\ell k,i}^*{\bm{u}}_{\ell k,i}\\
\label{eqn:Uprimedef}
{\bm{\mathcal{R}}}_i'&\define&{\mathrm{diag}}\left\{{\bm{R}}_{1,i}',\;{\bm{R}}_{2,i}',\;\dots,\;{\bm{R}}_{N,i}'\right\}\eq
It holds that \be \addbox{\;\Ex \R_{k,i}' = R_{k}'\;} \ee where \be
\addbox{\;R_k'\define \sum_{\ell\in{\cal N}_k}c_{\ell
k}\left[R_{u,\ell}\;+\;R_{v,\ell k}^{(u)}\right]\;}
  \ee
When there is no noise during the exchange of the regression data,
i.e., when $R_{v,\ell k}^{(u)}=0$, the expressions for
$\{\bm{R}_{k,i}',\;{\bm{\mathcal{R}}}_i',\;R_k'\}$ reduce to
expressions (\ref{defkajs812.ada})--(\ref{compadlkas.12}) and
(\ref{defkajs812}) for
$\{\bm{R}_{k,i},\;{\bm{\mathcal{R}}}_i,\;R_k\}$.

Likewise, we introduce (compare with (\ref{dkja891.lakdlk})): \bq
\label{eqn:zkiprimedef} {\bm{z}}_{k,i}&\define&\sum_{\ell\in{\cal
N}_k}c_{\ell
k}{\bm{u}}_{\ell k,i}^* \v_{\ell k}(i)\\
\label{eqn:ZZprimedef}
{\bm{z}}_i&\define&{\mathrm{col}}\left\{{\bm{z}}_{1,i},\;{\bm{z}}_{2,i},\dots,\;{\bm{z}}_{N,i}\right\}\eq
Compared with the earlier definition for $\s_i$ in
(\ref{dkja891.lakdlk}) when there is no noise over the exchange
links, we see that we now need to account for the various noisy
versions of the same regression vector $\u_{\ell,i}$ and the same
signal $\d_{\ell}(i)$. For instance, the vectors $\u_{\ell k,i}$ and
$\u_{\ell m,i}$ would denote two noisy versions received by nodes
$k$ and $m$ for the {\em same} regression vector $\u_{\ell,i}$ transmitted
from node $\ell$. Likewise, the scalars $\d_{\ell k}(i)$ and
$\d_{\ell m}(i)$ would denote two noisy versions received by nodes
$k$ and $m$ for the {\em same} scalar $\d_{\ell}(i)$ transmitted from node
$\ell$. As a result, the quantity $\z_{i}$ is not zero mean any
longer (in contrast to $\s_i$, which had zero mean). Indeed, note that \bq
\Ex\z_{k,i}&=& \sum_{\ell\in{\cal N}_k}c_{\ell k}\;\Ex{\bm{u}}_{\ell
k,i}^*
\v_{\ell k}(i)\nn\\
&=&\sum_{\ell\in{\cal N}_k}c_{\ell
k}\;\Ex\left(\left[{\bm{u}}_{\ell,i}+\v_{\ell
k,i}^{(u)}\right]^*\cdot\left[\v_{\ell}(i)\;+\;\v_{\ell
k}^{(d)}(i)\;-\;\v_{\ell k,i}^{(u)}\; w^o\right]\right)\nn\\
&=& -\left(\sum_{\ell\in{\cal N}_k}c_{\ell k}\;R_{v,\ell
k}^{(u)}\right)w^o
 \eq
It follows that \be
\Ex\z_i\;=\;-\ba{c}\displaystyle\sum_{\ell\in{\cal
N}_1}c_{\ell 1}\;R_{v,\ell 1}^{(u)}\\\\
\displaystyle\sum_{\ell\in{\cal N}_2}c_{\ell 2}\;R_{v,\ell
2}^{(u)}\\\\\vdots\\\\\displaystyle\sum_{\ell\in{\cal N}_N}c_{\ell
N}\;R_{v,\ell N}^{(u)} \ea\;w^o \ee Although we can continue our
analysis by studying this general case in which the vectors $\z_i$
do not have zero-mean (see \cite{zhaoxiao,zhaoxiao2abc}), we
shall nevertheless limit our discussion in the sequel to the case in
which there is no noise during the exchange of the regression data,
i.e., we henceforth assume that: \be \addbox{\;\v_{\ell
k,i}^{(u)}=0,\;\;\;R_{v,\ell k}^{(u)}\;=0,\;\;\;\u_{\ell
k,i}=\u_{\ell,i}\;}\;\;(\mbox{\rm assumption from this point
onwards}) \label{dlk9812.al}\ee We maintain all other noise sources, which
occur during the exchange of the weight estimates
$\{\w_{\ell,i-1},\bm{\psi}_{\ell,i}\}$ and the data
$\{\d_{\ell}(i)\}$. Under condition (\ref{dlk9812.al}),  we obtain \bq \Ex\z_i&=&0\\
\sigma_{\ell k}^2&=&\sigma_{v,\ell}^2+\sigma_{v,\ell k}^2\label{defl.saZZ.1}\\
R_{k}'&=&\sum_{\ell\in{\cal N}_k}\;c_{\ell
k}\;R_{u,\ell}\;\stackrel{(\ref{defkajs812})}{=}\;R_k
 \eq
Then, the covariance matrix of each term $\z_{k,i}$ is given by \be
\addbox{\;R_{z,k}\define  \sum_{\ell\in{\cal N}_k}c_{\ell
k}^2\;\sigma_{\ell k}^2 \;R_{u,\ell}\;}\ee  and the covariance matrix
of $\z_i$ is $N\times N$ block diagonal with blocks of size $M\times
M$:\be \addbox{\;{\cal Z}\define \Ex\z_i\z_i^*\;=\;\mbox{\rm diag}\{
R_{z,1},\;R_{z,2},\ldots,\;R_{z,N}
 \}\;} \label{defl.saZZ} \ee
Now repeating the argument that led to (\ref{like.lakd.ada}) we
arrive at the following recursion for the weight-error vector:
\begin{equation}
\addbox{\; \label{eqn:noisyerrorrecursion1}
{\widetilde{\bm{w}}}_i=\mathcal{A}_2^T\left(I_{NM}-{\mathcal{M}}{\bm{\mathcal{R}}}_i'\right){\mathcal{A}_1^T}{\widetilde{\bm{w}}}_{i-1}-{\mathcal{A}}_2^T{\mathcal{M}}{\bm{z}}_i
-{\mathcal{A}}_2^T\left(I_{NM}-{\mathcal{M}}{\bm{\mathcal{R}}}_i'\right){\bm{v}}_{i-1}^{(w)}
-{\bm{v}}_i^{(\psi)}\; } \;\;(\mbox{\rm noisy
links})\end{equation} For comparison purposes, we repeat recursion
(\ref{like.lakd.ada}) here (recall that this recursion corresponds
to the case when the exchanges over the links are not subject to
noise): \be \widetilde{\w}_i={\cal A}_2^T\left(I_{NM}-{\cal M}\bm
{\cal R}_i\right){\cal A}_1^T\;\widetilde{\w}_{i-1}\;-\;{\cal
A}_2^T{\cal M}{\cal C}^T\s_i\;\;\;\;\;(\mbox{\rm perfect links})
\label{like.lakd.perfect}\ee Comparing
(\ref{eqn:noisyerrorrecursion1}) and (\ref{like.lakd.perfect}) we
find that:
\begin{itemize}
  \item[(1)] The covariance matrix ${\bm{\mathcal{R}}}_i$ in (\ref{like.lakd.perfect}) is replaced by
  ${\bm{\mathcal{R}}}_i'$. Recall from (\ref{eqn:Uprimedef})
  that ${\bm{\mathcal{R}}}_i'$ contains the influence of the
  noises that arise during the exchange of the regression data,
  i.e., the $\{R_{v,\ell k}^{(u)}\}$. But since we are now assuming that $R_{v,\ell k}^{(u)}=0$, then ${\bm{\mathcal{R}}}_i'={\bm{\mathcal{R}}}_i.$

  \item[(2)] The term ${\cal C}^T\s_i$ in (\ref{like.lakd.perfect}) is replaced by $\z_i$. Recall
  from (\ref{eqn:zkiprimedef}) that $\z_i$ contains the
  influence of the noises that arise during the exchange of the
  measurement data and the regression data, i.e., the $\{\sigma_{v,\ell k}^2,\;R_{v,\ell
  k}^{(u)}\}$.

  \item[(3)] Two new driving terms appear involving ${\bm{v}}_{i-1}^{(w)}$ and
   ${\bm{v}}_i^{(\psi)}$. These terms reflect the influence of the
   noises during the exchange of the weight estimates
   $\{\w_{\ell,i-1},\bm{\psi}_{\ell,i}\}$.

\item[(4)] Observe further that:
\begin{enumerate}
\item[(4a)] The term involving ${\bm{v}}_{i-1}^{(w)}$ accounts for
noise introduced at the information-exchange step
(\ref{eqn:noisyDiffusionPriorDiff1}) {\em before}
adaptation.

\item[(4b)] The term involving $\z_i$ accounts for
noise introduced during the adaptation step (\ref{eqn:noisyDiffusionIncremental1}).

\item[(4c)] The term involving ${\bm{v}}_{i}^{(\psi)}$ accounts for
noise introduced at the information-exchange step (\ref{eqn:noisyDiffusionPostDiff1}) {\em after}
adaptation.

\end{enumerate}

\end{itemize}

\noindent Therefore, since we are not considering noise during the exchange of the regression data,
 the weight-error recursion (\ref{eqn:noisyerrorrecursion1}) simplifies to:
\begin{equation}
\addbox{\; \label{like.lakd.ada.yy}
{\widetilde{\bm{w}}}_i=\mathcal{A}_2^T\left(I_{NM}-{\mathcal{M}}{\bm{\mathcal{R}}}_i\right){\mathcal{A}_1^T}{\widetilde{\bm{w}}}_{i-1}-{\mathcal{A}}_2^T{\mathcal{M}}{\bm{z}}_i
-{\mathcal{A}}_2^T\left(I_{NM}-{\mathcal{M}}{\bm{\mathcal{R}}}_i\right){\bm{v}}_{i-1}^{(w)}
-{\bm{v}}_i^{(\psi)}\; } \;\;(\mbox{\rm noisy
links})\end{equation}
where we used the fact that  ${\bm{\mathcal{R}}}_i'= {\bm{\mathcal{R}}}_i$ under these conditions.

\subsection{Convergence in the Mean}
 Taking expectations of both sides of (\ref{like.lakd.ada.yy}) we find
that the mean error vector evolves according to the following  recursion:\be \addbox{\;\Ex\widetilde{\w}_i={\cal
A}_2^T\left(I_{NM}-{\cal M} {\cal R}\right){\cal
A}_1^T\cdot\Ex\widetilde{\w}_{i-1},\;\;i\geq 0\;}
\label{like.lakd.ada.2.yy}\ee with ${\cal R}$ defined by (\ref{r.dklajda}). This is the same recursion
encountered earlier in (\ref{like.lakd.ada.2}) during perfect data exchanges. Note that had we
considered noises during the exchange of the regression data, then the vector $\z_i$ in (\ref{like.lakd.ada.yy})
{\em would not} be zero mean and the matrix ${\bm{\mathcal{R}}}_i'$ will have to be used instead of
${\bm{\mathcal{R}}}_i$. In that
case, the recursion for $\Ex\widetilde{\w}_i$ will be different
from (\ref{like.lakd.ada.2.yy}); i.e., the presence of noise during the exchange of regression data alters the dynamics of
the mean error vector in an important way --- see \cite{zhaoxiao,zhaoxiao2abc} for details on how
to extend the arguments to this general case with a driving non-zero bias term. We can now extend
Theorem~\ref{kald8912.lkas} to the current scenario.

\begin{theorem} ({\rm {\bf Convergence in the Mean}}) \label{kald8912.lkas.yy} Consider
the problem of optimizing the global cost (\ref{opt.11}) with the
individual cost functions given by (\ref{gihjas.alk}). Pick a right
stochastic matrix $C$ and left stochastic matrices $A_1$ and $A_2$
satisfying (\ref{dlajs8.12}). Assume each node in the network runs
the perturbed adaptive diffusion algorithm
(\ref{eqn:noisyDiffusionPriorDiff1})--(\ref{eqn:noisyDiffusionPostDiff1}). Assume further that
the exchange of the variables
$\{\w_{\ell,i-1},\bm{\psi}_{\ell,i},\d_{\ell}(i)\}$ is subject to
additive noises as in (\ref{eqn:noisyw}), (\ref{eqn:noisypsi}), and
(\ref{eqn:noisyd}). We assume that the regressors are exchanged
unperturbed. Then, all estimators $\{\w_{k,i}\}$ across the network
will still converge in the mean to the optimal solution $w^o$ if the step-size
parameters $\{\mu_k\}$ satisfy \be \addbox{\;\mu_{k}<
\frac{2}{\lambda_{\max}(R_k)}\;} \label{KD8912.ALKD.ada.yy}\ee where
the neighborhood covariance matrix $R_k$ is defined by
(\ref{defkajs812}). That is, $\Ex\w_{k,i}\rightarrow w^o$ as $i\rightarrow \infty$.
\end{theorem}
\qd

\subsection{Mean-Square Convergence}
Recall from (\ref{kdjalk12}) that we introduced the matrix:
\be
{\cal B}\define
{\cal A}_2^T\left(I_{NM}-{\cal M} {\cal R}\right){\cal
A}_1^T\label{kdjalk12.yy}\ee
We further introduce the $N\times N$ block matrix with blocks of size $M\times M$ each:
 \be {\cal H}\define {\cal A}_2^T(I_{NM}-{\cal
M}{\cal R}) \ee Then, starting from (\ref{like.lakd.ada.yy}) and repeating
the argument that led to (\ref{kdl8912.12}) we can establish the validity of the
following variance relation: \be
\addbox{\;\Ex\|\widetilde{\w}_i\|_{\sigma}^2=
\Ex\|\widetilde{\w}_{i-1}\|_{{\cal F}\sigma}^2+\left[\mbox{\rm
vec}\left({\cal A}_2^T{\cal M}{\cal Z}^T{\cal M}{\cal
A}_2\right)+\mbox{\rm vec}\left(\left({\cal H}{\cal R}_{v}^{(w)T}{\cal
H}^*\right)^T\right)+\mbox{\rm vec}\left({\cal
R}_{v}^{(\psi)T}\right)\right]^T \sigma\;}\label{kdl8912.12.yy}\ee
for an arbitrary nonnegative-definite weighting matrix $\Sigma$ with $\sigma=\mbox{\rm vec}(\Sigma)$, and where
${\cal F}$ is the same matrix defined earlier either by
(\ref{jdk8912.l.kas}) or (\ref{ka8912.lkad}). We
can therefore extend the statement of Theorem~\ref{escxlasdk} to the present scenario.

\begin{theorem} ({\rm {\bf Mean-Square Stability}})
Consider the same setting of Theorem~\ref{kald8912.lkas.yy}.
Assume sufficiently small step-sizes to justify
ignoring terms that depend on higher powers of the step-sizes. The
perturbed adaptive diffusion algorithm
(\ref{eqn:noisyDiffusionPriorDiff1})--(\ref{eqn:noisyDiffusionPostDiff1})
is mean-square stable if, and only if, the matrix ${\cal F}$ defined by (\ref{jdk8912.l.kas}),
or its approximation (\ref{ka8912.lkad}),
 is stable (i.e., all its
eigenvalues are strictly inside the unit disc).
This condition is
satisfied by sufficiently small step-sizes $\{\mu_k\}$ that are also smaller than:\be \addbox{\;\mu_{k}< \frac{2}{\lambda_{\max}(R_k)}\;}
\label{KD8912.ALKD.ada.3.yy}\ee where the neighborhood covariance
matrix $R_k$ is defined by (\ref{defkajs812}). Moreover, the
convergence rate of the algorithm is determined by
$[\rho({\cal B})]^2$.
\end{theorem}

\qd

\noindent We conclude from the previous two theorems that the conditions for the mean
and mean-square convergence of the adaptive diffusion strategy are not
affected by the presence of noises over the exchange links (under the assumption that
the regression data are exchanged without perturbation; otherwise, the convergence conditions would be affected).
The mean-square performance, on the other hand, is affected as follows. Introduce the $N\times N$ block matrix:
\be \addbox{\;{\cal Y}_{\rm imperfect}\define {\cal A}_2^T{\cal M}{\cal Z}{\cal
M}{\cal A}_2\;+\;{\cal H}{\cal R}_{v}^{(w)}{\cal
H}^*\;+\;{\cal R}_{v}^{(\psi)}\;}\;\;\;\;(\mbox{\rm imperfect exchanges})\label{lkd9182}
 \ee
which should be compared with the corresponding quantity defined by (\ref{dlkas9812.1}) for
the perfect exchanges case, namely,
 \be {\cal Y}_{\rm perfect}= {\cal A}_2^T {\cal M} {\cal C}^T {\cal
S}{\cal C}{\cal M}{\cal A}_2 \label{dlkas9812.1.yy}\;\;\;\;(\mbox{\rm perfect exchanges})\ee
When perfect exchanges occur, the matrix ${\cal Z}$ reduces to ${\cal C}^T{\cal S}{\cal C}$.  We can relate ${\cal Y}_{\rm imperfect}$
 and ${\cal Y}_{\rm perfect}$ as follows. Let
 \be
{\cal R}^{(du)}\define \mbox{\rm diag}\left\{
\sum_{\ell\in{\cal N}_1} c_{\ell 1}^2\sigma_{v,\ell 1}^2R_{u,\ell},\;
\sum_{\ell\in{\cal N}_2} c_{\ell 2}^2\sigma_{v,\ell 2}^2R_{u,\ell},\;\ldots
\sum_{\ell\in{\cal N}_N} c_{\ell N}^2\sigma_{v,\ell N}^2R_{u,\ell}\right\}
 \ee
Then, using (\ref{defl.saZZ.1}) and (\ref{defl.saZZ}), it is straightforward to verify that
\be
{\cal Z}= {\cal C}^T{\cal S}{\cal C}\;+\;{\cal R}^{(du)}
\ee
and it follows that:
\bq
{\cal Y}_{\rm imperfect}&=&{\cal Y}_{\rm perfect}\;+\;
{\cal A}_2^T{\cal M}{\cal R}^{(du)}{\cal
M}{\cal A}_2\;+\;{\cal H}{\cal R}_{v}^{(w)}{\cal
H}^*\;+\;{\cal R}_{v}^{(\psi)}\nn\\
&\define& {\cal Y}_{\rm perfect}\;+\;\Delta{\cal Y}
\label{usekfdkj.14}
\eq
Expression (\ref{usekfdkj.14})
reflects the influence of the noises $\{R_{v}^{(w)},\;R_{v}^{(\psi)},\;\sigma_{v,\ell k}^2\}$.
Substituting the definition (\ref{lkd9182}) into (\ref{kdl8912.12.yy}), and taking the limit as
$i\rightarrow\infty$, we obtain from the latter expression that:
\be\addbox{\; \lim_{i\rightarrow\infty}\;\Ex\|\widetilde{\w}_{i}\|_{(I-{\cal
F})\sigma}^2\;=\; \left[\mbox{\rm vec}\left({\cal
Y}_{\rm imperfect}^T\right)\right]^T \sigma\;}\label{kdl8912.12.asas.yy}\ee
which has the same form as (\ref{kdl8912.12.asas}); therefore, we can proceed analogously to obtain:
\be \addbox{\;{\rm MSD}^{\rm
network}_{\rm imperfect}\;=\; \frac{1}{N}\cdot\left[\mbox{\rm vec}\left({\cal
Y}_{\rm imperfect}^T\right)\right]^T\cdot(I-{\cal F})^{-1}\cdot\mbox{\rm
vec}\left(I_{NM}\right)\;} \label{lkad8912.13.yy}\ee
and
\be \addbox{\;{\rm EMSE}^{\rm network}_{\rm imperfect}\;=\;
\frac{1}{N}\cdot\left[\mbox{\rm vec}\left({\cal
Y}_{\rm imperfect}^T\right)\right]^T\cdot(I-{\cal F})^{-1}\cdot\mbox{\rm
vec}\left({\cal R}_u\right)\;} \ee
Using (\ref{usekfdkj.14}), we see that the network MSD and EMSE deteriorate as follows:
\bq
{\rm MSD}^{\rm
network}_{\rm imperfect}&=&{\rm MSD}^{\rm
network}_{\rm perfect}\;+\;\frac{1}{N}\cdot\left[\mbox{\rm vec}\left(\Delta {\cal
Y}^T\right)\right]^T\cdot(I-{\cal F})^{-1}\cdot\mbox{\rm
vec}\left(I_{NM}\right)\\
{\rm EMSE}^{\rm network}_{\rm imperfect}&=&{\rm EMSE}^{\rm network}_{\rm perfect}\;+\;
\frac{1}{N}\cdot\left[\mbox{\rm vec}\left(\Delta{\cal
Y}^T\right)\right]^T\cdot(I-{\cal F})^{-1}\cdot\mbox{\rm
vec}\left({\cal R}_u\right)
\eq

\subsection{Adaptive Combination Weights}
We can repeat the discussion from Secs.~\ref{sec7.1} and
\ref{sec7.2} to devise one adaptive scheme to adjust the combination
coefficients in the noisy exchange case. We illustrate the construction by considering the ATC
strategy corresponding to $A_1=I_N, A_2=A, C=I_N$, so that only
weight estimates are exchanged and the update recursions are of the
form: \bq
\bm{\psi}_{k,i}&=&\w_{k,i-1}\;+\;\mu_k\u_{k,i}^*[\d_{k}(i)-\u_{k,i}\w_{k,i-1}]\label{xiao.1}\\
\w_{k,i}&=&\sum_{\ell\in{\cal N}_k}a_{\ell k}\;\bm{\psi}_{\ell k,i}
\eq where from (\ref{eqn:noisypsi}): \be \bm{\psi}_{\ell
k,i}\;=\;\bm{\psi}_{\ell,i}\;+\;\v_{\ell k,i}^{(\psi)}
\label{xiao.3}\ee In this case, the network MSD performance
(\ref{lkad8912.13.yy}) becomes \be {\rm MSD}^{\rm network}_{\rm
atc,C=I,imperfect}\;=\; \frac{1}{N}\sum_{j=0}^{\infty}\;\mbox{\rm
Tr}\left({\cal B}_{\rm atc,C=I}^j {\cal Y}_{\rm atc,imperfect}{\cal
B}_{\rm atc,C=I}^{*j}\right)\label{imajdka.act.yy}\ee where, since
now ${\cal Z}={\cal S}$ and ${\cal R}_{v}^{(w)}=0$, we have \bq
{\cal B}_{\rm
atc,C=I}&=&{\cal A}^T(I-{\cal M}{\cal R}_u)\\
{\cal Y}_{\rm
atc,imperfect}&=&{\cal A}^T{\cal M}{\cal S}{\cal M}{\cal A}\;+\;{\cal R}_{v}^{(\psi)}\label{usialdalk.yy}\\
R_v^{(\psi)}&=&{\mathrm{diag}}\left\{R_{v,1}^{(\psi)},\;R_{v,2}^{(\psi)},\;\dots,\;R_{v,N}^{(\psi)}\right\}\\
R_{v,k}^{(\psi)}&=&\sum_{\ell\in{\cal
N}_k}a_{\ell k}^2\;R_{v,\ell k}^{(\psi)}\\
{\cal R}_u&=&\mbox{\rm diag}\{R_{u,1},R_{u,2},\ldots,R_{u,N}\}\\
{\cal S}&=&\mbox{\rm diag}\{\sigma_{v,1}^2 R_{u,1},\;\sigma_{v,2}^2
R_{u,2},\ldots,\sigma_{v,N}^2 R_{u,N}\}\\
{\cal M}&=&\mbox{\rm diag}\{\mu_1 I_M,\;\mu_2 I_M,\ldots,\mu_N I_M\}\\
{\cal A}&=&A\otimes I_M \eq To
proceed, as was the case with (\ref{reduasa}), we
consider the following simplified optimization problem:
\be\label{reduasa.yy}
\begin{aligned}
    &\min_{A} \quad \text{Tr}({\cal Y}_{\rm atc,imperfect})\\
    &\text{subject to}\;\; A^T\mathds{1}=\mathds{1},\;\; a_{\ell k}\geq 0,\;\;
    a_{\ell k}=0 \text{ if $\ell \notin \mathcal{N}_k$}
\end{aligned}
\ee Using (\ref{usialdalk.yy}), the trace of ${\cal Y}_{\rm atc,imperfect}$ can be
expressed in terms of the combination coefficients as follows: \be
\mbox{\rm Tr}({\cal Y}_{\rm atc,imperfect})\;=\;\sum_{k=1}^N
\sum_{\ell=1}^N\;a_{\ell k}^2\;\left[\mu_\ell^2 \sigma_{v,\ell}^2\;\mbox{\rm
Tr}(R_{u,\ell})\;+\;\mbox{\rm Tr}\left(R_{v,\ell k}^{(\psi)}\right)\right] \ee so that problem (\ref{reduasa.yy}) can be decoupled
into $N$ separate optimization problems of the form:
\begin{equation}\label{eq20.yy}\boxed{\;\;\;
\begin{aligned}\\
    &\min_{\{a_{\ell k}\}_{\ell =1}^N} \quad \sum_{\ell =1}^Na^2_{\ell
    k}\left[\mu_\ell^2 \sigma_{v,\ell}^2\;\mbox{\rm
Tr}(R_{u,\ell})\;+\;\mbox{\rm Tr}\left(R_{v,\ell k}^{(\psi)}\right)\right]
    \text{,\;\;\;  $k=1,\ldots,N$}\\
    &\text{subject to}\\
    &\qquad  a_{\ell k}\geq 0,\qquad \sum_{\ell =1}^Na_{\ell k}=1,\quad
    a_{\ell k}=0 \text{ if $\ell \notin \mathcal{N}_k$\;\;\;}\\
\end{aligned}}
\end{equation}
With each node $\ell$, we associate the following nonnegative
variance products: \be \addbox{\;\gamma_{\ell k}^2\define
\mu_{\ell}^2\cdot\sigma^{2}_{v,\ell}\cdot \mbox{\rm Tr}(R_{u,\ell})\;+\;
\mbox{\rm Tr}\left(R_{v,\ell k}^{(\psi)}\right),\;\;k\in{\cal N}_{\ell}
\;} \label{deflak.d}\ee This
measure now incorporates information about the exchange noise covariances $\{R_{v,\ell k}^{(\psi)}\}$.
Then, the solution of
(\ref{eq20.yy}) is given by:
\begin{equation} \label{eq11.yy}
\addbox{\;    a_{\ell k}=
    \begin{cases}
    \frac{\gamma_{\ell k}^{-2}}{\sum_{m\in\mathcal{N}_k}\gamma_{m k}^{-2}},
    &\text{if $\ell \in\mathcal{N}_k$}\\
    0, &\text{otherwise}
    \end{cases}\;}\;\;\;\;\;(\mbox{\rm relative-variance
    rule})
\end{equation}
We continue to refer to this combination rule as the {\em relative-variance
combination rule} \cite{TS11}; it leads to a left-stochastic matrix
$A$. To evaluate the combination weights (\ref{eq11.yy}), the nodes need to
know the  variance products, $\{\gamma_{m k}^2\}$, of their
neighbors. As before, we can motivate one adaptive construction as follows.

We refer to the ATC recursion (\ref{xiao.1})--(\ref{xiao.3}) and use
the data model (\ref{lkad8912.lakd.ada}) to write for node $\ell$:
\be \boldsymbol{\psi}_{\ell k,i} =\w_{\ell,i-1}\;+\;\mu_{\ell}
\u_{\ell,i}^*\left[\u_{\ell,i}\widetilde{\w}_{\ell,i-1}+\v_{\ell}(i)\right]\;+\;\v_{\ell
k,i}^{(\psi)}\ee so that, in view of our earlier assumptions on the
regression data and noise in Secs.~\ref{sec.daada} and~\ref{sec9.1},
we obtain in the limit as $i\rightarrow\infty$: \be
\lim_{i\rightarrow\infty}\; \Ex\left\|\boldsymbol{\psi}_{\ell k,i}
-\w_{\ell,i-1}\right\|^2\;=\; \mu_{\ell}^2\cdot
\left(\lim_{i\rightarrow\infty}\Ex\|\widetilde{\w}_{i-1}\|^2_{\Ex\left(\u_{\ell,i}^*
\|\u_{\ell,i}\|^2\u_{\ell,i}\right)}\right)\;+\;\mu_{\ell}^2\cdot
\sigma_{v,\ell}^2\cdot \mbox{\rm Tr}(R_{u,\ell})\;+\; \mbox{\rm
Tr}\left(R_{v,\ell k}^{(\psi)}\right) \label{89ajd712.xiao}\ee

\noindent \\ \noindent In a manner similar to what was done before
for (\ref{89ajd712}), we can evaluate the limit on the right-hand
side by using the corresponding steady-state result
(\ref{kdl8912.12.asas.yy}). We select the vector $\sigma$ in
(\ref{kdl8912.12.asas.yy}) to satisfy: \be (I-{\cal
F})\sigma\;=\;\mbox{\rm vec}\left[\Ex\left(\u_{\ell,i}^*
\|\u_{\ell,i}\|^2\u_{\ell,i}\right)\right] \ee Then, from
(\ref{kdl8912.12.asas.yy}),
\be\lim_{i\rightarrow\infty}\;\Ex\|\widetilde{\w}_{i-1}\|_{\Ex\left(\u_{\ell,i}^*
\|\u_{\ell,i}\|^2\u_{\ell,i}\right)}^2\;=\; \left[\mbox{\rm
vec}\left({\cal Y}^T_{\rm atc, imperfect}\right)\right]^T \cdot
(I-{\cal F})^{-1}\cdot\mbox{\rm vec}\left[\Ex\left(\u_{\ell,i}^*
\|\u_{\ell,i}\|^2\u_{\ell,i}\right)\right]\ee Now recall from
expression (\ref{usialdalk.yy}) that the entries of ${\cal Y}_{\rm
atc,imperfect}$ depend on combinations of the squared step-sizes,
$\{\mu_m^2,\;m=1,2,\ldots,N\}$, and on terms involving
$\left\{\mbox{\rm Tr}\left(R_{v,m}^{(\psi)}\right)\right\}$.
This fact implies that the first term on the right-hand side of
(\ref{89ajd712.xiao}) depends on products of
 the form $\{\mu_{\ell}^2\mu_{m}^2\}$; these fourth-order factors can be ignored in comparison to the second-order factor $\mu_{\ell}^2$
 for small step-sizes. Moreover, the same first term on the right-hand side of
(\ref{89ajd712.xiao}) depends on products of the form
$\left\{\mu_{\ell}^2\mbox{\rm
Tr}\left(R_{v,m}^{(\psi)}\right)\right\}$, which can be ignored in
comparison to the last term, $\mbox{\rm Tr}\left(R_{v,\ell
k}^{(\psi)}\right),$ in (\ref{89ajd712.xiao}), which does not appear
multiplied by a squared step-size. Therefore, we can approximate:
 \bq
\lim_{i\rightarrow\infty}\; \Ex\left\|\boldsymbol{\psi}_{\ell k,i}
-\w_{\ell,i-1}\right\|^2&\approx& \mu_{\ell}^2\cdot
\sigma_{v,\ell}^2\cdot \mbox{\rm Tr}(R_{u,\ell}) \;+\; \mbox{\rm
Tr}\left(R_{v,\ell k}^{(\psi)}\right)
\nn\\
&=&\gamma_{\ell k}^2\eq in terms of the desired variance product,
$\gamma_{\ell k}^2$. Using the following instantaneous approximation
at node $k$ (where $w_{\ell,i-1}$ is replaced by $w_{k,i-1}$):
\begin{align}
   \Ex\|\boldsymbol{\psi}_{\ell k,i}-\w_{\ell,i-1}\|^2\;\approx\;\|\psi_{\ell k,i}-w_{k,i-1}\|^2
\end{align}
we can motivate an algorithm that enables node $k$ to estimate the
variance products $\gamma_{\ell k}^2$. Thus, let
$\widehat{\gamma}_{\ell k}^2(i)$ denote an estimate for
$\gamma_{\ell k}^2$ that is computed by node $k$ at time $i$. Then,
one way to evaluate $\widehat{\gamma}_{\ell k}^2(i)$ is through the
recursion:
\begin{equation} \label{eq20.yyz}
\begin{aligned}
 \addbox{\;   \widehat{\gamma}^2_{\ell k}(i)=(1-\nu_k)\cdot \widehat{\gamma}^2_{\ell
 k}(i-1)\;+\;\nu_k\cdot\|{\psi}_{\ell k,i}-w_{k,i-1}\|^2\;}
\end{aligned}
\end{equation}
where $\nu_k$ is a positive coefficient smaller than one. Indeed,
it can be verified that
\begin{equation}
    \lim_{i\rightarrow\infty}\Ex\boldsymbol{\widehat{\gamma}}^2_{\ell k}(i)
    \approx \gamma^2_{\ell k}
\end{equation}
so that the estimator $\bm{\widehat{\gamma}}^2_{\ell k}(i)$ converges on
average close to the desired variance product $\gamma^2_{\ell k}$. In this
way, we can replace the weights (\ref{eq11.yy}) by the adaptive
construction:
\begin{equation}\label{eq15}
\addbox{\;    a_{\ell k}(i)=
    \begin{cases}
    \frac{\widehat{\gamma}^{-2}_{\ell k}(i)}{\sum_{m\in\mathcal{N}_k}\widehat{\gamma}^{-2}_{m k}(i)},
    &\text{if $\ell\in\mathcal{N}_k$}\\
    0, &\text{otherwise}
    \end{cases}\;}
\end{equation}
Equations (\ref{eq20.yyz}) and (\ref{eq15}) provide one adaptive
construction for the combination weights $\{a_{\ell k}\}$.

\section{Extensions and Further Considerations}
Several extensions and variations of diffusion strategies are
possible. Among those variations we mention strategies that endow
nodes with temporal processing abilities, in addition to their spatial
cooperation abilities. We can also apply diffusion strategies to
solve recursive least-squares  and state-space estimation problems
in a distributed manner. In this section, we highlight select
contributions in these and related areas.

\subsection{Adaptive Diffusion Strategies with Smoothing Mechanisms}
In the ATC and CTA adaptive diffusion strategies (\ref{Equ:DiffusionAdaptation:ATC.adaptive})--(\ref{Equ:DiffusionAdaptation:CTA.adaptive}), each node in the network shares
information locally with its neighbors through a process of spatial
cooperation or combination. In this section, we describe briefly an
extension that adds a temporal dimension to the processing at the
nodes. For example, in the ATC implementation
(\ref{Equ:DiffusionAdaptation:ATC.adaptive}), rather than have each
node $k$ rely solely on current data,
$\{d_{\ell}(i),u_{\ell,i},\;\ell\in{\cal N}_k\}$, and on current
weight estimates received from the neighbors,
$\{{\psi}_{\ell,i},\;\ell\in{\cal N}_k\}$,  node $k$ can be allowed to store and process
present and past weight estimates, say, $P$ of them as in
$\{{\psi}_{\ell,j},\;j=i,i-1,\ldots,i-P+1\}$. In this way, previous
weight estimates can be smoothed and used more effectively to help enhance the
mean-square-deviation performance especially in the presence of noise over the communication links.

\begin{figure}[h]
\epsfxsize 11cm \epsfclipon
\begin{center}
\leavevmode \epsffile{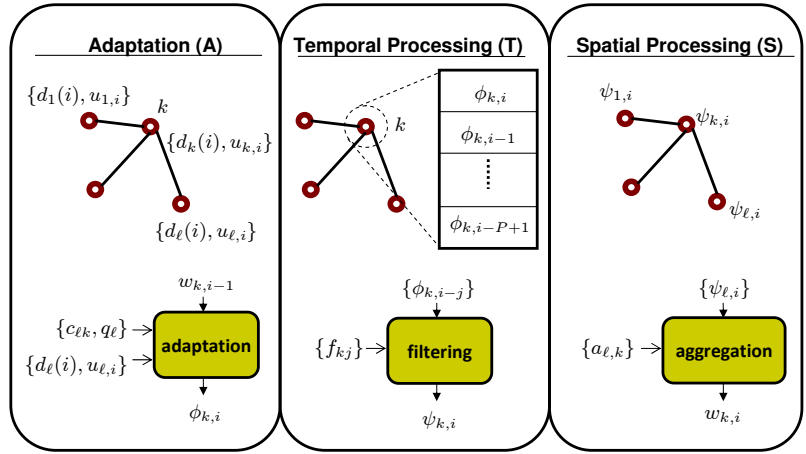} \caption{{\small Illustration of the three steps involved in an
ATS diffusion strategy: adaptation, followed by temporal processing or smoothing, followed by spatial
processing.}}\label{fig-XX.label}
\end{center}
\end{figure}

To motivate diffusion strategies with smoothing mechanisms, we continue to
assume that the random data $\{\d_k(i),\u_{k,i}\}$ satisfy the
modeling assumptions of Sec.~\ref{sec.daada}. The global cost
(\ref{opt.11}) continues to be the same but the individual cost
functions (\ref{gihjas.alk}) are now replaced by: \be
J_k(w)\;=\;\sum_{j=0}^{P-1}q_{kj}\; \Ex
\left|{\d}_{k}(i-j)-{\u}_{k,i-j}\;w\right|^{2} \label{eq2.leex} \ee
so that
\begin{equation}
J^{\rm glob}(w)=\sum_{k=1}^{N}\left(\sum_{j=0}^{P-1}q_{kj}\; \Ex
\left|{\d}_{k}(i-j)-{\u}_{k,i-j}\;w\right|^{2}\right)
\label{eq2.lee}
\end{equation}
 where each coefficient $q_{kj}$ is a non-negative scalar
representing the weight that node $k$ assigns to data from
time instant $i-j$. The coefficients $\{q_{kj}\}$ are assumed to satisfy the
normalization condition: \be q_{ko}>0,\;\;\;\;\;\sum_{j=0}^{P-1} q_{k
j}\;=\;1,\;\;\;k=1,2,\ldots,N \ee When the random processes
${\d}_{k}(i)$ and ${\u}_{k,i}$ are jointly wide-sense stationary, as
was assumed in Sec.~\ref{sec.daada}, the optimal solution $w^o$ that
minimizes (\ref{eq2.lee}) is still given by the same normal
equations (\ref{9s0lklad.as}). We can extend the arguments of
Secs.~\ref{sec.derivation.1} and~\ref{sec.adaptive.dd} to
(\ref{eq2.lee}) and arrive at the following version of a
diffusion strategy incorporating temporal processing (or smoothing) of the intermediate weight estimates\cite{Lee11,Lee12}:

\begin{eqnarray}
\phi_{k,i}&=&w_{k,i-1}\;+\;\mu_{k} \sum_{\ell \in  {\cal N}_{k}}
c_{\ell k}\;q_{\ell o}\;u^{*}_{\ell,i}
\left[d_{\ell}(i)\!-\!u_{\ell,i}w_{k,i-1}\right]\;\;\;\;\;(\mbox{\rm adaptation})\label{eq32a}\\
\psi_{k,i}&=&\sum^{P-1}_{j=0}f_{kj}\;\phi_{k,i-j}\label{eq32b}\;\;\;\;\;(\mbox{\rm temporal processing or smoothing})\\
w_{k,i}&=&\sum_{\ell \in  {\cal N}_{k}}a_{\ell
k}\;\psi_{\ell,i}\;\;\;\;\;\;\;\;\;(\mbox{\rm spatial processing})
\label{eq32c}
\end{eqnarray}
where the nonnegative coefficients $\{c_{\ell k}, a_{\ell k},
f_{kj}, q_{l o}\}$ satisfy: \bq
\mbox{\rm for}\;k=1,2,\ldots,N:&&\nn\\
        c_{\ell k}\geq 0,\;\;\;\sum_{k=1}^N c_{\ell k} = 1,\;\;\;\;  c_{\ell k}=0~\mathrm{if}~\ell \notin
        \mathcal{N}_{k}\\
        a_{\ell k}\geq 0,\;\;\sum_{\ell=1}^N a_{\ell k} = 1,\;\;\;\;  a_{\ell k}=0~\mathrm{if}~\ell \notin
        \mathcal{N}_{k}\\
        f_{kj}\geq 0,\;\;\;\;\;\sum_{j=0}^{P-1} f_{kj} = 1\\
        0<q_{\ell o} \leq 1\eq
Since only the coefficients $\{q_{\ell o}\}$ are needed, we
alternatively denote them by the simpler notation $\{q_{\ell}\}$ in
the listing in Table~\ref{table.spatio.label}. These are simply
chosen as nonnegative coefficients: \be 0<q_{\ell}\leq 1
,\;\;\;\ell=1,2,\ldots,N \ee Note that algorithm
(\ref{eq32a})-(\ref{eq32c}) involves three steps: (a) an adaptation
step (A) represented by (\ref{eq32a}); (b) a temporal filtering or smoothing step
(T) represented by (\ref{eq32b}), and a spatial cooperation step (S)
represented by (\ref{eq32c}). These steps are illustrated in Fig.~\ref{fig-XX.label}.
We use the letters (A,T,S) to label
these steps; and we use the sequence of letters (A,T,S) to designate
the order of the steps. According to this convention, algorithm
(\ref{eq32a})-(\ref{eq32c})  is referred to as the ATS diffusion
strategy since adaptation is followed by temporal processing, which
is followed by spatial processing. In total, we can obtain six
different combinations of diffusion algorithms by changing the order
by which the temporal and spatial combination steps are performed in
relation to the adaptation step. The resulting variations are
summarized in Table~\ref{table.spatio.label}. When we use only the
most recent weight vector in the temporal filtering step (i.e., set
$\psi_{k,i}=\phi_{k,i}$), which corresponds to the case $P=1$, the
algorithms of Table~\ref{table.spatio.label} reduce to the ATC and
CTA diffusion algorithms
(\ref{Equ:DiffusionAdaptation:ATC.adaptive}) and
(\ref{Equ:DiffusionAdaptation:CTA.adaptive}). Specifically, the
variants TSA, STA, and SAT (where spatial processing S precedes
adaptation A) reduce to CTA, while the variants TAS, ATS, and AST
(where adaptation A precedes spatial processing S) reduce to ATC.

The mean-square performance analysis of the smoothed
diffusion strategies can be pursued by extending the arguments of
Sec.~\ref{sec.mse}. This step is carried out in \cite{Lee11,Lee12} for doubly stochastic
combination matrices $A$ when the filtering coefficients $\{f_{kj}\}$ do not change with $k$.
For instance, it is shown in \cite{Lee12} that whether temporal processing is performed before or after adaptation, the strategy that performs adaptation before spatial
cooperation is always better. Specifically, the six diffusion variants can
be divided into two groups with the respective network MSDs satisfying the following relations:
\bq
\mbox{\rm Group \#1}:&\;\;& {\rm MSD}^{\rm network}_{\rm TSA}\;=\;
{\rm MSD}^{\rm network}_{\rm STA}\;\geq\;
{\rm MSD}^{\rm network}_{\rm TAS}\\\nn\\
\mbox{\rm Group \#2}:&\;\;& {\rm MSD}^{\rm network}_{\rm SAT}\;\geq\;
{\rm MSD}^{\rm network}_{\rm ATS}\;=\;
{\rm MSD}^{\rm network}_{\rm AST}\eq
Note that within groups 1 and 2, the order of the A and
T operations is the same: in group 1, T precedes A and in
group 2, A precedes T. Moreover, within each group, the order of the A and S operations determines performance; the strategy that performs A before S is
better. Note further that when $P = 1$, so that temporal processing is not performed, then TAS reduces to ATC and TSA reduces to CTA. This conclusion is
consistent with our earlier result (\ref{jakd.ka9831}) that ATC
outperforms CTA.

{\footnotesize
\begin{table}[h]
\begin{center}
\caption{\rm {\small Six diffusion strategies with temporal smoothing steps. } }
{\small
\begin{tabular}{|l|l|}\hline\hline &\\
      {\small {\bf TSA diffusion:}} &  {\small {\bf TAS diffusion:}}\\
      \normalsize $\phi_{k,i-1}=\displaystyle \sum^{P-1}_{j=0}f_{kj}\;w_{k,i-j-1}$ &       \normalsize $\phi_{k,i-1}=\displaystyle\sum^{P-1}_{j=0}f_{kj}\;w_{k,i-j-1}$ \\
      \normalsize $\psi_{k,i-1}=\displaystyle\sum_{\ell \in  {\cal N}_{k}}a_{\ell k}\;\phi_{\ell,i-1}$ &      \normalsize$\psi_{k,i}=\phi_{k,i-1}+\mu_{k}\displaystyle\sum_{\ell \in \normalsize {\cal N}_{k}}q_\ell\; c_{\ell
      k}\;
      u^{*}_{\ell,i}\left[d_{\ell}(i)-u_{\ell,i}\phi_{k,i-1}\right]$\\
      \normalsize$w_{k,i}=\psi_{k,i-1}+\mu_{k}\displaystyle\sum_{\ell \in \normalsize {\cal N}_{k}}q_\ell\; c_{\ell
      k}\;
      u^{*}_{\ell,i}\left[d_{\ell}(i)-u_{\ell,i}\;\psi_{k,i-1}\right]$&      \normalsize$w_{k,i}=\displaystyle\sum_{\ell \in  {\cal N}_{k}}a_{\ell,k}\psi_{\ell,i}$ \\
     \hline&\\
      {\small {\bf STA diffusion:}}&       {\small {\bf ATS diffusion:}}\\
      \normalsize $\phi_{k,i-1}=\displaystyle\sum_{\ell \in  {\cal N}_{k}}a_{\ell k}\;w_{\ell,i-1}$ &       \normalsize$\phi_{k,i}=w_{k,i-1}+\mu_{k}\displaystyle\sum_{\ell \in \normalsize {\cal N}_{k}}q_\ell\; c_{\ell
      k}\;
      u^{*}_{\ell,i}\left[d_{\ell}(i)-u_{\ell,i}w_{k,i-1}\right]$\\
      \normalsize $\psi_{k,i-1}=\displaystyle\sum^{P-1}_{j=0}f_{kj}\;\phi_{k,i-j-1}$ &      \normalsize$\psi_{k,i}=\displaystyle\sum^{P-1}_{j=0}f_{kj}\phi_{k,i-j}$ \\
      \normalsize$w_{k,i}=\psi_{k,i-1}+\mu_{k}\displaystyle\sum_{\ell \in \normalsize {\cal N}_{k}}q_\ell\; c_{\ell
      k}\;
      u^{*}_{\ell,i}\left[d_{\ell}(i)-u_{\ell,i}\psi_{k,i-1}\right]$&      \normalsize$w_{k,i}=\displaystyle\sum_{\ell \in  {\cal N}_{k}}a_{\ell k}\psi_{\ell,i}$\\
      \hline &\\
      {\small {\bf SAT diffusion:}}& {\small {\bf AST diffusion:}}\\
      \normalsize$\phi_{k,i-1}=\displaystyle\sum_{\ell \in  {\cal N}_{k}}a_{\ell k}w_{\ell,i-1}$ & \normalsize$\phi_{k,i}=w_{k,i-1}+\mu_{k}\displaystyle\sum_{\ell \in \normalsize {\cal N}_{k}}q_\ell\; c_{\ell
      k}\;
      u^{*}_{\ell,i}\left[d_{\ell}(i)-u_{\ell,i}w_{k,i-1}\right]$\\
      \normalsize$\psi_{k,i}=\phi_{k,i-1}+\mu_{k}\displaystyle\sum_{\ell \in \normalsize {\cal N}_{k}}q_\ell\; c_{\ell
      k}\;
      u^{*}_{\ell,i}\left[d_{\ell}(i)-u_{\ell,i}\phi_{k,i-1}\right]$&      \normalsize$\psi_{k,i}=\displaystyle\sum_{\ell \in  {\cal N}_{k}}a_{\ell k}\phi_{\ell,i}$\\
      \normalsize$w_{k,i}=\displaystyle\sum^{P-1}_{j=0}f_{kj}\psi_{k,i-j}$ &\normalsize$w_{k,i}=\displaystyle\sum^{P-1}_{j=0}f_{kj}\psi_{k,i-j}$\\
      \hline
\end{tabular} }
\label{table.spatio.label}
\end{center}
\end{table}
}

In related work, reference \cite{chouvar} started from the CTA algorithm (\ref{Equ:DiffusionAdaptation:CTA.adaptive.2}) without information exchange and added a useful projection step to it between the combination step and the adaptation step; i.e., the work considered an algorithm with an STA structure (with spatial combination occurring first, followed by a projection step, and then adaptation). The projection step uses the current weight
estimate, $\phi_{k,i}$, at node $k$ and projects it onto hyperslabs defined by the current and
past raw data. Specifically, the algorithm from \cite{chouvar} has the following form:
\bq
\phi_{k,i-1}&=&\displaystyle\sum_{\ell \in  {\cal N}_{k}}a_{\ell k}\;w_{\ell,i-1}\label{projdk.alkd.3}\\
\psi_{k,i-1}&=&{\cal P}_{k,i}^{'}[\phi_{k,i-1}]\label{projdk.alkd} \\
w_{k,i}&=&\psi_{k,i-1}-\mu_{k}\left\{\psi_{k,i-1}-
\sum_{j=0}^{P-1} f_{kj}\cdot{\cal P}_{k,i-j}[\phi_{k,i-1}]\right\} \label{projdk.alkd.2}
\eq
where the notation $\psi={\cal P}_{k,i}[\phi]$ refers to the act of projecting the vector $\phi$ onto the hyperslab $P_{k,i}$ that consists of all $M\times 1$ vectors $z$ satisfying (similarly for the projection ${\cal P}_{k,i}^{'}$):
\bq
P_{k,i}&\define& \left\{\;z\;\mbox{\rm such that}\;|d_k(i)-u_{k,i}z|\leq \epsilon_k\;\right\}\\
P_{k,i}^{'}&\define& \left\{\;z\;\mbox{\rm such that}\;|d_k(i)-u_{k,i}z|\leq \epsilon_k'\;\right\}
\eq
where $\{\epsilon_k,\epsilon_k'\}$ are positive (tolerance) parameters chosen by the designer to satisfy $\epsilon_k'>\epsilon_k$. For generic values $\{d,u,\epsilon\}$, where $d$ is a scalar and $u$ is a row vector, the projection operator is described analytically by the following expression \cite{slava10}:
\be
{\cal P}[\phi]\;=\;\phi\;+\;\left\{\begin{array}{ll}\frac{u^*}{\|u\|^2}\left[d-\epsilon-u\phi\right],&\mbox{\rm if}\;d-\epsilon>u\phi\\\\
0,&\mbox{\rm if}\;|d-u\phi|\leq\epsilon\\\\
\frac{u^*}{\|u\|^2}\left[d+\epsilon-u\phi\right],&\mbox{\rm if}\;d+\epsilon<u\phi
\end{array}\right.
\ee
The projections that appear in (\ref{projdk.alkd})--(\ref{projdk.alkd.2}) can be regarded as another example of a temporal processing step. Observe from the middle plot in Fig.~\ref{fig-XX.label} that the temporal step that we are  considering in the algorithms listed in Table~\ref{table.spatio.label}
 is based on each node $k$ using its current and past weight estimates, such as  $\{\phi_{k,i},\phi_{k,i-1},\ldots,\phi_{k,i-P+1}\}$, rather than only $\phi_{k,i}$ and current and past {\em raw} data
 $\{d_{k}(i),d_{k}(i-1),\ldots,d_k(i-P+1),u_{k,i},u_{k,i-1},\ldots,u_{k,i-P+1}\}$. For this reason,  the
 temporal processing steps in Table~\ref{table.spatio.label} tend to exploit information from across the network more broadly and the resulting mean-square error performance is generally improved relative to (\ref{projdk.alkd.3})--(\ref{projdk.alkd.2}).

\subsection{Diffusion Recursive Least-Squares}
Diffusion strategies can also be applied to recursive least-squares problems to enable distributed
solutions of least-squares designs \cite{Cattivelli07,Cattivelli08a}; see also \cite{ACL2006}. Thus, consider again
a set of $N$ nodes that are spatially distributed
over some domain. The objective of the network is to collectively
estimate some unknown column vector of
length $M$, denoted by $w^o$, using a least-squares criterion. At
every time instant $i$, each node $k$ collects a scalar measurement, $d_k(i)$, which
is assumed to be related to the unknown vector $w^o$ via the linear model:
\be d_k(i)\;=\;u_{k,i}w^o\;+\;v_k(i)\ee
In the above relation, the vector $u_{k,i}$ denotes a row regression vector of length $M$,
 and $v_k(i)$ denotes measurement noise. A snapshot of the data in the network at time $i$ can be captured by
 collecting the measurements and noise samples, $\{d_k(i),v_k(i)\},$ from across
all nodes into column vectors $y_i$ and $v_i$ of sizes $N\times 1$ each, and the regressors $\{u_{k,i}\}$
into a matrix $H_i$ of size $N\times M$:
\be
y_i\;=\;\ba{c}d_1(i)\\d_2(i)\\\vdots\\d_N(i)\ea\;(N\times 1),\;\;\;\;\;
v_i\;=\;\ba{c}v_1(i)\\v_2(i)\\\vdots\\v_N(i)\ea\;(N\times 1),\;\;\;\;\;
H_i\;=\;\ba{c}u_{1,i}\\u_{2,i}\\\vdots\\u_{N,i}\ea\;(N\times M)
\ee
Likewise, the history of the data across the network up to time $i$ can be
collected  into vector quantities as follows:
\be
{\cal Y}_i\;=\;\ba{c}y_i\\y_{i-1}\\\vdots\\y_0\ea,\;\;\;\;\;\;
{\cal V}_i\;=\;\ba{c}v_i\\v_{i-1}\\\vdots\\v_0\ea,\;\;\;\;\;\;
{\cal H}_i\;=\;\ba{c}H_i\\H_{i-1}\\\vdots\\H_0\ea
\ee
Then, one way to estimate $w^o$ is by formulating a global least-squares optimization problem of the form:
\be
\addbox{\;\min_{w}\;\;\|w\|^2_{\Pi_i}\;+\;\|{\cal Y}_i\;-\;{\cal H}_i w\|^2_{{\cal W}_i}\;}
\label{lkd912.12}\ee
where $\Pi_i>0$ represents a Hermitian regularization matrix and ${\cal W}_i\geq 0$ represents a Hermitian weighting matrix.
Common choices for $\Pi_i$ and ${\cal W}_i$ are
\bq
{\cal W}_i&=&\mbox{\rm diag}\{I_N,\;\lambda I_N,\;\ldots,\;\lambda^i I_N\}\\
\Pi_i&=&\lambda^{i+1}\delta^{-1}\eq
where $\delta>0$ is usually a large number and $0\ll\lambda\leq 1$ is a forgetting factor whose value is generally very close to
one. In this case, the global cost function (\ref{lkd912.12}) can be written in the equivalent form:
\be
\addbox{\;\min_{w}\;\;\lambda^{i+1}\|w\|^2\;+\;\sum_{j=0}^i\lambda^{i-j}\left(\sum_{k=1}^N |d_{k}(j)-u_{k,j} w|^2\right)\;}
\ee
which is an exponentially weighted least-squares problem.  We see that, for every time instant $j$,
 the squared errors, $|d_{k}(j)-u_{k,j} w|^2$, are summed across the network and scaled by the exponential weighting factor
 $\lambda^{i-j}$. The index $i$ denotes current time and the index $j$ denotes a time instant in the past. In this way,
 data occurring in the remote past are scaled more heavily than data occurring closer to present time.
 The global solution of (\ref{lkd912.12}) is given by \cite{Sayed08}:
\be
w_i\;=\;\left[\Pi_i+{\cal H}_i{\cal W}_i{\cal H}_i\right]^{-1}{\cal H}_i^*{\cal W}_i{\cal Y}_i
\label{doas.1}\ee
and the notation $w_i$, with a subscript $i$, is meant to indicate that the estimate $w_i$ is based on all
data collected from across the network up to time $i$. Therefore, the $w_i$ that is
computed via (\ref{doas.1}) amounts to a global construction.

In \cite{Cattivelli07,Cattivelli08a} a diffusion strategy was developed that allows nodes to approach the
global solution $w_i$ by relying solely on local interactions. Let $w_{k,i}$ denote a local estimate for $w^o$ that is computed by node $k$ at time $i$.
The diffusion recursive-least-squares (RLS) algorithm takes the following form. For every node $k$, we
start with the initial conditions $w_{k,-1}=0$ and $P_{k,-1}=\delta I_M,$ where $P_{k,-1}$ is an $M\times M$ matrix.
Then, for every time instant $i$, each node
$k$ performs an incremental step followed by a diffusion step as follows:

\be
\begin{array}{l}
\textrm{\small {\bf Diffusion RLS.}}\\\hline\hline\\
\textrm{\small {\bf Step 1} (incremental update)}\\
\;\;\;\; {\psi}_{k,i}\leftarrow w_{k,i-1}\\
\;\;\;\; P_{k,i}\leftarrow \lambda^{-1}P_{k,i-1}\\
\;\;\;\; \textrm{for every neighboring node }\ell \in\mathcal{N}_k\textrm{, update}:\\
\;\;\;\;\;\;\; {\psi}_{k,i}\leftarrow  {\psi}_{k,i} \;+\;\dfrac{c_{\ell k}P_{k,i}u_{\ell,i}^*}{1+c_{\ell k}u_{\ell,i}P_{k,i}u_{\ell,i}^*}\left[d_{\ell,i}-u_{\ell,i}{\psi}_{k,i}\right] \\\\
\;\;\;\;\;\;\; P_{k,i}\leftarrow  P_{k,i}-\dfrac{c_{\ell k}P_{k,i}u_{\ell,i}^*u_{\ell,i}P_{k,i}}{1+c_{\ell k}u_{\ell,i}P_{k,i}u_{\ell,i}^*}\\
\;\;\;\; \textrm{end}\\\smallskip\\
\textrm{\small {\bf Step 2} (diffusion update)}\\
\;\;\;\; w_{k,i}=  \displaystyle \sum_{\ell\in\mathcal{N}_k}a_{\ell k}{\psi}_{\ell,i}\\\\\hline
\end{array}
\label{diff.rls}\ee
\noindent \\

\noindent where the symbol $\leftarrow$ denotes a sequential assignment,
and where the scalars $\{a_{\ell k},c_{\ell k}\}$ are nonnegative coefficients
satisfying:
\bq
\mbox{\rm for}\;k=1,2,\ldots,N:&&\nn\\
        c_{\ell k}\geq 0,\;\;\;\sum_{k=1}^N c_{\ell k} = 1,\;\;\;\;  c_{\ell k}=0~\mathrm{if}~\ell \notin
        \mathcal{N}_{k}\\
        a_{\ell k}\geq 0,\;\;\sum_{\ell=1}^N a_{\ell k} = 1,\;\;\;\;  a_{\ell k}=0~\mathrm{if}~\ell \notin
        \mathcal{N}_{k}
\eq
The above algorithm  requires that at every instant $i$, nodes communicate
to their neighbors their measurements $\{d_{\ell}(i),u_{\ell,i}\}$ for
the incremental update, and the intermediate estimates $\{{\psi}_{\ell,i}\}$
for the diffusion update. During the incremental update, node $k$ cycles through its neighbors and incorporates their
data contributions represented by $\{d_{\ell}(i), u_{\ell,i}\}$ into $\{{\psi}_{k,i},P_{k,i}\}$. Every other node in the
network is performing similar steps. At the end of the incremental step, neighboring nodes share their
 intermediate estimates $\{{\psi}_{\ell,i}\}$ to undergo diffusion. Thus, at the end of both steps, each node
$k$ would have updated the quantities $\{w_{k,i-1}, P_{k,i-1}\}$ to $\{w_{k,i}, P_{k,i}\}$. The
quantities $P_{k,i}$ are matrices of size $M\times M$ each. Observe that the diffusion RLS implementation
(\ref{diff.rls}) does not require the nodes to share their matrices $\{P_{\ell,i}\}$, which would amount to a substantial
burden in terms of communications resources since each of these matrices has $M^2$ entries. Only the quantities $\{d_{\ell}(i),u_{\ell,i},\psi_{\ell,i}\}$ are shared.
The mean-square performance and convergence of the
diffusion RLS strategy are studied in some detail in \cite{Cattivelli08a}.

The incremental step of the diffusion RLS strategy (\ref{diff.rls}) corresponds to performing a number of
$|{\cal N}_k|$ successive least-squares updates starting from the initial conditions $\{w_{k,i-1},P_{k,i-1}\}$ and
ending with the values $\{\psi_{k,i},P_{k,i}\}$ that move on to the diffusion update step. It can be verified
from the properties of recursive least-squares solutions \cite{Sayed03,Sayed08} that these variables satisfy the
following equations at the {\em end} of the incremental stage (step 1):
\bq
P_{k,i}^{-1}&=&\lambda P_{k,i-1}^{-1}\;+\;\sum_{\ell\in{\cal N}_k} c_{\ell k} u_{\ell,i}^*u_{\ell,i}\label{pp.11}\\
P_{k,i}^{-1}\psi_{k,i}&=&\lambda P_{k,i-1}^{-1}w_{k,i-1}\;+\;
\sum_{\ell\in{\cal N}_k} c_{\ell k} u_{\ell,i}^*d_{\ell}(i)
\eq
Introduce the auxiliary $M\times 1$ variable:
\be
q_{k,i}\define P_{k,i}^{-1}\psi_{k,i}
\ee
Then, the above expressions lead to the following alternative form of the diffusion RLS strategy (\ref{diff.rls}).\\
\be
\begin{array}{l}
\textrm{\small {\bf Alternative form of diffusion RLS.}}\\\hline\hline\\
\;\;\;\;\;\;w_{k,i-1}\;=\;\displaystyle\sum_{\ell\in{\cal N}_k} a_{\ell k} \psi_{\ell,i-1}\\\\
\;\;\;\;\;\;P_{k,i}^{-1}\;=\;\lambda P_{k,i-1}^{-1}\;+\;\displaystyle\sum_{\ell\in{\cal N}_k} c_{\ell k}u_{\ell,i}^*u_{\ell,i}\\\\
\;\;\;\;\;\;q_{k,i}\;=\; \lambda P_{k,i-1}^{-1}w_{k,i-1}\;+\;
\displaystyle\sum_{\ell\in{\cal N}_k} c_{\ell k} u_{\ell,i}^*d_{\ell}(i)\\\\
\;\;\;\;\;\;\psi_{k,i}\;=\;P_{k,i}q_{k,i}\\\\\hline
\end{array}
\label{eq712.13}\ee
\noindent \\

Under some approximations, and for the special choices $A=C$ and $\lambda=1$, the diffusion RLS strategy (\ref{eq712.13})
 can be reduced to a form given in \cite{Xiao06} and which is described by the following equations:
\bq
P_{k,i}^{-1}&=&\sum_{\ell\in{\cal N}_k} c_{\ell k} \left[P_{\ell,i-1}^{-1}\;+\;u_{\ell,i}^*u_{\ell,i}\right]\label{xiaolall.1}\\
q_{k,i}&=&\sum_{\ell\in{\cal N}_k} c_{\ell k}\left[q_{\ell,i-1}\;+\;u_{\ell,i}^*d_{\ell}(i)\right]\\
\psi_{k,i}&=&P_{k,i}q_{k,i}\label{xiaolall.3}
\eq
Algorithm (\ref{xiaolall.1})--(\ref{xiaolall.3}) is motivated in \cite{Xiao06} by using consensus-type arguments. Observe
that the algorithm requires the nodes to share the variables $\{d_{\ell}(i),u_{\ell,i},q_{\ell,i-1},P_{\ell,i-1}\}$, which
corresponds to more
communications overburden than required by diffusion RLS; the latter only requires that nodes share
$\{d_{\ell}(i),u_{\ell,i},\psi_{\ell,i-1}\}$. In order to illustrate how a special case of diffusion RLS (\ref{eq712.13})
can be related to this scheme, let us set
\be A=C\;\;\;\;\mbox{\rm and}\;\;\;\;\lambda=1\ee
Then, equations (\ref{eq712.13}) give:
\be
\begin{array}{l}
\textrm{\small {\bf Special form of diffusion RLS when $A=C$ and $\lambda=1$.}}\\\hline\hline\\
\;\;\;\;\;\;w_{k,i-1}\;=\;\displaystyle\sum_{\ell\in{\cal N}_k} c_{\ell k} \psi_{\ell,i-1}\\\\
\;\;\;\;\;\;P_{k,i}^{-1}\;=\;P_{k,i-1}^{-1}\;+\;\displaystyle\sum_{\ell\in{\cal N}_k} c_{\ell k}u_{\ell,i}^*u_{\ell,i}\\\\
\;\;\;\;\;\;q_{k,i}\;=\; P_{k,i-1}^{-1}w_{k,i-1}\;+\;
\displaystyle\sum_{\ell\in{\cal N}_k} c_{\ell k} u_{\ell,i}^*d_{\ell}(i)\\\\
\;\;\;\;\;\;\psi_{k,i}\;=\;P_{k,i}q_{k,i}\\\\\hline
\end{array}
\label{eq712.13b}\ee
\noindent \\
\noindent Comparing these equations with (\ref{xiaolall.1})--(\ref{xiaolall.3}), we find that
algorithm (\ref{xiaolall.1})--(\ref{xiaolall.3}) of \cite{Xiao06} would relate to the diffusion RLS algorithm
(\ref{eq712.13}) when the following approximations are justified:
\bq
\sum_{\ell\in{\cal N}_k}c_{\ell k} P_{\ell,i-1}^{-1}&\approx& P_{k,i-1}^{-1}\label{approx.1}\\
\sum_{\ell\in{\cal N}_k}c_{\ell k} q_{\ell,i-1}&=&
\sum_{\ell\in{\cal N}_k}c_{\ell k} P_{\ell,i-1}^{-1} \psi_{\ell,i-1}\nn\\
&\approx&\sum_{\ell\in{\cal N}_{k}} c_{\ell k}P_{k,i-1}^{-1}\psi_{\ell,i-1}\nn\\
&=&
P_{k,i-1}^{-1}\sum_{\ell\in{\cal N}_{k}} c_{\ell k}\psi_{\ell,i-1}\\
&=&
P_{k,i-1}^{-1}w_{k,i-1}
\label{approx.2}
\eq
It was indicated in \cite{Cattivelli08a} that the diffusion RLS
implementation (\ref{diff.rls}) or (\ref{eq712.13}) leads to enhanced performance in comparison to the consensus-based
update (\ref{xiaolall.1})--(\ref{xiaolall.3}).

\subsection{Diffusion Kalman Filtering}
Diffusion strategies can also be applied to the solution of
distributed state-space filtering and smoothing problems
\cite{Cattivelli08b,Cattivelli08c,Cattivelli10b}. Here, we describe briefly the diffusion version of
the Kalman filter; other variants and smoothing filters can be found in
\cite{Cattivelli10b}. We assume that some system of interest is
evolving according to linear state-space dynamics, and that every
node in the network collects measurements that are linearly related
to the unobserved state vector. The objective  is for every node to
track the state of the system over time based solely on local
observations and on neighborhood interactions.

Thus, consider a network consisting of $N$ nodes observing the state
vector, $\x_i,$ of size $n\times 1$ of a linear state-space model.
At every time $i$, every node $k$ collects a measurement vector
$\y_{k,i}$ of size $p\times 1$, which is related to the state vector
as follows: \bq
\label{equ:modelnodek}
\x_{i+1}&=&F_i\x_i+G_i\n_i\\
\y_{k,i}&=&H_{k,i}\x_i+\v_{k,i},\;\;\;k=1,2,\ldots,N \eq
 The signals $\n_i$ and $\v_{k,i}$ denote state and measurement
noises of sizes $n\times 1$ and $p\times 1$, respectively, and they
are assumed to be zero-mean, uncorrelated and white, with covariance
matrices denoted by
\begin{equation}\label{equ:modelnoisecov}
\Ex\left[\begin{array}{c}
\n_i\\
\v_{k,i}
\end{array}\right]
\left[\begin{array}{c}
\n_j\\
\v_{k,j}
\end{array}\right]^*
\define \left[\begin{array}{cc}
Q_i & 0\\
0 & R_{k,i}
\end{array}\right]\delta_{ij}
\end{equation}
The initial state vector, $\x_o,$ is assumed to be zero-mean with
covariance matrix \be \Ex\x_o\x_o^*\;=\;\Pi_o>0\label{eq.pi}\ee
 and is
uncorrelated with $\n_i$ and $\v_{k,i}$, for all $i$ and $k$. We
further assume that $R_{k,i}>0$. The
parameter matrices $\{F_i,G_i,H_{k,i},Q_i,R_{k,i},\Pi_o\}$ are assumed to be known by
node $k$.

Let $\widehat{\x}_{k,i|j}$ denote a local estimator for $\x_i$ that is computed by node $k$ at time $i$ based solely
on local observations and on neighborhood data up to time $j$. The following diffusion strategy was proposed in
\cite{Cattivelli08b,Cattivelli08c,Cattivelli10b} to evaluate approximate predicted and filtered versions of these local estimators
in a distributed manner for data satisfying
model (\ref{equ:modelnodek})--(\ref{eq.pi}).  For every node $k$, we
start with $\widehat{\x}_{k,0|-1}=0$ and $P_{k,0|-1}=\Pi_o$, where $P_{k,0|-1}$ is an $M\times M$ matrix.
At every time instant $i$, every node $k$ performs an incremental step followed by a diffusion step:
\be
\begin{array}{l}
\textrm{\small {\bf Time and measurement-form of the diffusion Kalman filter.}}\\\hline\hline\\
\textrm{\small {\bf Step 1} (incremental update)}\\
\;\;\;\; \bm{\psi}_{k,i}\leftarrow \widehat{\x}_{k,i|i-1}\\
\;\;\;\; P_{k,i}\leftarrow P_{k,i|i-1}\\
\;\;\;\; \textrm{for every neighboring node }\ell \in\mathcal{N}_k\textrm{, update}:\\
\;\;\;\;\;\;\; R_{e}\leftarrow  R_{\ell,i}+H_{\ell,i}P_{k,i}H_{\ell,i}^* \\
\;\;\;\;\;\;\; \bm{\psi}_{k,i}\leftarrow  \bm{\psi}_{k,i} +P_{k,i}H_{\ell,i}^*R_{e}^{-1}\left[\y_{\ell,i}-H_{\ell,i}\bm{\psi}_{k,i}\right] \\
\;\;\;\;\;\;\; P_{k,i}\leftarrow  P_{k,i}-P_{k,i}H_{\ell,i}^*R_{e}^{-1}H_{\ell,i}P_{k,i}\\
\;\;\;\; \textrm{end}\\\smallskip\\
\textrm{\small {\bf Step 2} (diffusion update)}\\
\;\;\;\; \widehat{\x}_{k,i|i}=  \displaystyle \sum_{\ell\in\mathcal{N}_k}a_{\ell k}\bm{\psi}_{\ell,i}\\
\;\;\;\; P_{k,i|i}=  P_{k,i}\\
\;\;\;\; \widehat{\x}_{k,i+1|i}=F_i\widehat{\x}_{k,i|i}\\
\;\;\;\; P_{k,i+1|i}= F_iP_{k,i|i}F_i^*+G_iQ_iG_i^*.\\\\\hline
\end{array}
\ee
\noindent \\

\noindent where the symbol $\leftarrow$ denotes a sequential assignment, and where the scalars $\{a_{\ell k}\}$ are nonnegative coefficients
satisfying:
\bq
\mbox{\rm for}\;k=1,2,\ldots,N:&&\nn\\
a_{\ell k}\geq 0,\;\;\;\;\sum_{\ell=1}^N a_{\ell k} = 1,\;\;\;\;  a_{\ell k}=0~\mathrm{if}~\ell \notin
        \mathcal{N}_{k}
\eq
The above algorithm  requires that at every instant $i$, nodes communicate
to their neighbors their measurement matrices $H_{\ell,i}$, the noise
covariance matrices $R_{\ell,i}$, and the measurements $\y_{\ell,i}$ for
the incremental update, and the intermediate estimators $\bm{\psi}_{\ell,i}$
for the diffusion update. During the incremental update, node $k$ cycles through its neighbors and incorporates their
data contributions represented by $\{\y_{\ell,i}, H_{\ell,i}, R_{\ell,i}\}$ into $\{\bm{\psi}_{k,i},P_{k,i}\}$. Every other node in the
network is performing similar steps. At the end of the incremental step, neighboring nodes share their
updated intermediate estimators $\{\bm{\psi}_{\ell,i}\}$ to undergo diffusion. Thus, at the end of both steps, each node
$k$ would have updated the quantities $\{\widehat{\x}_{k,i|i-1}, P_{k,i|i-1}\}$ to $\{\widehat{\x}_{k,i+1|i}, P_{k,i+1|i}\}$. The
quantities $P_{k,i|i-1}$ are $n\times n$ matrices. It is important to note that even though the notation $P_{k,i|i}$
and $P_{k,i|i-1}$ has been retained for these variables, as in the standard Kalman filtering notation \cite{Sayed08,Kai00},
these matrices do \emph{not} represent any longer the covariances of the state
estimation errors, $\tilde{\x}_{k,i|i-1}=\x_i-\widehat{\x}_{k,i|i-1},$ but can be related to them \cite{Cattivelli10b}.

An alternative representation of the diffusion Kalman filter may be obtained in information form by
further assuming that $P_{k,i|i-1}>0$ for all $k$ and $i$; a sufficient condition for this fact to hold is
to requires the matrices $\{F_i\}$ to be invertible \cite{Kai00}. Thus, consider again data satisfying
model (\ref{equ:modelnodek})--(\ref{eq.pi}). For every node $k$, we start with
$\widehat{\x}_{k,0|-1}=0$ and $P_{k,0|-1}^{-1}=\Pi_o^{-1}$.
At every time instant $i$, every node $k$ performs an incremental step followed by a diffusion step:

\be\begin{array}{l}
\textrm{\small {\bf Information form of the diffusion Kalman filter.}}\\\hline\hline\\
\textrm{\small {\bf Step 1} (incremental update)}\\
\;\;\;\;\;\; S_{k,i}=\displaystyle \sum_{\ell\in\mathcal{N}_k}H_{\ell,i}^*R_{\ell,i}^{-1}H_{\ell,i}\\\\
\;\;\;\;\;\; \q_{k,i}=\displaystyle \sum_{\ell\in\mathcal{N}_k}H_{\ell,i}^*R_{\ell,i}^{-1}\y_{\ell,i}\\\\
\;\;\;\;\;\; P_{k,i|i}^{-1}=P_{k,i|i-1}^{-1}+S_{k,i}\\
\;\;\;\;\;\;
\bm{\psi}_{k,i}=\widehat{\x}_{k,i|i-1}+P_{k,i|i}\left[\q_{k,i}-S_{k,i}\widehat{\x}_{k,i|i-1}\right]\\\bigskip\\
\textrm{\small {\bf Step 2:} (diffusion update)}\\
\;\;\;\;\;\; \widehat{\x}_{k,i|i}=\displaystyle \sum_{\ell\in\mathcal{N}_k}a_{\ell k}\bm{\psi}_{\ell, i}\\
\;\;\;\;\;\; \widehat{\x}_{k,i+1|i}=F_i\widehat{\x}_{k,i|i}\\
\;\;\;\;\;\; P_{k,i+1|i}=F_iP_{k,i|i}F_i^*+G_iQ_iG_i^*\\\\\hline
\end{array}
\label{increald.1l2k}\ee
\noindent \\

\noindent The incremental update in (\ref{increald.1l2k}) is similar to the update used
in the distributed Kalman filter derived in \cite{Saber07}. An important difference in the
algorithms is in the diffusion step. Reference \cite{Saber07} starts from a continuous-time consensus implementation and
discretizes it to arrive at the following update relation: \bq
\widehat{\x}_{k,i|i}&=&\bm{\psi}_{k,i}+\epsilon\sum_{\ell\in\mathcal{N}_k}(\bm{\psi}_{\ell,i}-\bm{\psi}_{k,i})\eq
which, in order to facilitate comparison with (\ref{increald.1l2k}), can be equivalently rewritten as:
\bq
\label{equ:olfaticomb}
\widehat{\x}_{k,i|i}&=&(1+\epsilon-n_k\epsilon
)\cdot\bm{\psi}_{k,i}\;+\;\sum_{\ell \in\mathcal{N}_{k}\backslash\{ k\}}\epsilon\cdot\bm{\psi}_{\ell,i} \label{consskjd.a}\eq
where $n_k$ denotes the degree of node $k$ (i.e., the size of its neighborhood, ${\cal N}_k$). In comparison,
the diffusion step in (\ref{increald.1l2k}) can be written as:
\bq
\widehat{\x}_{k,i|i}&=&a_{k k}\cdot \bm{\psi}_{k,i}\;+\;\sum_{\ell\in\mathcal{N}_k\backslash\{k\}}a_{\ell k}\cdot\bm{\psi}_{\ell, i}\label{increald.1l2k.xx}\eq
Observe that the weights used in (\ref{equ:olfaticomb}) are  $(1+\epsilon-n_k\epsilon)$ for the node's estimator,
$\bm{\psi}_{k,i}$, and
$\epsilon$ for all other estimators, $\{\bm{\psi}_{\ell,i}\}$, arriving from the neighbors of node $k$. In
contrast, the diffusion step (\ref{increald.1l2k.xx}) employs a convex combination of the estimators $\{\bm{\psi}_{\ell,i}\}$ with generally different weights $\{a_{\ell k}\}$ for different neighbors; this choice is motivated by the
desire to employ combination coefficients that enhance the fusion of information at node $k$,
 as suggested by the discussion in App.~D of  \cite{Cattivelli10b}. It was verified in \cite{Cattivelli10b} that the diffusion
implementation (\ref{increald.1l2k.xx}) leads to enhanced performance in comparison to the consensus-based
update (\ref{consskjd.a}). Moreover, the weights $\{a_{\ell k}\}$ in
(\ref{increald.1l2k.xx}) can also be adjusted over time in order to
 further enhance performance, as discussed in \cite{asilomar09}. The mean-square performance and convergence of
 the diffusion Kalman filtering implementations are studied in some detail in \cite{Cattivelli10b}, along with
other diffusion strategies for smoothing problems including fixed-point and fixed-lag smoothing.

\subsection{Diffusion Distributed Optimization}\label{sec.diskd8192}
The ATC and CTA steepest-descent diffusion strategies (\ref{Equ:DiffusionAdaptation:ATC}) and (\ref{Equ:DiffusionAdaptation:CTA}) derived earlier in Sec.~\ref{sec.derivation.1} provide distributed mechanisms for the solution of global optimization problems of the form:
\be
\min_{w}\;\sum_{k=1}^N J_{k}(w) \label{opt.11.now}\ee where the individual costs,
$J_{k}(w)$, were assumed to be quadratic in $w$, namely,
\be
J_{k}(w)\;=\;\sigma_{d,k}^2 -w^* r_{du,k} - r_{du,k}^*\; w
+ w^*R_{u,k}\; w\;\label{a90ks.alk.now}\ee
for given parameters $\{\sigma_{d,k}^2, r_{du,k}, R_{u,k}\}$. Nevertheless, we remarked in that section that similar diffusion strategies can be applied to more general cases involving individual cost functions, $J_k(w)$, that are not necessarily quadratic in $w$ \cite{Chen10,Chen10b,ChenSSP2012}. We restate below, for ease of reference, the general  ATC and CTA diffusion strategies (\ref{Equ:DiffusionAdaptation:ATC.2}) and (\ref{Equ:DiffusionAdaptation:CTA.2}) that can be used for the
distributed solution of global optimization problems of the form (\ref{opt.11.now})  for more general convex functions $J_k(w)$:

\be \mbox{\rm (ATC strategy)}\;\;\;
\addbox{\;
                \label{Equ:DiffusionAdaptation:ATC.2.now}
                \begin{array}{l}
                    \psi_{k,i}  =   \displaystyle w_{k,i-1} - \mu_k\; \sum_{\ell \in \mathcal{N}_k}
c_{\ell k}\;\left[\nabla_{w}J_{\ell} (w_{k,i-1})\right]^*\\
                    w_{k,i} =   \displaystyle \sum_{\ell \in \mathcal{N}_k} a_{\ell k}\; \psi_{\ell,i}
                \end{array}\;}
       \ee
and
\be \mbox{\rm (CTA strategy)}\;\;\;
\addbox{\;
                \label{Equ:DiffusionAdaptation:CTA.2.now}
                \begin{array}{l}
                    \psi_{k,i-1}    =   \displaystyle \sum_{\ell \in \mathcal{N}_k} a_{\ell k} \;w_{\ell,i-1}
                    \\
                    w_{k,i} =    \displaystyle\psi_{k,i-1}
                                - \mu_k \sum_{\ell \in \mathcal{N}_k} c_{\ell
                                k}\;\left[\nabla_{w}J_{\ell} (\psi_{k,i-1})\right]^*
                \end{array}\;}
   \ee
   \noindent \\

\noindent for positive step-sizes $\{\mu_k\}$ and for nonnegative coefficients $\{c_{\ell k},a_{\ell k}\}$ that satisfy:
\be
\begin{array}{r}
\mbox{\rm for}\;k=1,2,\ldots,N:\\
c_{\ell k}\geq 0,\;\;\;\;\displaystyle\sum_{k=1}^N c_{\ell k} = 1,\;\;\;\;  c_{\ell k}=0~\mathrm{if}~\ell \notin \mathcal{N}_{k}\\
a_{\ell k}\geq 0,\;\;\;\;\displaystyle\sum_{\ell=1}^N a_{\ell k} = 1,\;\;\;\;  a_{\ell k}=0~\mathrm{if}~\ell \notin \mathcal{N}_{k}
\end{array}\label{run.run.now}
\ee
That is, the matrix $A=[a_{\ell k}]$ is left-stochastic while the matrix $C=[c_{\ell k}]$ is right-stochastic:
\be
C\mathds{1}=\mathds{1},\;\;\;\;\;A^T\mathds{1}=\mathds{1}
\label{satisyud.now}\ee
We can again regard the above ATC and CTA strategies as special cases of the following general diffusion scheme:
\bq
				\phi_{k,i-1}	&=&	 \displaystyle\sum_{\ell\in{\cal N}_k} a_{1,\ell k}\; w_{\ell,i-1}		 \label{gen1xx}\\
				\psi_{k,i}		&=&	 \displaystyle\phi_{k,i-1} - \mu_k \sum_{\ell\in{\cal N}_k} c_{\ell k} \left[\nabla_w J_{\ell}(\phi_{k,i-1})\right]^*\label{gen1zz}\\
				w_{k,i}		&=&	 \displaystyle\sum_{\ell\in{\cal N}_k} a_{2,\ell k}\; \psi_{\ell,i}
	\label{gen1yy}\eq
where the coefficients $\{a_{1,\ell k}, a_{2,\ell k}, c_{\ell k}\}$ are nonnegative coefficients
corresponding to the $(l,k)$-th entries of combination matrices $\{A_1,A_2,C\}$ that satisfy:
\be A_1^T\mathds{1}=\mathds{1},\;\;\;\;
A_2^T\mathds{1}=\mathds{1},\;\;\;\;C\mathds{1}=\mathds{1}\ee
The convergence behavior of these diffusion strategies can be examined under both conditions of noiseless updates (when the gradient vectors are available) and noisy updates (when the gradient vectors are subject to gradient noise). The following properties can be proven for the diffusion strategies (\ref{gen1xx})--(\ref{gen1yy})\cite{Chen10,Chen10b,ChenSSP2012}. The statements that follow assume, for convenience of presentation, that all data are {\em real-valued}; the conditions would need to be adjusted for complex-valued data. \\

\noindent {\bf {\em Noiseless Updates}}\\
Let
\be
J^{\rm glob}(w)\;=\;\sum_{k=1}^N J_{k}(w)\label{alkd812.as}
\ee
denote the global cost function that we wish to minimize. Assume $J^{\rm glob}(w)$
is strictly convex so that its minimizer $w^o$ is unique. Assume further that each individual cost function $J_{k}(w)$ is convex and has a minimizer at the {\em same} $w^o$. This case is common in practice; situations abound where nodes in a network need to work cooperatively to attain a common objective (such as tracking a target, locating the source of
chemical leak, estimating a physical model, or identifying
a statistical distribution). The case where the $\{J_{k}(w)\}$ have different
individual minimizers is studied in \cite{Chen10,ChenSSP2012}, where it is shown that
the same diffusion strategies of this section are still applicable and nodes would converge instead to a Pareto-optimal
solution.\\

\begin{theorem} ({\rm {\bf Convergence to Optimal Solution: Noise-Free Case}}) Consider
the problem of minimizing the strictly convex global cost (\ref{alkd812.as}), with the
individual cost functions $\{J_{k}(w)\}$ assumed to be convex with each having a minimizer at the same $w^o$. Assume that all data are real-valued and suppose the Hessian matrices of the individual costs are bounded from below and from above as follows:
\be
				\label{Equ:ConvergenceAnalysis:Thm:Cond_Hessian}
				\lambda_{\ell,\min} I_M \;\le\; \nabla_w^2 J_{\ell}(w) \;\le\; \lambda_{\ell,\max} I_M,
				\quad \ell=1,2,\ldots,N
			\ee
		for some positive constants $\{\lambda_{\ell,\min},\lambda_{\ell,\max}\}$. Let
 \bq \sigma_{k,\min}\define \sum_{\ell\in {\cal N}_k} c_{\ell k} \lambda_{\ell,\min},\;\;\;\;\;\;\sigma_{k,\max}\define \sum_{\ell\in {\cal N}_k} c_{\ell k} \lambda_{\ell,\max}\eq
Assume further that $\sigma_{k,\min}>0$ and that the positive step-sizes are chosen such that:
			\be
				\label{Equ:ConvergenceAnalysis:Thm:Cond_mu}
				\mu_k
				\le
				\frac{2}{\sigma_{k,\max}},\;\;
				\quad
				k=1,\ldots,N
				\ee
Then, it holds that $w_{k,i}\rightarrow w^o$ as $i\rightarrow\infty$. That is, the weight estimates generated by (\ref{gen1xx})--(\ref{gen1yy}) at all nodes will tend towards the desired global minimizer.

\qd
	\end{theorem}

\smallskip
\noindent  We note that in works on distributed sub-gradient methods (e.g., \cite{ramdistributed,NedicOzdal}),
the norms of the sub-gradients are usually required to be uniformly bounded. Such a requirement is restrictive in the unconstrained optimization of differentiable functions. Condition (\ref{Equ:ConvergenceAnalysis:Thm:Cond_Hessian}) is more relaxed since it allows the gradient vector $\nabla_{w} J_{\ell}(w)$ to have {\em unbounded} norm. This extension is important because requiring bounded gradient norms, as opposed to bounded Hessian matrices, would exclude the possibility of using quadratic costs for the $J_{\ell}(w)$ (since the gradient vectors would then be unbounded). And, as we saw in the body of the chapter, quadratic costs play a critical role in adaptation and learning over networks.\\

\noindent {\bf {\em Updates with Gradient Noise}}\\
It is often the case that we do not have access to the exact gradient vectors to use in (\ref{gen1zz}), but to noisy versions of them, say,
\be
\widehat{\nabla_w J_{\ell}(\bm{\phi}_{k,i-1})}\define
\nabla_w J_{\ell}(\bm{\phi}_{k,i-1})\;+\;\v_{\ell}(\widetilde{\bm{\phi}}_{k,i-1})
\ee
where the random vector variable $\v_{\ell}(\cdot)$ refers to gradient noise; its value is generally dependent on the weight-error vector realization,
 \be \widetilde{\bm{\phi}}_{k,i-1}\define w^o-\bm{\phi}_{k,i-1}\ee
 at which the gradient vector is being evaluated. In the presence of gradient noise, the weight estimates at the various nodes become random quantities and we denote them by the boldface notation $\{\w_{k,i}\}$. We assume that, conditioned on the past history of the
 weight estimators at all nodes, namely,
  \be {\cal F}_{i-1}\define \left\{\w_{m,j},\;m=1,2,\ldots,N,\;j<i\right\}\ee
the gradient noise has zero mean and its variance is
 upper bounded as follows:
 \bq
 \Ex\left\{\v_{\ell}(\widetilde{\bm{\phi}}_{k,i-1})\;|\;{\cal F}_{i-1}\right\}&=&0\\
 \Ex\left\{\|\v_{\ell}(\widetilde{\bm{\phi}}_{k,i-1})\|^2\;|\;{\cal F}_{i-1}\right\}&\leq&\alpha \|\widetilde{\bm{\phi}}_{k,i-1}\|^2\;+\;\sigma_v^2\label{cond43.43}
 \eq
 for some $\alpha>0$ and $\sigma_v^2\geq 0$. Condition (\ref{cond43.43}) allows
the variance of the gradient noise to be time-variant, so long as it does not grow faster than
$\Ex\|\widetilde{\bm{\phi}}_{k,i-1}\|^2$. This
condition on the noise is more general than the ``uniform-bounded assumption'' that appears in
\cite{ramdistributed}, which required instead:
 \bq
 \Ex\left\{\|\v_{\ell}(\widetilde{\bm{\phi}}_{k,i-1})\|^2\right\}\leq\sigma_v^2,\;\;\;\;\;\;\;\;
 \Ex\left\{\|\v_{\ell}(\widetilde{\bm{\phi}}_{k,i-1})\|^2\;|\;{\cal F}_{i-1}\right\}\leq\sigma_v^2
 \eq
 These two requirements are special cases of (\ref{cond43.43}) for $\alpha= 0$. Furthermore, condition (\ref{cond43.43}) is similar to
condition (4.3) in \cite{berst00}, which requires the noise variance to satisfy:
\be
\Ex\left\{\|\v_{\ell}(\widetilde{\bm{\phi}}_{k,i-1})\|^2\;|\;{\cal F}_{i-1}\right\}\leq\alpha\left[ \|\nabla_{w} J_{\ell}(\bm{\phi}_{k,i-1})\|^2\;+\;1\right]\ee
This requirement can be verified to be a combination of the ``relative random noise'' and the ``absolute random noise'' conditions defined in \cite{Poljak87} --- see \cite{Chen10b}.

Now, introduce the column vector:
\be
\z_i\define \sum_{\ell=1}^N\;\mbox{\rm col}\left\{c_{\ell 1}\v_{\ell}(w^o),\;c_{\ell 2}\v_{\ell}(w^o),\;\ldots,\;
c_{\ell N}\v_{\ell}(w^o)
\right\}
\ee
and let
\be
{\cal Z}\define \Ex\z_i\z_i^*
\label{docid12}\ee
Let further
 \be
 \widetilde{\w}_{i}\define\mbox{\rm col}\left\{\widetilde{\w}_{i,1},\;\widetilde{\w}_{i,2},\ldots,\widetilde{\w}_{i,N}\right\}
 \ee
where
\be
\widetilde{\w}_{k,i}\define w^o-\w_{k,i}
\ee
Then, the following result can be established \cite{Chen10b}; it characterizes the network mean-square deviation in steady-state, which is defined as
\be
{\rm MSD}^{\rm network}\define \lim_{i\rightarrow\infty}\left(\frac{1}{N}\sum_{k=1}^{N}\Ex\|\widetilde{\w}_{k,i}\|^2\right)
\ee

\begin{theorem} ({\rm {\bf Mean-Square Stability: Noisy Case}})
Consider the problem of minimizing the strictly convex global cost (\ref{alkd812.as}), with the
individual cost functions $\{J_{k}(w)\}$ assumed to be convex with each having a minimizer at the same $w^o$. Assume all data are real-valued and suppose the Hessian matrices of the individual costs are bounded from below and from above as stated in (\ref{Equ:ConvergenceAnalysis:Thm:Cond_Hessian}). Assume further that the diffusion strategy
(\ref{gen1xx})--(\ref{gen1yy}) employs noisy gradient vectors, where the noise terms are zero mean and satisfy conditions (\ref{cond43.43}) and (\ref{docid12}). We select the positive step-sizes to be sufficiently small and to satisfy:
\be
\mu_k\;<\;\min\left\{\frac{2\sigma_{k,\max}}{\sigma_{k,\max}^2\;+\;\alpha\|C\|^2_{1}},
\frac{2\sigma_{k,\min}}{\sigma_{k,\min}^2\;+\;\alpha\|C\|^2_{1}}
\;\frac{}{}\right\}
\ee
for $k=1,2,\ldots,N$. Then, the diffusion strategy (\ref{gen1xx})--(\ref{gen1yy}) is mean-square stable and the mean-square-deviation of the network is given by:
\be
{\rm MSD}^{\rm network}\;\approx\;\frac{1}{N}\left[\mbox{\rm vec}\left({\cal A}_2^T{\cal M} {\cal Z}^T {\cal M} {\cal A}_2\right)\right]^T \cdot(I-{\cal F})^{-1}\cdot\mbox{\rm vec}(I_{NM})
\ee
where
\bq
{\cal A}_2&=&A_2\otimes I_M\\
{\cal M}&=&\mbox{\rm diag}\{\mu_1 I_M,\;\mu_2 I_M,\;\ldots,\mu_N I_M\}\\
{\cal F}&\approx&{\cal B}^T\otimes {\cal B}^*\\
{\cal B}&=&{\cal A}_2^T\left(I-{\cal M}{\cal R}\right){\cal A}_1^T\\
{\cal R}&=&\sum_{\ell=1}^N\mbox{\rm diag}\left\{c_{\ell 1}\nabla_{w}^2 J_{\ell}(w^o),\;c_{\ell 2}\nabla_{w}^2 J_{\ell}(w^o),\;\ldots,\;
c_{\ell N}\nabla_{w}^2 J_{\ell}(w^o)
\right\}
\eq
	\end{theorem}
\qd

\bigskip
\bigskip

\begin{center}
{\bf Acknowledgements}
\end{center}
The development of the theory and applications of diffusion adaptation over networks has benefited greatly from the insights and contributions of several UCLA PhD students, and several visiting graduate students to the UCLA Adaptive Systems Laboratory (http://www.ee.ucla.edu/asl). The assistance and contributions of all
students are hereby gratefully acknowledged, including Cassio G. Lopes, Federico S. Cattivelli, Sheng-Yuan Tu, Jianshu Chen, Xiaochuan Zhao, Zaid Towfic, Chung-Kai Yu, Noriyuki Takahashi, Jae-Woo Lee,
Alexander Bertrand, and Paolo Di Lorenzo. The author is also particularly thankful to
S.-Y. Tu, J. Chen, X. Zhao, Z. Towfic, and C.-K. Yu for their assistance in reviewing an earlier draft of this article.

\clearpage
\appendix

\section{Appendix: Properties of Kronecker Products}\label{appendix.KronProp}
For ease of reference, we collect in this appendix some useful properties of Kronecker products. All matrices are assumed to be of compatible dimensions; all inverses are assumed to exist whenever necessary. Let $E=[e_{ij}]_{i,j=1}^n$ and $B=[b_{ij}]_{i,j=1}^m$ be $n\times n$ and $m\times m$ matrices, respectively. Their Kronecker product is denoted by $E\otimes B$ and is defined as the $nm\times nm$ matrix whose entries are given by \cite{horn}:
\be
E\otimes B=\ba{cccc}e_{11} B&e_{12}B&\ldots&e_{1n} B\\e_{21}B&e_{22} B&\ldots&e_{2n} B\\\vdots&&\vdots\\e_{n1}B&e_{n2} B&\ldots&e_{nn} B\ea
\ee
In other words, each entry of $E$ is replaced by a scaled multiple of $B$. Let $\{\lambda_i(E), i=1,\ldots,n\}$ and $\{\lambda_j(B), j=1,\ldots,m\}$ denote the eigenvalues of $E$ and $B$, respectively. Then, the eigenvalues of $E\otimes B$ will consist of all $nm$ product combinations $\{\lambda_i(E)\lambda_j(B)\}$. Table~\ref{Tabla.kdonr} lists some well-known properties of Kronecker products.

{\small
\begin{table}[h]
\begin{center}
\caption{\rm {\small Properties of Kronecker products}.} {\small
\begin{tabular}{l}\hline\hline \\
$\;\;\;\;(E+ B)\otimes C=(E\otimes C)\;+\;(B\otimes C)\;\;\;\;$\\
$\;\;\;\;(E\otimes B)(C\otimes D)=(EC\otimes BD)\;\;\;\;$\\\\
$\;\;\;\;(E\otimes B)^T = E^T \otimes B^T\;\;\;\;$\\
$\;\;\;\;(E\otimes B)^* = E^* \otimes B^*\;\;\;\;$\\
$\;\;\;\;(E\otimes B)^{-1} = E^{-1} \otimes B^{-1}\;\;\;\;$\\
$\;\;\;\;(E\otimes B)^{\ell} = E^{\ell} \otimes B^{\ell}\;\;\;\;$\\\\
$\;\;\;\;\{\lambda(E\otimes B)\}=\{\lambda_i(E)\lambda_j(B)\}_{i=1,j=1}^{n,m}\;\;\;\;$\\
$\;\;\;\;\det(E\otimes B)=(\det E)^m(\det B)^n\;\;\;\;$\\\\
$\;\;\;\;\mbox{\rm Tr}(E\otimes B)=\mbox{\rm Tr}(E)\mbox{\rm Tr}(B)\;\;\;\;$\\
$\;\;\;\;\mbox{\rm Tr}(E B)=\left[\mbox{\rm
vec}(B^T)\right]^T\mbox{\rm vec}(E)\;\;\;\;$\\
$\;\;\;\;\mbox{\rm vec}(EC B) = (B^T\otimes E)\mbox{\rm vec}(C)\;\;\;\;$\\
\\\hline
\end{tabular} }
\label{Tabla.kdonr}
\end{center}
\end{table}
}

\section{Appendix: Graph Laplacian and Network
Connectivity}\label{appendix.B} Consider a network consisting of $N$
nodes and $L$ edges connecting the nodes to each other. In the
constructions below, we only need to consider the edges that connect
distinct nodes to each other; these edges do not contain any
self-loops that may exist in the graph and which connect nodes to
themselves directly. In other words, when we refer to the $L$ edges of a
graph, we are excluding self-loops from this set; but we are still allowing
loops of at least length $2$ (i.e., loops generated by paths covering at least $2$ edges).

\begin{figure}[h]
\epsfxsize 10cm \epsfclipon
\begin{center}
\leavevmode \epsffile{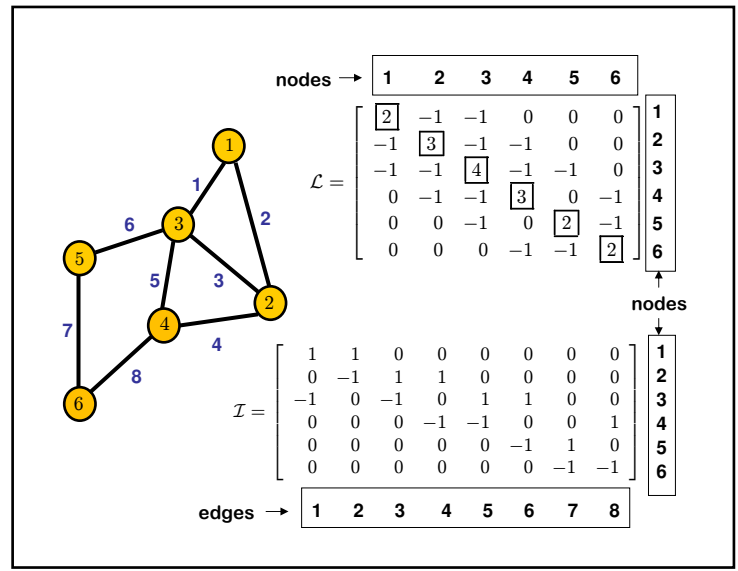} \caption{{\small A network with
$N=6$ nodes and $L=8$ edges. The nodes are marked $1$ through $6$
and the edges are marked $1$ through $8$. The corresponding
Laplacian and incidence matrices ${\cal L}$ and ${\cal I}$ are
$6\times 6$ and $6\times 8$.}}\label{fig-I.label}
\end{center}
\end{figure}

The neighborhood of any node $k$ is denoted by ${\cal N}_k$ and it
consists of all nodes that node $k$ can share information with;
these are the nodes that are connected to $k$ through edges, in
addition to node $k$ itself. The degree
of node $k$, which we denote by $n_k$, is  defined as the
positive integer that is equal to the size of its neighborhood: \be
n_k\define |{\cal N}_k| \ee
 Since $k\in{\cal N}_k$, we
always have $n_k\geq 1$. We further associate with the network an
$N\times N$ Laplacian matrix,  denoted by ${\cal L}$. The matrix
${\cal L}$ is symmetric and its entries are defined as follows
\cite{cve98,bel98,kocay}: \be \left[{\cal
L}\right]_{k\ell}=\left\{\begin{array}{rl}n_k-1,&\mbox{\rm if}\;
k=\ell\\
-1,&\mbox{\rm if}\;k\neq \ell\;\mbox{\rm and nodes}\;k\;{\rm
and}\;\ell\;\mbox{\rm are neighbors}\\ 0,&\mbox{\rm
otherwise}\end{array}\right. \label{define.laplacian}\ee Note that
the term $n_k-1$ measures the number of edges that are incident on
node $k$, and the locations of the $-1's$ on row $k$ indicate the
nodes that are connected to node $k$. We also associate with the
graph an $N\times L$ incidence matrix, denoted by ${\cal I}$. The
entries of ${\cal I}$ are defined as follows. Every column of ${\cal
I}$ represents one edge in the graph. Each edge connects two nodes
and its column will display two nonzero entries at the rows
corresponding to these nodes: one entry will be $+1$ and the other
entry will be $-1$. For directed graphs, the choice of which entry
is positive or negative can be used to identify the nodes from which
edges emanate (source nodes) and the nodes at which edges arrive
(sink nodes). Since we are dealing with undirected graphs, we shall
simply assign positive values to lower indexed nodes and negative
values to higher indexed nodes: \be \left[{\cal
I}\right]_{ke}=\left\{\begin{array}{rl}+1,&\mbox{\rm if node $k$
is the lower-indexed node connected to edge $e$}\\
-1,&\mbox{\rm if node $k$ is the higher-indexed node connected to
edge $e$}\\
0,&\mbox{\rm otherwise}\end{array}\right. \ee
Figure~\ref{fig-I.label} shows the example of a network with $N=6$
nodes and $L=8$ edges. Its Laplacian and incidence matrices are also
shown and these have sizes $6\times 6$ and $6\times 8$,
respectively. Consider, for example, column $6$ in the incidence
matrix. This column corresponds to edge $6$, which links nodes $3$
and $5$. Therefore, at location ${\cal I}_{36}$ we have a $+1$ and
at location ${\cal I}_{56}$ we have $-1$. The other columns of
${\cal I}$ are constructed in a similar manner.

Observe that the Laplacian and incidence matrices of a graph are
related as follows: \be {\cal L}={\cal I}\;{\cal I}^T \ee
 The
Laplacian matrix conveys useful information about the topology of
the graph. The following is a classical result from graph theory
\cite{cve98,bel98,kocay,fie73}.

\begin{lemma} {\rm {\bf (Laplacian and Network Connectivity)}} \label{laplaoruais.as}
Let \be \theta_1\geq\theta_2\geq \ldots\geq \theta_N \ee denote the
ordered eigenvalues of ${\cal L}$. Then the following properties
hold:
\begin{enumerate}
\item[(a)] ${\cal L}$ is symmetric nonnegative-definite so that $\theta_i\geq 0$.

\item[(b)] The rows of ${\cal L}$ add up to zero so that ${\cal
L}\mathds{1}=0$. This  means that $\mathds{1}$ is a right
eigenvector of ${\cal L}$ corresponding to the eigenvalue zero.

\item[(c)] The smallest eigenvalue is always zero, $\theta_N=0$. The
second smallest eigenvalue, $\theta_{N-1}$, is called the algebraic
connectivity of the graph.

\item[(d)] The number of times that zero is an eigenvalue of ${\cal L}$ (i.e., its
multiplicity) is equal to the number of connected subgraphs.

\item[(e)] The algebraic connectivity of a connected graph is nonzero, i.e., $\theta_{N-1}\neq 0$. In other words, a graph is connected if, and only if, its
algebraic connectivity is nonzero.

\end{enumerate}

\end{lemma}
\bp Property (a) follows from the identity ${\cal L}={\cal I}\;{\cal
I}^T$. Property (b) follows from the definition of ${\cal
L}$. Note that for each row of ${\cal L}$, the entries on the row
add up to zero. Property (c) follows from properties (a) and (b) since ${\cal
L}\mathds{1}=0$ implies that zero is an eigenvalue of ${\cal L}$. For
part (d), assume the network consists of two separate connected
subgraphs. Then, the Laplacian matrix would have a block diagonal
structure, say, of the form ${\cal L}=\mbox{\rm diag}\{{\cal
L}_1,{\cal L}_2\}$, where ${\cal L}_1$ and ${\cal L}_2$ are the
Laplacian matrices of the smaller subgraphs. The smallest eigenvalue
of each of these Laplacian matrices would in turn be zero and unique by property (e). More
generally, if the graph consists of $m$ connected subgraphs, then
the multiplicity of zero as an eigenvalue of ${\cal L}$ must be
$m$. To establish property (e), first observe that if the algebraic connectivity is nonzero then it is obvious that the graph must be connected. Otherwise, if the graph were disconnected, then its Laplacian matrix would be block diagonal and the algebraic multiplicity of zero as an eigenvalue of ${\cal L}$ would be larger than one so that $\theta_{N-1}$ would be zero, which is a contradiction.  For the converse statement, assume the graph is connected and let $x$ denote an arbitrary eigenvector of ${\cal L}$ corresponding to the eigenvalue at zero, i.e., ${\cal L}x=0$. We already know that ${\cal L}\mathds{1}=\mathds{1}$ from property (b). Let us verify that $x$ must be proportional to the vector $\mathds{1}$ so that the algebraic multiplicity of the eigenvalue at zero is one. Thus note that $x^{T}{\cal L}x=0$. If we denote the individual entries of $x$ by $x_k$, then this identity implies that for each node $k$:
 \[
\sum_{\ell\in{\cal N}_k} (x_k-x_{\ell})^2\;=\;0
 \]
It follows that the entries of $x$ within each neighborhood have equal values. But since the graph is connected, we conclude that all entries of $x$ must be equal. It follows that the eigenvector $x$ is proportional to the vector $\mathds{1}$, as desired.
\ep

\noindent

\section{Appendix: Stochastic Matrices}\label{app.D} Consider $N\times N$ matrices $A$ with
nonnegative entries, $\{a_{\ell k}\geq 0\}$. The matrix $A=[a_{\ell k}]$ is said to be right-stochastic if it satisfies
\be
A\mathds{1}=\mathds{1}\;\;\;\;\;\;\;\;\;\;(\mbox{\rm right-stochastic})
\ee
in which case each row of  $A$ adds up to one.  The matrix $A$ is said to be left-stochastic if it satisfies
\be
A^T\mathds{1}=\mathds{1}\;\;\;\;\;\;\;(\mbox{\rm left-stochastic})
\ee
in which case each column of $A$ adds up to one. And the matrix is said to be
doubly stochastic if both conditions hold so that
both its columns and rows add up to one:
\be
A\mathds{1}=\mathds{1},\;\;\;\;\;A^T\mathds{1}=\mathds{1}\;\;\;\;\;\;\;(\mbox{\rm doubly-stochastic})
\ee
Stochastic matrices arise frequently in the study of adaptation over networks. This appendix lists some
of their properties.

\begin{lemma} ({\rm {\bf Spectral Norm of Stochastic Matrices}}) \label{thm.Casaadasa}
Let $A$ be an $N\times N$ right or left or doubly stochastic matrix. Then, $\rho(A)=1$ and, therefore,
all eigenvalues of $A$ lie inside the unit disc, i.e., $|\lambda(A)|\leq 1$.
\end{lemma}
\bp We prove the result for right stochastic matrices; a similar argument applies to left
or doubly stochastic matrices. Let $A$ be a right-stochastic matrix. Then,
$A\mathds{1}=\mathds{1}$, so that $\lambda=1$ is one of the eigenvalues of $A$. Moreover,
for any matrix $A$, it holds that $\rho(A)\leq\|A\|_{\infty}$,
where $\|\cdot\|_{\infty}$ denotes the maximum absolute row sum of its matrix argument. But since all rows
of $A$ add up to one, we have $\|A\|_{\infty}=1$. Therefore, $\rho(A)\leq 1$. And
since we already know that $A$ has an eigenvalue at $\lambda=1$, we conclude that $\rho(A)=1$.

\ep

\smallskip

\noindent The above result asserts that the spectral radius of a stochastic matrix is unity
and that $A$ has an eigenvalue at $\lambda=1$. The result, however, does not rule out the
possibility of multiple eigenvalues at $\lambda=1$, or even other eigenvalues
with magnitude equal to one. Assume, in addition, that the stochastic matrix $A$ is {\em regular}. This
means that there exists an integer power
$j_o$ such that all entries of $A^{j_o}$ are {\em strictly} positive, i.e.,
\be
\mbox{\rm for all $(\ell,k)$, it holds that}\;\left[A^{j_o}\right]_{\ell k}\;>\;0,\;\mbox{\rm for some $j_o>0$}
\label{kad9012.}\ee
Then a result in matrix theory known as the Perron-Frobenius Theorem \cite{horn} leads to
the  following stronger characterization of the eigen-structure of $A$.

\begin{lemma} \label{perrona.da}({\rm {\bf Spectral Norm of Regular Stochastic Matrices}})
Let $A$ be an $N\times N$ right stochastic and regular matrix. Then:
\begin{enumerate}
\item[(a)] $\rho(A)=1$.
\item[(b)] All other eigenvalues of $A$ are strictly inside the unit circle (and, hence,
have magnitude strictly less than one).
\item[(c)] The eigenvalue at $\lambda=1$ is simple, i.e., it has multiplicity one. Moreover, with proper sign scaling,
all entries of the corresponding eigenvector are positive. For a right-stochastic $A$, this eigenvector
is the vector $\mathds{1}$ since $A\mathds{1}=\mathds{1}$.
\item [(d)] All other eigenvectors associated with the other
eigenvalues will have at least one negative or complex entry.
\end{enumerate}
\end{lemma}
\bp Part (a) follows from Lemma~\ref{thm.Casaadasa}. Parts (b)-(d) follow from the
Perron-Frobenius Theorem  when $A$ is regular \cite{horn}.

\ep

\bigskip

\begin{lemma} ({\rm {\bf Useful Properties of Doubly Stochastic Matrices}}) \label{thm.adasxxa}
Let $A$ be an $N\times N$ doubly stochastic matrix. Then the following properties hold:
\begin{enumerate}
\item[(a)] $\rho(A)=1$.
\item[(b)] $AA^T$ and $A^TA$ are doubly stochastic as well.
\item[(c)] $\rho(AA^T)=\rho(A^TA)=1$.
\item[(d)] The eigenvalues of $AA^T$ or $A^TA$ are real and lie inside the interval $[0,1]$.
\item[(e)] $I-AA^T\geq 0$ and $I-A^TA\geq 0$.
\item[(f)] $\mbox{\rm Tr}(A^T H A)\leq\mbox{\rm Tr}(H)$, for any $N\times N$ nonnegative-definite Hermitian matrix $H$.
\end{enumerate}
\end{lemma}
\bp Part (a) follows from Lemma~\ref{thm.Casaadasa}. For part (b), note that $AA^T$ is symmetric and
$AA^T\mathds{1}=A\mathds{1}=\mathds{1}.$ Therefore, $AA^T$ is doubly stochastic. Likewise for $A^TA$.
Part (c) follows from part (a) since $AA^T$ and $A^TA$ are themselves doubly stochastic matrices.
For part (d), note that $AA^T$ is symmetric and nonnegative-definite. Therefore, its eigenvalues are
real and nonnegative. But since $\rho(AA^T)=1$, we must have $\lambda(AA^T)\in[0,1]$. Likewise for the
matrix $A^TA$. Part (e) follows from part (d). For part (f), since $AA^T\geq 0$ and its eigenvalues lie within $[0,1]$, the matrix
$AA^T$ admits an eigen-decomposition of the form: \[AA^T
\;=\;U\Lambda U^T \] where $U$ is orthogonal (i.e., $U^{-1}=U^T$) and
$\Lambda$ is diagonal with entries in the range $[0,1]$. It then
follows that \bq \mbox{\rm Tr}(A^THA)&=&
\mbox{\rm Tr}(AA^TH)\nn\\
&=&\mbox{\rm Tr}(U\Lambda U^T H)\nn\\
&=&\mbox{\rm Tr}(\Lambda U^T H U)\nn\\
&\stackrel{(*)}{\leq} &\mbox{\rm Tr}(U^THU)\nn\\
&= &\mbox{\rm Tr}(UU^T H)\nn\\
&= &\mbox{\rm Tr}(H)\nn\eq
where step $(*)$ is because $U^THU=U^{-1}HU$ and, by similarity, the matrix $U^{-1}HU$ has the same eigenvalues as $H$.
Therefore, $U^THU\geq 0$. This means that the diagonal entries of $U^THU$ are nonnegative. Multiplying $U^THU$
by $\Lambda$ ends up scaling
the nonnegative diagonal entries to smaller values so that $(*)$ is justified.

\ep

\section{Appendix: Block Maximum Norm}\label{app.a}
Let $x = \mathrm{col}\{x_1,x_2,\ldots,x_N\}$ denote an $N\times 1$
block column vector whose individual entries are of size $M\times 1$
each. Following \cite{Ber97,Takahashi10,Tak08}, the block
maximum norm of $x$ is denoted by $\|x\|_{b,\infty}$ and is defined
as \be
 \addbox{       \|x\|_{b,\infty} \define  \max_{1 \le k \le N}
 \|x_k\|}
\ee where $\| \cdot \|$ denotes the Euclidean norm of its vector
argument. Correspondingly, the induced block maximum norm of an arbitrary $N
\times N$ block matrix ${\cal A}$, whose individual block entries are of size
$M\times M$ each, is defined as \be \addbox{\;\|{\cal A}\|_{b,\infty}
\define \max_{x \neq 0}
                            \frac{\|{\cal A}x\|_{b,\infty}}{\|x\|_{b,\infty}}\;}
\label{defioa.slak}\ee The block maximum norm inherits the unitary
invariance property of the Euclidean norm, as the following result
indicates \cite{Takahashi10}.

\begin{lemma} {\rm {\bf (Unitary Invariance)}}
    \label{Lemma:BlockMaximumNorm}
    Let ${\cal U}=\mathrm{diag}\{U_1,U_2,\ldots,U_N\}$ be an $N \times N$ block diagonal matrix with $M \times M$
    unitary blocks $\{U_k\}$. Then, the following properties hold:
        \begin{enumerate}
            \item[(a)]
                $\|{\cal U}x\|_{b,\infty}=\|x\|_{b,\infty}$
            \item[(b)]
                $\|{\cal U}{\cal A}{\cal U}^*\|_{b,\infty} = \|{\cal A}\|_{b,\infty}$
        \end{enumerate}
        for all block vectors $x$ and block matrices ${\cal A}$ of appropriate dimensions.
        \qd
\end{lemma}

\bigskip

\noindent The next result provides useful bounds for the block maximum norm of a block matrix.\\

\begin{lemma} {\rm {\bf (Useful Bounds)}}
    \label{Lemma:BlockMaximumNorm.232.alkjdlak}
    Let ${\cal A}$ be an arbitrary $N\times N$ block matrix with blocks $A_{\ell k}$ of size $M\times M$ each.
    Then, the following results hold:
    \begin{enumerate}
     \item[(a)]  The norms of ${\cal A}$ and its complex conjugate are related as follows:
     \be \|{\cal A}^*\|_{b,\infty}\;\leq\; N\cdot \|{\cal A}\|_{b,\infty}\label{89akd.uasuyef}\ee

     \item[(b)]  The norm of ${\cal A}$ is bounded as follows:
     \be \max_{1\leq\ell,k\leq N}\|A_{\ell k}\|\;\leq\;\|{\cal A}\|_{b,\infty}\;\leq\; N\cdot
\left(\max_{1\leq\ell,k\leq N}\|A_{\ell k}\|\right)\label{kdl9812.}\ee
 where $\|\cdot\|$ denotes the $2-$induced norm (or maximum
singular value) of its matrix argument.

     \item[(c)]  If ${\cal A}$ is Hermitian and nonnegative-definite (${\cal A}\geq 0$), then there exist finite positive constants $c_1$ and $c_2$
      such that
     \be
c_1\cdot\mbox{\rm Tr}({\cal A})\;\leq\;\|{\cal A}\|_{b,\infty}\;\leq\;c_2\cdot\mbox{\rm Tr}({\cal A})
\label{use.kajd}     \ee

    \end{enumerate}
\end{lemma}

\bp Part (a) follows directly from part (b) by noting that
\bqn
\|{\cal A}^*\|_{b,\infty}&\leq& N\cdot
\left(\max_{1\leq\ell,k\leq N}\|A_{\ell k}^*\|\right)\\
&=& N\cdot
\left(\max_{1\leq\ell,k\leq N}\|A_{\ell k}\|\right)\\
&\leq& N\cdot
\|{\cal A}\|_{b,\infty}
\eqn
where the equality in the second step is because $\|A_{\ell k}^*\|=\|A_{\ell k}\|$; i.e.,
complex conjugation does not alter the $2-$induced norm of a matrix.

To establish part (b), we consider arbitrary $N\times 1$ block vectors $x$ with entries
$x=\mbox{\rm col}\{x_1,x_2,\ldots,x_N\}$ and where each $x_k$ is $M\times 1$. Then, note that
\bq \|{\cal
A}x\|_{b,\infty}&=&\max_{1\leq \ell \leq
N}\left\|\sum_{k=1}^N A_{\ell k} x_{k}\right\|\nn\\
&\leq&\max_{1\leq \ell \leq
N}\left(\sum_{k=1}^N \|A_{\ell k}\|\cdot \|x_{k}\|\right)\nn\\
&\leq&\left(\max_{1\leq \ell \leq N}\sum_{k=1}^N
\|A_{\ell k}\|\right)\cdot \max_{1\leq k\leq N}\|x_{k}\|
\nn\\
&\leq&\left(\max_{1\leq \ell \leq N}\sum_{k=1}^N
\max_{1\leq k\leq N}\|A_{\ell k}\|\right)\cdot
\|x\|_{b,\infty}
\nn\\
&\leq&N\cdot \left(\max_{1\leq \ell, k \leq N}\|A_{\ell k}\|\right) \cdot
\|x\|_{b,\infty}
\nn\eq
so that
\bq \|{\cal
A}\|_{b,\infty}&\define&\max_{x\neq 0}\frac{\|{\cal
A}x\|_{b,\infty}}{\|x\|_{b,\infty}}\;\leq\; N\cdot \left(\max_{1\leq \ell, k \leq N}\|A_{\ell k}\|\right)\nn\eq
which establishes the upper bound in (\ref{kdl9812.}).

To establish the lower bound, we assume without loss of generality
that $\max_{1\leq\ell, k\leq N} \|A_{\ell k}\|$ is attained at $\ell=1$ and $k=1$. Let $\sigma_1$ denote the
largest singular value of $A_{11}$ and let $\{v_1,u_1\}$ denote the corresponding $M\times 1$ right and left singular vectors. That is,
\be
\|A_{11}\|=\sigma_1,\;\;\;\;\;A_{11}v_1=\sigma_1 u_1
\ee
where $v_1$ and $u_1$ have unit norms. We now construct an $N\times 1$ block vector $x^o$ as follows:
\be
x^o\define \mbox{\rm col}\{v_1,\;0_M,\;0_M,\;\ldots,\;0_M\}
\ee
Then, obviously,
\be
\|x^o\|_{b,\infty}=1
\ee
and
\be
{\cal A}x^o\;=\;\mbox{\rm col}\{A_{11}v_1,\;A_{21}v_1,\;\ldots,\;A_{N1}v_1\}
\ee
It follows that
\bq
\|{\cal A} x^o\|_{b,\infty}&=& \max \;\{\;\|A_{11}v_1\|,\;\|A_{21}v_1\|,\;\ldots,\;\|A_{N1}v_1\|\;\}\nn\\
&\geq &\|A_{11}v_1\|\nn\\
&=&\|\sigma_1 u_1\|\nn\\
&=&\sigma_1\nn\\
&=&\|A_{11}\|\nn\\
&=&\max_{1\leq \ell,k\leq N}\;\|A_{\ell k}\|
\eq
Therefore, by the definition of the block maximum norm,
\bq \|{\cal
A}\|_{b,\infty}&\define&\max_{x\neq 0}\left(\frac{\|{\cal
A}x\|_{b,\infty}}{\|x\|_{b,\infty}}\right)\nn\\&\geq&
\frac{\|{\cal
A}x^o\|_{b,\infty}}{\|x^o\|_{b,\infty}}\nn\\&=&\|{\cal
A}x^o\|_{b,\infty}\nn\\
&\geq &\max_{1\leq \ell,k\leq N}\;\|A_{\ell k}\|\eq
which establishes the lower bound in (\ref{kdl9812.}).

To establish part (c), we start by recalling that all norms on finite-dimensional vector spaces are equivalent \cite{horn,kre}. This means
that if $\|\cdot\|_a$ and $\|\cdot\|_d$ denote two different matrix norms, then there exist positive
 constants $c_1$ and $c_2$ such that for any matrix $X$,
 \be
c_1\cdot \|X\|_{a}\;\leq\;\|X\|_d\;\leq\;c_2\cdot\|X\|_a
\label{equikadj.12} \ee
Now, let $\|{\cal A}\|_{*}$ denote the nuclear norm of the square matrix ${\cal A}$. It is defined as the sum of its
singular values:
\be
\|{\cal A}\|_{*}\define \sum_{m} \sigma_m ({\cal A})
\ee
Since ${\cal A}$ is Hermitian and nonnegative-definite, its eigenvalues coincide with its singular values and, therefore,
\[ \|{\cal A}\|_{*}\;=\;\sum_{m} \lambda_m ({\cal A})\;=\;\mbox{\rm Tr}({\cal A})
\]
Now applying result (\ref{equikadj.12}) to the two norms $\|{\cal A}\|_{b,\infty}$ and
$\|{\cal A}\|_{*}$ we conclude that
\be
c_1\cdot \mbox{\rm Tr}({\cal A})\;\leq\;\|{\cal A}\|_{b,\infty}\;\leq\;c_2\cdot \mbox{\rm Tr}({\cal A})
\ee
as desired.
\ep

\bigskip

\noindent The next result relates the block maximum norm of an
extended matrix to the $\infty-$norm (i.e., maximum absolute row
sum) of the originating matrix. Specifically, let $A$ be an $N\times
N$ matrix with bounded entries and introduce the block matrix \be
{\cal A}\define A\otimes I_M\label{90dakjsd}\ee
 The extended matrix ${\cal
A}$ has blocks of size $M\times M$ each.\\

\begin{lemma} {\rm {\bf (Relation to Maximum Absolute Row Sum)}}
    \label{Lemma:BlockMaximumNorm.232}
    Let ${\cal A}$ and $A$ be related as in (\ref{90dakjsd}). Then, the following properties hold:
        \begin{enumerate}
            \item[(a)]
                $\|{\cal A}\|_{b,\infty}=\|A\|_{\infty}$, where the
                notation
                $\|\cdot\|_{\infty}$ denotes the maximum absolute
                row sum of its argument.

            \item[(b)]  $\|{\cal A}^*\|_{b,\infty}\;\leq\; N\cdot \|{\cal
                A}\|_{b,\infty}$.
        \end{enumerate}
\end{lemma}

\bp The results are obvious for a zero matrix $A$. So assume $A$ is
nonzero. Let $x=\mbox{\rm col}\{x_1,x_2,\ldots,x_N\}$ denote an
arbitrary
  $N\times 1$ block vector whose individual entries $\{x_k\}$ are
vectors of size $M\times 1$ each. Then,
\bq \|{\cal
A}x\|_{b,\infty}&=& \max_{1\leq k\leq
N}\left\|\sum_{\ell=1}^N a_{k\ell} x_{\ell}\right\|\nn\\
&\leq&\max_{1\leq k\leq
N}\left(\sum_{\ell=1}^N |a_{k\ell}|\cdot \|x_{\ell}\|\right)\nn\\
&\leq&\left(\max_{1\leq k\leq N}\sum_{\ell=1}^N
|a_{k\ell}|\right)\cdot
\max_{1\leq\ell\leq N}\|x_{\ell}\|\nn\\
&=&\|A\|_{\infty}\cdot \|x\|_{b,\infty}\eq
so that
\bq \|{\cal
A}\|_{b,\infty}&\define&\max_{x\neq 0}\frac{\|{\cal
A}x\|_{b,\infty}}{\|x\|_{b,\infty}}\;\leq\;
 \|A\|_{\infty} \label{adcasl.12}\eq The argument so far
establishes that $\|{\cal A}\|_{b,\infty}\leq\|A\|_{\infty}$. Now,
let $k_o$ denote the row index that corresponds to the maximum
absolute row sum of $A$, i.e.,
\[
\|A\|_{\infty}\;=\;\sum_{\ell=1}^N|a_{k_o\ell}|
\]
We construct an $N\times 1$ block vector $z=\mbox{\rm
col}\{z_1,z_2,\ldots,z_N\}$, whose $M\times 1$ entries $\{z_{\ell}\}$ are
chosen as follows:
\[
z_{\ell}\;=\;\mbox{\rm sign}(a_{k_o\ell})\cdot e_1
\]
where $e_1$ is the $M\times 1$ basis vector:
\[
e_1=\mbox{\rm col}\{1,0,0,\ldots,0\}
\]
and the sign function is defined as
\[
\mbox{\rm sign}(a)\;=\;\left\{\begin{array}{rl}1,&\mbox{\rm {\small
if}}\;a\geq0\\
-1,&\mbox{\rm {\small otherwise}}\end{array}\right.
\]
Then, note that $z\neq 0$ for any nonzero matrix $A$, and
\[
\|z\|_{b,\infty}\;=\;\max_{1\leq\ell\leq N}\;\|z_{\ell}\|\;=\;1\]
Moreover,
 \bq \|{\cal
A}\|_{b,\infty}&\define& \max_{x\neq 0}\frac{\|{\cal
A}x\|_{b,\infty}}{\|x\|_{b,\infty}}\nn\\
&\geq& \frac{\|{\cal
A}z\|_{b,\infty}}{\|z\|_{b,\infty}}\nn\\
&=& \|{\cal A}z\|_{b,\infty}\nn\\
&=& \max_{1\leq k\leq
N}\left\|\sum_{\ell=1}^N  a_{k\ell} z_{\ell}\right\| \nn\\
&\geq& \left\|\sum_{\ell=1}^N  a_{k_o\ell} z_{\ell}\right\| \nn\\
&=& \left\|\sum_{\ell=1}^N  a_{k_o\ell}\cdot \mbox{\rm {\small sign}}(a_{k_o\ell})e_1\right\| \nn\\
&=& \sum_{\ell=1}^N |a_{k_o\ell}|\cdot\|e_1\|\nn \\
&=&\sum_{\ell=1}^N |a_{k_o\ell}| \nn\\
&=&\|A\|_{\infty}  \label{adcasl.123}\eq Combining this result with
(\ref{adcasl.12}) we conclude that $\|{\cal A}\|_{b,\infty}=\|A\|_{\infty}$,
which establishes part (a). Part (b) follows from the statement of part (a) in Lemma~\ref{Lemma:BlockMaximumNorm.232.alkjdlak}.

 \ep

 \noindent
The next result establishes a useful property for the block maximum
norm of right or left stochastic matrices; such matrices arise as
combination matrices for distributed processing
over networks as in (\ref{dlajs8.12}) and (\ref{defioas.a}).\\

\begin{lemma} {\rm {\bf (Right and Left Stochastic Matrices)}}
    \label{Lemma:BlockMaximumNorm.right}
    Let $C$ be an $N\times N$  right stochastic matrix, i.e., its entries are nonnegative and it satisfies $C\mathds{1}=\mathds{1}$.
    Let $A$ be an $N\times N$ left stochastic matrix, i.e., its entries are nonnegative and it satisfies $A^T\mathds{1}=\mathds{1}$.
Introduce the block matrices \be {\cal A}^T\define A^T\otimes
I_M,\quad\quad{\cal C}\define C\otimes I_M \ee The matrices ${\cal
A}$ and ${\cal C}$ have blocks of size $M\times M$ each. It holds
that \be \addbox{\;\|{\cal A}^T\|_{b,\infty}\;=\;1,\quad\quad
\|{\cal C}\|_{b,\infty}\;=\;1\;} \label{kajd8912.}\ee
\end{lemma}
\bp Since $A^T$ and $C$ are right stochastic matrices, it holds that
$\|A^T\|_{\infty}=1$ and $\|C\|_{\infty}=1$. The desired result then follows from part (a) of
Lemma~\ref{Lemma:BlockMaximumNorm.232}.

\ep

\noindent The next two results establish useful properties for
the block maximum norm of a block diagonal matrix transformed by stochastic matrices; such
transformations arise as coefficient matrices that control the evolution of weight
error vectors over networks, as in (\ref{like.lakd}).\\

\begin{lemma} {\rm {\bf (Block Diagonal Hermitian Matrices)}}
    \label{Lemma:BlockMaximumNorm.block}
    Consider an $N\times N$ block diagonal Hermitian matrix
${\cal D}=\mbox{\rm diag}\{D_1,\;D_2,\;\ldots,D_N\}$, where each
$D_k$ is $M\times M$ Hermitian. It holds that \be
\addbox{\;\rho({\cal D})\;=\; \max_{1\leq k\leq N}\;\rho(D_k)\;=\;\|{\cal
D}\|_{b,\infty}}
\label{hdk912l}\ee where $\rho(\cdot)$ denotes the spectral radius
(largest eigenvalue magnitude) of its argument. That is, the
spectral radius of ${\cal D}$ agrees with the  block maximum norm of
${\cal D}$, which in turn agrees with the largest spectral radius of
its block components.
\end{lemma}
\bp We already know that the spectral radius of any matrix ${\cal X}$ satisfies
$\rho({\cal X})\leq \|{\cal X}\|$, for any induced matrix norm \cite{golub,horn}. Applying this result to ${\cal D}$ we readily get that
$\rho({\cal D})\leq \|{\cal D}\|_{b,\infty}$. We now establish the reverse inequality, namely, $\|{\cal D}\|_{b,\infty}\leq \rho({\cal D})$. Thus, pick an arbitrary $N\times 1$ block vector $x$ with entries
$\{x_1,x_2,\ldots,x_N\}$, where each $x_k$ is $M\times 1$.
From definition (\ref{defioa.slak}) we have \bq \| {\cal
D}\|_{b,\infty}&\define&\max_{x\neq 0}\frac{\|{\cal
D}x\|_{b,\infty}}{\|x\|_{b,\infty}}\nn\\&=& \max_{x\neq
0}\left(\;\frac{1}{\|x\|_{b,\infty}}\cdot \max_{1\leq k\leq
N}\left\|D_k x_k\right\|\;\right)\nn\\&\leq & \max_{x\neq
0}\left(\;\frac{1}{\|x\|_{b,\infty}}\cdot \max_{1\leq k\leq N}
\left(\left\|D_k\right\|\cdot \left\|x_{k}\right\|\right)\;\right)\nn\\&=&
\max_{x\neq 0}\max_{1\leq k\leq
N}\left(\left\|D_k\right\|\cdot\frac{
\left\|x_{k}\right\|}{\|x\|_{b,\infty}}\right)\;\nn\\
&\leq &\max_{1\leq k\leq N}\|D_k\|\nn\\
&=&\max_{1\leq k\leq N}\rho(D_k) \label{comibne.4}\eq where the
notation $\|D_k\|$ denotes the $2-$induced norm of $D_k$ (i.e., its
largest singular value). But since $D_k$ is assumed to be Hermitian,
its $2-$induced norm agrees with its spectral radius, which explains
the last equality.

\ep

\begin{lemma} {\rm {\bf (Block Diagonal Matrix Transformed by Left Stochastic Matrices)}}
    \label{Lemma:BlockMaximumNorm.block.left}
    Consider an $N\times N$ block diagonal Hermitian matrix
${\cal D}=\mbox{\rm diag}\{D_1,\;D_2,\;\ldots,D_N\}$, where each
$D_k$ is $M\times M$ Hermitian.
    Let $A_1$ and $A_2$ be $N\times N$ left stochastic matrices, i.e., their entries are nonnegative and they
    satisfy $A_1^T\mathds{1}=\mathds{1}$ and $A_2^T\mathds{1}=\mathds{1}.$ Introduce the block matrices \be {\cal A}_1^T= A_1^T\otimes
I_M,\quad\quad{\cal A}_2^T\define A_2^T\otimes I_M \ee The matrices
${\cal A}_1$ and ${\cal A}_2$ have blocks of size $M\times M$ each.
Then it holds that \be \addbox{\;\rho\left({\cal A}_2^T\cdot {\cal
D}\cdot {\cal A}_1^T\;\right)\;\leq\;\rho({\cal D})\;}
\label{kajd8912.xxas}\ee
\end{lemma}
\bp Since the spectral radius of any matrix never exceeds any
induced norm of the same matrix, we have that \bq \rho\left({\cal
A}_2^T\cdot {\cal D}\cdot {\cal A}_1^T\;\right)&\leq& \left\|\;
{\cal A}_2^T\cdot {\cal D}\cdot {\cal A}_1^T\;
\right\|_{b,\infty}\nn\\
&\leq &\left\|{\cal A}_2^T\right\|_{b,\infty}\cdot \left\|{\cal
D}\right\|_{b,\infty}\cdot \left\|{\cal A}_1^T\;
\right\|_{b,\infty}\nn\\
&\stackrel{(\ref{kajd8912.})}{=} &\left\|{\cal
D}\right\|_{b,\infty}\nn\\
&\stackrel{(\ref{hdk912l})}{=} &\rho({\cal D}) \label{askialkjd812}\eq
 \ep

\noindent In view of the result of
Lemma~\ref{Lemma:BlockMaximumNorm.block}, we also conclude from
(\ref{kajd8912.xxas}) that
 \be \addbox{\;\rho\left({\cal A}_2^T\cdot {\cal
D}\cdot {\cal A}_1^T\;\right)\;\leq\;\max_{1\leq k\leq
N}\rho(D_k)\;} \label{kajd8912.xxas.2}\ee It is worth noting that
there are choices for the matrices $\{{\cal A}_1,{\cal A}_2,{\cal
D}\}$ that would result in strict inequality in
(\ref{kajd8912.xxas}). Indeed, consider the special case:
\[
{\cal D}=\ba{cc}2&0\\0&1\ea,\;\;\;\;{\cal
A}_1^T=\ba{cc}\frac{1}{3}&\frac{2}{3}\\\frac{2}{3}&\frac{1}{3}\ea,\;\;\;\;
{\cal
A}_2^T=\ba{cc}\frac{1}{3}&\frac{2}{3}\\\frac{2}{3}&\frac{1}{3}\ea
\]
This case corresponds to $N=2$ and $M=1$ (scalar blocks). Then,
\[
{\cal A}_2^T {\cal D} {\cal A}_1^T =
\ba{cc}\frac{2}{3}&\frac{2}{3}\\\frac{2}{3}&1\ea
\]
and it is easy to verify that
\[
\rho({\cal D})\;=\;2,\quad\quad\rho({\cal A}_2^T {\cal D} {\cal
A}_1^T)\;\approx\;1.52
\]
The following conclusions follow as corollaries to the statement of
Lemma~\ref{Lemma:BlockMaximumNorm.block.left}, where by a stable
matrix $X$ we mean one whose eigenvalues lie strictly inside the
unit circle.

\begin{corollary}  {\rm {\bf (Stability Properties)}} \label{coruais.as}Under the same
setting of Lemma~\ref{Lemma:BlockMaximumNorm.block.left}, the
following conclusions hold:
\begin{enumerate}
\item[(a)] The matrix ${\cal A}_2^T {\cal D} {\cal A}_1^T$ is stable
whenever ${\cal D}$ is stable.

\item[(b)] The matrix ${\cal A}_2^T {\cal D} {\cal A}_1^T$ is stable
for all possible choices of left stochastic matrices ${\cal A}_1$
and ${\cal A}_2$ if, and only if, ${\cal D}$ is stable.
\end{enumerate}

\end{corollary}

\bp Since ${\cal D}$ is block diagonal, part (a) follows immediately from (\ref{kajd8912.xxas}) by
noting that $\rho({\cal D})<1$ whenever ${\cal D}$ is stable. [This statement fixes the argument that appeared in App.~I of \cite{Cattivelli10} and Lemma 2 of \cite{Cattivelli10b}. Since the matrix $X$ in App.~I of \cite{Cattivelli10} and the matrix ${\cal M}$ in Lemma 2 of \cite{Cattivelli10b} are block diagonal, the $\|\cdot\|_{b,\infty}$ norm should replace the $\|\cdot\|_{\rho}$ norm used there, as in the proof that led to (\ref{askialkjd812}) and as already done in \cite{Takahashi10}.] For
part (b), assume first that ${\cal D}$ is stable, then ${\cal A}_2^T {\cal
D} {\cal A}_1^T$ will also be stable by part (a) for any
left-stochastic matrices ${\cal A}_1$ and ${\cal A}_2$. To prove the
converse, assume that ${\cal A}_2^T {\cal D} {\cal A}_1^T$ is stable
for any choice of left stochastic matrices ${\cal A}_1$ and ${\cal A}_2$. Then,
${\cal A}_2^T {\cal D} {\cal A}_1^T$ is  stable
for the particular choice
${\cal A}_1=I={\cal A}_2$ and it follows that ${\cal D}$
must be stable.

\ep

\section{Appendix: Comparison with Consensus Strategies}\label{app.C} Consider a connected
network consisting of $N$ nodes. Each node has a state or measurement value $x_k$, possibly a vector of
size $M\times 1$. All nodes in the network are interested in evaluating the average value of their
states, which we denote by
\be
{w}^o\define \frac{1}{N}\sum_{k=1}^N x_k\label{lkad9012.}
\ee
A centralized solution to this problem would require each node to transmit its
measurement $x_k$ to a fusion center. The central processor would then compute
${w}^o$ using (\ref{lkad9012.}) and transmit it back to all nodes.
This centralized mode of operation suffers from at least two limitations. First,
 it requires communications and power resources to transmit the data
back and forth between the nodes and the central processor; this problem is compounded if the
fusion center is stationed at a remote location. Second, the architecture has a
critical point of failure represented by the
central processor; if it fails, then operations would need to be halted.\\

\noindent {\bf {\em Consensus Recursion}}\\
The consensus strategy provides an elegant distributed solution to the same problem, whereby
nodes interact locally with their neighbors and are able to converge to $w^o$ through these interactions.
Thus, consider an arbitrary node $k$ and
assign nonnegative weights $\{a_{\ell k}\}$ to the edges linking $k$ to its neighbors
$\ell\in{\cal N}_k$. For each node $k$, the weights $\{a_{\ell k}\}$ are assumed to
add up to one so that
 \bq\mbox{\rm for}\;k=1,2,\ldots,N:&&\nn\\
        a_{\ell k}\geq 0,\;\;\;\;\;\sum_{\ell=1}^N a_{\ell k} = 1,\;\;\;\;  a_{\ell k}=0~\mathrm{if}~\ell \notin
        \mathcal{N}_{k}
 \label{A.cond} \eq
The resulting combination matrix is denoted by $A$ and its $k-$th column consists of the
entries $\{a_{\ell k},\ell=1,2,\ldots,N\}$. In view of (\ref{A.cond}), the combination matrix $A$
is
seen to satisfy $A^T\mathds{1}=\mathds{1}$. That is, $A$ is left-stochastic. The consensus strategy can be described as follows.
Each node $k$ operates repeatedly on the
data from its
neighbors and updates its state iteratively according to the rule:
\be
\addbox{\;w_{k,n} \;=\;\sum_{\ell\in{\cal N}_k} a_{\ell k}\; w_{\ell,n-1},\;\;\;n> 0\;}
\label{consensus}\ee
where $w_{\ell,n-1}$ denotes the state of node $\ell$ at iteration $n-1$, and $w_{k,n}$
denotes the updated state of node $k$ after iteration $n$. The initial conditions are
\be
w_{k,o}\;=\;x_k,\;\;\;\;k=1,2,\ldots,N
\ee
If we collect the states of all nodes at iteration $n$ into a column vector, say,
\be
z_n\define \mbox{\rm col}\{w_{1,n},\;w_{2,n},\;\ldots,w_{N,n}\}
\ee
Then, the consensus iteration (\ref{consensus}) can be equivalently rewritten in vector form as follows:
\be
\addbox{\;z_{n}\;=\;{\cal A}^T z_{n-1},\;\;n> 0\;}\label{kald9012.}
\ee
where
\be
{\cal A}^T=A^T\otimes I_M
\ee
The initial condition is
\bq
z_{o}&\define& \mbox{\rm col}\{x_{1},\;x_{2},\;\ldots,x_{N}\}\nn\\
\eq
\bigskip
\noindent {\bf {\em Error Recursion}}\\
Note that we can express the average value, ${w}^o$, from (\ref{lkad9012.}) in the form
\be
\addbox{\;{w}^o= \frac{1}{N}\cdot(\mathds{1}^T\otimes I_M)\cdot z_{o}\;}
\label{dqoid.as}\ee
where $\mathds{1}$ is the vector of size $M\times 1$ and whose entries are all equal to one. Let
\be
\widetilde{w}_{k,n}=w^o-w_{k,n}
\ee
denote the weight error vector for node $k$ at iteration $n$; it measures how far the iterated state
is from the desired average value $w^o$. We collect all error vectors across the network into an
$N\times 1$ block column vector whose entries are of size $M\times 1$ each:
\be
\widetilde{w}_n\define\ba{c}\widetilde{w}_{1,n}\\\widetilde{w}_{2,n}\\\vdots\\\widetilde{w}_{N,n}\ea
\ee
Then,
\be
\addbox{\;\widetilde{w}_n \;=\; (\mathds{1}\otimes I_M)w^o\;-\;z_n\;}\label{jakd.9812}
\ee

\bigskip

\noindent {\bf {\em Convergence Conditions}}\\
The following result is a classical result on consensus strategies \cite{groot74,berger,Tsi84}. It provides conditions under which
the state of all nodes will converge to the desired average, ${w}^o$, so
that $\widetilde{w}_n$ will tend to
zero.

\begin{theorem} ({\rm {\bf Convergence to Consensus}}) \label{thm.Casaa} For any initial states
$\{x_k\}$, the successive iterates
$w_{k,n}$ generated by the consensus iteration
(\ref{consensus}) converge to the network average value $w^o$ as $n\rightarrow\infty$ if, and only if,
the following three conditions are met:
\bq
A^T\mathds{1}&=&\mathds{1}\label{cond.11xx}\\
A\mathds{1}&=&\mathds{1}\\
\rho\left(A^T-\frac{1}{N}\mathds{1}\mathds{1}^T\right)&<&1\label{cond.13xx}
\eq
That is, the combination matrix $A$ needs to be doubly stochastic, and the matrix
$A^T-\frac{1}{N}\mathds{1}\mathds{1}^T$ needs to be stable.
\end{theorem}
\bp ({\em Sufficiency}). Assume first that the three conditions stated in the theorem hold. Since $A$ is doubly stochastic,
then so is any power of $A$, say, $A^{n}$ for any $n\geq 0$, so that
\be
\left[A^{n}\right]^T\mathds{1}=\mathds{1},\;\;\;\;\;\;
A^{n}\mathds{1}=\mathds{1}
\ee
Using this fact, it is straightforward to verify by induction the validity of the following equality:
\bq
\left(A^T-\frac{1}{N}\mathds{1}\mathds{1}^T\right)^{n}&=&
\left[A^{n}\right]^T-\frac{1}{N}\mathds{1}\mathds{1}^T
\label{89akljkda}
\eq
Likewise, using the Kronecker product identities
\bq
(E+ B)\otimes C&=&(E\otimes C)\;+\;(B\otimes C)\\
(E\otimes B)(C\otimes D)&=&(EC\otimes BD)\\
(E\otimes B)^n &=& E^n \otimes B^n
\eq
for matrices $\{E,B,C,D\}$ of compatible dimensions, we observe that
\bq \left({\cal A}^{n}\right)^{T}\;-\;\frac{1}{N}\cdot(\mathds{1}\otimes I_M)\cdot
(\mathds{1}^T\otimes I_M)&=&
\left[\left({A}^{n}\right)^{T}\otimes I_M\right] \;-\;\frac{1}{N}\cdot(\mathds{1}
\mathds{1}^T\otimes I_M)\nn\\
&=&
\left[\left({A}^{n}\right)^{T}\;-\;\frac{1}{N}\cdot\mathds{1}
\mathds{1}^T\right]\otimes I_M\nn\\
&\stackrel{(\ref{89akljkda})}{=}&\left(A^T-\frac{1}{N}\mathds{1}\mathds{1}^T\right)^{n}\otimes I_M\nn\\
&=&\left[\left(A^T-\frac{1}{N}\mathds{1}\mathds{1}^T\right)\otimes I_M\right]^{n} \label{9a089das}\eq
 Iterating (\ref{kald9012.}) we find that
\be z_n=\left[{\cal A}^{n}\right]^T z_{o}\label{908as1.lk12}\ee
and, hence, from (\ref{dqoid.as}) and (\ref{jakd.9812}),
\bq
\widetilde{w}_n&=&
-\left[\left({\cal A}^{n}\right)^{T}\;-\;\frac{1}{N}\cdot(\mathds{1}\otimes I_M)\cdot(\mathds{1}^T\otimes I_M)\right]\cdot z_{o}\nn\\
&\stackrel{(\ref{9a089das})}{=}&
-\left[\left(A^T-\frac{1}{N}\mathds{1}\mathds{1}^T\right)\otimes I_M\right]^{n}\cdot z_{o}\label{k8012a.lkad}
\eq
Now recall that, for two arbitrary matrices $C$ and $D$ of
compatible dimensions, the eigenvalues of the Kronecker product
$C\otimes D$ is formed of all product combinations
$\lambda_i(C)\lambda_j(D)$ of the eigenvalues of $C$ and $D$
\cite{golub}. We conclude from this property, and from the fact that
$A^T-\frac{1}{N}\mathds{1}\mathds{1}^T$ is stable, that the coefficient matrix
\[
\left(A^T\;-\;\frac{1}{N}\cdot\mathds{1}\mathds{1}^T\right)\otimes I_M\]
is also stable. Therefore,
\be
\widetilde{w}_n\;\;\rightarrow\;0\;\;\;\mbox{\rm as}\;n\rightarrow\infty
\ee
\noindent ({\em Necessity}). In order for $z_n$ in (\ref{908as1.lk12}) to converge to
$(\mathds{1}\otimes I_M)w^o$, for any initial state $z_o$, it must hold that
\be
\lim_{n\rightarrow\infty} \left({\cal A}^{n}\right)^T\cdot z_{o}\;=\;
\frac{1}{N}\cdot (\mathds{1}\otimes I_M)\cdot(\mathds{1}^T\otimes I_M)\cdot z_{o}
\ee
for any $z_{o}$. This implies that we must have
\be\lim_{n\rightarrow\infty} \left({\cal A}^{n}\right)^T\;=\;
\frac{1}{N}\cdot (\mathds{1}\mathds{1}^T\otimes I_M)
\ee
or, equivalently,
\be\lim_{n\rightarrow\infty} \left(A^{n}\right)^T\;=\;
\frac{1}{N}\mathds{1}\mathds{1}^T\label{conds.1}\ee
This in turn implies that we must have
\be\lim_{n\rightarrow\infty} A^T\cdot \left(A^{n}\right)^T\;=\;
A^T\cdot\frac{1}{N}\mathds{1}\mathds{1}^T\label{conds.2}\ee
But since
\be
\lim_{n\rightarrow\infty} A^T\cdot \left(A^{n}\right)^T\;=\;
\lim_{n\rightarrow\infty} \left(A^{n+1}\right)^T\;=\;
\lim_{n\rightarrow\infty} \left(A^{n}\right)^T
\ee
we conclude from (\ref{conds.1}) and (\ref{conds.2}) that it must hold that
\be
\frac{1}{N}\mathds{1}\mathds{1}^T
\;=\;\frac{1}{N}A^T\cdot \mathds{1}\mathds{1}^T
\ee
That is,
\be\frac{1}{N}\left(A^T\mathds{1}\;-\;
\mathds{1}\right)\cdot \mathds{1}^T=0
\ee
from which we conclude that we must have $A^T\mathds{1}=\mathds{1}$. Similarly, we can show that
$A\mathds{1}=\mathds{1}$ by studying the limit of $\left(A^{n}\right)^T A^T$. Therefore,
$A$ must be a doubly stochastic matrix.  Now using the fact that $A$ is doubly stochastic,
we know that (\ref{89akljkda}) holds. It follows
that in order for condition (\ref{conds.1}) to be satisfied, we must have
\be
\rho\left(A^T-\frac{1}{N}\mathds{1}\mathds{1}^T\right)<1\label{lkad9812.12}
\ee
\ep

\bigskip

\noindent {\bf {\em Rate of Convergence}}\\
\noindent From (\ref{k8012a.lkad}) we conclude that the rate of convergence of the error vectors
$\{\widetilde{w}_{k,n}\}$ to zero is determined by the spectrum of the matrix
\be A^T-\frac{1}{N}\mathds{1}\mathds{1}^T\ee
Now since $A$ is a doubly stochastic matrix, we know that it has an eigenvalue at $\lambda=1$.
Let us denote the eigenvalues of $A$ by
$\lambda_k(A)$ and let us order them in terms of their magnitudes as follows:
\be
0\leq \;|\lambda_M(A)|\;\leq\;\ldots\;\leq\;|\lambda_{3}(A)|\;\leq\;|\lambda_{2}(A)|\;\leq\; 1
\label{eq.570}\ee
where $\lambda_{1}(A)=1$. Then, the eigenvalues of the coefficient matrix $(A^T-\frac{1}{N}\mathds{1}\mathds{1}^T)$ are equal to
\be
\left\{\;\lambda_M(A),,\ldots,\;\lambda_3(A),\;\lambda_2(A),0\;\right\}
\ee
It follows that the magnitude of
$\lambda_{2}(A)$ becomes the spectral radius of $A^T-\frac{1}{N}\mathds{1}\mathds{1}^T$.
Then condition (\ref{lkad9812.12}) ensures that $|\lambda_2(A)|<1$. We therefore arrive at the following conclusion.\\

\begin{corollary} ({\rm {\bf Rate of Convergence of Consensus}}) Under conditions (\ref{cond.11xx})--(\ref{cond.13xx}),
the rate of convergence of the successive iterates $\{w_{k,n}\}$ towards the network average $w^o$ in the
consensus strategy (\ref{consensus}) is determined by the
second largest eigenvalue magnitude of $A$, i.e., by $|\lambda_2(A)|$ as defined in (\ref{eq.570}).\end{corollary}

\qd

\smallskip

\noindent It is worth noting that doubly stochastic matrices $A$ that are also {\em regular} satisfy
conditions (\ref{cond.11xx})--(\ref{cond.13xx}). This is because, as we already know from
Lemma~\ref{perrona.da}, the eigenvalues of such matrices satisfy $|\lambda_m(A)|<1$, for
$m=2,3,\ldots,N$, so that condition (\ref{cond.13xx}) is automatically satisfied. \\

\begin{corollary} ({\rm {\bf Convergence for Regular Combination Matrices}})\label{codia.dalk} Any doubly-stochastic and
regular matrix $A$ satisfies the three conditions (\ref{cond.11xx})--(\ref{cond.13xx}) and, therefore,
ensures the convergence of the consensus iterates $\{w_{k,n}\}$ generated
by
(\ref{consensus}) towards $w^o$ as $n\rightarrow\infty$.\end{corollary}

\qd

\smallskip

\noindent A regular combination matrix $A$ would result when the two conditions listed below
are satisfied by the graph connecting the nodes over which the consensus iteration is applied.

\begin{corollary} ({\rm {\bf Sufficient Condition for Regularity}}) Assume the
combination matrix $A$ is doubly stochastic and
that the graph over which the consensus iteration (\ref{consensus}) is applied satisfies the following two conditions:
\begin{enumerate}
\item[(a)] The graph is connected. This means that there exists a path connecting
any two arbitrary nodes in the network. In terms of the Laplacian matrix that is associated
with the graph (see Lemma~\ref{laplaoruais.as}), this means that the second smallest
eigenvalue of the Laplacian is nonzero.
\item[(b)] $a_{\ell k}=0$ if, and only if, $\ell\notin {\cal N}_k$. That is, the combination weights
are strictly positive between any two neighbors, including $a_{kk}>0$.
\end{enumerate}
Then, the corresponding matrix $A$ will be regular and, therefore,
the consensus iterates $\{w_{k,n}\}$ generated by
(\ref{consensus}) will converge towards $w^o$ as $n\rightarrow\infty$.
\end{corollary}
\bp We first establish that conditions (a) and (b) imply that $A$ is a regular matrix, namely,
that
there should exist an integer $j_o>0$ such that
\be
\left[A^{j_o}\right]_{\ell k}\;>\;0
\label{hdka8912}\ee
for all $(\ell,k)$. To begin with,
by the rules of matrix multiplication, the $(\ell,k)$ entry of the $i-$th power of
$A$ is given by:
\be
\left[A^i\right]_{\ell k}\;=\;\sum_{m_1=1}^N\sum_{m_2=1}^N\ldots\sum_{m_{i-1}=1}^N
\;a_{\ell m_1}a_{m_1 m_2}\ldots
a_{m_{i-1} k}
\label{k89a.dlkasd}\ee
The summand in (\ref{k89a.dlkasd}) is nonzero if, and only if, there is some sequence of indices
$(\ell,m_1,\ldots,m_{i-1},k)$ that forms a path from node $\ell$ to node $k$.
Since the network is assumed to be
connected, there exists a minimum (and finite) integer value $i_{\ell k}$ such that a path exists
from node $\ell$ to node $k$ using $i_{\ell k}$ edges and that
\[\left[A^{i_{\ell k}}\right]_{ \ell k}\;>\;0
\]
In addition, by induction, if $\left[A^{i_{\ell k}}\right]_{\ell k}>0$, then
\bqn
\left[A^{i_{\ell k}+1}\right]_{\ell k}&=&\sum_{m=1}^N\;\left[A^{i_{\ell k}}\right]_{\ell m}\;a_{mk}\\
&\geq &\left[A^{i_{\ell k}}\right]_{\ell k}\;a_{kk}\\
&>&0
\eqn
Let
\[
j_o\;=\;\max_{1\leq k,\ell\leq N}\;\{i_{\ell k}\}
\]
Then, property (\ref{hdka8912}) holds for all $(\ell,k)$. And we conclude
from (\ref{kad9012.}) that $A$ is a regular matrix.
It then follows from Corollary~\ref{codia.dalk} that the consensus iterates $\{w_{k,n}\}$ converge to
the average network value $w^o$.

\ep

\bigskip

\noindent {\bf {\em Comparison with Diffusion Strategies}}\\
\noindent Observe that in comparison to diffusion strategies, such as the ATC strategy (\ref{Equ:DiffusionAdaptation:ATC.adaptive}),
the consensus iteration (\ref{consensus}) employs the same quantities $w_{k,\cdot}$ on both sides of the
iteration. In other words, the consensus construction keeps iterating on the same set of vectors until they
converge to the average value $w^o$. Moreover, the index $n$ in the consensus algorithm is an iteration index.
In contrast, diffusion strategies employ different quantities on both sides of the combination step
in  (\ref{Equ:DiffusionAdaptation:ATC.adaptive}), namely, $w_{k,i}$ and $\{\psi_{\ell,i}\}$; the latter
variables have been processed through an information exchange step and are updated (or filtered) versions of
the $w_{\ell,i-1}$. In addition, each step of the diffusion strategy
(\ref{Equ:DiffusionAdaptation:ATC.adaptive}) can incorporate new data, $\{d_{\ell}(i),u_{\ell,i}\}$, that are
collected by the nodes at every time instant. Moreover, the index $i$ in the diffusion
implementation is a time index (and not an iteration index); this is because diffusion strategies are
inherently adaptive and perform online learning. Data keeps streaming in and diffusion incorporates the
new data into the update equations at every time instant. As a result, diffusion strategies are able to
respond to data in an adaptive manner, and they are also able to solve general optimization problems: the vector
$w^o$ in adaptive diffusion iterations is the minimizer of a global cost function (cf.~(\ref{opt.11})),
while the vector $w^o$ in consensus iterations is the average value of the initial states of the nodes (cf.~(\ref{lkad9012.})).

Moreover, it turns out that diffusion strategies influence the evolution of the network dynamics in an interesting and advantageous manner in comparison to consensus strategies. We illustrate this point by means of an example. Consider initially the ATC strategy (\ref{Equ:DiffusionAdaptation:ATC.adaptive.2})  without information exchange,  whose update equation we repeat below for ease of reference:
\bq
          \bm{\psi}_{k,i}  &=&   \displaystyle \w_{k,i-1} + \mu_k \u_{k,i}^*\left[\d_{k}(i)-\u_{k,i}
                    \w_{k,i-1}\right]\label{comadjl12.1}\\
                    \w_{k,i} &=&   \displaystyle \sum_{\ell \in \mathcal{N}_k} a_{\ell k}\; \bm{\psi}_{\ell,i}\;\;\;\;\;\;\;\;\;\;\quad\quad\quad\quad\quad\quad
                    \quad\quad\quad\quad\quad\quad\quad\;\;\;\;\;\;\;\;\;\;\;(\mbox{\rm ATC diffusion })\label{comadjl12.2}
\eq
These recursions were derived in the body of the article as an effective distributed solution for optimizing (\ref{opt.11})--(\ref{gihjas.alk}).  Note that they involve two steps, where the weight estimator $\w_{k,i-1}$ is first updated to the intermediate estimator $\bm{\psi}_{k,i}$, before  the intermediate estimators from across the neighborhood are combined to obtain $\w_{k,i}$. Both steps of ATC diffusion (\ref{comadjl12.1})--(\ref{comadjl12.2}) can be combined into a single update as follows:
\be
\addbox{\;\w_{k,i} \;=\;   \displaystyle \sum_{\ell \in \mathcal{N}_k} a_{\ell k} \left[\w_{\ell,i-1}\;+\;\mu_{\ell} \u_{\ell,i}^*\left(\d_{\ell}(i)-\u_{\ell,i}
                    \w_{\ell,i-1}\right)\right]\;}
                    \;\;\;\;\;(\mbox{\rm ATC diffusion})\label{comadjl12.2.atc}
\ee
Likewise, consider the CTA strategy (\ref{Equ:DiffusionAdaptation:CTA.adaptive.2})  without information exchange,  whose update equation we also repeat below:
\bq
          \bm{\psi}_{k,i-1} &=&   \displaystyle \sum_{\ell \in \mathcal{N}_k} a_{\ell k}\; \w_{\ell,i-1}\;\;\;\;\;\;\;\;\;\;\quad\quad\quad\quad\quad\quad
                    \quad\quad\quad\quad\quad\quad\;\;\;(\mbox{\rm CTA diffusion })\label{comadjl12.3.cta}\\
                    \w_{k,i}  &=&   \displaystyle \bm{\psi}_{k,i-1} + \mu_k \u_{k,i}^*\left[\d_{k}(i)-\u_{k,i}
                    \bm{\psi}_{k,i-1}\right]
                    \label{comadjl12.2.cta}
\eq
Again, the CTA strategy involves two steps: the weight estimators $\{\w_{\ell,i-1}\}$ from the neighborhood of node $k$ are first combined to yield the intermediate estimator $\bm{\psi}_{k,i-1}$,  which is subsequently updated to $\w_{k,i}$. Both steps of CTA diffusion can also be combined into a single update as follows:
\be\addbox{\;
\w_{k,i} \;=\;   \displaystyle \sum_{\ell \in \mathcal{N}_k} a_{\ell k}\; \w_{\ell,i-1}\;+\; \mu_{k} \u_{k,i}^*\left[\d_{k}(i)-\u_{k,i}
                    \displaystyle \sum_{\ell \in \mathcal{N}_k}a_{\ell k}\w_{\ell,i-1}\right]\;}
                    \quad(\mbox{\rm CTA diffusion})\label{comadjl12.2.cta.abc}
\ee
Now, motivated by the consensus iteration (\ref{consensus}), and based on a procedure for distributed optimization suggested in \cite{Ber97} (see expression (7.1) in that reference), some works in the literature (e.g., \cite{moore03,barba07,OlfatiCDC,Nedic2010,giannakis09,Mat2009,Kar2009,moura10,dimakis10}) considered distributed strategies that correspond to the following form for the optimization problem under consideration (see, e.g., expression (1.20) in \cite{Nedic2010} and expression (9) in \cite{moura10}):
\be
                    \addbox{\;\w_{k,i} =   \displaystyle \sum_{\ell \in \mathcal{N}_k} a_{\ell k}\; \w_{\ell,i-1}\;+\; \mu_k \u_{k,i}^*\left[\d_{k}(i)-\u_{k,i}
                    \w_{k,i-1}\right]\;}\;\;\;\;\;\;(\mbox{\rm consensus strategy})
\label{dja8912.a}\ee
This strategy can be derived by following the same argument we employed earlier in Secs.~\ref{sec.steepest.sec} and~\ref{sec.adaptive.dd} to arrive at the diffusion strategies, namely,  we replace $w^o$ in \eqref{Equ:DiffusionAdaptation:Combination_intermediate} by $w_{\ell,i-1}$ and then apply the instantaneous approximations (\ref{instaldlakd.laprad}).
Note that the {\em same} variable $\w_{k,\cdot}$ appears on both sides of the equality in (\ref{dja8912.a}). Thus, compared with the ATC diffusion strategy (\ref{comadjl12.2.atc}),  the update from $\w_{k,i-1}$ to $\w_{k,i}$ in the consensus implementation (\ref{dja8912.a}) is only influenced by data $\{\d_k(i),\u_{k,i}\}$ from node $k$. In contrast, the ATC diffusion structure (\ref{comadjl12.1})--(\ref{comadjl12.2}) helps incorporate the influence of the data $\{\d_{\ell}(i),\u_{\ell,i}\}$ from across the neighborhood of node $k$ into the update of $\w_{k,i}$, since these data are reflected in the intermediate estimators $\{\bm{\psi}_{\ell,i}\}$. Likewise, the contrast with the CTA diffusion strategy (\ref{comadjl12.2.cta.abc}) is clear, where the right-most term in (\ref{comadjl12.2.cta.abc}) relies on a combination of all estimators from across the neighborhood of node $k$, and not only on $\w_{k,i-1}$ as in the consensus strategy (\ref{dja8912.a}). These facts have desirable implications on the evolution of the weight-error vectors across diffusion networks. Some simple algebra, similar to what we did in Sec.~\ref{sec.mse}, will show that the mean of the extended error vector for the consensus strategy (\ref{dja8912.a}) evolves according to the recursion:
\be\addbox{\;\Ex\widetilde{\w}_i=\left({\cal A}^T-{\cal M}
{\cal R}_u\right)\cdot\Ex\widetilde{\w}_{i-1},\;\;i\geq
0\;}\;\;\;\;\;\;\;\;\;\;\;\;\quad\quad(\mbox{\rm consensus strategy})
\label{conshdj9812}\ee
where ${\cal R}_u$ is the block diagonal covariance matrix defined by (\ref{replaced}) and
$\widetilde{\w}_i$ is the aggregate error vector defined by (\ref{ahdak819aldka}). We can  compare the above mean error dynamics with the ones that correspond to the ATC and CTA diffusion strategies (\ref{comadjl12.1})--(\ref{comadjl12.2}) and
(\ref{comadjl12.3.cta})--(\ref{comadjl12.2.cta.abc}); their error dynamics follow as special cases from (\ref{like.lakd.ada.2}) by setting $A_1=I=C$ and $A_2=A$ for ATC and
$A_2=I=C$ and $A_1=A$ for CTA:
\be\addbox{\;
\Ex\widetilde{\w}_i={\cal A}^T\left(I_{NM}-{\cal M}
{\cal R}_u\right)\cdot\Ex\widetilde{\w}_{i-1},\;\;i\geq
0\;}\;\;\;\;\;\quad\quad (\mbox{\rm ATC diffusion})
\label{conshdj9812.2}\ee
and
\be\addbox{\;
\Ex\widetilde{\w}_i=\left(I_{NM}-{\cal M}
{\cal R}_u\right){\cal A}^T\cdot\Ex\widetilde{\w}_{i-1},\;\;i\geq
0\;}\;\;\;\;\;\quad\quad (\mbox{\rm CTA diffusion})
\label{conshdj9812.3}\ee
We observe that the coefficient matrices that control the evolution of $\Ex\widetilde{\w}_i$ are different in all three cases. In particular,
 \bq
 \mbox{\rm consensus strategy (\ref{conshdj9812}) is stable in the mean}\;&\Longleftrightarrow&\;\rho\left({\cal A}^T-{\cal M}
{\cal R}_u\right)<1\\
 \mbox{\rm ATC diffusion (\ref{conshdj9812.2}) is stable in the mean}\;&\Longleftrightarrow&\;\rho\left[{\cal A}^T\left(I_{NM}-{\cal M}
{\cal R}_u\right)\right]<1\\
 \mbox{\rm CTA diffusion (\ref{conshdj9812.3}) is stable in the mean}\;&\Longleftrightarrow&\;\rho\left[\left(I_{NM}-{\cal M}
{\cal R}_u\right){\cal A}^T\right]<1
 \eq
 It follows that the mean stability of the consensus network is sensitive to the choice of the combination matrix $A$. This is not the case for the diffusion strategies. This is because from property (\ref{kajd8912.xxas}) established in App.~\ref{app.a}, we know that the matrices ${\cal
A}^T\left(I_{NM}-{\cal M}{\cal R}_u\right)$ and $\left(I_{NM}-{\cal M}{\cal R}_u\right){\cal
A}^T$ are stable if
 $\left(I_{NM}-{\cal M}{\cal R}_u\right)$ is
stable. Therefore, we can select the step-sizes to satisfy $\mu_k<2/\lambda_{\max}(R_{u,k})$ for the ATC or CTA diffusion strategies and ensure their mean stability regardless of the combination matrix $A$. This also means that the diffusion networks will be mean stable whenever the individual nodes are mean stable, regardless of the topology defined by $A$. In contrast, for consensus networks, the network can exhibit unstable mean behavior even if all its individual nodes are stable in the mean. For further details and other results on the mean-square performance of diffusion networks in relation to consensus networks, the reader is referred to \cite{yusayed12,yusayed12AA}.\\

\noindent {\bf Acknowledgement}. The development of the theory and applications of diffusion adaptation over networks has benefited greatly from the insights and contributions of several UCLA Ph.D. students, and several visiting graduate students to the UCLA
Adaptive Systems Laboratory (http://www.ee.ucla.edu/asl). The assistance and contributions of all students are hereby gratefully acknowledged, including Cassio G. Lopes, Federico S. Cattivelli, Sheng-Yuan Tu, Jianshu Chen, Xiaochuan Zhao, Zaid Towfic, Chung-Kai Yu, Noriyuki Takahashi, Jae-Woo Lee, Alexander Bertrand, and Paolo Di Lorenzo. The author is also particularly thankful to S.-Y. Tu, J. Chen, X. Zhao, Z. Towfic, and C.-K. Yu for their
assistance in reviewing an earlier draft of this chapter.\\

\end{document}